\documentclass[natbib,smallextended]{svjour3}
\journalname{Astronomy Astrophysics Review}

\smartqed  

\usepackage{graphicx}
\usepackage{xcolor}
\usepackage{txfonts}
\usepackage{fancyhdr}
\usepackage{lastpage}
\usepackage{lmodern}

\usepackage[colorlinks=true,linkcolor=blue,citecolor=blue,urlcolor=teal]{hyperref}
\usepackage{natbib}
\usepackage[utf8]{inputenc}

\DeclareRobustCommand{\ion}[2]{\textup{#1\,\textsc{\lowercase{#2}}}}
\def\teff{$T\rm_{eff}$}
\newcommand{\logg}{\ensuremath{\log\,g}}

\newcommand{\cdt}{$\rm^{12}C/^{13}C~$}

\begin{document}

\title{The most metal poor stars}

\author{Piercarlo Bonifacio$^1$, Elisabetta Caffau$^1$, Patrick Fran{\c c}ois$^{2,3}$ and Monique Spite$^1$}

\institute{$^1$ LIRA, Observatoire de Paris, Universit{\'e} PSL, Sorbonne Universit{\'e}, Universit{\'e} Paris Cité, CY Cergy Paris Universit{\'e}, CNRS,92190 Meudon, France
\\
\email{Piercarlo.Bonifacio@observatoiredeparis.psl.eu}\\
\email{Elisabetta.Caffau@observatoiredeparis.psl.eu}\\
\email{Monique.Spite@observatoiredeparis.psl.eu}\\
$^2$ LIRA, Observatoire de Paris, Universit{\'e} PSL, Sorbonne Universit{\'e}, Universit{\'e} Paris Cité, CY Cergy Paris Universit{\'e}, CNRS,75014 Paris, France\\
\email{Patrick.Francois@observatoiredeparis.psl.eu}\\
$^3$UPJV, Universit\'e de Picardie Jules Verne,33 rue St Leu, Amiens, 80080, France}

\maketitle

\abstract{
The most metal-poor stars found in the Galaxy and in nearby galaxies are witnesses of the early evolution of the Universe. 
In a general picture in which we expect the metallicity to increase monotonically with time, as a result of the
metal production in stars, we also expect the most metal-poor stars to be the most primitive objects accessible
to our observations. The abundance ratios in these stars provide us important information on the first generations
of stars that synthesised the nuclei that we observe in these stars. Because they are so primitive the modelling
of their chemical inventory can be often satisfactorily achieved by assuming that all the metals were
produced in a single supernova, or just a few. This is simpler than modelling the full chemical evolution,
using different sources, that is necessary at higher metallicity. The price to pay for this relative
ease of interpretation is that these stars are extremely rare and require specifically tailored observational
strategies in order to assemble statistically significant samples of stars. 
In this review we try to summarise the main observational results that have been obtained in the last ten years.
}



\section{Introduction}

Analysing the difference in the color magnitude diagrams of  the solar neighbourhood stars and the stars belonging to globular clusters,  \citet{baade_ngc_1944, baade_resolution_1944}
came to the conclusion that stars could be divided into two broad families, Population I for the solar neighbourhood stars and Population II stars for the globular cluster stars. 
It appeared later that  Population II stars were old metal-poor stars.
\citet{osterbrock_walter_1995} gives an interesting historical summary of the discovery of these two populations.
Currently the concept has evolved, since the solar neighbourhood hosts stars of many different
origins. Part have been formed in the disc at different radii and migrated to the solar neighbourhood \citep[see e.g. ][and references therein]{minchev_estimating_2018}, part have been formed in external galaxies that have merged with the Milky Way. The most relevant major
merger that we can trace is the Gaia Sausage Enceladus (GSE) event \citep{belokurov_co-formation_2018,haywood_disguise_2018,helmi_merger_2018}.
First spectroscopic confirmations of the low abundances in metal poor stars were done by \citet{chamberlain_atmospheres_1951} who showed that
the weakness of metallic lines in HD\,19445 and HD\,140283 were due to a low metal abundance and not to a high effective
temperature. \citet{helfer_abundances_1959} by the analysis of four stars,  among them two belonging to globular clusters and
one high velocity star, further confirmed the low metallicity of some stars.  High velocity stars were soon associated with the metal-poor halo stars, thought to be the witness  of the early galactic history. 
In the search of the elusive Population III \footnote{We refer to Population III as the generation of stars formed from primordial 
material, that is, with no metals.}, the first systematic search for metal-poor stars was done by  \citet{beers_search_1985}  analysing the  strength of the Ca {\sc II} H\&K lines on spectra taken on photographic plates. From their first data release of the survey, they successfully discovered  134 stars with [Fe/H]\footnote{For any pair of elements [X/Y] = log(X/Y)+12, where the argument of the logarithm is the ratio of the abundance by number of element X to Y.} $\le  -2 $.
The following release of Beers' catalogue  \citep{beers_search_1992} led the discovery 70 extremely metal poor (hereafter EMP, see Table \ref{tab:emp_class} for a formal definition) 
stars. High resolution high S/N ratio spectroscopy was then used to confirm the metallicity and the detailed chemical composition of these EMP stars as described in Sec. \ref{sec:abundance_patterns} . 

High resolution,  high SNR spectroscopy follow-up observations revealed the diversity of the chemical composition of these EMP  stars challenging 
the models of formation and evolution of the first supernovae but also the early galactic chemical evolution. From the lithium  Spite's plateau \citep{spite_lithium_1982,spite_abundance_1982} and the lithium  ``melt down'' at very low metallicity  \citep{sbordone_metal-poor_2010}, to the very rare EMP stars,  to the  $\alpha$-poor EMP stars, to  EMP stars highly enriched in r-process and/or s-process, these stars have revealed a wide diversity of chemical composition, witness of the nucleosynthesis of the first stars. 
Nevertheless, the vast majority of EMP stars are  high in [$\alpha$/Fe] where the $\alpha$ elements are produced by type II supernovae \citep[and references therein]{matteucci_modelling_2021}.
All EMP stars also show the presence of neutron capture elements 
formed, mostly, by the  r-process \citep{matteucci_modelling_2021}. 

In this review we concentrate on the lowest metallicity stars found in our Galaxy and in Local Group galaxies.
We tried to cover all the relevant literature subsequent to the review of \citet{frebel_near-field_2015} and up
to June 2024, although some more recent papers are also included, but not in a systematic way.
This review covers the observational picture, for a review on the formation of low-mass low-metallicity
stars we refer the reader to the review by \citet{klessen_first_2023} and for
the nucleosynthesis in stars to the review by \citet{arcones_origin_2023}.
For a discussion on   3D-NLTE line formation we refer the reader to the review by \citet{lind_three-dimensional_2024}.
In writing this review we made multiple times use of the SAGA database\footnote{\url{http://sagadatabase.jp/}} \citep{suda_stellar_2008}.

\section{The initial mass function of the first generation of stars}

\citet{salpeter_luminosity_1955} noted that the
luminosity function of a set of stars of different ages 
must depend on three factors:
{\em (i)} $\xi(M)$, the relative probability of creation
of a star of mass $M$; {\em (ii)} the rate of creation of stars;
{\em (iii)} the evolution of stars off the Main Sequence.  
From the study of the luminosity function
of Galactic stars and some simple hypothesis on the creation
rate of stars and their evolutionary properties he then derived what he 
called the ``original mass function'':

\begin{equation}
\xi(M) \approx 0.03 (M/M_\odot)^{-1.35}
\end{equation}

This is what is currently called the Salpeter Initial Mass Function (hereafter IMF),
it is usually generalised to Salpeter-like IMFs, where the exponent
is a real value $-\alpha$.
This power-law form of the IMF is often used \citep[see e.g.][]{kroupa_variation_2001},
perhaps using different values of $\alpha$ for different mass intervals.
Other functional forms, in particular log-normal and exponential,
have been invoked to model the IMF \citep[see e.g.][]{larson_early_1998,chabrier_galactic_2001}.
Since the discovery that the Population II stars are metal deficient,
it was argued that the Population II IMF was ``top heavy'', i.e.
had a larger fraction of massive stars than the present-day IMF.
Probably the first such claim dates back to \cite{schwarzschild_evolution_1953} who relied
also on the high frequency of white dwarfs, and on the excess of red stars
in distant elliptical galaxies. 

There is a physical reason why one should expect that at low
metallicities the formation of higher mass stars is favoured.
Based on the work of \citet{jeans_stability_1902}, on the stability
of a spherical nebula one can define the ``Jeans mass'', that is the
mass above which the nebula will start collapsing under
the effect of gravity. One can show that
$M_J \propto T^{3/2}$ thus the higher the temperature,
the higher $M_J$.
A nebula, under the effect of gravity, develops a pressure and temperature
gradient, that try to compensate the effect of gravity and try to find
an equilibrium configuration. There are several physical
mechanisms that can contribute to cool the nebula, thus lowering the
Jeans mass. The two most relevant are: 
{\em (i)} collisional excitation of ions or
molecules, followed by radiative recombination, this is often inaccurately
referred to as ``line cooling''; {\em (ii)} collisional excitation
of rotation or vibration modes of dust particles, followed by emission
of far-infrared photons, this is collectively referred to as ``dust cooling''. 
Both mechanisms are expected to become less efficient, the lower the metallicity.
Since the most efficient transitions to cool a collapsing nebula
are transitions of \ion{O}{i} and \ion{C}{ii} \citep{bromm_formation_2003}, line cooling is
less efficient the lower the metallicity.
The most abundant elements, H and He, do not have any
low-lying levels that can be collisionally excited. 
Molecules, such as H$_2$, CO and OH  
play a role at  lower temperatures
than found in collapsing metal-poor
gas clouds \citep[see e.g.][]{omukai_protostellar_2000}.
The very  existence of dust requires the presence of metals and 
thus  dust cooling becomes less efficient at low metallicity
\citep[see e.g.][]{chon_transition_2021}.
It has to born in mind that the Jeans mass is at all metallicities much larger than the typical mass of stars, thus the collapsing cloud must fragment into smaller fragments at some stage. This also implies that stars are not formed as isolated objects, but are formed in groups. The fragmentation at small scales ($M < 0.01 -- 0.1 M_\odot$) is favoured by the presence of dust \citep{chon_transition_2021}.

The situation becomes more extreme for a gas that is devoid of metals, i.e. stars formed from primordial gas, that is constituted only of H and He isotopes and traces of $^7$Li. In this situation it has been argued that only the formation of stars of mass 
larger than 10 $M_\odot$ is possible \citep[e.g.][]{larson_early_1998}, thus making the IMF of the first stellar generation (Pop III) very different from what it is in the current Galaxy (Pop I) or even in the metal-poor regime (Pop II). Theoretical considerations lead to postulate the existence of a ``critical'' metallicity ($Z_{cr}$) above which the formation of low-mass stars ($M< 1M_\odot$) is possible and  the star formation mode transitions from Pop III to Pop II \citep[see e.g.][and references therein]{omukai_thermal_2005,chon_transition_2021}.
The value of $Z_{cr}$ is observationally constrained by the metal-weak tail of the metallicity distribution functions of the Milky Way and other galaxies where it can be determined. Since low-mass stars have lifetimes longer than the age of the 
Universe, no star with $Z< Z_{cr}$ should be observable.
At present the most metal-poor object known in the Universe is the star  SDSS\,J102915.14+172927.9  \citep{caffau_extremely_2011,caffau_primordial_2012,caffau_sdss_2024} with $Z\le 6.9\times10^{-7}$, implying that 
$Z_{cr}$ is certainly below this value.
The topic of the formation of Pop III stars is extensively reviewed by \citet{klessen_first_2023}  and we refer
the reader to that review for further details. We simply point out that our current understanding is that  a top-heavy IMF is not
necessarily required, some simulations predict that fragmentation may form low mass stars even at zero metallicity \citep{greif_simulations_2011,stacy_building_2016}.

\begin{figure}
\centering
\includegraphics[width=11.5cm,clip=true]{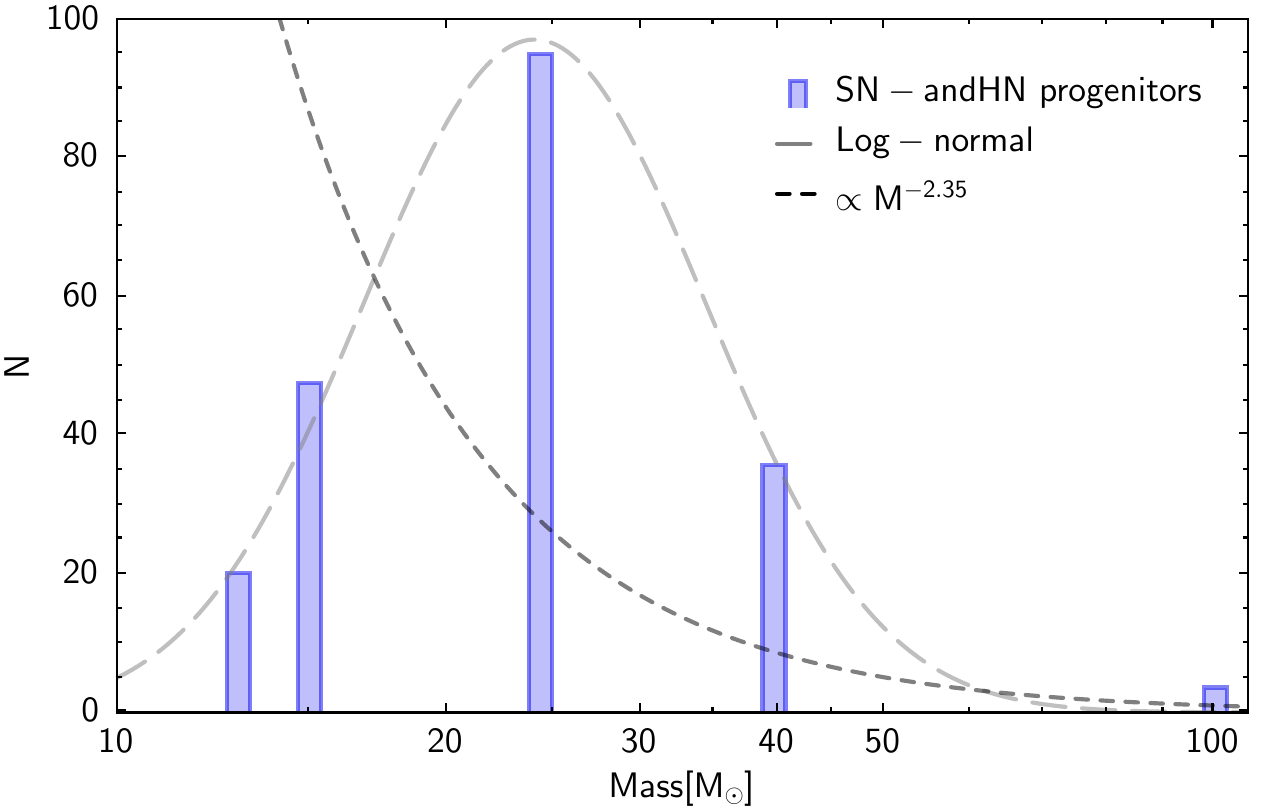}
\caption{Observed and theoretical high-mass portion of the IMF, adapted from \citet{ishigaki_initial_2018}.
}
\label{fig:IMF_Ishigaki}
\end{figure}

Observationally the high-mass portion of the IMF 
can be constrained by observing the metals produced
by these stars, that govern the chemical composition
of the following generation of stars.
The  fate of a Pop III massive star depends on the mass of its He core \citep{heger_how_2003}.
We refer the reader to the detailed discussion on this point in the review
of \citet{klessen_first_2023}. For our purpose we simply recall
that there are three main physical quantities that govern the final
outcome: stellar rotation, magnetic fields, energy of explosion.
The effect of rotation is to produce larger He cores and may result
in an asymmetric explosion, with the presence of a jet \citep{grimmett_chemical_2021}.
Even a weak initial magnetic field can be enhanced by orders of magnitudes during the explosion, 
and can affect the explosion energy \citep{nakamura_three-dimensional_2025}. The energy of
the explosion greatly depends on the amount of material that is not ejected, but falls back
onto the remnant. When the explosion energy is much smaller than $10^{51}$\,erg one usually calls
it a faint SNe. Finally, as we shall further detail below, not all massive stars explode as SNe,
some may collapse directly to a black hole and some may become a ``failed'' SN with a neutron star,
or black hole remnant.
In this case the star does not
eject chemically enriched material
\citep{heger_nucleosynthetic_2002,zhang_fallback_2008}.
By making the assumption that each EMP star was formed from material enriched by a single SN,
whose mass can be estimated by fitting the observed abundance pattern
to theoretical yields of SNe, 
one can derive indirect information on the masses of Pop III stars. This approach has been used by
\citet{ishigaki_initial_2018} and their result is shown in
Fig.\,\ref{fig:IMF_Ishigaki}. 
\citet{jiang_modified_2024} have revised the above approach, by taking
into account the criterion of explodability of a star of a given mass. Their derived IMF continues
to increase for masses below 25 $M_\odot$, at variance with the results
of \citet{ishigaki_initial_2018}. The two results are not in contradiction,
because in the picture of 
\citet{jiang_modified_2024} the population of stars below  25 $M_\odot$  
 is dominated by failed SNe, while
\citet{ishigaki_initial_2018} take into account only successful explosions.

The yields of Pop III stars with masses in excess of 100 $M_\odot$ were not
included in the model of \citet{ishigaki_initial_2018}, because the abundance ratios in the sample
of stars analysed  are better fit by ordinary core collapse SNe.
Stars with such a large mass are expected to end their
lives as Pair Instability Supernovae (hereafter PISN) that have
a very distinct nucleosynthetic signature with
a marked odd-even effect \citep[][]{heger_nucleosynthetic_2002,takahashi_stellar_2018}.
Although there are a few stars whose abundance patterns can be interpreted
as formed from material that has been polluted by a PISN, as well as lower mass core-collapse SNe
\citep[see e.g.][]{aoki_chemical_2014,salvadori_probing_2019,aguado_pisn-explorer_2023,caffau_pristine_2023}
they are still few and the interpretation is not unique. 
\citet{xing_metal-poor_2023} claimed that the abundance pattern of LAMOST\,J1010+2358 ([Fe/H]=--2.4)
can be interpreted as having formed from material that has been polluted only by
a PISN of 260\,$M_\odot$. This result was however confuted
by \citet{skuladottir_pair-instability_2024} who, based on a more complete
abundance inventory, favour material polluted by a Pop II 13 $M_\odot$
SNe and a 39 $M_\odot$ Pop III SNe. 
\citet{thibodeaux_lamost_2024} also determined a new chemical inventory for this star, 
from a different spectrum than the one used by 
\citet{skuladottir_pair-instability_2024}, in good agreement with the former and
favour a solution in which the material has been polluted by an 11\,$M_\odot$ SN.

There have been claims of observed super luminous SNe in the local Universe that could indeed be PISNe;
a good candidate for this is SN 2007bi \citep{gal-yam_supernova_2009}. However, subsequent
analysis  of its nebular spectra ruled out this interpretation
\citep{mazzali_nature_2019}.
We can thus say that, in spite of some circumstantial evidence in favour, there is no compelling evidence that PISN ever
existed. 
Thanks to the abundance patterns in extremely metal-poor stars, we have a reasonably good understanding of the 
IMF of Pop III in the range 10-100\,$M_\odot$, but we have very few constraints both on the low-mass and
the high-mass end of the IMF.

\citet{jiang_modified_2024} derive IMFs under two different assumptions:
either each EMP star has been formed out of gas
enriched by a single SN (mono-enrichment) or by
two SNe (dual-enrichment).
In the latter hypothesis they predict the existence of
a sizeable fraction of PISNs (see their figure 4, panel b).
The analysis of \citet{hartwig_machine_2023} concludes that only
about one third of the EMP stars can be classified as mono-enriched, therefore
the dual-enriched case should apply to the majority of the stars.
Alexander Heger and collaborators have provided a code, 
{\tt STARFIT\footnote{\url{https://2sn.org/starfit/}}},
that fits a given abundance pattern to the yields of zero metallicity SNe of \citet{heger_nucleosynthetic_2002}
and \citet{heger_nucleosynthesis_2010}. The code provides a set of best fitting  models for the star
that has polluted the gas to provide the observed abundance pattern.
The underlying assumption is the mono-enrichment. Because of its ease
of use the code has been used in many publications, it is very difficult
to provide an exhaustive list.
As an example we suggest the reader can consider
the papers of 
\citet{ito_bd44493_2009,placco_metal-poor_2015,lombardo_chemical_2022,mardini_chemical_2022,ji_spectacular_2024}.

.

\section{Searching for extremely metal-poor stars} \label{sec:surveys}

We propose  a new definition of the classes of metal-poor stars, that is 
summarised in Table\,\ref{tab:emp_class}.
The most metal-poor bin,
UMP stars, contains stars with [Fe/H] $<-4.0$.
The vast majority of stars found in this [Fe/H] regime have [C/Fe]$> 1$  (the carbon enhanced metal-poor stars CEMP discussed in Sec.\,\ref{sec:C-rich}). This group of stars
may testify  a different mode of star formation.
Next are the 
EMP stars,  those
with $\rm -4.0 <[Fe/H] < -2.8$. This classification is based
on the lithium abundances: below --2.8 one sees an increased scatter
in the abundances and many Li-poor stars, well below the Spite plateau (see Sec.\,\ref{sec_libeb} and Fig.\,\ref{Fig:LiFe}).
We shall refer to very metal-poor stars (VMP) as those
with $\rm -2.8 \le [Fe/H] < -1.5$.
Finally we call  metal-poor (MP) the stars 
with $\rm -1.5 \le [Fe/H] \le -0.5$. This latter definition has the virtue of
aligning the definition of MP between the exoplanet community and the stellar 
astronomers community.
In this section we shall review the main searches that have been
conducted in the last 20 years to find EMP and also UMP stars.

\begin{table}[thb]
    \centering
    \caption{Adopted classification of metal-poor stars based on their [Fe/H].}
    \begin{tabular}{rr}
    \hline
       $\rm [Fe/H] <-4.0$  &  UMP \\
    $\rm -4.0 \le [Fe/H] < -2.8$ &    EMP \\ 
    $\rm -2.8 \le [Fe/H] < -1.5$ & VMP \\
    $\rm -1.5 \le [Fe/H] \le -0.5$ & MP\\
    \hline
    \end{tabular}
    \label{tab:emp_class}
\end{table}

\subsection{Spectroscopic Surveys\label{subsec:spectro}}

A very efficient way to select metal-poor stars is in spectroscopic databases. 
In fact, in the case the quality of the spectra is good enough \citep[see e.g. the Gaia-ESO Survey,][]{gilmore_gaia-eso_2012,hourihane_gaia-eso_2023} 
they offer immediately all the information (detailed abundances) needed to derive the characteristics of the stars themselves 
and to put constraints on the masses of the previous stellar generation.

At present there are some medium- to high-resolution on-going spectroscopic surveys that were successful in  detecting  and confirming EMP stars.
Galactic Archaeology with HERMES  \citep[GALAH,][]{de_silva_galah_2015} provided many EMP stars (see e.g.  the confirmation of the most iron-poor star known, \citealt{keller_single_2014} selected for its photometry obtained with the SkyMapper telescope, see Sec.\,\ref{subbsec:photom}, and EMP in the Bulge \citealt{howes_extremely_2015}). 
This survey in its phase 1 was a magnitude limited survey and in phase 2 it is particularly focused on deriving ages, so the primary targets are main-sequence turn-off stars.
Apache Point Observatory Galactic Evolution Experiment
\citep[APOGEE,][]{majewski_apache_2017} 
is an unbiased sample, fact that is very important to investigate the chemical evolution of the Galaxy,
but the strategy does not favour the detection of EMP stars that are hardly present in the sample.
GALAH and overall APOGEE are providing the community with large sample of stars with detailed chemical inventory.
The H3 survey \citet{conroy_mapping_2019} takes advantage of the wide field (about $1^\circ$ diameter) of the 
6.5\,m MMT telescope to conduct a survey at  resolving power $R\approx 23000$ over a limited spectral range
(15\,nm) down to $r=18$. The data of this survey shall eventually be available
through its web site\footnote{\url{http://h3survey.rc.fas.harvard.edu/}}.
The Gaia satellite holds a special place among the surveys.
The resolving power of the Radial Velocity Spectrometer (RVS) is $R=11\,500$ and 
the spectral coverage is limited (27\,nm), but strategically centred to include the strong
\ion{Ca}{ii} IR triplet lines. While in ground-based observations this region
is plagued by numerous OH telluric emission lines. This requires an accurate sky subtraction
in order to be confidently used, and this is not the case for a space-based instrument like RVS.
The region has very few measurable lines at low metallicity, however
the \ion{Ca}{ii} IR triplet can be measured even for EMP stars. In fact
\citet{recio-blanco_gaia_2023} have shown that the instrument is capable
of recovering two well-known metal-poor  stars: HD\,140283 and HD\,200654.
In the future data releases that will benefit of more transits for each star and therefore
higher signal-to-noise ratio, we can expect that new EMP stars shall be discovered.

The disadvantage of the low-resolution, and sometimes also medium-resolution, spectroscopic surveys is that they lack a detailed chemical inventory and they
rely on high-resolution follow-up observations to obtain it.
The advantage is the shorter observing time that implies larger amount of spectra.
Radial Velocity Experiment RAVE \citet{steinmetz_radial_2006} is an extremely successful project but mainly observing at high metallicities.
The HK Survey \citep{beers_search_1985,beers_search_1992} has been very successful in the discovery of EMP stars \citep[see e.g.][]{cayrel_first_2004}. 
The Hamburg/ESO Survey HES \citep{christlieb_finding_2003} observed the first ultra Fe-poor star \citep{christlieb_stellar_2002}.
The Large Sky Area Multi-Object Fiber Spectroscopic Telescope LAMOST \citep{deng_lamost_2012,liu_k_2014} observed many EMP, confirmed at high-resolution \citep[see e.g.][]{li_four-hundred_2022}.
Sloan Digital Sky Survey SDSS \citep{york_sloan_2000} has been very successful in the selection of EMP stars \citep[see e.g.][]{aguado_j00230307_2018,jeong_search_2023}.
The HALO7D survey \citep{cunningham_halo7d_2019} covers a very specific niche:
it targets halo TO stars observed at low resolution with DEIMOS at the 10\,m Keck telescope
in the fields of the extra-galactic survey CANDELS
\citep{grogin_candels_2011,koekemoer_candels_2011} for which multi-epoch imaging
from the Hubble Space Telescope is available. In this way proper motions
are available for targets as faint as $m_{F606W}=23.5$. HALO7D is a deep pencil-beam survey
that is complementary to other wider and shallower surveys.
The high multiplex Dark Energy Spectroscopic Instrument (DESI) on the 4\,m Mayall telescope at
the Kitt Peak National Observatory, although primarily devoted to extragalactic observations
is conducting a Milky Way Survey \citep{cooper_overview_2023}. A first catalogue
has already been released \citep{koposov_desi_2024} and follow-up observations
have already started \citep{allende_prieto_gtc_2023}.

\subsection{Photometric Surveys\label{subbsec:photom}}

Photometric databases contain much larger samples of stars than spectroscopic databases.
They are then the ideal place to search for rare objects as EMP stars.
To give an example, Gaia\,DR3 \citep{gaia_collaboration_gaia_2023}
contains $1.46\times 10^9$ sources to be compared to 999\,645 RVS spectra.
It is not trivial to select metal-poor  candidates from photometry, also if we know that they are characterised by being bluer than solar-metallicity stars.
With wide-band photometry blue stars can be selected in the search of EMP candidates.
Classical wide band photometric metallicity estimates use the $U-B$ colour
\citep[see e.g.][]{wallerstein_abundances_1962}, while 
modern versions rely on colours like $u-g$ \citep[see e.g.][]{bonifacio_topos_2021}. It has to be kept in mind that such colours
are a measure of the Balmer jump, and thus they are sensitive to both metallicity
and surface gravity.

An example of the use of wide-band photometry is \citet{schlaufman_best_2014} who
made use of the all-sky APASS optical,  Two Micron All Sky Survey \citep[2MASS, ][]{skrutskie_two_2006} near-infrared, and WISE mid-infrared photometry to identify bright metal-poor star candidates through their lack of molecular absorption near 4.6 microns (Best and Brightest metal-poor stars B \& B). 
 They identified seven previously unknown stars with $\rm [Fe/H]\leq -3.0$.
  \citet{placco_r-process_2019} observed, at resolving power 1200-2000, 857 stars selected in the B \& B survey. Out of these  133 happen to be CEMP stars, 18 have $\rm [Fe/H < -3.0$ and 39 fulfil our interest with $\rm [Fe/H < -2.8$.
\citet{limberg_targeting_2021} selected stars in B \& B survey and observed with GEMINI and SOAR 1896 stars and 35 are EMP stars.
\citet{xu_stellar_2022} used Gaia, 2MASS and ALLWISE photometry, to search for relatively bright VMP giants using three different criteria. They discovered also few stars with $\rm [Fe/H]<-3$.

The use of a narrow-band photometry allows to select stars with a small flux in a small wavelength range where strong lines fall in a way to select EMP candidates.
Narrow-band photometry is generally combined with broad/medium-band photometry for the stellar parameters determination.

Pristine \citep{starkenburg_pristine_2017} combines a wide-band photometry (e.g. SDSS or Gaia) with a narrow-band photometry centred on the \ion{Ca}{ii} H and K lines.
These lines are so strong to be detectable at the lowest metallicities in the cool stars (K, G and F stars, the stars of still sufficient low-mass to have a main-sequence life of the order of or longer than the age of the universe), the interesting candidates to be formed from a gas enriched by just one or few stellar generations. The Pristine filter is mounted on the wide-field imager MegaCam at the 3.6\,m CFHT 
and observed a considerable fraction of the sky.
The follow-up observations allowed to confirm several EMP stars, among which one of the most metal-poor star known \citep[see][]{starkenburg_pristine_2018}. 
An artificial Pristine filter has been defined by \citet{martin_pristine_2023} from the spectro-photometric Gaia\,DR3 data and a catalogue of metal-poor candidates is provided.

 SkyMapper Southern Survey \citep[SMSS,][]{keller_skymapper_2007} observed in the southern hemisphere with a narrow-band filter centred, as Pristine, on the \ion{Ca}{ii} H and K lines, but the filter is slightly larger than the Pristine filter. The Survey is currently at its fourth data release \citep{onken_skymapper_2024}.
\citet{casagrande_skymapper_2019} complemented the SkyMapper photometry with 2MASS to derive a metallicity
calibration.
 \citet{huang_beyond_2022} derived metallicities (also distance and ages) for 20 million stars SMSS and Gaia EDR3, including 25000 candidate EMP stars.

On-going multi-band photometric observations on narrow filters are providing the tools to select stars with specific and also multiple characteristic, such as metal-poor (with a filter centred on the \ion{Ca}{ii}-K and -H lines) and carbon enhanced (with a filter centred on the molecular G-band hosting CH lines).
The S-PLUS Southern Photometric Local Universe Survey \citep[S-PLUS, ][]{mendes_de_oliveira_southern_2019} is imaging $\rm\sim 9300\, deg^2$ of the sky in 12 optical bands and provided the first data release with $\rm\sim 336\, deg^2$ with limit magnitude $r=21$. 
\citet{almeida-fernandes_data_2022} presented the second data release of S-PLUS covering $\rm\sim 950\, deg^2$ in the sky.
\citet{herpich_fourth_2024} presented the S-PLUS 4th data release, covering $\rm\sim 3000\, deg^2$.
Metal-poor candidates selected from S-PLUS have been confirmed as EMP stars. \citet{placco_mining_2022} with low-resolution spectroscopy confirmed that 15\% of the 522 stars selected from S-PLUS have $\rm [Fe/H]<-3.0$. \citet[][]{placco_splus_2021} present the chemical inventory of the evolved star SPLUS\,J210428.01--004934.2 (\teff=4812\,K and \logg=1.95) providing $\rm [Fe/H]=-4.03$ and derived also carbon ($\rm [C/Fe]=-0.06$).
\citet{whitten_photometric_2021} derived a metallicity estimations and A(C) on 700000 stars from  S-PLUS photometry, using artificial neural network methodology SPHINX.

The Javalambre Photometric Local Universe Survey \citep[J-PLUS, ][]{cenarro_j-plus_2019} is a photometric survey observing in the northern hemisphere 12 bands that allows the selection of metal-poor stars. 
\citet{whitten_j-plus_2019} developed the pipeline SPHINX, based on neural-network, to derive stellar effective temperature and metallicity from J-PLUS photometry.
\citet{galarza_j-plus_2022}, with the machine learning pipeline SPEEM, investigated the J-PLUS Data Release 2. Of the 177 candidates selected with $\rm [Fe/H]<-2.5$, they obtained spectra for 11 stars and confirmed the low metallicity in 64\% of them, finding also a star with $\rm [Fe/H]<-3$.

The mini-JPAS survey \citep{bonoli_minijpas_2021}, that uses 56 filters to take photometry from celestial objects, has a great potentiality in selecting a large number of metal-poor candidates.
Stellar Abundances and Galactic Evolution Survey \citep[SAGES][]{fan_stellar_2023} is a multi-band photometric survey with the goal to provide accurate stellar parameters. The first release covered $\rm\sim 9960\, deg^2$ in the sky.

The availability of the Gaia prism spectra, collectively referred to as XP spectra, represents
a real revolution for photometry. In fact, as described by \citet{montegriffo_gaia_2023}, it is possible
to use the spectra to compute magnitudes in any desired system, wide, intermediate or narrow.
As above-mentioned \citet{martin_pristine_2023} took advantage of this fact to extend the 
Pristine photometry to the whole sky.
The number of papers that have produced large catalogues of
metallicities using the Gaia XP spectra is too large to 
be completely reviewed and we cite only a personal selection
of the available catalogues.
To start with the astrophysical parameter inference system \citep[Apsis][]{fouesneau_gaia_2023}
that includes in the Gaia DR3 catalogue estimates of metallicity, effective temperature
and surface gravity for 470 million stars based on the XP spectra.
\citet{andrae_robust_2023} publish metallicities for 175 million stars based on XP spectra.
\citet{zhang_parameters_2023} publish stellar parameters, including metallicities for 220 million spectra.
\citet{li_aspgap_2024} concentrate on red giant stars and use XP spectra
to provide atmospheric parameters, but also [$\rm \alpha/M$] values for 27 million stars.
\citet{xylakis-dornbusch_metallicities_2024} provide metallicities for 10 million stars, but
with special attention to metal-poor stars. 
To test the ability of their method to select metal-poor stars they selected 26 stars and observed
them at high resolution, all the stars have [Fe/H] $< -2.0$, as expected, 15 have
[Fe/H]$< -2.5 $ and two [Fe/H]$< -3.0$ confirming the high efficiency of this
catalogue to select metal-poor stars.
Finally \citet{khalatyan_transferring_2024} provide metallicities for 217 million stars based on XP spectra.
Since all these catalogues have only recently been made available their exploitation has
only recently started. One approach to select truly metal poor stars is to cross-match
several catalogues and select the stars that are below a given metallicity threshold
in more than one catalogue (or more than two or three...).
The underlying data, the XP spectra are the same, but the algorithms to derive
metallicities are different.

An ingenious method to select metal poor stars has been devised by
\citet{melendez_2mass_2016}, it is in between spectroscopy and photometry.
They cross match  large catalogues that provide spectral types with  photometric catalogues
and select stars with a large discrepancy between spectral type and colours.
This allowed them to identify the bright extremely metal poor star 2MASS\,J18082002-5104378
(V=11.9, [Fe/H]=--4.1).

\subsection{Kinematical selections\label{subsec:kinem}}

\citet{roman_correlation_1950} suggested that high-speed stars are metal-poor.
\citet{schwarzschild_spectroscopic_1950} suggested that high-velocity stars have a C/Fe higher and a Fe strength (abundance) lower than low-velocity stars.
\citet{chamberlain_atmospheres_1951} analysed two high-speed stars (HD 19445 and HD 140283) and concluded that they are poor in Ca and Fe.
\citet{roman_group_1954} investigated a sample of fast stars (selected from absolute radial velocity $>\pm 75$\,km/s or proper-motion $>100$\,km/s) and concluded that they appear as F stars but too blue.
\citet{roman_catalogue_1955} presented a catalogue of fast stars because these objects are interesting for the Galactic structure and evolution.
It is stated in that paper that for the F- and G-type stars, the weakening of the metallic lines 
is correlated to an
ultraviolet excess.
Stars with weakest lines have the largest ultraviolet excess and the largest velocities.
From this statement by \citet{roman_catalogue_1955}, we can conclude that the fastest stars are the most metal-poor.
This is also the point of \citet{eggen_evidence_1962} who find that the stars with the largest  
ultraviolet excess\footnote{The ultraviolet excess is defined as $\delta(U-B) = (U-B)_{Hyades} - (U-B)_*$.
The Hyades is an open cluster of solar metallicity. \citet{sandage_new_1969} noted
that at any given metallicity the ultraviolet excess reaches a maximum value at $(B-V)=+0.6$ and proposed
to normalise the index to that of a star of this colour. This normalised ultraviolet excess
is usually denoted as $\delta(U-B)_{0.6}$.}
have the highest eccentricity, velocity and angular momentum.
\citet{wallerstein_abundances_1962}, investigating the chemical content of a sample of fast stars,
derived some interesting and, at the time, innovative conclusions: 
(i) a correlation between the stellar Fe abundance and the velocity parameter; 
(ii) a correlation between ultraviolet excess and metallicity;
(iii) the fact that the ratios in the abundances is not the same for all stars in the sample; 
(iv) the velocity dispersion increases with decreasing metallicity.

Selecting high-speed stars is still a way to select metal-poor stars and in the recent years, thanks to Gaia, it has been possible
to select stars for their speed, using Vr, parallax and proper motions from the Gaia DR2 and DR3.
But the goal has been more the quest on what are these high-speed stars than searching for EMP stars.
And in fact, in the chemical investigations based on stars selected for their high velocity, very few stars happen to be EMP.
Some of the investigations of the stars selected by their  high velocity are:
 \citet{matas_pinto_detailed_2022}; 
 \citet{caffau_high-speed_2020};
 \citet{bonifacio_high-speed_2024};
 \citet{caffau_high_2024};
 \citet[][using Gaia for the selection and S-PLUS for the parameter determination]{quispe-huaynasi_characterization_2024}, but few stars  are below $-3.0$.
 A similar study was done in \citet{quispe-huaynasi_j-plus_2023} using  J-PLUS.
\citet{quispe-huaynasi_high-velocity_2022} analysed a sample of fast stars (PM and distances from Gaia, $\rm V_r$ and metallicity from APOGEE) and they are MP $[-2.5,-0.5]$ Halo giant stars.
 \citet{marchetti_predicting_2018} in their paper is searching for fast stars, but no investigation in the metallicity is provided.
 \citet{hattori_old_2018} also select high-speed stars in Gaia-DR2, but claim that most are MP just from photometry compared to isochrones.
 \citet{li_origins_2023} investigates fast stars selected in surveys, so they have the chemical analysis and conclude they are all MP old stars.
 \citet{reggiani_chemical_2022} provide a chemical investigation of 15 late type fast stars: $\rm -2.5<[Fe/H]<-0.9$.
 \citet{li_591_2021} selected fast stars in LAMOST and ended with 591 stars. 86\% have $\rm [Fe/H]<-1$. Few stars (1 to 3) have $\rm [Fe/H]<-3$.
 \citet{hawkins_fastest_2018} computed a chemical investigation of 5 fast stars \citep[from][]{marchetti_predicting_2018}: the stars are giants $\rm -2<[Fe/H]<-1$ with no peculiarity.
 \citet{du_new_2019} select fast stars but no chemical investigation as the papers by \citet{marchetti_predicting_2018,de_la_fuente_marcos_flying_2019}. 
 \citet{du_origin_2018} select 24 high-velocity ($\rm Vgc>0.85$\,Vesc) stars with chemistry from LAMOST, mainly metal-poor alpha-enhanced, no EMP ($\rm -2.2<[Fe/H]<0$).
 \citet{du_high-velocity_2018} select local high-speed ($\rm V>220$\,km/s with respect to local standard rest) stars, 16 are high-velocity stars; they are metal-poor ($\rm -3<[Fe/H]<-0.3$) no EMP.
 \citet{nelson_detailed_2024} investigated a sample of 16 high-speed stars that happen to be all bounded to the Galaxy. All the stars happen to be metal-poor with just one very metal-poor star. 

\subsection{The metal-poor tail of the metallicity distribution function of the Milky Way}

One of the side-products of searches for metal-poor stars is that one can obtain 
information on the metallicity distribution function (hereafter MDF), or at least on its
low-metallicity tail. From the theoretical point of view the MDF is an output
of any galaxy evolution model that takes chemical composition into account. The comparison
between observed and theoretical MDF has, potentially, the power to exclude some
classes of models. The metal-poor tail of the MDF is of particular interest
since it provides information on the first steps of the evolution of the galaxy under study.
From the observational point of view the main difficulties are twofold: to obtain reliable
metallicity estimates for a statistically significant number of stars and to understand
the bias inherent in the selection of stars for which metallicity is estimated. 
The techniques and endeavours to solve the first issue have been discussed in subsections \ref{subsec:spectro} to
\ref{subsec:kinem}. In this section we shall describe efforts to determine
the metal-poor tail of the MDF of the Galaxy. 
Since the metal-poor stars are mainly found in the Galactic halo, 
in the literature most papers refer to the ``halo'' MDF, this use
mainly reflects a definition of the halo based on metallicity, rather
than on dynamical quantities, that were largely unavailable until recently. 
For a thorough discussion of the implications
of the metal-poor tail of the MDF we refer the reader to \citet{salvadori_cosmic_2007}.

The Hamburg-ESO objective prism survey \citep[hereafter HES][]{reimers_hamburgeso_1997}, designed
to select bright QSOs, turned out to have an extremely interesting stellar  content
\citep{christlieb_finding_2003,christlieb_stellar_2004,christlieb_stellar_2008}. Coupled to follow-up spectroscopic
observations it led to the determination of an MDF \citep{schorck_stellar_2009}, based on
a sample of 1638 stars.  \citet{schorck_stellar_2009} corrected their MDF for the selection 
function of the HES as a function of $(B-V)_0$ colour.
One of the prominent characteristics of this MDF was a sharp drop at metallicity $\sim -3.6$.
\citet{li_stellar_2010} derived the MDF by using unevolved stars selected from HES.

\citet{yong_most_2013-1} proposed an MDF based on a sample of 190 stars all observed
spectroscopically at high resolution and included a correction for the selection bias. 
With respect to the HES MDF, their MDF shows a smooth decrease below metallicity --3.6
and no sharp drop.

\citet{allende_prieto_deep_2014} derived an MDF from a sample of 5100 F and G stars
observed as spectro-photometric standard stars with the spectrographs \citep{smee_multi-object_2013}
designed and built for the Baryon Oscillation Spectroscopic Survey \citep[BOSS][]{dawson_baryon_2013}.
They corrected the MDF for the colour selection that was used to select the 
target stars. This MDF shows a very smooth metal-poor tail that extends down to --4.00.

Taking advantage of the large number of photometric
metallicities available from the
Pristine survey \citep{starkenburg_pristine_2017},
\citet{youakim_pristine_2020} selected a sample of about 80\,000 Turn-off stars
to determine the MDF. In order to correct for the bias they used a Gaussian mixture model
that takes into account the photometric errors.
This MDF is based on a much larger number of stars than the ones previously discussed
and, as pointed out by the authors themselves,
it provides a larger fraction of stars with [Fe/H] $< -2.0$ than other
published MDFs.

The H3 survey \citep{conroy_mapping_2019} is conducted at
relatively high resolving power ($R\sim 23\,000$) over a small spectral
range ($513-530$\,nm) with the MMT 6\,m telescope and a multi-fibre instrument.
\citet{naidu_evidence_2020} provide an MDF based on a sample of 5684 giants
observed in the H3 survey. They also compute dynamical quantities
and classify their stars that belong to different dynamical sub-structures. 

\citet{carollo_nature_2021} follow a different approach from other investigations, 
because they begin by defining the halo on the basis
of integrals of motion, the main focus being the Milky Way dynamics. 
However, using metallicities from
SDSS-SEGUE DR7 \citep{yanny_segue_2009}
they provide MDFs for different ranges in angular momentum ($L_z$).
What is mostly relevant to the current discussions is that
none of these MDFs extend to metallicity below $-3.00$ and for the
highly retrograde sample  ($L_z < -1000 ~{\rm kpc\,km\,s^{-1}}$)
the MDF peaks at metallicity $-2.2$, much more metal-poor than the other samples.

\begin{figure}
\centering
\includegraphics[width=11.5cm,clip=true]{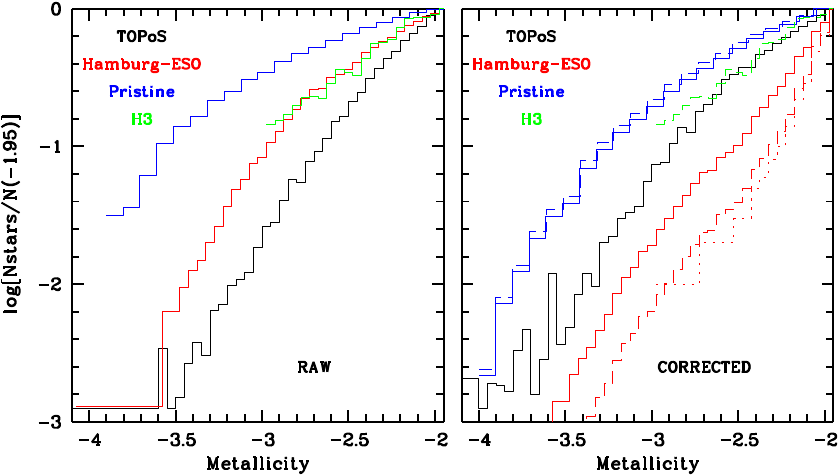}
\caption{The logarithmic ratio of the number of stars in each metallicity bin
over the number of stars in the metallicity bin at --1.95 from  \citet{bonifacio_topos_2021} (black),
\citet{schorck_stellar_2009}  (red), \citet{youakim_pristine_2020} (blue) and \citet{naidu_evidence_2020} (green),
both raw (left panel) and bias-corrected (right panel). For the HES data \citet{schorck_stellar_2009} provide three possible
bias corrections and they are depicted in red with  solid, dotted and dashed lines. Figure adapted from 
\citet{bonifacio_topos_2021}.
}
\label{fig:MDF}
\end{figure}

\begin{figure}
\centering
\includegraphics[width=10cm,clip=true]{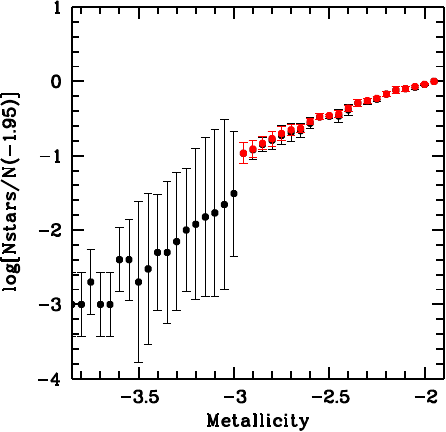}
\caption{The average logarithmic ratio of the number of stars in each metallicity bin
over the number of stars in the metallicity bin at --1.95 obtained by averaging 
the  bias-corrected ratio of \citet{bonifacio_topos_2021} and 
the uncorrected ratio of \citet{schorck_stellar_2009}  (black). 
For the metallicity range above --3.0 also the ratio
of \citet{naidu_evidence_2020} are added to the average (red).
Figure adapted from 
\citet{bonifacio_topos_2021}.
}
\label{fig:MDF_aver}
\end{figure}

The TOPoS project \citep{caffau_topos_2013}, performed an independent analysis of a large number of low resolution
spectra of TO stars observed with both SDSS and BOSS spectrographs in order to select extremely metal-poor stars
for high resolution follow-up. In \citet{bonifacio_topos_2021} they used this sample of data to derive the MDF
of the halo and, after cleaning, their sample consists of 139\,493 unique stars, the largest sample
among the above-cited studies.
To correct for the bias in the sample \citet{bonifacio_topos_2021} used as reference
a sample of over $24\times 10^{6}$ stars extracted from the SDSS photometric
catalogue, with the same selection criteria as the stars in the spectroscopic sample.
For these stars they estimated the metallicity from the  reddening-free
index $p = (u-g)-0.5885914(g-z)$. By comparing the photometric and spectroscopic MDFs they derived
a bias function assuming that the two have to be identical since they sample the same
underlying population.
The bias-corrected spectroscopic MDF confirms that the SDSS spectroscopic sample
is heavily biased in favour of metal-poor stars.

To depict the metal-poor tail of the MDF
we prefer to use the logarithmic ratio
of the number of stars in a  given metallicity bin to the number of stars
in the bin at metallicity --1.95.
In Fig.\,\ref{fig:MDF} we show the comparison of the metal-poor tail 
of four of the MDFs discussed above.
In the left panel the published bias corrections have been taken into account, except
for the H3 MDF, that is supposed to be unbiased. 
It is apparent that, while the H3 MDF does not extend to extremely low
metallicities, it is in good agreement with the HES raw MDF and
with the bias-corrected TOPoS MDF. 
For this reason \citet{bonifacio_topos_2021} suggested that it is
reasonable to average these three MDFs, thus obtaining an error
estimate in each metallicity bin from the variance among the three.
The result of this averaging process is shown in Fig.\,\ref{fig:MDF_aver}.
The error increases below metallicity --3.0, but the number of stars in each  bin decreases with decreasing the metallicity.
The apparent change in slope at --3.0 is not statistically significant.

The metal-poor tail of the MDF places  constraints on the properties of the first stars, 
even if some of these stars were formed in dwarf galaxies that later merged to form
the Milky Way \citep[see][and references therein]{bonifacio_topos_2021}.
For this reason it is clearly important to be able to reduce the error bars on the MDF.
In our opinion the way forward is to derive MDFs by using larger samples of stars
with different, but well understood,  observational biases, so that they can
ultimately be averaged to provide an error estimate.
The future looks promising, the Gaia final data release should provide large
samples of metal poor stars both with spectroscopic and photometric metallicity
estimates. The wide field spectroscopic surveys like WEAVE \citep{jin_wide-field_2024}
and 4MOST \citep{de_jong_4most_2019} also hold the promise to provide accurate metallicities
for large samples of stars and a known selection function.

\section{Abundance patterns of EMP stars in our Galaxy} \label{sec:abundance_patterns}

The main purpose of determining the chemical composition of the atmosphere of the EMP stars is to determine the chemical composition of the gas from which these stars formed at the very beginning of the Galaxy. To date we know of no star with a primordial composition, with only hydrogen, helium and some Li. The analysis of the atmosphere of the EMP stars should allow us to figure out what type of stars existed in the early days of the Galaxy and are responsible for the enrichment of the primordial matter.

\subsection{Abundances of the light elements Li, Be and B}
\label{sec_libeb}
\citet{burbidge_synthesis_1957} already noted that, in the Universe, Li, Be and B are extremely rare as compared with their neighbours in the periodic table of the elements: He, C and N. In fact these elements are very fragile, since they are destroyed as soon as the temperature reaches $2.5 \times 10^6$ K for Li,  $3.5 \times 10^6$ K for Be, and about $5 \times 10^6$ K for B.
As a consequence a process to create Li, Be, and B could not be sustained by nuclear fusion reactions inside stars \citep[see also][]{boesgaard_intriguing_2023}. 
Lithium has many possible sources:  the big bang (primordial nucleosynthesis for $\rm^7Li$), novae (where it is expelled as soon as it is  built),
AGB stars, RGB stars,  and spallation. We refer the reader to \citet{romano_gaia-eso_2021} and references
therein, for details on these possible sources. By contrast Be\footnote{Only the unstable isotope $^7$Be is formed in the big bang, but it decays to $^7$Li with a half life of 53.22 days.} and B cannot be made in the big bang nor in stars, the only
source of production is spallation \citep[see][and references therein]{prantzos_production_2012}. 

Moreover, if the convection zone in the stellar atmosphere is deep enough, as in cool stars (even in dwarf stars), these fragile elements are swept along to hot deep layers where they are destroyed and  little by little they are depleted  by dilution in the atmosphere of the stars.\\
The ``primitive'' abundance of Li can (a priori) be only observed in warm metal-poor unevolved (dwarf, turn-off and subgiant) stars with effective temperatures higher than about 5900\,K. 
In these stars indeed, the convective zone is supposed to be high enough to prevent these elements from being destroyed. 
Be and B are harder to destroy and their ``primitive'' abundance can be observed in stars as cool as 5200\,K.
But if one wants a simultaneous evaluation of the abundances of the three elements one has to target unevolved
stars with effective temperatures higher than 5900\,K.
In this section we will discuss only the abundances of the elements in stars found not to be carbon-rich. The C-rich stars are indeed very common at low metallicity and will be discussed in section \ref{sec:C-rich}. To be sure that a star is not C-rich we adopted a very strict limit $\rm [C/Fe] < +0.7$ (see section \ref{sec:C-rich}).

\begin{figure}
\centering
\includegraphics[width=8cm,clip=true]{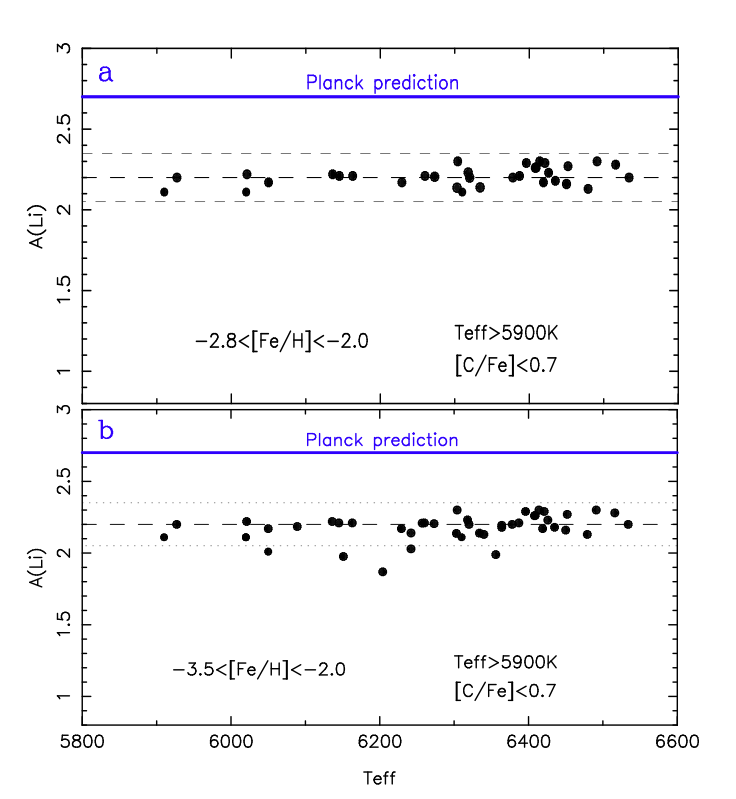}
\caption{A(Li) versus \teff\ in warm dwarfs or turnoff stars with:  (panel a) $\rm -2.8<[Fe/H]<-2$  and  (panel b) including the more metal-poor stars with $\rm -3.5<[Fe/H]<-2.8$. For all the stars in this figure  
[C/Fe]$<+0.7$, a very strict limit to be sure that the stars appearing in this figure are not carbon-rich (see Fig. \ref{Fig-CFeH} and section \ref{defCEMP}.)
}
\label{Fig:LiT}
\end{figure}

$\bullet$ Li \\
In Fig.\,\ref{Fig:LiT}a we have plotted A(Li) vs. \teff\ for dwarfs and turnoff metal-poor stars with 
$\rm T_{eff} > 5900 K$ and  $\rm [Fe/H] \ge -2.8$, following 
\citet{bonifacio_first_2007}, \citet{hosford_lithium_2010}, \citet{melendez_observational_2010}, \citet{sbordone_metal-poor_2010}, \citet{spite_lithium_2015}, \citet{reggiani_constraining_2017}, \citet{matas_pinto_metal-poor_2021}. 
Many of these papers contain C-rich stars, however in Fig.\,\ref{Fig:LiT} we decided not to show the  C-rich stars, 
nor the blue-stragglers. The latter are expected to decrease  their lithium abundance 
during the process of formation of the blue straggler \citep{glaspey_lithium_1994,glebbeek_blue_2010}.
In these stars the abundance of Li is constant  independently of the temperature of the star.
This Li abundance defines a "plateau" called "Spite plateau" after \citet{spite_lithium_1982}, and it was thought that this abundance ($\rm A(Li) \approx 2.2 $) was the primordial abundance of $\rm^{7}Li$, as it is built during the big bang. 
Moreover
the value of this observed plateau of A(Li) is the same for stars formed in other galaxies
\citep{matteucci_evolution_2021}. Observationally this rests on the fact that it is
observed in the Globular Cluster $\omega$ Cen \citep{monaco_lithium_2010}, 
that is believed to be the nucleus
of an accreted galaxy, and
M\,54 \citep{mucciarelli_cosmological_2014}, that belongs to the Sgr dSph.
The plateau has also been found among disrupted accreted galaxies such as GSE \citep{molaro_lithium_2020,simpson_galah_2021}
and Sequoia \citep{molaro_lithium_2020} as well as in the halo stream S2 \citep{aguado_s2_2021}, that is believed to be a disrupted galaxy.
\\
However, the data of the Planck satellite \citep{planck_collaboration_planck_2016} allow a precise prediction of the quantity of $\rm^{7}Li$ produced by the big bang in the frame of the standard model, and this value ($\rm A(Li) \approx 2.7$, see Fig.\,\ref{Fig:LiT}) is about 0.5 dex higher than the level of the plateau.  
This is known as the cosmological ``Li-problem'', that we refer to as the first ``Li-problem''.
A thorough discussion of this is beyond the scope of this review.
We only mention here that there are two classes of solutions of the problem proposed:
a depletion of Li in stars due to stellar phenomena
\citep[see, e.g.][]{korn_probable_2006,fu_lithium_2015,boesgaard_lithium_2023,boesgaard_primordial_2024,borisov_coherent_2024,nguyen_combined_2024}
or a revision of the big bang nucleosynthesis including new physics
\citep[see e.g][]{salvati_breaking_2016,luo_big_2019,talukdar_big_2024,singh_re-examining_2024}.
The recent finding that the Li abundance in the metal poor gas
of the Small Magellanic cloud matches the abundance in metal-poor dwarf stars
\citep{molaro_extragalactic_2024} would rule out all ``stellar'' explanations,
if confirmed along other lines of sight.
\\

There is a second "Li-problem". With the advent of more efficient spectrographs and receptors, 
the Li abundance could be measured in more and more metal-poor stars. 
If, like in Fig.\,\ref{Fig:LiT}b, we include  stars with a metallicity lower than $\rm [Fe/H] = -2.8$ the spread of the Li abundance becomes much larger.

\begin{figure}
\centering
\includegraphics[width=8cm,clip=true]{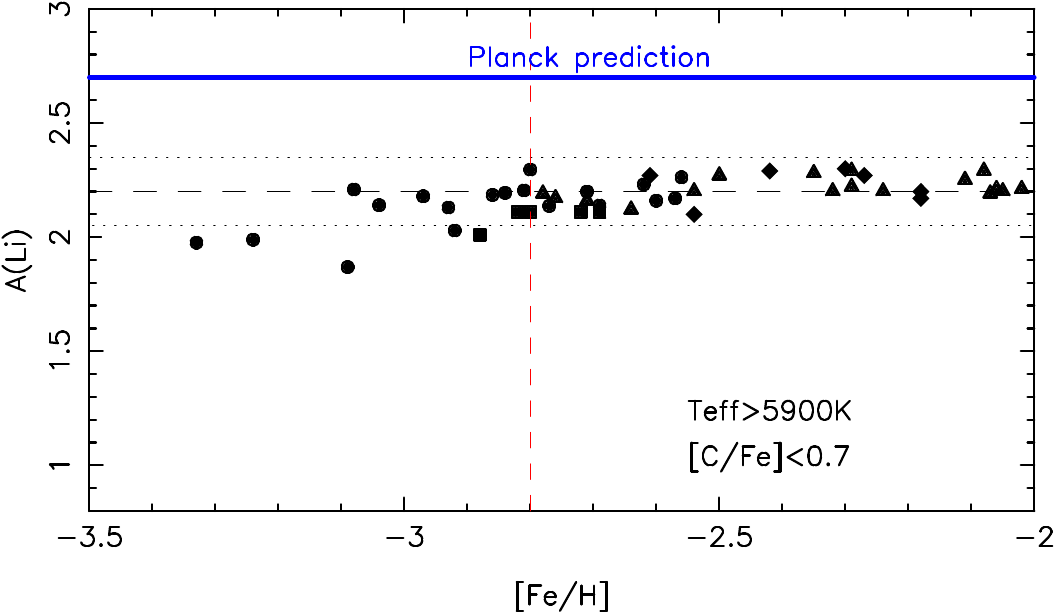}
\caption{A(Li) versus [Fe/H]  in warm dwarfs or turnoff stars, where stars with $\rm [Fe/H] < -2.8$ are included: 
\citep[black dots:][]{bonifacio_first_2007,sbordone_metal-poor_2010,spite_lithium_2015,matas_pinto_metal-poor_2021}, \citep[black squares:][]{roederer_search_2014},
\citep[black triangles:][]{melendez_observational_2010},
\citep[black diamonds:][]{reggiani_constraining_2017}.
We insist on the fact that for all the stars in this figure  [C/Fe]$<+0.7$, 
a very strict limit (see Fig. \ref{Fig-CFeH} and Sec.\,\ref{defCEMP}).
At very low metallicity from $\rm[Fe/H] \lesssim -2.8$ (red dashed line) a melt-down of the "plateau" is  observed.}
\label{Fig:LiFe}
\end{figure}

If we now plot for all the stars  A(Li) vs. [Fe/H] (Fig.\,\ref{Fig:LiFe}), we first observe a plateau in the interval $\rm -2.8<[Fe/H]<-2.0$, but then, at lower metallicity, a melt-down of this plateau is observed  \citep{sbordone_metal-poor_2010}. \\
Moreover, let us note that  in Fig.\,\ref{Fig:LiFe}, even for $\rm[Fe/H]>-2.8$, A(Li),  seems to be increasing  slightly with [Fe/H] \citep[e.g.][]{norris_critique_2023}. It was first suggested that this trend could be explained by a slight enrichment  of the matter in $\rm^{6}Li$ by cosmic rays \citep{fields_evolution_1999}. But it seems now that, in very metal-poor stars,  the contribution of $\rm^{6}Li$ in the total lithium abundance is negligible \citep[e.g.][]{wang_non-detection_2022}.

Since the carbon-rich stars almost appear  at the same metallicity as the melt-down of the Li-plateau (see section \ref{sec:C-rich}), \citet{norris_critique_2023} tried to explain this melt-down by a link between the formation of the C-rich stars and the Li deficiency. However, from our Fig.\,\ref{Fig:LiFe}, which does not contain C-rich stars, a melt-down of the Li plateau appears in stars without carbon enrichment.
The three stars with the lowest Li abundance are, in order of decreasing
metallicity CS\,22966-011, CS\.22948-093 and CS\,22988-031 that have [C/Fe]= +0.45, +0.6 and +0.38, 
respectively.
\\
On the other hand \citet{norris_critique_2023} also explain the slight slope of A(Li) vs. [Fe/H] by a merging of two populations of C-rich and C-normal stars 
in the region $\rm -2.8<[Fe/H]<-2$, but this slight trend is observed even when only C-normal stars are taken into account (see Fig.\,\ref{Fig:LiFe}).

Note that in this section we have only considered stars with $\rm [Fe/H] > -3.5$. At lower metallicity, in warm dwarfs and turnoff stars, the CH band is so weak that it is very difficult to ascertain that $\rm [C/Fe]<+0.7$. The most metal-poor stars are generally C-rich stars and are discussed in section \ref{sec:C-rich}. 

Up to now there is no clear explanation of the difference between the lithium abundance based on stellar and cosmological endeavours: depletion of Li in the stars or uncertainty in the predictions of the big bang model \citep[see in particular][]{norris_critique_2023}.

$\bullet$ Be and B\\
These elements are not supposed to be formed during the big bang and thus no plateau with [Fe/H], is expected. Following \citet{reeves_galactic_1970} and \citet{meneguzzi_production_1971} these elements are built by spallation in the interstellar medium: energetic neutrons and protons bombard mainly C, N, and O atoms, and break them into  $\rm^{6}Li$, Be, and B. The reverse process is also possible, by which fast C, N and O nuclei in the cosmic rays break up after collision with H atoms in
the interstellar medium.

\begin{figure}
\centering
\includegraphics[width=6cm,clip=true]{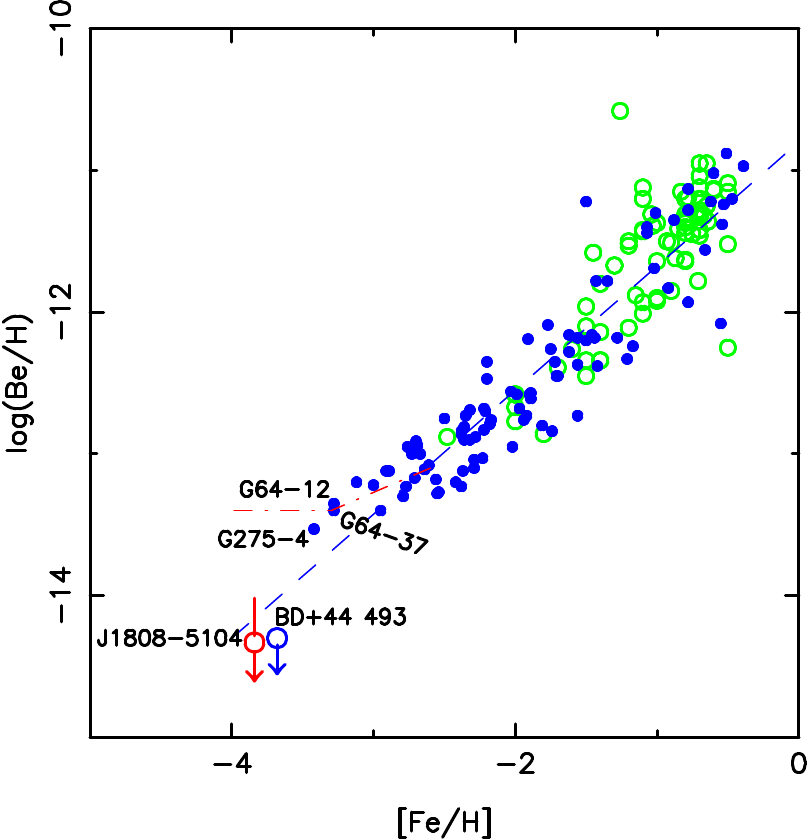}
\caption{A(Be) versus [Fe/H]  in warm dwarfs or turnoff stars. 
The green open circles are from \citet{smiljanic_beryllium_2009} and the blue filled circles from \citet{boesgaard_beryllium_2011}.
 The upper limit of the abundance of Be in 2MASS J1808-5104 and BD +44~493 are indicated with big red and blue open circles. The blue dashed straight line represents the mean relation.
At very low metallicity the Be abundance continues to decrease, there is no indication of a plateau.}
\label{fig:Be-Fe}
\end{figure}

The Be abundance can be measured in the near UV from ground-based telescopes, using the resonance lines of Be\,II at 313\,nm.

In metal-poor stars B\,I is only measured in the UV at 249.7\,nm \citep{duncan_evolution_1997,duncan_boron_1998,garcia_lopez_boron_1998,primas_new_1999,boesgaard_boron_2005}.
The B II and B III resonance lines \citep{boesgaard_boron_1981,mendel_testing_2006} 
   can only be measured  in relatively hot  and thus young, metal-rich stars.

Already \citet{molaro_upper_1984}, thanks to a low upper limit in the Be abundance in an old metal-poor star, put forward the existence of a slope in the Be abundance as a function of metallicity. The slope was confirmed by \citet{gilmore_is_1992}. In Fig.\,\ref{fig:Be-Fe} the abundance of Be is plotted as a function of [Fe/H], following \citet{spite_be_2019}. Their upper-limit in the Be abundance was in particular derived in two extremely metal-poor stars BD\,$+44~493$ \citep{ito_bd44493_2009,placco_hubble_2014} and 2MASS\,J1808-5104 \citep{spite_be_2019,mardini_chemical_2022}, 
and their very low Be abundance (see Fig.\,\ref{fig:Be-Fe}) confirms the linear decrease of the Be abundance with metallicity  \citep[see e.g.][]{boesgaard_intriguing_2023}, 
down to $\rm [Fe/H]\sim -4.0$.\\
 
A similar trend was also found for the relation of the  boron abundance vs. [Fe/H] \citep[see][]{duncan_evolution_1997,duncan_boron_1998,primas_galactic_2000,boesgaard_intriguing_2023}.

\subsection{Elements from C to Zn in normal EMP stars}

\begin{figure}
\centering
\includegraphics[width=8cm,clip=true]{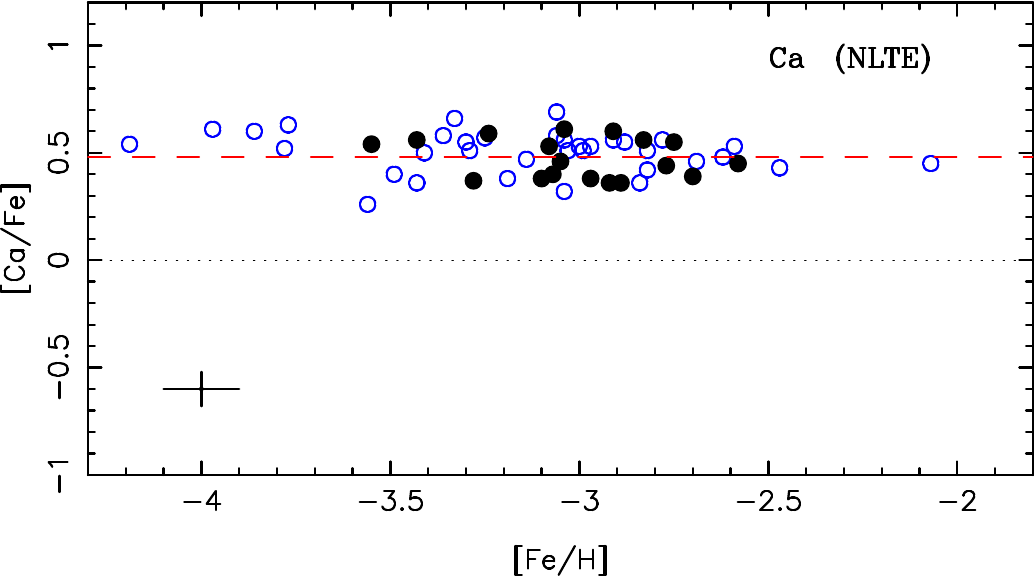}
\caption{\footnotesize{[Ca/Fe] vs. [Fe/H]  computed for the 54 VMP and EMP dwarfs (black dots) and giants (open blue circles) studied very homogeneously and corrected for NLTE \citep{bonifacio_first_2009,spite_nlte_2012}.}}.
\label{Fig-NLTECaFeH}
\end{figure}

In the infancy of the Universe only massive stars of previous generations had time to explode as SN type II and enrich the gas from where the old EMP stars we observed formed.
C is produced by the triple $\alpha$ reaction during He fusion, but is normally not considered as an $\alpha$ element.
The C abundance will be discussed in section \ref{sec:C-rich}.\\
The $\alpha$ elements from O to Ca  (O, Mg, Si, S and Ca) are the result of $\alpha$-particle captures, during the fusion of  C, Ne and O in very massive stars. In the literature Ti is sometimes also considered an $\alpha$ elements, but we prefer not to consider
it among the $\alpha$ elements since, besides being formed through $\alpha$ captures, it is also partly synthesised in nuclear statistical equilibrium, along with
iron-peak elements.
Since these stars were the first to enrich the interstellar matter, these elements appear to be more abundant compared to iron in the oldest, more metal-poor  galactic stars.
As an example we give the trend of [Ca/Fe] vs. [Fe/H] in Fig.\ref{Fig-NLTECaFeH}. All these 54 stars were studied very homogeneously by  \citet{cayrel_first_2004}, \citet{bonifacio_first_2009}  
and corrected for NLTE by \citet{spite_nlte_2012}. 
A very similar result was obtained by \citet{sneden_iron-peak_2023} from a sample of 37 warm metal-poor stars. 

\begin{figure}
\centering
\includegraphics[width=8cm,clip=true]{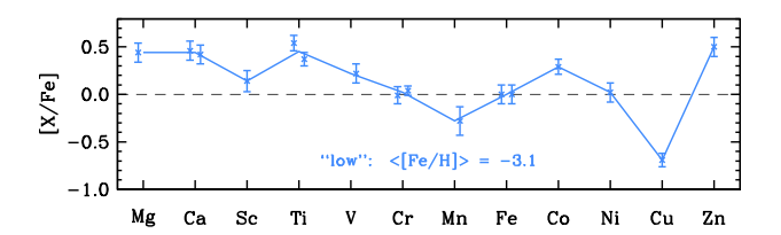}
\caption{Mean elemental abundances for low metallicity stars following \citet{sneden_iron-peak_2023}. Figure reproduced with permission.}
\label{Fig-snedenFegroup}
\end{figure}

Based on a pure LTE analysis  \citet{sneden_iron-peak_2023} give an updated trend of the Fe-group elements as a function of [Fe/H] in the most metal-poor dwarf stars. Generally speaking they are in good agreement with the results of \citet{bonifacio_first_2009}. The mean values are given in Fig.\,\ref{Fig-snedenFegroup}. 
Following \citet{sneden_iron-peak_2023}, the $\alpha$ elements Mg and Ca are overabundant relative to iron by about 0.4 dex; the lightest iron peak elements Sc, Ti and V are also overabundant relative to Fe; the heavier elements of the iron peak have a ratio [X/Fe] close to zero except for Cu, that seems to be strongly under-abundant while zinc is significantly over-abundant relative to Fe.
NLTE effects could probably explain the low [Cu/Fe] ratio at low metallicity 
\citep[see e.g.][]{andrievsky_galactic_2018,roederer_new_2018}, however it seems that  NLTE effects and  granulation (3D effects) are not able to compensate for the large over-abundance of Zn at low metallicity \citep[e.g.][]{bonifacio_first_2009,roederer_new_2018}, and this effect 
is thus probably  real.

\begin{figure}
\centering
\includegraphics[width=8cm,clip=true]{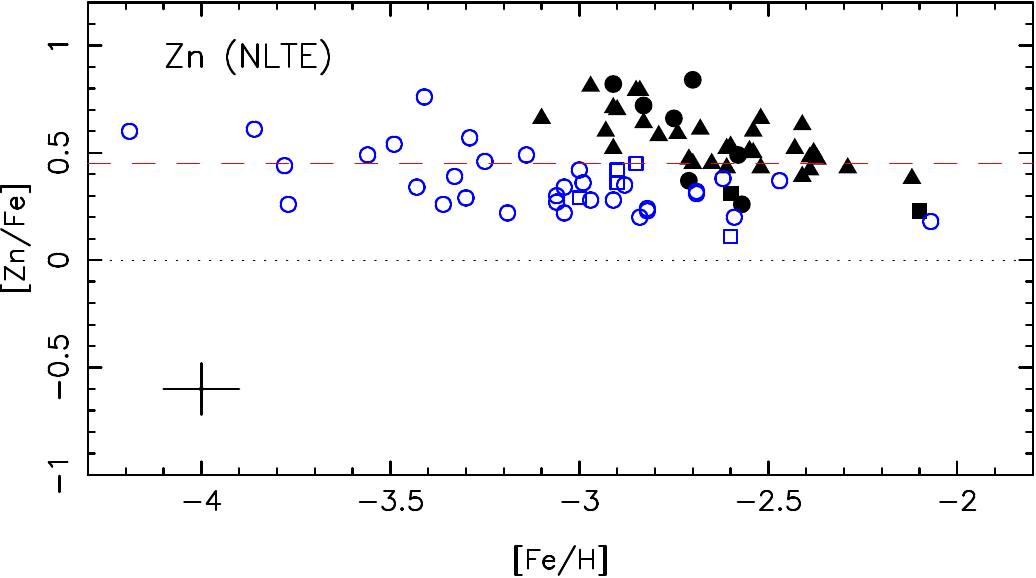}
\caption{[Zn/Fe] vs. [Fe/H]  computed for  VMP and EMP dwarfs (black filled symbols)  and giants  (blue open symbols) from  \citet{cayrel_first_2004}, \citet{bonifacio_first_2009}, \citet{matas_pinto_metal-poor_2021}: circles,  or  \citet{lai_detailed_2008}: squares, and \citet{sneden_iron-peak_2023}: triangles.}
\label{Fig-FeZn}
\end{figure}

The Fe-group elements are mostly made during the explosion of supernovae, in the deepest part of the ejecta. 
Core-collapse supernovae (CCSNe) in the mass range of 20-40 $M_{\odot}$  are generally considered  to be the main contributors \citep{klessen_first_2023}. But they are supposed to form very little Zn  \citep[see e.g.][]{grimmett_chemical_2021,prantzos_galactic_2019}.\\
 However, in the EMP stars the mean value of [Zn/Fe] is close to +0.4 dex (see Fig.\,\ref{Fig-snedenFegroup}).
 As a consequence \citet{grimmett_chemical_2021} propose that Zn be produced by jet-driven hypernovae. These hypernovae produce very little Fe, and the matter, in the early Galaxy, would be  enriched by a mix of CCSNe and hypernovae. 
 In Fig  \ref{Fig-FeZn} we have plotted [Zn/Fe] vs. [Fe/H] for a sample of EMP stars studied in \citet{cayrel_first_2004}, \citet{bonifacio_first_2009},  \citet{lai_detailed_2008}  \citet{matas_pinto_metal-poor_2021} and \citet{sneden_iron-peak_2023}. None of the stars in this figure is carbon-rich: they have all a ratio $\rm[C/Fe] < 0.6 $. The abundance of Zn has been corrected for NLTE effects following the tables of \citet{takeda_non-lte_2005}, we could not use the recent values of \citet{sitnova_non-lte_2022} since they have not computed the corrections for dwarf stars below [Fe/H]=--2.
  There is an important bias in Fig. \ref{Fig-FeZn}: below [Fe/H]=--2.7, in turnoff stars, the Zn lines are very weak and can be measured only if the Zn abundance is relatively large. 
  In Fig. \ref{Fig-FeZn} the spread  minimum to maximum  in [Zn/Fe] is about 0.7\,dex,
  which is large when compared to e.g. [Ca/Fe] (see Fig. \ref{Fig-NLTECaFeH}, the spread is 
  about 0.4\,dex),  and cannot be explained by measurement errors. 
It could be the result of a variable contribution of  jet-driven hypernovae to the medium enrichment.

\subsection{$\alpha$-poor stars} \label{sec:alpha-poor-stars}
EMP stars are, as expected, enhanced in $\alpha$-elements, but there are exceptions.
A famous exception is the most metal-poor star known SDSS\,J102915.14 +172927.9 \citep{caffau_extremely_2011,caffau_sdss_2024} whose Mg and Ca are not enhanced with respect to Fe.
But not always Mg and Ca scale in a  consistent way with respect to Fe in these non-$\alpha$-enhanced or $\alpha$-poor stars. There are some stars in which both Mg and Ca are depleted or not-enhanced but in some stars just one of them is low.
Four $\alpha$-poor stars at metal-poor regime ([Fe/H] around $-2.0$) have already been investigated by \citet{ivans_chemical_2003} 
and \citet{sakari_r-process_2019}.  The proposed  explanation is that this peculiar chemical composition can be the result of a larger contribution from SN\,Ia.
We can find $\alpha$-poor stars also among EMP stars, and the list is quite long. 
\citet{li_four-hundred_2022} analysed $\alpha$-poor stars, including some EMP and in particular J1458+1128 with $\rm [Mg/Fe]=-0.23$ and $\rm [Ca/Fe]=-1.06$.
\citet{aoki_high-resolution_2013,venn_pristine_2020,purandardas_observational_2021,yong_most_2013,caffau_x-shooter_2013,hansen_exploring_2014,matsuno_high-resolution_2017,francois_chemical_2018,bonifacio_chemical_2012} also investigated EMP, non-$\alpha$-enhanced or $\alpha$-poor stars also at extremely low [Fe/H].

 In the dwarf galaxies the star formation is slow or it proceeds in bursts, giving the time to type Ia SNe 
to produce iron. For this reason the knee corresponding to the enhancement in the $\alpha$-elements is found at a lower metallicity than in the Galaxy  \citep[see][]{matteucci_metallicity_1990}.
According to both \citet{bonifacio_topos_2018} and 
\citet{sakari_r-process_2019} the low-$\alpha$ stars could have formed in low-mass dwarf galaxies and subsequently accreted to the Milky Way Halo, carrying memory of a different chemical evolution, as has been claimed by \citet{hayes_disentangling_2018}.  
But this interpretation can be challenged by a star like SDSS\,J102915.14+172927.9, that has a Galactic disc orbit, not easily reconciled with an accretion event.
\citet{caffau_sdss_2024} suggested that this star may be a true Pop\,III star whose atmosphere has been polluted by metal-rich gas during its encounters
with Galactic gas clouds with the mechanism described by \citet{yoshii_metal_1981}. 

\subsection{The neutron-capture elements}
The elements heavier than the iron group ($Z>30$ or so) are mainly formed by neutron capture on iron peak elements. Neutron capture occurs mainly in three locations: in the envelopes of evolved low or intermediate mass stars in their AGB phase \citep[the slow or "s-process", see][]{arcones_origin_2023}, or in some sort of explosive event likely a core-collapse supernova or a merging of neutron stars \citep[the rapid "r-process" see][ for more details]{cowan_origin_2021}. Moreover an intermediate process "i-process" is sometimes evoked. It would be a n-capture process triggered by the rapid ingestion of a substantial quantity of H in He-burning convective regions in 
for example super-AGB or He shell flash in low metallicity stars, which would lead to the formation of $\rm^{13}C$ and then to the reaction $\rm^{13}C(\alpha,n)^{16}O$  
(\citealt{cowan_production_1977} see also \citealt{hampel_intermediate_2016,roederer_diverse_2016,choplin_intermediate_2021}). 
Another possible site for the "i-process" are rapidly accreting white
dwarfs \citep{stephens_3d1d_2021} and proton injection in a He burning shell in massive stars \citep{banerjee_new_2018}.\\
Since the s-process occurs during the evolution of relatively low mass stars with a very long life time \citep{arcones_origin_2023}, it is not supposed to contribute to the enrichment of the matter in the early Galaxy.
There is however the possibility of an s-process taking place in massive rotating stars, 
and the products should be ejected by the stars through winds, prior to their explosion as SN
\citep{frischknecht_non-standard_2012,frischknecht_s-process_2016,choplin_non-standard_2018,banerjee_new_2019}.
In the following we shall refer to the s-process occurring in AGB stars as the main s-process.

Three large projects were or still are dedicated to the study of neutron-capture elements in metal-poor stars. Let us cite:\\
$\bullet$ the HERES project \\
The aim of the HERES project(Hamburg ESO R enhanced Stars) was to select and study  metal-poor stars selected in the Hamburg ESO survey and enhanced in neutron-capture elements,  to possibly identify the sites for the nucleosynthesis processes \citep[let us cite in particular][]{christlieb_hamburgeso_2004,barklem_hamburgeso_2005,hayek_hamburgeso_2009,mashonkina_hamburgeso_2014}.\\
$\bullet$ the R-process Alliance\\
The goal of the R-process Alliance was again to 
precise the main production site and the possible secondary sites of the neutron capture elements in the early Galaxy, among which are neutron star mergers, jets in rotational supernovae and neutrino driven winds  \citep[see in particular][]{hansen_r-process_2018,sakari_r-process_2018,ezzeddine_r-process_2020,holmbeck_r-process_2020,bandyopadhyay_r-process_2024}.  \\
$\bullet$ the CERES project\\
The project CERES (Chemical Evolution of R-process Elements in Stars) aims to provide a homogeneous analysis of a sample of metal-poor stars with 
$ \rm [Fe/H] < -1.5.$ However many stars studied in this survey are EMP stars \citep{lombardo_chemical_2022,lombardo_chemical_2025,alencastro_puls_chemical_2025}.
\\

The neutron-capture elements are generally divided into three peaks, according to their atomic number \citep[see e.g.][]{sneden_neutron-capture_2008,frebel_nuclei_2018}. For the s-process the centre of each peak corresponds to elements with 
very small cross section for neutron-capture reaction
(elements with a magic number of neutrons).  For the r-process the centre of each peak corresponds to the decay of unstable neutron-rich elements with magic neutron numbers too. These peaks are slightly shifted relative to the peaks of the s-process because the proton number (or the mass) are not the same for the two processes. Generally speaking, we consider that
Sr, Y, Zr,  belong to the first peak; 
elements from Ba to Hf belong to the second peak; and the heavier elements from Os to U to the third peak.\\

The first extensive study of the behaviour of a rather large number of heavy elements from Sr (Z=38) to Yb (Z=70) in an homogeneous sample of EMP giant stars is probably the paper of \citet{francois_first_2007} after the papers of \citet{mcwilliam_spectroscopic_1995}, 
\citet{ryan_extremely_1996} and \citet{christlieb_hamburgeso_2004}. 
All these papers underline the large spread of the heavy elements in EMP stars.  For example, in stars with $\rm[Fe/H] \approx -3.$,  the ratio Eu/Fe varies, from star to star,  by more than a factor of 100. 

An extensive review of the behaviour of the heavy elements in the early Galaxy,  mainly based on the "HERES" and the "R process Alliance" surveys, has been done by \citet{sneden_neutron-capture_2008} and \citet{frebel_nuclei_2018}. \\

\begin{figure}
\centering
\includegraphics[width=5.5cm,clip=true]{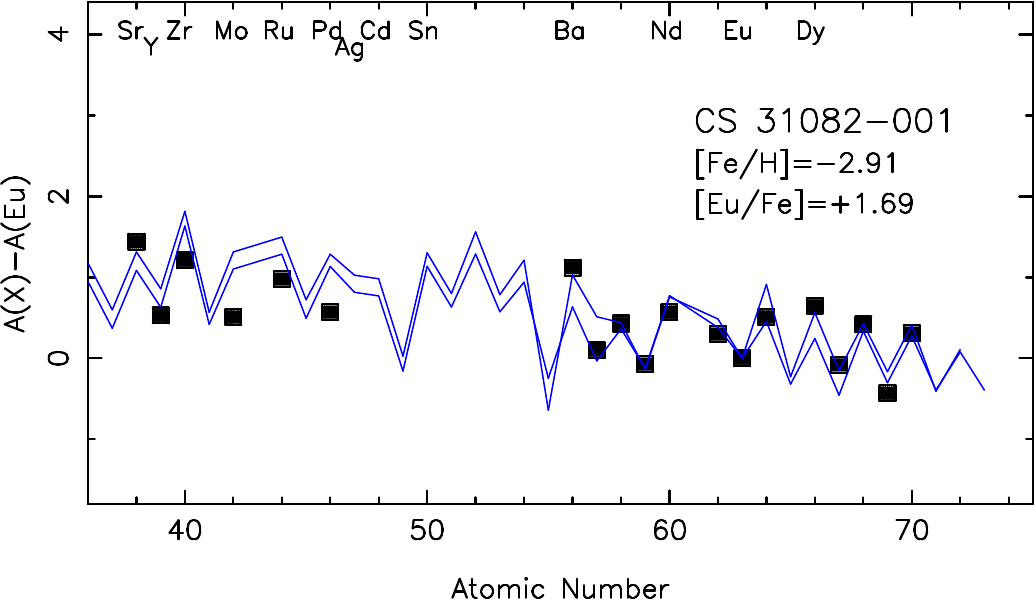}
\includegraphics[width=5.5cm,clip=true]{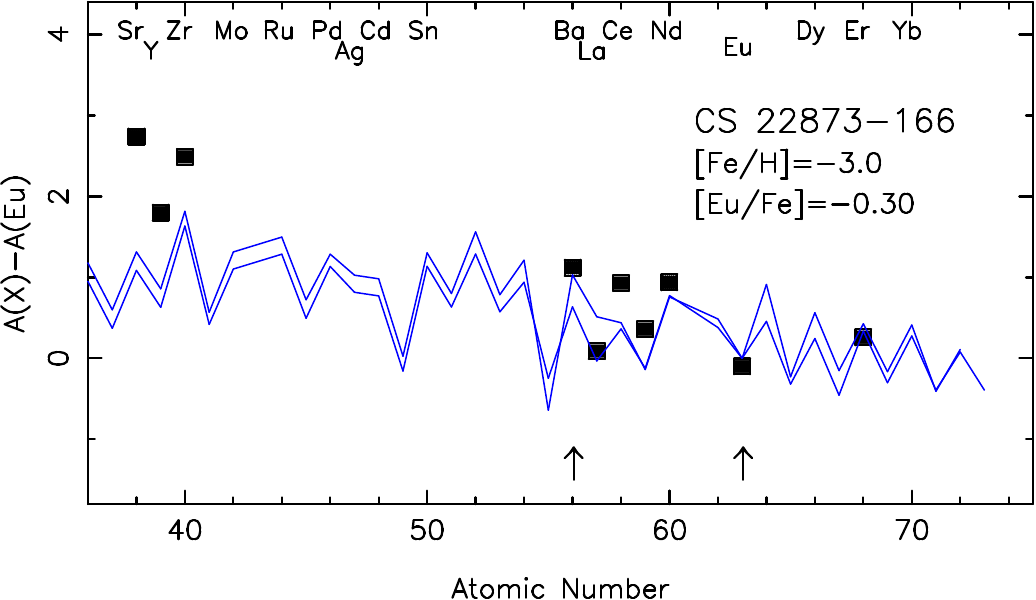}
\caption
{ Abundance patterns for neutron capture elements in two EMP stars with $\rm [Fe/H] \approx -3$ compared to the theoretical r-process patterns computed by \citet{wanajo_cold_2007}  for a hot and a cold model (blue solid lines). 
CS 31082-001 is a classical r-rich star  \citep[left panel,][]{cayrel_measurement_2001,hill_first_2002,siqueira_mello_first_2013,ernandes_reanalysis_2023} and CS 22873-166 \citep[right panel,][]{francois_first_2007} is on the contrary an r-poor star.
From Ba to Yb, in both stars, there is a rather good agreement between the theoretical and the observed patterns, but in CS 22873-166 the lighter elements(here Sr Y Zr) are more abundant than expected by the r-process
}
\label{Fig-heavy-patterns}
\end{figure}

\noindent $\bullet$ Theoretical patterns of the r-process elements\\
\citet{arlandini_neutron_1999} and \citet{simmerer_rise_2004} determined for each  element in the Sun, the quantity formed by the different processes (s-process and r+i processes). In the Sun, the isotopic abundances are known (meteorites) but matter was enriched with heavy elements formed in massive stars (r-process) and in low and intermediate-mass stars  (s-process). By calculating the products of the s-process in low and intermediate-mass stars, \citet{arlandini_neutron_1999} were able to deduce by subtraction the distribution of elements produced by the r-process. This distribution is known as the solar r-process pattern,
but it may contain also elements formed by an i-process.\\
On the other hand \citet{wanajo_cold_2007} and \citet{wanajo_electron-capture_2011} directly computed 
the distribution of the elements produced by the r-process. From Sr to heavier elements, both distributions are very similar.\\

\noindent $\bullet$ Second peak elements - definition of the r-rich stars\\
In Fig. \ref{Fig-heavy-patterns} we compare the abundance pattern of the heavy elements of two EMP stars with $\rm [Fe/H] \approx -3$  which differ by a factor of 100 in Eu/Fe. In both cases the observed pattern of the elements from Ba to Yb is rather well represented by the predictions of a pure r-process as computed by \citet{wanajo_cold_2007}.  For example, the ratio [Eu/Ba] is almost constant in the EMP stars, at least between [Fe/H]=--4 and [Fe/H]=--2, and close to +0.5  \citep{spite_abundance_2018}.\\
In the early Galaxy, Eu and Ba are both formed by the r-process and the proportion of these elements is conformed to the predictions of this process. As expected,  the main s-process does not contribute to the enrichment of second peak elements in  the early Galaxy.\\
Following  \citet{beers_discovery_2005} a star is called  r-rich if $\rm [Eu/Fe] \geq +0.3$ ("r-I"  for $\rm +0.3 <[Eu/Fe]< +1$, and "r-II" if  $\rm [Eu/Fe] > 1.0 $).\\

Moreover, it is interesting to remark that the distribution of the second peak r-process elements, is the same in the Sun and in the EMP stars formed less than 1\,Gyr after the big bang. According  to \citet{frebel_nuclei_2018}, the constancy of this distribution over time suggests that the r-process is universal.

Recently \citet{roederer_element_2023} have suggested that the elements Ru to Ag in r-enhanced stars display abundance patterns that
can be interpreted as an r-process that proceeds to the production of elements heavier than U, which
then fission, populating this interval of atomic numbers.  
\\

\noindent $\bullet$  First peak elements\\
On the other hand, in CS\,22873-166 (r-poor star), unlike CS\,31082-001 (Fig. \ref{Fig-heavy-patterns}) the abundances of the first peak elements Sr, Y, Zr are not compatible with the abundances expected by a pure main r-process. These stars are sometimes called Sr-rich stars, because they contain more strontium than expected.  Other examples can be find, for example, in HD\,122563 and HD\,88609 \citep{honda_neutron-capture_2006,honda_neutron-capture_2007}. All these stars are "r-poor". In fact the scatter of the ratio [Sr/Ba] increases when [Ba/Fe] (or [Eu/Fe]) decreases \citep{spite_high_2014,spite_abundance_2018}.
Generally speaking, when a star is r-rich, the main r-process is sufficient to explain all the heavy elements from Sr to Yb, but when the interstellar gas out of which  the star was formed, is poor in heavy elements, then  appears another type of enrichment superposed to the main r-process often called weak-r process, which would build mainly first peak elements \citep{cowan_origin_2021}. \\
The elements Ge, As and Se (atomic numbers 32, 33, 34) are very difficult to measure since 
the use of lines in the UV is needed and the only one accessible from the ground is Ge. 
\citet{roederer_germanium_2012} and \citet{peterson_trans-iron_2020} measured these elements in metal-poor stars, 
however, to our knowledge, 
the only EMP stars for which one of these elements, Ge, has been 
measured are CS\,31082-001 \citep{siqueira_mello_first_2013,ernandes_reanalysis_2023}, and HD\,115444 \citep{westin_r-process-enriched_2000}.  Both stars are  r-II stars with respectively [Eu/Fe]=+1.69 and +0.85,  and in both cases Ge is underabundant with $\rm [Ge/Fe]=-0.55$ and $-0.47$
\\

\noindent $\bullet$ Extremely r-rich stars\\
\citet{francois_first_2007} and \citet{hill_first_2002} (ESO large program "First Stars First Nucleosynthesis") studied the abundance of the heavy elements in 26 red giants with a metallicity lower than [Fe/H]=--2.8, and found only three stars with  [Eu/Fe]$>+0.7$, the characteristic of the r-II stars following \citet{holmbeck_r-process_2020}. Over the past decades efforts have been made to search for r-process-enhanced stars. \citet{roederer_nine_2014} reported the discovery of new r-process enhanced stars, but none of them has a metallicity lower than [Fe/H]=--2.8. Later \citet{hansen_r-process_2018} selected relatively bright metal-poor stars in the RAVE catalogue and obtained for these stars high resolution spectra. Among the 107 stars they studied only 19 have a metallicity below --2.8 and finally among these 19 stars only three have a 
ratio [Eu/Fe]$>0.7.$ The r-II stars are very rare, their ratio [Eu/Fe] is larger than +0.7, but generally it is lower than about +1.8. \\
However \citet{cain_r-process_2020} reported the discovery of  star  2MASS J15213995-3538094 (hereafter J1521-3538) with [Fe/H]=--2.8 and an extremely large Eu enhancement: [Eu/Fe]=+2.23. It is to date the EMP star with the largest r-process enhancement in our Galaxy. Since such stars were found in the ultra faint dwarf galaxy (UFD) Reticulum II  (see section \ref{sec:UFDgal} ) and that  J1521-3538 shows a bound, prograde orbit around the Galaxy with a high eccentricity, unlike the other known r-II stars, \citet{cain_r-process_2020} suggest that J1521-3538 has been accreted from a low mass dwarf galaxy.
The  star 2MASS J22132050-5137385  with an even higher Eu  enhancement, [Eu/Fe]=+2.45 has been found by
\citet{roederer_r-process_2024}, however it has [Fe/H]=--2.2 and does not qualify as EMP, according
to our criteria.
\\

\noindent $\bullet$ Third peak elements - Ages determination\\
In the most r-rich stars it is possible to measure the abundance of some of the third peak elements from Z=76 to Z=92 \citep[see e. g. ][]{barbuy_first_2011,frebel_discovery_2007}, and in particular the abundance of the radioactive elements Th and U which both belong to the actinide group. 
If we know the quantity of these elements formed by the r-process from theoretical predictions then from the half-lives of $\rm ^{232}Th$ and $\rm ^{238}U$  (14.05 and 4.468 Gyr, respectively) it is possible to deduce directly the time elapsed since the production of these elements i.e. the age of the star. 
This age determination is independent of Galactic chemical evolution and stellar internal theories but it is highly dependent on the r-process production ratios used \citep[see for example][]{schatz_thorium_2002,farouqi_charged-particle_2010}. Moreover 
the stars for which actinides can be measured are a handful and are called ``actinide boost stars'', with $\rm 0.14 \leq Th/Eu \leq 1.1$, the known ones are listed at the end of this section.

\begin{figure}
\centering
\includegraphics[width=8cm,clip=true]{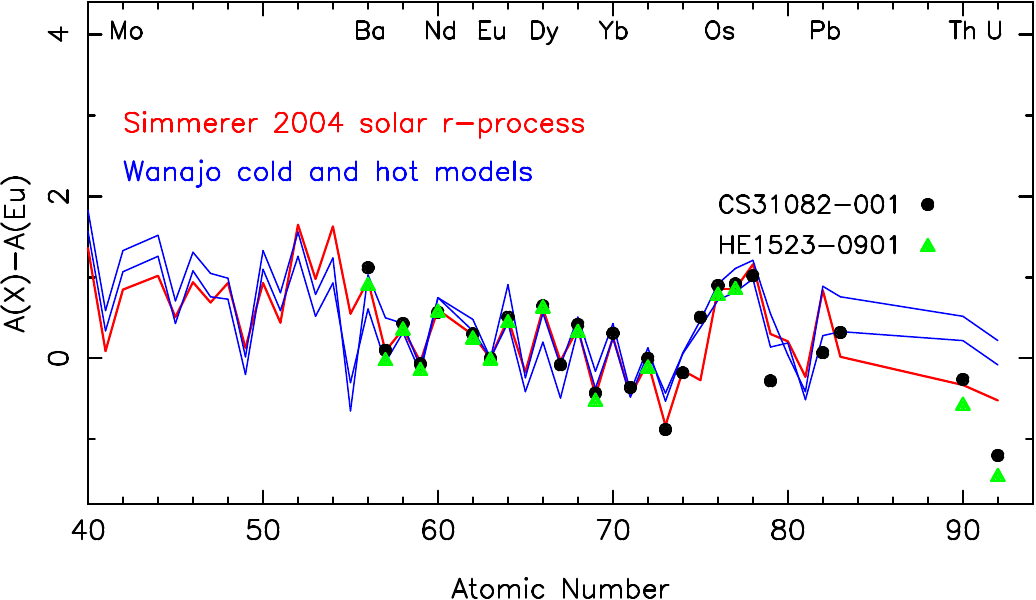}
\caption
{Abundance patterns for neutron capture elements scaled to Eu in two EMP r-II stars with $\rm [Fe/H] \approx -3$ compared to the theoretical r-process patterns computed by \citet{wanajo_cold_2007}  for a hot and a cold model (solid blue lines) and to the solar r-process pattern following 
\citet[][solid red line]{simmerer_rise_2004}. 
It is interesting to note that the hot and cold models of \citet{wanajo_cold_2007} lead to about the same pattern of the second peak of n-capture elements but for the same production of Eu (second peak element), the hot model produces more Th and U (actinides) than the cold model.
}
\label{Fig-thirdpeak}
\end{figure}

In Fig. \ref{Fig-thirdpeak}, we compare the abundance pattern of second and third peak n-capture elements in CS 31082-001 \citep{barbuy_first_2011} and HE 1523-0901 \citep{frebel_discovery_2007} to the prediction of the solar r-process pattern and to the hot and cold models of  \citet{wanajo_cold_2007}. These models correspond to different sites of formation: the cold model is associated to lower mass supernovae or binary neutron star mergers and the hot model to massive supernovae.
The figure shows that the theoretical predictions of the abundance pattern of the second peak elements (from Ba to Pb) are in rather good agreement with all the models,
but the theoretical predictions diverge, for the  actinide group elements like Th and U. In fact the  actinides  are overproduced by the main r-process, but this production is highly sensitive to the distribution of the electron fraction and the velocity of the dynamical ejecta  \citep[e.g.][]{thielemann_sources_2023}.\\
\citet{holmbeck_actinide_2019,holmbeck_actinide-rich_2019} consider that neutron stars mergers (NSM) are the main site of the r-process production, but during this merging the ejecta would be diluted with the products of, for example, an NSM accretion disk wind.\\
\citet{wanajo_actinide-boosting_2024} study the formation of the n-capture elements in Black Hole-Neutron Star mergers, and they show that the combination of dynamical and post-merger ejecta can reproduce the observations is every cases (actinide boosted or not boosted stars). 

However,  uncertainties in the production ratio U/Th are expected to largely cancel out because these elements have nearly the same atomic mass. The more precise estimation of the age of a star is expected to be deduced from the comparison of the observed and predicted ratio  U/Th.
 The origin of the boost of the actinides (compared to the lanthanides) is still debated. 
\\
From the U/Th ratio  an age of about 14 Gyr was found for CS 31082-001 \citep{cayrel_measurement_2001,hill_first_2002,barbuy_first_2011} and 13.2 Gyr for HE 1523-0901 with an error of about 3 Gyr in both cases. These values are in good agreement with the Planck predictions of the age of the Universe ($13.7\pm  0.13$ Gyr). 

Up to today U has been measured only in a handful of stars, besides the two above mentioned it has been measured in
CS\,29497-004 \citep{hill_hamburgeso_2017}, RAVE\,J2038-0023 \citep{placco_rave_2017}, 2MASS\,J09544277+5246414
\citep{holmbeck_r-process_2018} and J1521-3538 \citep{cain_r-process_2020}. In addition
there is a tentative detection in BD+17\,3248 \citep{cowan_chemical_2002}.
It is worth mentioning that \citet{shah_uranium_2023} were able to measure U in four of these stars using
two new U{\sc II} lines at 405\,nm and 409\,nm besides the usual one at 385.9\,nm.

\section{Carbon enhanced metal-poor stars, the most pristine objects?}  \label{sec:C-rich}

For a long time it is known that some stars are carbon-rich. In his attempt of a general spectral classification of the stars, Angelo Secchi in 1868 created a special ``type''  for these C-rich stars (\citealt{secchi_sugli_1868} see also \citealt{bonifacio_riscoperta_2018} for an historical account). The stars he observed with his objective-prism telescope had  a very strong C{\small 2} band, and later, \citet{mccarthy_angelo_1994} 
showed that they were in fact asymptotic giant branch (AGB) stars.
During the AGB phase of the stellar evolution, carbon and heavy elements formed by nucleosynthesis inside the star are brought to the surface, and ejected by stellar winds. 
Many of these stars were observed  by N.C. Dun\'er in 1893 at the Uppsala refractor equipped with direct-vision prism, and their spectra were described in \citet{duner_spectra_1899}.\\
The metal-poor carbon-rich stars were called by \citet{keenan_spectra_1942} ``CH stars''  since the CH band appears to be very strong in their spectrum.
Later \citet{bidelman_carbon_1956} classifies these carbon-rich stars into three different groups (Hydrogen deficient stars, CH stars, and Ba stars) the largest group being that of the CH stars.
Most of these stars are now known binaries  \citep{lucatello_binary_2006} and are also rich in neutron-capture elements like Sr, Y, Zr and Ba 
as  \citet{wallerstein_chemical_1964} pointed out sixty years ago.
An  historical perspective of the topic is also provided in the introduction by
\citet{caffau_investigation_2018}.\\
They are too old to be massive AGB stars, some of them are even dwarfs or sub-giants,
thus it is generally admitted that they are members of binary systems where the former primary star transferred matter during its AGB phase onto the atmosphere of the presently observable companion 
\citep[see e.g. ][] {masseron_holistic_2010,lugaro_s_2012,abate_carbon-enhanced_2015-1,abate_carbon-enhanced_2015,karinkuzhi_chemical_2015}.

\subsection{Definition of the Carbon Enhanced Metal Poor stars }   \label{Paragraph-Cdef}

To our knowledge the first time the term Carbon Enhanced Metal Poor, CEMP, appeared is
in the paper by \citet{lucatello_stellar_2003} and in the review by \citet{christlieb_finding_2003}.
In the literature the evaluation of the C abundance in EMP and VMP is based on the CH G-band, with very few exceptions
\citep[see e.g][]{takeda_carbon_2013,amarsi_carbon_2019}. Again with few exceptions the synthesis of the G-band is done using 1D model atmospheres
assuming LTE.
Large surveys like the HK survey \citep{beers_search_1985,beers_search_1992} and the Hamburg-ESO survey \citep[HES/HERES,][]{christlieb_hamburgeso_2004}
have revealed the existence of an important population of carbon-rich stars \citep{beers_discovery_2005,lucey_carbon-enhanced_2023},  moreover it could be shown that at low metallicity the number of CEMP stars strongly increases \citep {lucatello_frequency_2006}.
 
\begin{figure}
\centering
\includegraphics[width=8cm,clip=true]{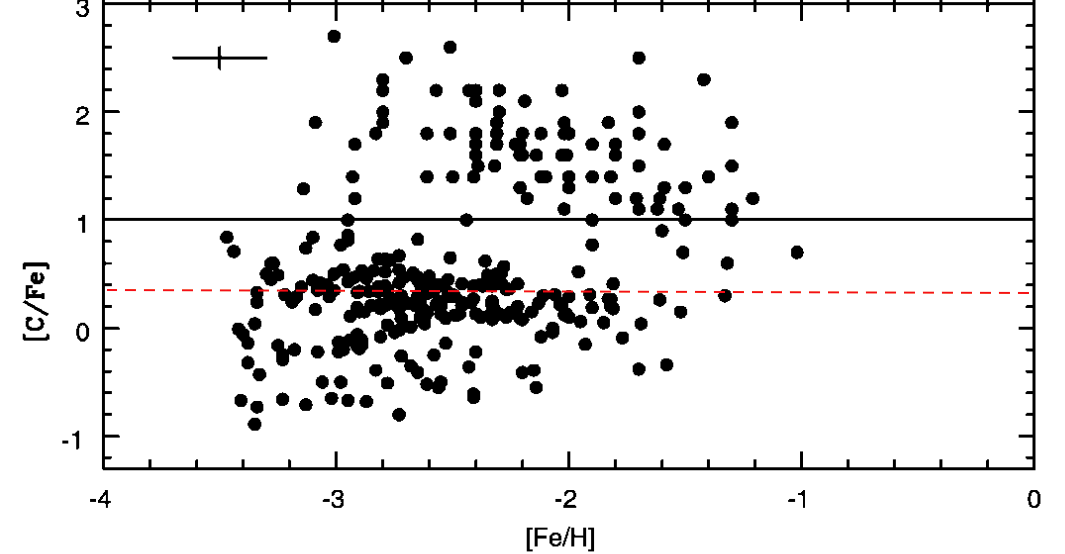}
\caption
{\footnotesize{[C/Fe] vs. [Fe/H]  for the  HERES sample following  \citet{lucatello_frequency_2006}. The black line indicates the original cutoff for considering a star to be CEMP star, $\rm[C/Fe] \ge +1.0$}. The red dashed line represents the mean value of [C/Fe] in EMP stars \citet{bonifacio_first_2009}. (The C-poor stars with $\rm [C/Fe] < 0$ are mainly mixed giants where C has been transformed into N).Figure reproduced from
 \citet{lucatello_frequency_2006} with permission.}
\label{Fig-CFeH}
\end{figure}

In Fig. \ref{Fig-CFeH} the [C/Fe] ratio of the stars of the HES survey is plotted vs. [Fe/H] following  \citet {lucatello_frequency_2006}. At low metallicity  ($ -3.5 <\rm[Fe/H]< -1.5$), the carbon abundance of the majority of the stars is close to [C/Fe]=+0.4,   \citep[see also][]{bonifacio_first_2009},
 but more than 20\% of the stars exhibit $\rm[C/Fe] \ge +1.0$. \\
 Comparing the results from low resolution surveys (HES survey; Sloan Digital Sky Survey, SDSS) and high resolution surveys (e.g. follow-up of SkyMapper \citealt{yong_high-resolution_2021} or
 Pristine \citealt{starkenburg_pristine_2017,aguado_pristine_2019}),
 \citet{arentsen_inconsistency_2022} found large differences (up to +0.4\,dex) in the determination of [C/Fe] depending on the adopted method for the abundances determination.\\
 The  threshold adopted for the definition of the CEMP stars is sometimes [C/Fe]=+1, as
 suggested by \citet{beers_discovery_2005}, but sometimes only [C/Fe]=+0.7 \citep[e.g.][]{aoki_carbonenhanced_2007}.  
The latter value has been justified by \citet{yoon_observational_2016} by 
noticing that the distribution of [C/Fe] is bimodal (see their figure 1) and +0.7 nicely
separates the two peaks. It should however be kept in mind that this bimodality
only appears on applying the corrections of \citet{placco_carbon-enhanced_2014} to the carbon
abundances, the uncorrected data show a single peak and an extended tail. 
The corrections of \citet{placco_carbon-enhanced_2014} take into account the evolutionary status
of the star through its surface gravity. 
 The value +0.7 is not recommended \citep[see also][]{bonifacio_topos_2015}  since it is too close to the mean value of the normal metal-poor stars compared to the uncertainty on the determination of [C/Fe].\\ 
 To be sure that a star is C-rich, based on 1D spectral synthesis of the G-band, we selected stars with $\rm [C/Fe]>+1.0$, and on the contrary to select stars with no carbon enhancement we adopted the threshold $\rm [C/Fe] < +0.7$.

\subsection{Frequency of the CEMP stars}  \label{defCEMP}

It is clear that the frequency of CEMP stars in a given interval of metallicity depends on the threshold ([C/Fe]=0.7 or [C/Fe]=1) adopted for defining a CEMP star. But the general trend should be about the same.\\
\citet{lee_carbon-enhanced_2014} estimate the frequency of the CEMP stars as a function of the metallicity and of the position of the star in the Galaxy, using low-resolution (R= 2000) stellar spectra from the Sloan Digital Sky Survey (SDSS) and its Galactic sub-survey, the Sloan Extension for Galactic Understanding and Exploration (SEGUE). A little later \citet{placco_carbon-enhanced_2014} from [C/Fe] ratios collected in the literature and based on high resolution spectra, also studied the frequency of the CEMP stars as a function of the metallicity. They corrected the [C/Fe] values measured in giants for the mixing effect. Both studies confirm an increase in CEMP proportion when the metallicity decreases. 
Moreover \citet {beers_bright_2017} reanalysed the 1777 spectra of the of the HES survey, with new estimates of the atmospheric parameters. They confirm the increase of the CEMP stars with the decrease of the metallicity: we computed that in their sample, in the interval $\rm -3.0 < [Fe/H]< 2.5$, only 15\% (12 stars over 80) have a [C/Fe] ratio above +1, but at lower metallicity, in the interval  $\rm -3.5 < [Fe/H] < -3.0$ they are 26\% (six  stars over 23).\\
\citet {beers_bright_2017}  also observe an increase in the fraction of CEMP stars with distance from the Galactic plane and moreover a large number of the observed CEMP stars have kinematics consistent with the metal-weak thick-disk population. Moreover \citet{lee_chemical_2017} show that  the carbon-enhanced metal-poor (CEMP)
stars in the outer-halo region exhibit a higher frequency of CEMP-no stars  than of CEMP-s stars, whereas the stars in the inner-halo region exhibit a higher frequency of CEMP-s stars (see Sec.\,\ref{sec-classCEMP} for the definition of the different classes of CEMP stars).
The fraction of CEMP stars as a function of metallicity is discussed in detail in \citet{arentsen_inconsistency_2022}.

In the most metal-poor stars known today, (below $\rm[Fe/H]\le -4.5$), almost all the stars have a $\rm[C/Fe]$ ratio above this limit. The only known exception is: SDSS\,J102915.14+172927.9  \citep{caffau_primordial_2012} which, despite its very low Fe abundance ([Fe/H]=--4.7),  seems to also have a low C abundance (see Fig. 2).  
Pristine\,J221.8781+09.7844
\citep{starkenburg_pristine_2018,lardo_pristine_2021} with [Fe/H]=--4.8 and a CH band not visible on the spectrum (but corresponding to $\rm[C/Fe] < +2.3$)  is also a good candidate. 

\subsection{3D NLTE corrections}

 It is well known that molecular bands, including the G-band are strongly affected by granulation
effects, that are stronger in extremely metal-poor stars \citep{collet_three-dimensional_2007,gallagher_-depth_2017}
and can be as large as --0.9\,dex for the UMP giant  HE 0107-5240 \citep{collet_chemical_2006}.
The NLTE effects on molecular bands have been less explored, although according to \citet{popa_non-local_2023} they are
non-negligible and can attain +0.2\,dex for a giant star at metallicity --4.0.\\
This large correction of  [C/Fe] in metal-poor stars should also affect the threshold of the definition of CEMP stars.
At similar stellar parameters and chemical composition, stars with a large difference in the strength of the G-band 
(therefore difference in [C/Fe]) exist, and some differences have to persist also after applying a 3D, or 3D-NLTE, correction.
In our definition of the CEMP stars (Sec.\,\ref{Paragraph-Cdef}) we adopted, in fact, a threshold 0.6\,dex 
above the mean value of the [C/Fe] ratio ($\rm \overline{[C/Fe]} \approx 0.4$) .\\

As an example of the limitations of our current understanding of
the line formation in the atmospheres of metal-poor stars we take
the three stars G64-12, G64-37 and G275-4
that have been found to be C-rich with $\rm [Fe/H] \approx -3.2$ and $\rm [C/Fe] \approx 1.1$ by \citet{placco_g64-12_2016}  and \citet{jacobson_cd_2015}, all based on 1D spectral synthesis of the G-band.
However, for the three stars the 1D LTE analysis of the atomic carbon lines by \citet{amarsi_carbon_2019} 
provides [C/Fe] $\approx  0.2, 0.3, 0.4$ respectively; the 3D NLTE analysis of the three stars
gives $[\rm C/Fe] \approx 0.0$. That two different abundance indicators, provide abundances that differ by 
more than 0.6\,dex is clearly unsatisfactory.
Unfortunately a 3D NLTE computation of the G-band is not yet available, to see if such
approach would bring the different indicators in agreement.
 
The modelling of the molecular bands needs to be improved.
To start with it is well known that 3D and NLTE have generally  opposite directions and need to be
treated together. In the second place the molecule formation equilibrium are all interconnected,
mainly through CO formation, which requires a simultaneous modelling of all the most relevant
molecules. In the third place \citet{gallagher_-depth_2017} have shown that while the structure of
1D model atmospheres is, by and large, the same for a solar-scaled  chemical composition and
for a CEMP chemical composition, the structure of a 3D model can be significantly different, especially for high
values of the C/O ratio. This requires that the 3D abundance determination needs to be done iteratively, 
computing at each step a model with the current abundances of C,N and O.
In this situation we decided to consider, in the following, for the star classification the abundances
derived from 1D LTE modelling of the CH band, although we are aware that in the future a more physically motivated
modelling may change the classification of some stars that are currently borderline CEMP.

\subsection{The different classes of CEMP stars as a function of [Fe/H]}  \label{sec-classCEMP}
Different classes of CEMP stars can be defined, depending on the enrichment in neutron-capture elements. Here we consider that a CEMP star is rich in neutron capture element if $\rm[Ba/Fe]>1.0$ or $\rm[Eu/Fe]>1.0$ \citep{beers_discovery_2005}.
 
When a CEMP star does not present any significant enrichment in neutron-capture elements, it is called CEMP-no.
But CEMP stars enriched in neutron-capture elements only formed by the s-process are called CEMP-s,  and 
 the rare CEMP stars enriched in s- but also in r-process elements are called CEMP-r/s  
\footnote{Many papers use this notation, although CEMP-r+s has been used by \citet{aoki_carbonenhanced_2007} and \citet{karinkuzhi_low-mass_2021} and even CEMP-sr by \citet{lugaro_s_2012} }. Some CEMP stars, whose prototype is CS\,22892-52, show apparently a pure r-process pattern and are 
called CEMP-r \citep{aoki_carbonenhanced_2007,shank_r-process_2023}.
Moreover it has been shown that in stars previously classified CEMP-r/s  
the overabundance of the neutron-capture elements could be 
explained by a production of these elements by the i-process and these stars were called CEMP-i \citep{goswami_spectroscopic_2021,goswami_peculiar_2022}. 
For a detailed  classification of the CEMP-s and -r/s stars based on the relative abundances of the neutron-capture elements, see e.g. \citet{abate_how_2016,hansen_abundances_2019,karinkuzhi_low-mass_2021,goswami_spectroscopic_2021}.

\begin{figure}
\centering
\includegraphics[width=10cm,clip=true]{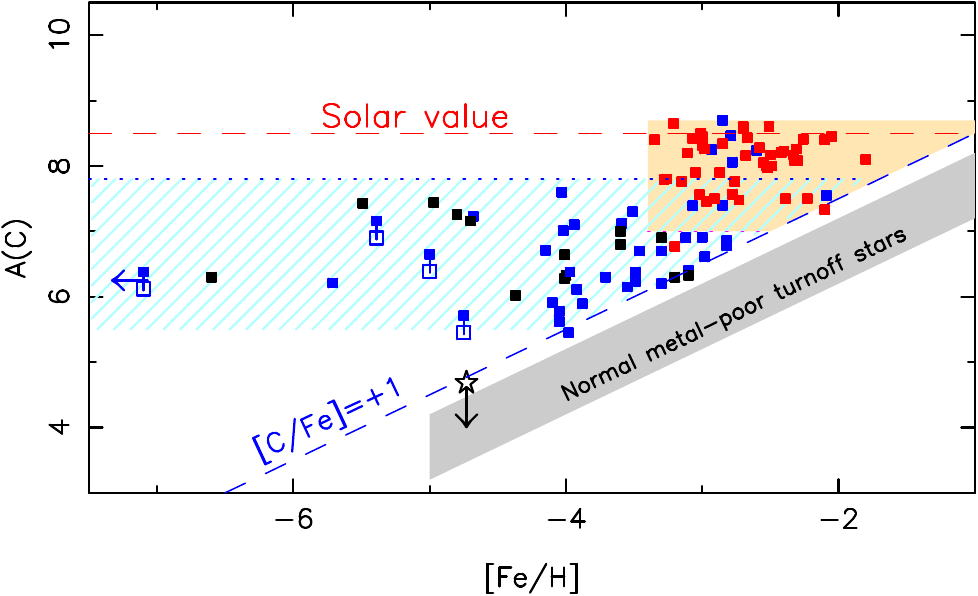}
\caption{\footnotesize   Carbon abundance A(C) of CEMP stars as a function of [Fe/H] for stars with $\rm[Fe/H] < -2.0$ \citep{bonifacio_topos_2015,spite_cemp-no_2018}. The CEMP-no stars are represented by blue squares (giants are represented by two symbols: an open blue square for the measured value of the C abundance, and a filled blue square for the empirically corrected (for the first dredge-up) A(C) value). The CEMP-s, -r/s, or -i stars are indistinctly represented by red squares. When the Ba abundance is unknown the star is marked with a black square. 
The upper limit marked as a black star symbol, represents the upper limit of A(C) in  SDSS\,J102915.14+172927.9 the most metal-poor star with a normal carbon abundance  \citep{caffau_primordial_2012}.
The red dashed line represents the C abundance in the Sun. The normal stars (not enriched in C), are located in the grey zone. All the stars above the blue dashed line are  CEMP with $\rm[C/Fe] \ge +1$. 
}
\label{Fig-ACFe}
\end{figure}

In  Fig. \ref{Fig-ACFe} the carbon abundance A(C) is plotted vs. [Fe/H] for all the known CEMP stars with $\rm[C/Fe] \ge +1.0$ and $\rm [Fe/H] < -2.0$ (updated data from  \citealt{bonifacio_topos_2015,spite_cemp-no_2018}).
It can be seen that:\\
\noindent $\bullet$ No CEMP star has a A(C) value significantly higher than the solar value.\\
$\bullet$ When  $\rm [Fe/H] > -3.4$: \\  
${}$\hspace{1cm}if $\rm A(C) \lesssim 7.0$  the stars are CEMP-no;\\ 
${}$\hspace{1cm}if $\rm A(C) \gtrsim  7.0$  there is a mix of CEMP-no and CEMP-s or -r/s stars.\\
$\bullet$ When  $\rm [Fe/H] < -3.4$ (the most metal-poor stars):\\
${}$\hspace{1cm} presently all the CEMP stars with known Ba abundance are confirmed CEMP-no\\
${}$\hspace{1cm} 
and $\rm A(C) < 7.8$.\\

In the orange zone of the figure (the "upper A(C) band" with $\rm[Fe/H] > -3.4$ and $\rm A(C) > 7.0$, the large majority of the CEMP stars are Ba-rich, they are mostly CEMP-s or -r/s, but a few stars are CEMP-no.  On the contrary, in the part of the figure hatched in cyan ("lower A(C) band"), A(C) is between 5.5 and 7.8  and as soon as  $\rm[Fe/H] < -3.4$ all the CEMP stars are CEMP-no. \\
However, there is an exception in Fig. \ref{Fig-ACFe}, following \citet{matsuno_high-resolution_2017} the star SDSS\,J1036+1212 with [Fe/H] = --3.6 and [C/Fe] = +1.2 and thus A(C)= 6.1, has a very high Ba abundance ([Ba Fe] = +1.68). It would be a CEMP-r/s star inside the low A(C) band. But \citet{behara_three_2010} using a very similar method to determine the main atmospheric parameters of this star had found a temperature 500 K higher and thus a higher metallicity and carbon abundance ([Fe/H]=--3.2, A(C)=6.8) and with these parameters SDSS J1036 +1212 has a quasi-normal position in Fig. \ref{Fig-ACFe}, at the lower limit of the upper A(C) band. 

With the advent of very large surveys of metal-poor stars, it is possible that the limits of the different sub-classes of CEMP stars in Fig. \ref{Fig-ACFe} change, but the probability of finding a CEMP-s star with $\rm [Fe/H]< -3.4 ~and~ A(C) < 7$ is, a priori, very low.

In their study of CEMP stars from the literature,
\citet{yoon_observational_2016} confirmed the general trends presented in \citet{bonifacio_topos_2015}.
Up to now no star has been found with  a carbon abundance lower than A(C) = 5.5 if $\rm [Fe/H] < -5.0$  but this can be due to the difficulty of measuring the CH band at this very low metallicity.\\ 

\subsubsection{CEMP-s and CEMP-r/s stars}
Many CEMP stars are enriched in neutron-capture elements. These elements are mainly formed by the slow or the rapid processes ``s-process'', occurring at neutron densities $\lesssim 10^{11}$ \,cm$^{-3}$ or ``r-process'', occurring at neutron densities $\gtrsim 10^{24}$\,cm$^{-3}$ \citep[see e.g.][]{sneden_neutron-capture_2008,frebel_nuclei_2018,cowan_origin_2021}.    But if the main s-process 
is expected to happen inside the AGB stars, it is not the case for the r-process that requires a very high neutron density.

Most of the CEMP stars enriched in neutron-capture elements, are in fact only enriched in s-process elements (CEMP-s), and a pollution by the ejecta of a past thermally-pulsing AGB companion (now a white dwarf) can explain this enrichment \citep[see e.g.][]{campbell_evolution_2008,abate_carbon-enhanced_2015,placco_hubble_2015}. \\
From a comparison of CEMP-s stars to AGB model yields \citep[e.g.][]{cristallo_evolution_2011}, \citet{hansen_abundances_2016} and \citet{goswami_spectroscopic_2023} suggest that the AGB stars progenitors of the CEMP-s stars were primarily of the lower mass variety \citep[in agreement with e.g. ][]{kennedy_ofe_2011,bisterzo_s-process_2012}.

Some of the CEMP stars, are enriched in s-elements and also in elements generally formed by the r-process, thus, when $\rm [Ba/Eu]>0.5 $ they are called by \citet{beers_discovery_2005} CEMP-r/s. 
Several stars of this type are now known \citep[e.g.][]{gull_r_2018,goswami_peculiar_2022}.
\citet{hollek_chemical_2015}, measuring also the [Y/Ba] ratio in the CEMP-s and the CEMP-r/s stars, show that there is a continuum between the CEMP-s and the CEMP-r/s stars, rather than a distinct cut off separating the two groups of objects, and they propose a new more progressive classification based on the [Y/Ba] ratio.
Several papers tried to find clear criteria to disentangle the CEMP-s and the CEMP-r/s stars \citep[see][ and references therein]{goswami_spectroscopic_2021}, and it turns out not to be easy because of the observed continuity between these classes of CEMP  stars. \\

The AGB nucleosynthesis fails to reproduce the heavy elements abundances in the CEMP-r/s stars \citep[e.g.][]{abate_carbon-enhanced_2015-1}. 
Thus it was first supposed that these  CEMP-r/s stars were formed from a gas already enriched in r-process elements, and were later polluted by the ejecta of a past AGB companion.

However, it seems that a process intermediate between the s- and the r-process (``i-process'', with densities in the range
$ 10^{14}\rm\,cm^{-3}\le \rm n_{density} \le 10^{16}\, cm^{-3}$), could  happen in low-mass low-metallicity AGB stars  \citep{hampel_intermediate_2016,karinkuzhi_low-mass_2021,goswami_peculiar_2022}. 
This process could be responsible for the formation of elements generally attributed to the r-process, like Eu.
In low-mass, low-metallicity AGB stars, 
indeed, protons could be brought by mixing to the hot C-rich layer and this ingestion of protons would result in the reaction $\rm ^{13}C(\alpha,n)^{16}O$, 
that would produce high neutron densities up to  $\rm 10^{14} ~to~ 10^{15}$\,    
$\rm cm^{-3}$ \citep[see e.g.][]{caffau_cemp_2019}. 

\subsubsection{CEMP-no stars} 

This class of CEMP stars concerns stars not enriched in neutron-capture elements. 
As we have seen, they dominate at low metallicity. 
Below $\rm [Fe/H]=-3.4$ (Fig. \ref{Fig-ACFe}), 
all CEMP stars belong to this class. Below $\rm[Fe/H]\approx -5$ it 
seems that all the stars are enriched in C and all the stars where the abundance of 
Ba could be measured (blue squares in Fig.\,\ref{Fig-ACFe}) are CEMP-no. 
As a consequence the CEMP-no stars are believed to be the direct descendants of  
the first-generation stars and provide a unique opportunity to probe the early Galactic nucleosynthesis \citep{hartwig_descendants_2018}. 

As we saw above, the CEMP-s stars are almost always in binary systems but the CEMP-no stars more often appear to be single stars \citep{starkenburg_binarity_2014,hansen_abundances_2016}. 
The binarity frequency among CEMP-no stars should be compared to
the binarity frequency expected at a given metallicity of all stars.
There are theoretical expectations
that Pop III stars have a high binary fraction, e.g. \citet{stacy_constraining_2013}
find 35\%. 
The dependence of the binary fraction is always difficult to establish
if one wants to include all possible periods, since very long period
binaries are  difficult to detect and characterise observationally.
\citet{moe_close_2019} restricted their study to ``close'' binaries, 
defined as those with periods less than $10^4$ days (28 years) and semi-major
axis less than 10 $au$, among stars of type FGK. These limits include probably also
most of the other existing surveys on binary fraction, that are 
very well reviewed in section two of the above-mentioned paper.
According to their analysis \citet{moe_close_2019} conclude that
the fraction of binary systems increases with decreasing metallicity,
being about 55\% at [Fe/H]=--3.0 and decreasing to
20\% at solar metallicity.
While this result supports the theoretical expectations one should be
aware that it relies on the correction for completeness of the various
surveys analysed by \citet{moe_close_2019}. The uncorrected fractions of
the different surveys show no trend for metallicity.
Among the stars  with [Fe/H] below --4.0, little is known on the binary fraction, 
since in any case the number involved is small.

\citet{arentsen_binarity_2019} combining the radial velocities variation in the samples of \citet{starkenburg_binarity_2014}, \citet{hansen_abundances_2016} and in their own sample of CEMP-no stars, found that 32\% of the CEMP-no stars (11 out of 34) are binaries. 
\citet{arentsen_binarity_2019} conclude that this is consistent with the general
binary fraction at this metallicity, and in fact this is even lower than the fraction
predicted by \citet{moe_close_2019} at metallicity --3.0.\\
\citet{roederer_neutron-capture_2014} measured the radial velocity of a sample of 16 CEMP-no stars, 
and could not detect variations however  they point out that some of them were observed only once.\\
\citet{caffau_topos_2016} detected that SDSS J092912.32+023817.0 ([Fe/H]=--4.97, [C/Fe]=+3.91)
is a double-lined spectroscopic binary, although they were unable to determine the orbital parameters.\\
\citet{schlaufman_ultra_2018} reported that the single-lined binary 2MASS J18082002-5104378, a CEMP-no star with $\rm[Fe/H]\approx -4$\,,
has a circular orbit with a short orbital period P=34.757d, and they show that the mass of the secondary star is about $\rm 0.14 M_\odot$. From these observations they deduce that in the early Galaxy, low mass stars were formed and they survived to the present day.\\
\citet{aguado_espresso_2022} monitored the radial velocity of eight stars with $\rm [Fe/H]\le -4.5$
and determine a tentative orbit for HE0107-5240 with a very long period, of the order of 36 years,
confirming the binary nature of this star, that was already highlighted by \citet{arentsen_binarity_2019}
and \citet{bonifacio_espresso_2020}. They highlighted also radial velocity variations in SMSS 1605-1433,
although they could not convincingly ascribe them to a binary companion.

Putting all the sparse information available in the literature, we conclude that 
assuming that the fraction of binary systems among CEMP-no stars is the same as that of
the general population in the same metallicity range is a reasonable working hypothesis.
The corollary is that the CEMP-no nature should not be connected to the presence
or absence of a companion.
The origin of the CEMP-no stars is still under debate \citep{chiaki_seeding_2020}. It is generally assumed that their carbon abundance is intrinsic and that they were formed from a cloud of carbon-rich gas, in spite of the fact that
some pollution by an AGB star cannot be excluded at least for the binary component of the CEMP-no stars.

\subsubsection{Abundance pattern of the CEMP stars}
The first CEMP star studied in detail was probably CS\,22949-037. This CEMP giant with [Fe/H] $\approx -4.0$, was discovered in the HK objective prism survey of \citet{beers_search_1992} and \citet{beers_estimation_1999}. The CNO abundances were first estimated by \citet{mcwilliam_spectroscopic_1995} and \citet{norris_extremely_2002}, then \citet{depagne_first_2002} measured the abundance of 21 elements from C to Ba in this star  and they showed that CS\,22949-037 is a CEMP-no. 
But, in a giant with $\rm log g \approx 1.5$ the CNO abundance can be affected by mixing inside the star that induces a decrease of C and O and an increase of N. To study the original abundance pattern of the CEMP stars it is more secure to analyse CEMP dwarf or subgiant stars.

\begin{figure}
\centering
\includegraphics[width=11.5cm,clip=true]{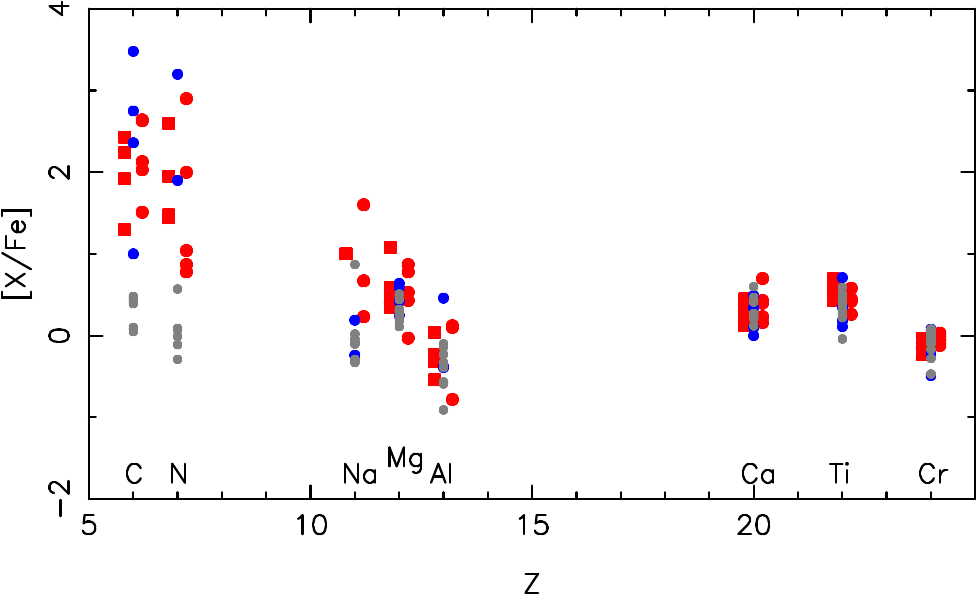}
\caption{\footnotesize [X/Fe] vs. the atomic number Z of  elements between C to Cr for CEMP-no (blue dots) and CEMP-s or CEMP-r/s (red dots and red squares) following \citet{hansen_elemental_2015}. The [X/Fe] ratios of the CEMP-s and CEMP-r/s stars are slightly shifted to the right and to the left for a better visibility. Reference normal stars are represented by grey dots.
}
\label{Fig-ACFe-hansen}
\end{figure}

\citet{hansen_elemental_2015} studied, in the same interval of metallicity ($\rm-4.7 <[Fe/H]< -2$), a sample of 8 normal metal-poor stars and 39 CEMP stars belonging to the different classes of CEMP in the hope of finding differences in abundance ratios in the different classes of CEMP stars to constrain the possible astrophysical sites of element production. They found that in all these stars C and N and also Na are strongly enhanced, often  Mg and Al are also enhanced. They did not find clear differences between the behaviour of the elements ratios in the different classes of CEMP stars. This is illustrated in Fig.\,\ref{Fig-ACFe-hansen} where [X/Fe] is plotted vs. the atomic number of the element for stars enriched in neutron-capture elements (red symbols CEMP-s or CEMP-r/s) or without enrichment (blue symbols CEMP-no). The reference stars are plotted in grey, since C and N abundances could be measured in very few normal stars by \citet{hansen_elemental_2015}, we added the normal stars analysed in \citet{spite_detailed_2022} in about the same interval of metallicity.

Taking data from the literature, \citet{frebel_near-field_2015} also showed that in the CEMP-no stars not only C but also O and N are enhanced and frequently but not always Na, Mg and Al,  but the elements like Si, Ca and heavier elements have a normal abundance compared to the normal metal-poor stars as it can be seen in Fig. \ref{Fig-ACFe-hansen}.
They suggested that the interstellar medium from which the CEMP-no stars were formed has been enriched by material partially processed by nucleosynthetic H burning into regions experiencing He burning in a progenitor star.

\citet{roederer_neutron-capture_2014} carefully studied the abundance pattern of the neutron-capture elements in 11 CEMP-no stars and 5 nitrogen-rich   stars (NEMP-no). They show that  the three groups of stars
CEMP-no,  NEMP-no, and EMP that show no enhancement in C or N,  do not show different distributions of [Sr/Fe] or [Ba/Fe]. The pattern of the heavy elements when it can be defined, is generally compatible with an r-process or the sum of an r- and a weak-r process, however CS 22878-101 with its very high [Ba/Eu] ratio shows evidence for at least a partial s-process origin.

\subsubsection{Origin of the CEMP-no stars}
Since among the CEMP-no stars there are the most iron-poor stars still shining in our Galaxy, then, if we assume that their abundance pattern is intrinsic, this surface abundance pattern can be used to constrain the very first generation of Population III massive stars, that were able to enrich the interstellar medium very early in the universe.

\begin{figure}
\centering
\includegraphics[width=11.0cm,clip=true]{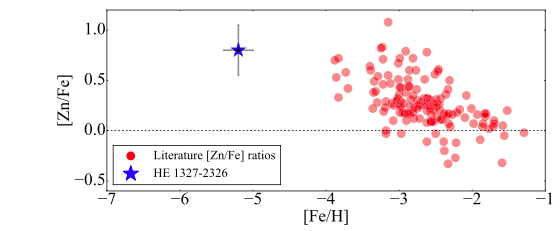}
\caption{\footnotesize [Zn/Fe] vs. [Fe/H] for EMP normal stars  from the literature (red dots), and for HE\,1327-2326 (blue star).
Figure reproduced from \citet{ezzeddine_evidence_2019}.
}
\label{Fig:HE1327-2326Zn}
\end{figure}

\begin{figure}
\centering
\includegraphics[width=10.0cm,clip=true]{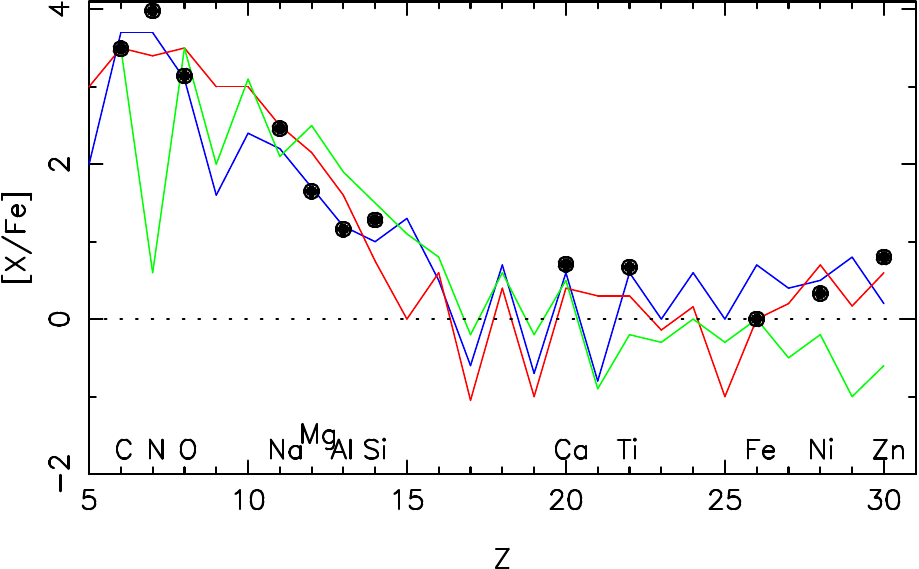}
\caption{\footnotesize [X/Fe] vs. the atomic number Z in HE\,1327-2326 (black dots), compared to theoretical patterns from \citet{ezzeddine_evidence_2019} (red line, for aspherical SNe), and \citet{jeena_rapidly_2023} (blue line, for rapidly rotating massive stars). The predictions of \citet{vanni_characterizing_2023} (medium polluted by low energy Pop\,III SNe) for stars with $\rm[Fe/H]<-4$ and $\rm [C/Fe] > 2.5 $ have been added  in green. (Note that in this last simulation, the uncertainty in the nitrogen production is extremely large, about 2\,dex). 
}
\label{Fig:compar_theo}
\end{figure}

Different models of progenitors were proposed to explain the high C abundance in CEMP-no stars.

\noindent $\bullet$ First of all, models using the "mixing and fallback" mechanism were suggested. In these models large fraction of the inner ejecta fall back onto the remnant neutron star or black hole. As a consequence the SNe ejects mostly the lighter elements of the external layers \citep{umeda_first-generation_2003,umeda_variations_2005,heger_nucleosynthesis_2010,tominaga_abundance_2014}.\\
 \citet{meynet_are_2010}  and \citet{chiappini_first_2013} proposed that mixing in rapidly rotating massive stars leads to an enhancement of CNO later ejected during the explosion.
 
 Unfortunately in this  kind of models the level of dilution of the SN ejecta, required to explain the high level of the carbon abundance A(C) in CEMP-no stars (up to A(C)=7.8), does not seem to be compatible with the dilution found in simulations of SN ejecta from PopIII stars \citep[see e.g. ][]{magg_minimum_2020} and, as a consequence, other explanations were also explored.\\
\noindent $\bullet$ \citet{placco_observational_2016}
analysed the CEMP-no star HE 0020-1741 ([Fe/H] = --4.1, [C/Fe]=+1.7) and compared the
pattern of this and other UMP stars with a large grid of SN models, using the {\tt STARFIT} code
and concluded that the pattern cannot be easily reproduced by a single polluting SN, but
at least two polluting SNe. This is in line with the findings
of \citet{hartwig_machine_2023}.
\\
\noindent $\bullet$ \citet{ezzeddine_evidence_2019}  analysed UV spectra of HE\,1327-2326 and they measured the abundance of several key elements in this extremely iron-poor CEMP-no star with [Fe/H]=--5.6 and [C/Fe]=+3.5. In particular they found that in this star $\rm [Zn/Fe] \approx +0.8$ (Fig. \ref{Fig:HE1327-2326Zn}).\\ 
When compared to normal EMP stars this very high value of [Zn/Fe] is almost  aligned with the values measured for EMP stars \citep[see ][]{cayrel_first_2004,bonifacio_first_2009} 
but it seems that at such a  low metallicity, the spherical models of SNe are not able to reproduce the elements pattern of HE 1327-2326 and in particular the high [Zn/Fe] ratio. 
The best fit is obtained for an aspherical SNe explosion model with bipolar outflows of a first star progenitor with $\rm 25 M\odot$ and an explosion energy of $10^{51}$~erg (Fig. \ref{Fig:compar_theo}, red line).

\noindent $\bullet$ \citet{jeena_rapidly_2023} proposed that rapidly rotating massive Pop III stars with masses between 20 and 35 $M_{\odot}$ and initial equatorial velocities between 40 and 70\,\% of the critical value, could also explain the high C abundance in CEMP-no stars. Massive zero-metals Pop III stars  would undergo a very efficient mixing, become quasi-chemically homogeneous, and eject very large amount of C, N, O in the wind which pollutes the interstellar medium. \citet{jeena_rapidly_2023}  were able to reasonably reproduce the abundance pattern of the elements in 14 CEMP-no stars with different values of A(C) (Fig. \ref{Fig:compar_theo}, blue line)  .

\noindent $\bullet$ \citet{vanni_characterizing_2023} suggest that low-energy Pop III SNe (with $E_{SN} < 2 \times 10^{51}$ erg) would be entirely responsible for the pollution of the cloud forming the CEMP halo stars  with $\rm[C/Fe]> +2.5$, and that CEMP stars with $\rm[C/Fe] < +2.5$ would be born in environments polluted by both Pop III and Pop II stars.
HE\,1327-2326 with its  very high [C/Fe] belongs to the first group, and it seems (Fig. \ref{Fig:compar_theo}, green line) that 
also this process produces too little zinc to explain the complete abundance pattern.\\

In fact in these computations, it is supposed that the gas from which the CEMP stars have been formed has been enriched by only one type of supernova, and this is unlikely to be the case.
\citet{bonifacio_astrophysics_2003} had in fact proposed a two SN model to explain the abundances
of HE 0107-5240, but that model was ruled out by the subsequent measurement of the oxygen abundance in the star 
by \citet{bessell_oxygen_2004}.
Accounting for metal contributions from several types of SNe would require more free parameters, thus a first approach is to recognise and select only mono-enriched second-generation stars \citep{hartwig_descendants_2018}. \\
In a new approach \citet{hartwig_machine_2023,hartwig_span_2024} use a semi analytic model A-SLOTH \citep{hartwig_public_2022} 
based on nine independent observables (like the stellar mass of the Galaxy, the relative fraction of EMP stars and the SFR at high redshift) can compute up to 11 observables.
In \citet{hartwig_span_2024} they conclude that IMF in the very early times should be around  $\propto{M^{-1.77}}$,
with masses between 13.6 and 197 $M_\odot$.

\subsubsection{Lithium abundance in the CEMP stars and comparison with EMP stars} \label{sec:LiCEMP}

\begin{figure}
\centering
\includegraphics[width=10.0cm,clip=true]{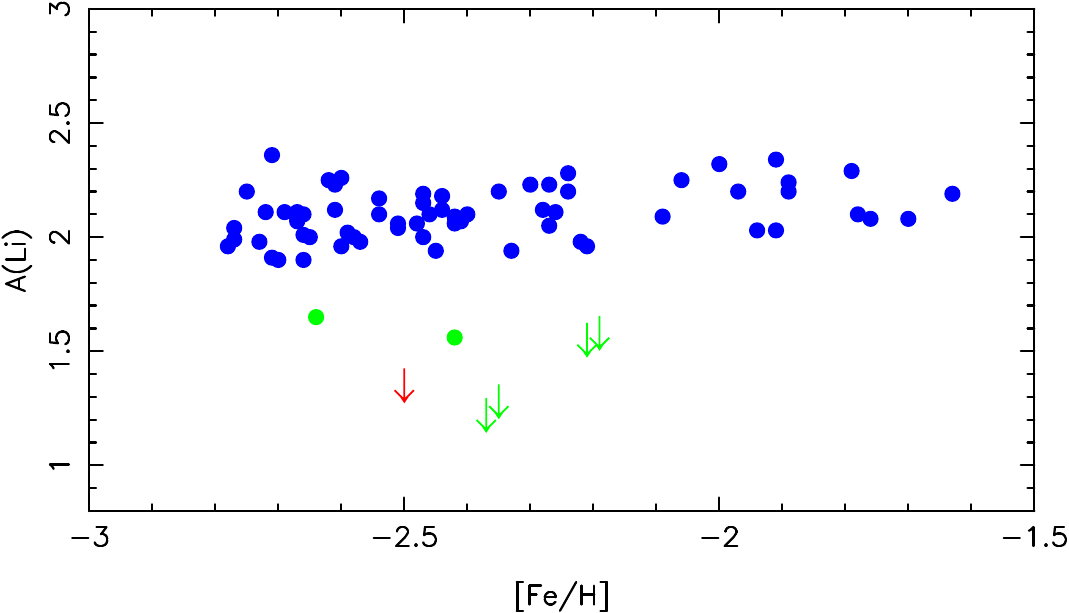}
\caption{\footnotesize A(Li) vs. [Fe/H] for the  dwarfs or turnoff stars with $\rm T_{eff} \geq 5900 K$ and $\rm [Fe/H] > -2.8$  following \citet{roederer_search_2014}.  The `normal' metal-poor stars are marked with blue dots and the CEMP stars with green dots or green arrows. The red arrow represents G122-069 a blue straggler following Matas-Pinto 2021 (PhD-thesis).
}
\label{Fig:li-roederer}
\end{figure}

\begin{figure}
\centering
\includegraphics[width=8cm,clip=true]{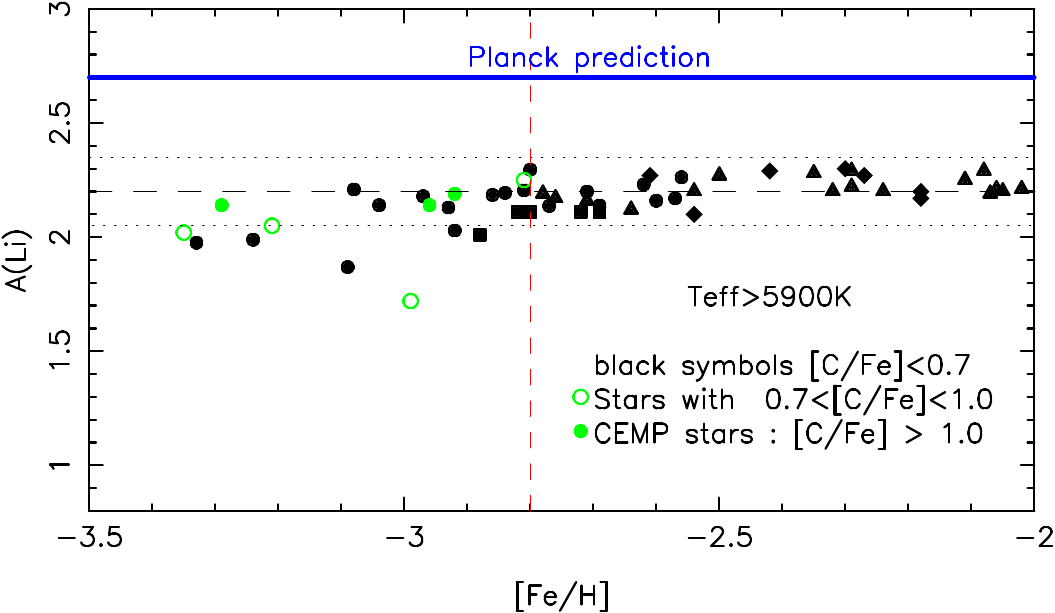}
\caption{A(Li) versus [Fe/H]  in warm dwarfs or turnoff stars, where are included CEMP stars with [C/Fe]$>1$ and stars in the "grey zone" (which are often considered as CEMP stars) with $\rm 0.7 < [C/Fe] < +1$.}
\label{Fig:LiEMPCEMPFe}
\end{figure}

When the lithium abundance is compared in unevolved `normal' and CEMP stars, the CEMP stars display a larger fraction of stars with Li abundances below the `Spite plateau' although some of them do lie on this plateau \citep{sivarani_first_2006,behara_three_2010,hansen_elemental_2015,matsuno_lithium_2017}. However, in the latter papers the iron abundance of the majority of the stars studied was less than $\rm [Fe/H]=-2.8$ and thus the deficiency of the lithium abundance could also be due to the large iron deficiency. However, this deficiency of the Li abundance is also found in the CEMP dwarfs and turnoff stars ($\rm T_{eff} \geq 5900 K$) studied by e.g.  \citet{roederer_search_2014} in the interval $\rm-2.5<[Fe/H]<-1.5$ (Fig. \ref{Fig:li-roederer}). 
Note that all the stars in this figure where  \citet{roederer_search_2014} could only measure an upper limit of the Li abundance are CEMP-s but the two lithium deficient stars represented by green dots are CEMP-no with [Ba/Fe]=--0.5 and --1.2.
The situation is still unclear,
if the Li deficiency were connected to the C overabundance one would still have to explain the
CEMP stars that lie on the Spite plateau.
To try to help to disentangle these different interpretations, Fig. \ref{Fig:LiEMPCEMPFe} is identical to Fig. \ref{Fig:LiFe}, but the stars with 
[C/Fe]$ > +0.7$ are included with an indication of the [C/Fe] value. 

\subsubsection{~Ratio \cdt in CEMP stars}
\citet{romano_evolution_2017} computed the theoretical Galactic evolution of the ratio $\rm^{12}C/^{13}C$ based on different models of  C production, including or not super-AGB stars, and fast rotating massive stars (FRMS). These models show an extremely rapid decrease of the \cdt ratio in the very early Galaxy. The models which include FRMS reach the ratio observed in a sample of 'unmixed' giants: \cdt $\sim$ 30 \citep{spite_first_2006} in less than 0.5 Gyr then the \cdt ratio slowly increases toward the solar value. About the same value, \cdt = 31, was measured in HD\,140283 a star known to be extremely  old  but not extremely metal-poor and not C-rich \citep{spite_12_2021}.\\
More recently \citet{molaro_12c13c_2023} measured this ratio  in five CEMP stars, two  dwarfs and three giants in the lower RGB. 
These CEMP stars, with $\rm [Fe/H]<-4.7$, 
are supposed to be the most pristine galactic stars. In these stars the \cdt~ratio varies from 39 to 78. They could represent the rapid decrease of the ratio \cdt at the very beginning of the Galaxy.
The high $\rm^{12}C/^{13}C$ ratio of GHS143 \citep[a young Galactic metal-poor stars not enhanced in carbon,][]{caffau_high_2024}, on the other side, could represent the increase with time of the C isotopic ratio.

\section{EMP stars in external galaxies}

Among the external galaxies close enough from the  Milky Way to be  resolved into stars and for which we can obtain spectra 
of sufficient quality (resolution, S/N ratio, wavelength range)  for metallicity and/or
abundance determination, the Magellanic Clouds  are the largest galaxies  visible at naked eye in the southern hemisphere.
Dwarf  galaxies (Carina, Draco, Fornax, Sagittarius, Sculptor and Sextans) discovered during the  20th 
century showed a different chemical history from our Galaxy as clearly demonstrated by \citet{venn_stellar_2004}.  
These first results reveal that these galaxies   are metal-poor with a   lower [$\alpha$/Fe] ratios at a given metallicity when compared to galactic  stars.  Many more fainter/smaller galaxies, nicknamed  as ultra faint dwarf spheroidal galaxies (UfdSph) or  ultra faint dwarf galaxies (UFD),  have been found since thanks to large photometric and spectroscopic surveys 
 like the  Sloan Digital Sky Survey   \citep[SDSS,][]{york_sloan_2000}, the Dark Energy Survey  \citep[DES][]{collaboration_dark_2005}, and others. \citet{mcconnachie_revised_2020}  built a catalogue with the positional, structural, and dynamical parameters for all dwarf galaxies in and around the Local Group up to a distance $\simeq$ 3 Mpc from the Sun.  We used  the revised public version of their catalogue\footnote{https://www.cadc-ccda.hia-iha.nrc-cnrc.gc.ca/en/community/nearby/}   to select the galaxies with a distance modulus closer  than 22.5. This choice corresponds to the range of galaxies  
for which individual information based on medium to high resolution spectroscopy on stars along the giant branch (radial velocities, metallicities) has been obtained with the current instrumentation as it will be shown in the following subsections. 
 
\begin{figure}
\centering
\includegraphics[width=12.0cm,clip=true]{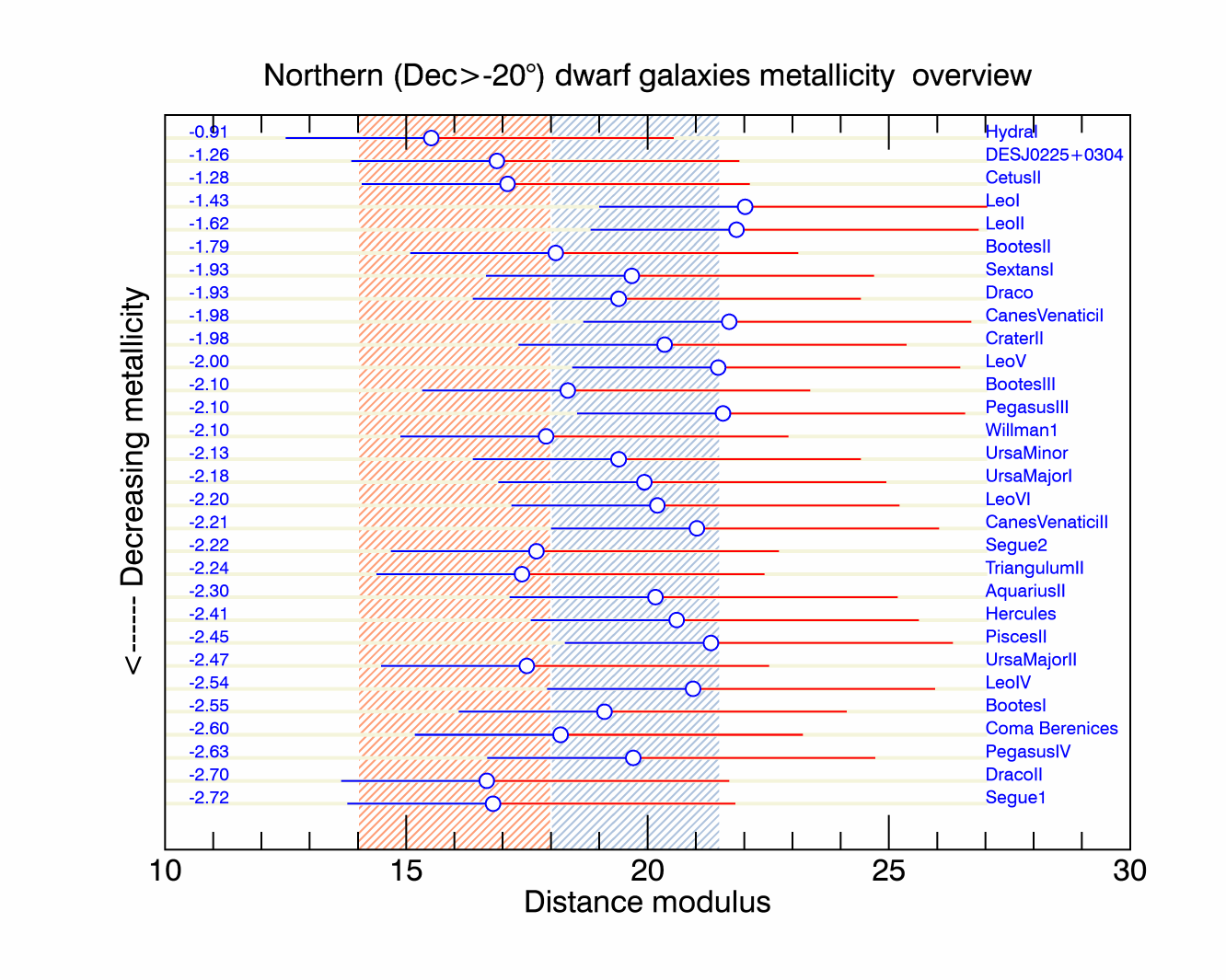}
\caption{\footnotesize   Dwarf and ultra-faint dwarf galaxies observed from northern observatories (CFHT, Subaru, Keck, ORM/La Palma, ...). 
The galaxies are ranked by decreasing  mean metallicity. The blue horizontal lines show a 3 magnitude range above the horizontal branch, illustrating the tip of red giant branch. The red horizontal lines represent a 5 magnitude range illustrating the location of the  red giant branch  from the horizontal branch down to the Turn-off. The light blue dashed area shows the typical magnitude interval where medium resolution abundance analyses have been made (i.e. with Keck DEIMOS). The light salmon dashed area shows the typical magnitude interval where high  resolution abundance analyses have been made (Keck HIRES, HDS).  }
\label{Fig:Galaxies_North_overview}
\end{figure}

\begin{figure}
\centering
\includegraphics[width=12.0cm,clip=true]{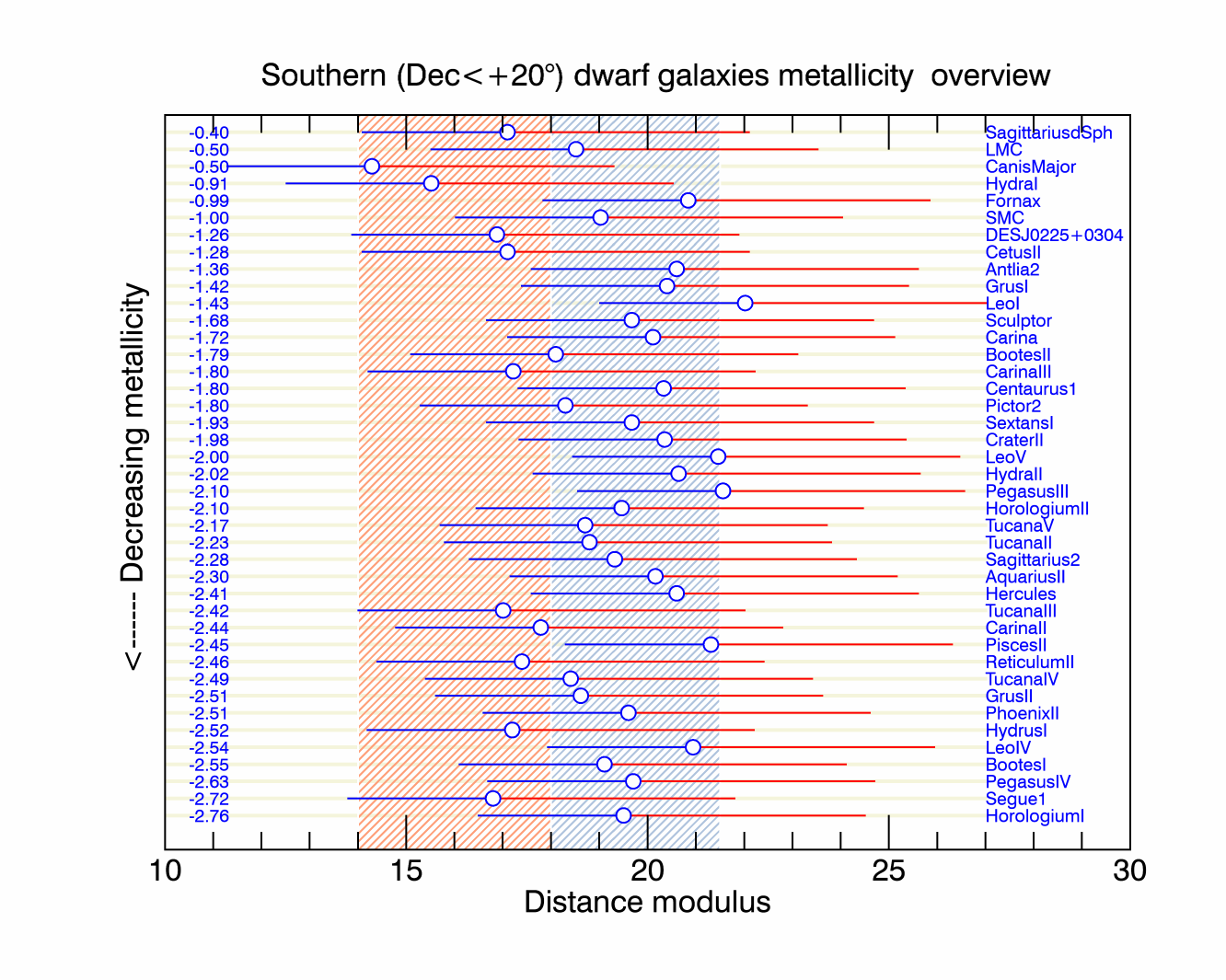}
\caption{\footnotesize  Dwarf and ultra-faint dwarf galaxies observed from southern observatories (Paranal, La Silla, CTIO, ...). 
The galaxies are ranked by decreasing mean metallicity. The blue horizontal lines show a 3 magnitude range above the horizontal branch, illustrating the tip of red giant branch. The red horizontal lines represent a 5 magnitude range illustrating the location of the  red giant branch  from the horizontal branch down to the Turn-off. The light blue dashed area shows the typical magnitude interval where medium resolution abundance analyses have been made (FORS2, FLAMES, X-Shooter, Magellan/IMACS). The light salmon dashed area shows the typical magnitude interval where high  resolution abundance analyses have been made (UVES, MIKE).}
\label{Fig:Galaxies_South_overview}
\end{figure}

 Fig. \ref{Fig:Galaxies_North_overview}  and Fig. \ref{Fig:Galaxies_South_overview} show respectively,  the  distance modulus versus mean metallicity  of the revised public version of  the catalogue of \citet{mcconnachie_observed_2012}, the list of dwarf and  ultra faint galaxies limited to a distance modulus of $\le$ 22.5
visible from northern observatories (defined  here as $\delta \ge   -20$, and respectively visible from  southern observatories (defined  here as $\delta \le $  +20) .
For each galaxy, the blue  line indicates the region  of magnitude where stars are located  approximately above the horizontal branch up to the tip of the red giant branch. The red line represents the range of magnitude from  the approximate location of  the horizontal branch 
 down to the turn-off.

 Unfortunately, these figures do not show  that around their mean metallicity,  most of these UFD galaxies exhibit a large metallicity spread, giving the possibility  to find a significant number of  extremely metal poor stars.

 The galaxies are sorted by  decreasing metallicity from the top to the base of each plot. 
 In both figures, we have highlighted in salmon (resp. blue) dashed areas representing the typical magnitude ranges  where high resolution spectroscopy (resp. medium-low resolution) has been used to derive  stellar metallicities  and/or detailed abundance ratios. 
 It is interesting to note that high resolution spectroscopy is only  possible for a handful of  stars generally located along the tip of the red giant branch.   For fainter stars, the difficulty resides on the low S/N ratio that can be achieved  even after several hours of accumulated data with the most efficient spectrographs on 10m class telescopes. 

\subsection {The Magellanic Clouds}

\begin{itemize}

\item {} Small Magellanic Cloud \\
Among the first high resolution spectroscopic studies of individual stars in the SMC, 
\citet{hill_chemical_1997} studied six SMC cool stars (K supergiants) thanks to high resolution spectra obtained with  the ESO CASPEC/3.6\,m spectrograph. They found that  the mean metallicity of the young population of the SMC was around $\rm [Fe/H] = -0.7$ in agreement with previous studies \citep{russell_abundances_1989,spite_chemical_1989,luck_chemical_1992}.

\begin{figure}
\centering
\includegraphics[width=10.0cm,clip=true]{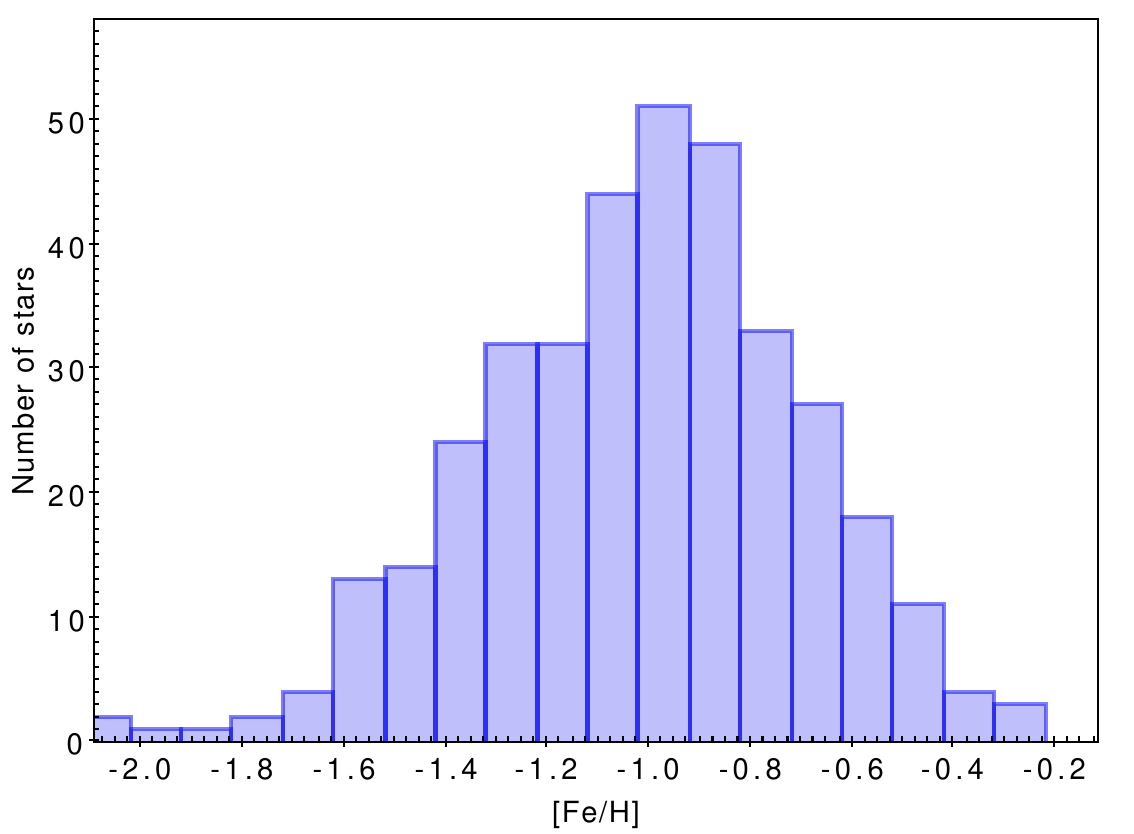}
\caption{\footnotesize  Small Magellanic Cloud metallicity histogram of the sample of stars from \citet{parisi_ca_2010}. 
Metallicity has been determined using calibration relations of the Ca{\sc II} infrared triplet}
\label{Fig:smc-parisi}
\end{figure}

\citet{parisi_ca_2010}  have obtained metallicities for $\sim$360 red giant stars distributed in 15 Small Magellanic Cloud (SMC) fields from near-infrared spectra covering the Ca{\sc II} triplet lines using the VLT + FORS2. The metallicity distribution (MD) of the whole sample shows a mean value of [Fe/H] =$-1.00 \pm 0.02$, with a dispersion of $0.32 \pm 0.01$, in agreement with global mean [Fe/H] values found in previous studies. The most metal poor star of their sample has a metallicity of [Fe/H] =--2.03. 
Fig. \ref{Fig:smc-parisi} presents the metallicity histogram  of the sample of stars studied by \citet{parisi_ca_2010} showing a rather moderate low metallicity of the SMC.
\citet{reggiani_most_2021} obtained high resolution MIKE/Magellan spectra of four giants  and found  SMC metal poor stars  down to --2.6.
So far, no EMP star has been discovered in the SMC. \\

\item{} Large Magellanic Cloud \\
The first  studies of F supergiants using high resolution spectroscopy have been performed in the
early '90s \citep{russell_abundances_1989,mcwilliam_abundance_1991,luck_chemical_1992,barbuy_oxygen_1994,hill_chemical_1995}.  They showed that  young stars in the LMC were metal-poor with a moderate deficiency
[Fe/H]  of the order of --0.3 dex. They also found that heavy s-process and r-process elements were slightly overabundant with   a [X/Fe] of the order of +0.3\,dex on average. 

For the older LMC stellar population, one of the first study of the detailed  chemical composition of metal-poor red giants in  the LMC based on high resolution spectra was done by 
\citet{pompeia_chemical_2008}. They  used FLAMES/VLT to obtain high resolution spectra  of 59 red giant stars from the inner disk of the LMC, ~2 kpc from the centre of the galaxy. They found moderately low metallicities  ranging from $\rm [Fe/H]= -0.28$ to $-1.7$. 
\citet{lapenna_tagging_2012} determined the detailed composition of  89 stars in the disk of the LMC and derived metallicities ranging from $-0.09$ to $-1.51$. 
Thanks to high resolution spectra obtained with FLAMES/VLT,  \citet{van_der_swaelmen_chemical_2013} determined the detailed chemical composition of  a  sample of 106 LMC red giants located in the bar and 58   red giants in the disc of the LMC. They obtained moderately low metallicities ranging from $\rm [Fe/H] = -0.65$ to +0.37 for their whole sample. They also measured  abundances of several $\alpha$ and neutron-capture elements. 
 A more extensive work has been done by \citet{nidever_lazy_2020}  with  spectra from   APOGEE metallicities and $\alpha$-element abundances were measured for a large sample of LMC red giants spanning a range of metallicity from $\rm[Fe/H]= -0.2$ to   rather metal poor values with $\rm [Fe/H] \simeq -2.5$ although no EMP star has been found in their sample. 
Using  the mid-infrared metal-poor star selection of \citet{schlaufman_best_2014}  and their own analysis of archival data, \citet{reggiani_most_2021} obtained high resolution MIKE/Magellan spectra of nine giants 
and found  LMC metal poor stars down to $\simeq -2.5$ and SMC stars down to $-2.6$.
A recent article from  \citet{oh_high-resolution_2024}  presented a high-resolution spectroscopic study of  seven extremely metal-poor stars in the Large Magellanic Cloud. They confirmed that all seven stars, two of which with $\rm [Fe/H] \le -2.8$, 
are the most
metal-poor stars found so far in the Magellanic Clouds.

\begin{table}
\centering
\caption{Abundances ratios for the metal-poor stars discovered by \citet{chiti_enrichment_2024}. [C/Fe], [Mg/Fe], [Ca/Fe] and [Eu/Fe] are the  carbon,
magnesium, calcium, and europium abundances, respectively. [C/Fe]c is the carbon abundance after correcting 
for the evolutionary state of the star as described in \citet{chiti_enrichment_2024}.}
\begin{tabular}{lcccccc}
\hline
      Name		&       [Fe/H]	&  [C/Fe]    	&   [C/Fe]c	   &    [Mg/Fe]	 &  [Ca/Fe]  &      [Eu/Fe]   	\\
\hline
 $\rm LMC-003 $	&	-2.97	&	 -0.38	&   0.35	   &     0.38	 &    0.19   &	 $<$  0.24  \\
 $\rm LMC-100 $	&	-2.67	&	 -0.28	&   0.44	   &     0.39	 &    0.25   &	$<$    0.37  \\
 $\rm LMC-104 $	&	-2.56	&	 -0.62	&   0.15	   &     0.34	 &    0.19   &	          0.18  \\
 $\rm LMC-109 $	&	-2.85	&	 -0.55	&   0.19	   &     0.43	 &    0.23   &	          0.48  \\
 $\rm LMC-119 $	&	-4.15	& $<$-0.35	&  $<$ 0.30	   &     0.42	 &    0.47   &	 $<$    1.04  \\
 $\rm LMC-124 $	&	-2.97	&	 -0.33	&   0.40	   &     0.49	 &    0.37   &	$<$   0.47  \\
 $\rm LMC-204 $	&	-2.83	&	 -0.15	&   0.59	   &     0.31	 &    0.09   &	 $<$   0.63  \\
 $\rm LMC-206 $	&	-2.56	&	 -0.34	&   0.05	   &    -0.05	 &    0.02   &	 $<$   0.36  \\
 $\rm LMC-207 $	&	-3.34	&	 -0.12	&   0.65	   &     0.16	 &    0.48   &	 $<$    0.54  \\
 $\rm LMC-215 $	&	-3.09	&	-0.14	&   0.61	   &     0.48	 &    0.28   &	  $<$   0.24  \\
    \hline
    \end{tabular}
    \label{tab:chiti_2024}
\end{table}

The elemental abundance ratios are generally consistent with Milky Way halo stars of similar [Fe/H] values.  This work has been extended by the study of  \citet{chiti_enrichment_2024} who  discovered a 
set of stars with metallicities ranging  from $-2.5$ to $-4.15$. The star LMC-119 is  the most metal-poor stars discovered so far in the Magellanic Clouds with $\rm [Fe/H] = -4.15$. Their detailed abundance results are shown in Table\,\ref{tab:chiti_2024}.

\begin{figure}
\centering
\includegraphics[width=10.0cm,clip=true]{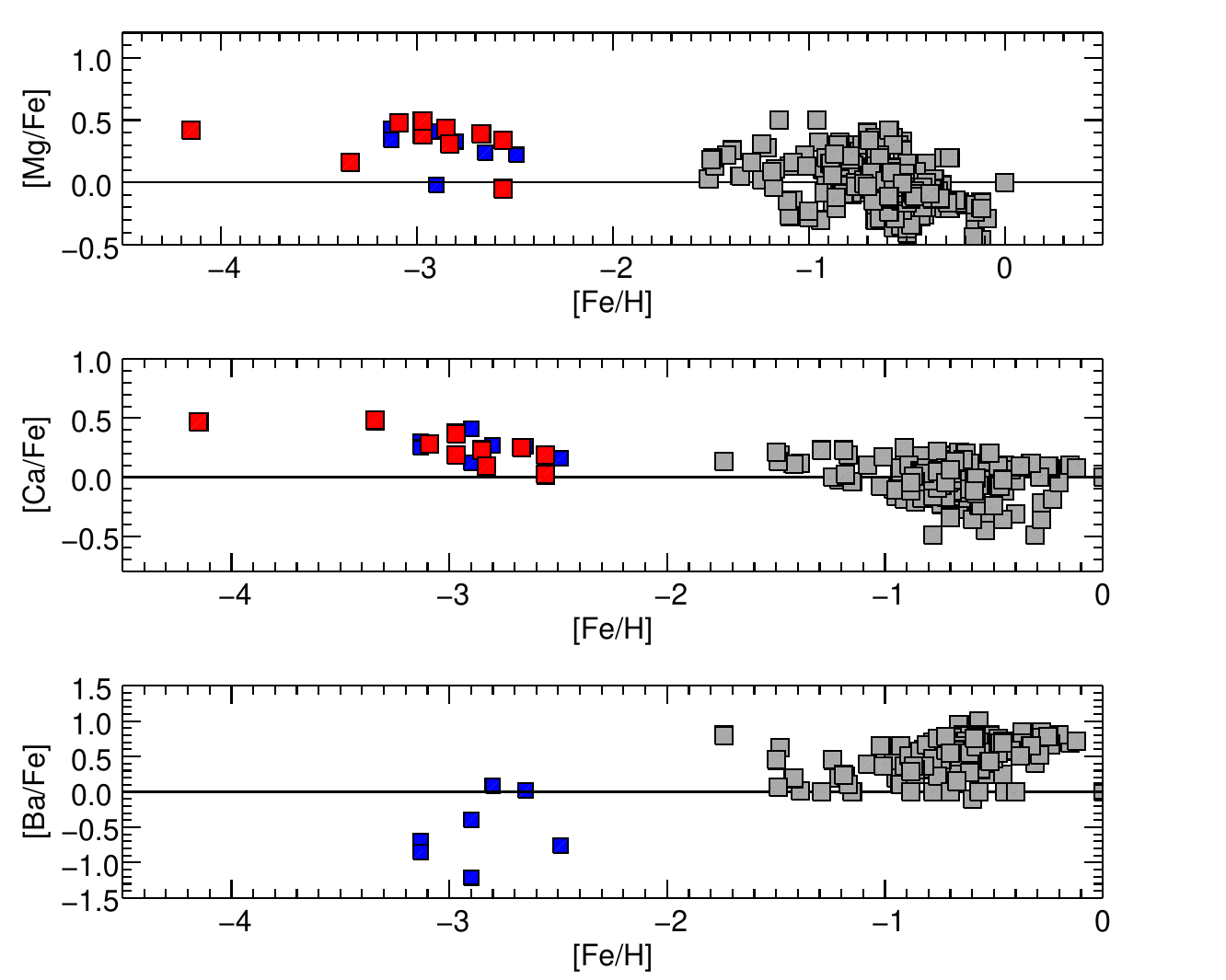}
\caption{\footnotesize [Mg/Fe], [Ca/Fe] and [Ba/Fe] abundance  ratio as a function of [Fe/H] of LMC stars. 
Blue symbols represent the most metal poor stars discovered so far in the LMC  \citep{oh_high-resolution_2024}.
Red symbols are data from \citet{chiti_enrichment_2024}.
Black symbols are literature data from 
\citet{pompeia_chemical_2008}, \citet{lapenna_tagging_2012} and \citet{van_der_swaelmen_chemical_2013}.  }
\label{Fig:lmc-mgfe}
\end{figure}

Fig. \ref{Fig:lmc-mgfe} shows the abundance of [Mg/Fe], [Ca/Fe] and [Ba/Fe]  measured in  the LMC stars where only literature data aiming at studying the low metallicity sample 
has been selected. From this literature data,  four stars found by \citet{oh_high-resolution_2024}  and seven stars found by \citet{chiti_enrichment_2024} can be considered as  true EMP stars,  the most metal poor star found by \citet{chiti_enrichment_2024} being classified as an UMP star.

\begin{figure}
\centering
\includegraphics[width=10.0cm,clip=true]{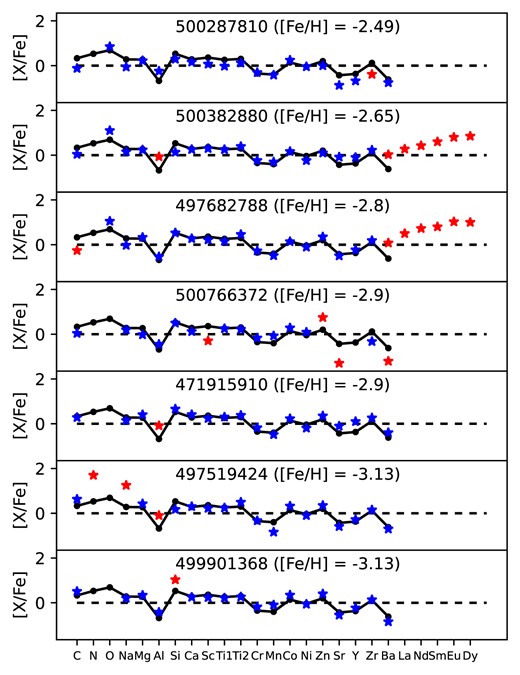}
\caption{\footnotesize Overview of the abundance measurements for the  stars studied by \citet{oh_high-resolution_2024}   with literature values. The star symbols represent 
the abundance measurements, and the black dots represent the mean MW values from the literature. The star colours indicate if the abundance is within 
0.5\,dex of the MW average (blue) or not (red).  Figure reproduced from  \citet{oh_high-resolution_2024}, with permission. }
\label{Fig:lmc-oh_stars}
\end{figure}

Fig. \ref{Fig:lmc-oh_stars} shows a comparison between the abundances of the most metal poor LMC  stars found by \citet{oh_high-resolution_2024}  and literature data for the Milky way halo. They found that their LMC results are consistent with that of the MW halo for most of the elements measured, although with
some discrepancies by at least 0.5\,dex compared to the MW, in at least one element.

\end{itemize}

\subsection {Dwarf Galaxies}  

\begin{itemize}

 \item{} Carina \\
  Carina was discovered by \citet{cannon_new_1977} by visual inspection of a plate
  in the ESO/SRC Southern Sky Survey.
  Thanks to FLAMES observations of a sample of stars belonging to Carina, 
  \citet{koch_complexity_2006} determined the metallicity in a sample of Carina giants and obtained a mean metallicity of $\rm [Fe/H] \simeq -1.7$  for Carina,  with a  the full range of metallicities  spanning approximately $\rm -3.0 < [Fe/H] < 0.0$.
 \citet{koch_complexity_2008}  obtained high-resolution spectroscopy of ten red giants in the Carina dwarf spheroidal (dSph) galaxy with UVES and determined their detailed chemical composition. Two of the stars had metallicities below $\rm [Fe/H] = -2.5$. 
 Nine red giants were studied by \citet{venn_nucleosynthesis_2012} using high resolution spectra obtained with 
 FLAMES/UVES and MIKE. Two stars were found to have very low  metallicites [Fe/H] of $-2.81$  and $-2.86$.
\citet{lemasle_vltflames_2012} used  FLAMES/VLT  in high-resolution mode  to measure the abundances of
several chemical elements, including Fe, Mg, Ca and Ba, in a sample of 35 individual Red Giant Branch stars in Carina. 
They found metallicities ranging from $\rm [Fe/H]= -1.18$ to $-2.51$.  

\citet{susmitha_abundance_2017} identified a CEMP-no star with a metallicity $\rm [Fe/H]= - 2.5$ thanks to the analysis of UVES spectra. 
\citet{norris_populations_2017} obtained R=47000  spectra of 63 stars with  the FLAMES/ESO spectrograph linked to UVES. 
They derived  detailed abundances in a sample of stars with  metallicities ranging from $\rm [Fe/H] = -2.68$ to $-0.64$. 

 \citet{hansen_evidence_2023} obtained high resolution  MIKE spectra  of six CEMP stars identified in the Carina dSph.
 Three of their stars had a metallicity below $\rm [Fe/H]  = -2.6$. \\
  
\begin{figure}
\centering
\includegraphics[width=12.0cm,clip=true]{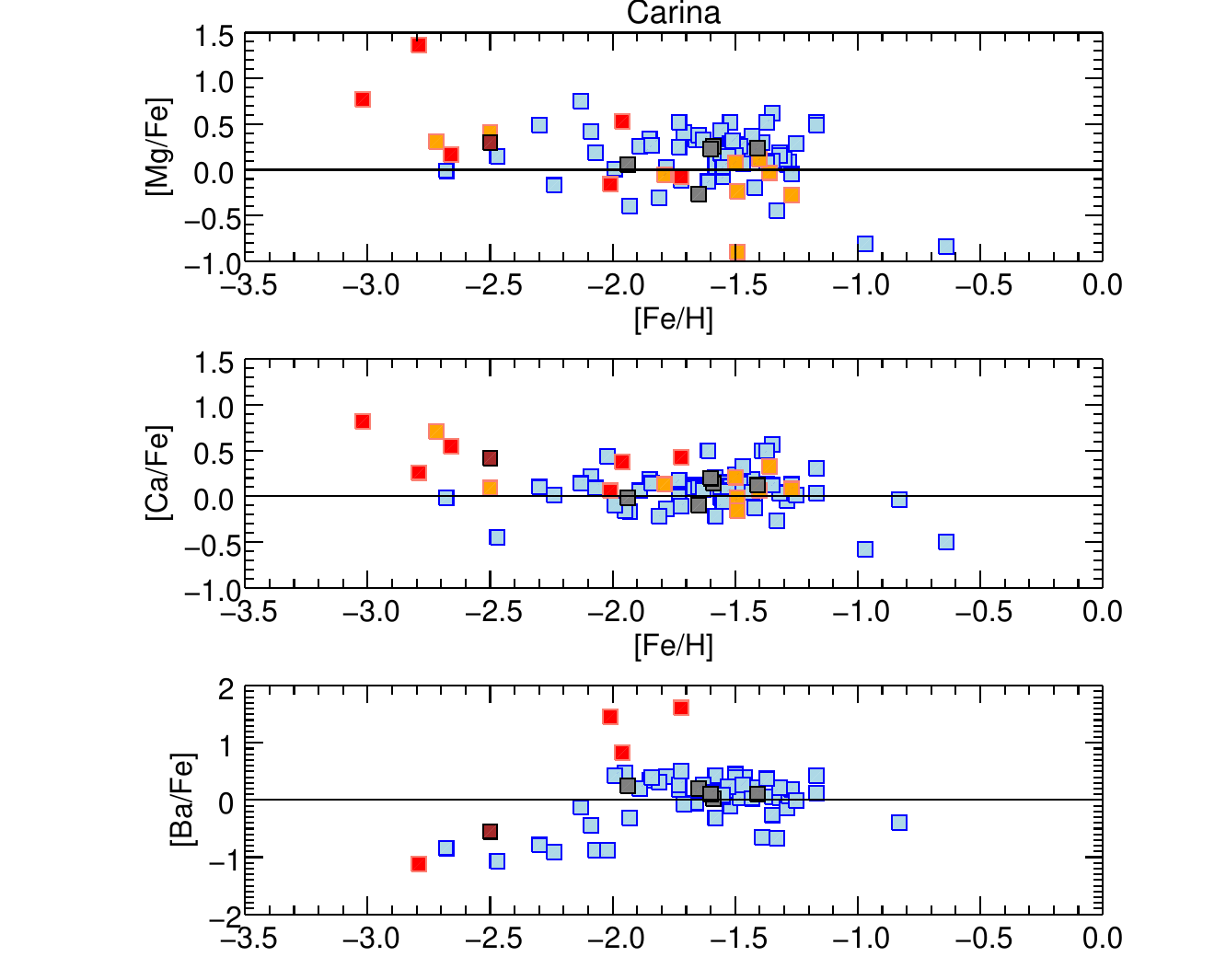}
\caption{\footnotesize The abundances for Carina stars (grey filled squares) are mainly from \citet{norris_populations_2017}, which includes a reanalysis of the data from \citet{shetrone_vltuves_2003}, \citet{venn_nucleosynthesis_2012}, and \citet{lemasle_vltflames_2012}. The stars represented as red symbols are the CEMP stars analysed by  \citet{hansen_evidence_2023}.  }
\label{Fig:carina_abundance}
\end{figure}

 In Fig. \ref{Fig:carina_abundance} are shown the results of the detailed abundances found in Carina dwarf galaxy. This figure shows that  only one of the  stars analysed by  \citet{hansen_evidence_2023} can be considered as an EMP star. 
The  Carina metallicity histogram based on the published data is shown in Fig. \ref{Fig:dSph_mdf}. \\
 
 \begin{figure}
\centering
\includegraphics[width=5.5cm,clip=true]{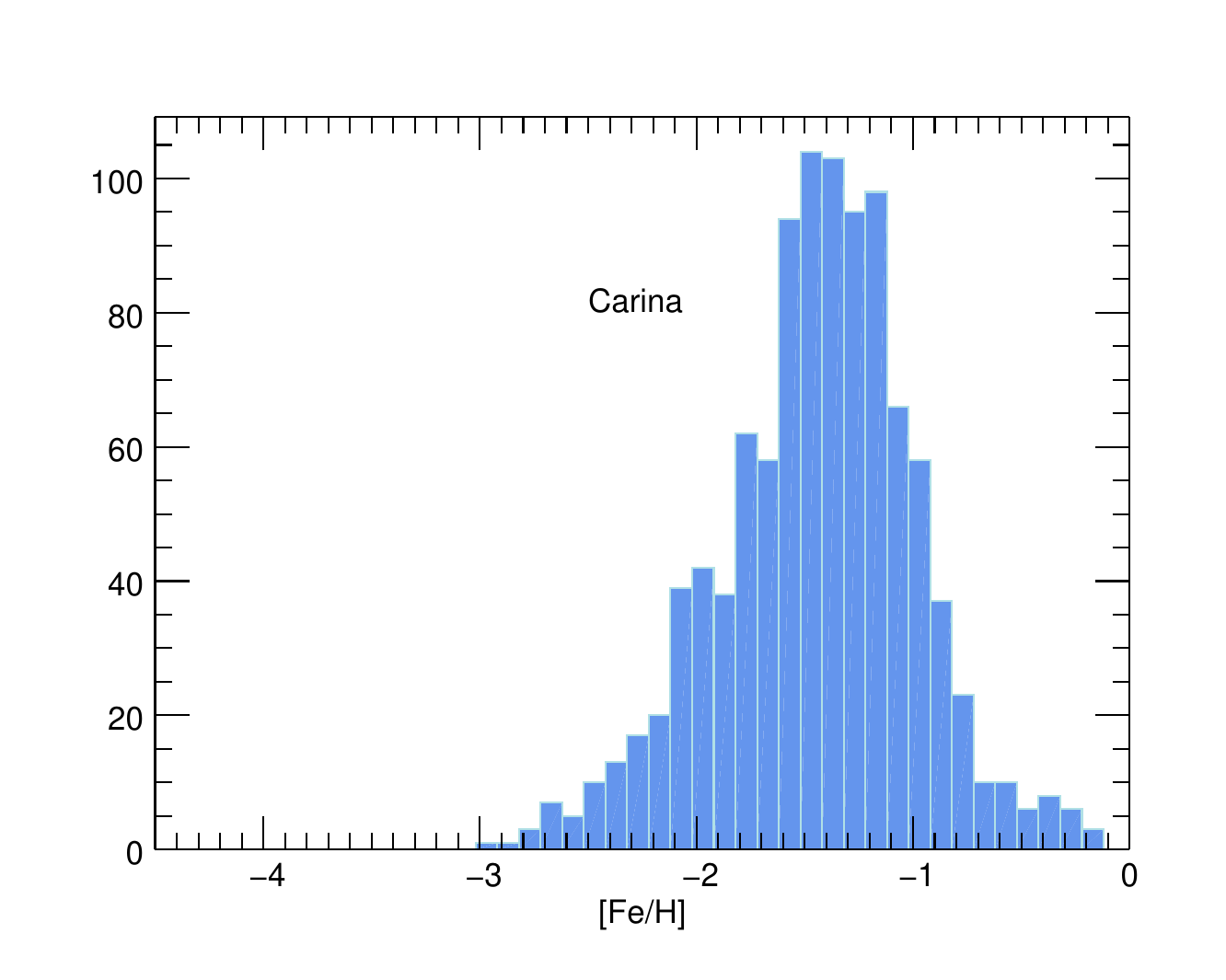}
\includegraphics[width=5.5cm,clip=true]{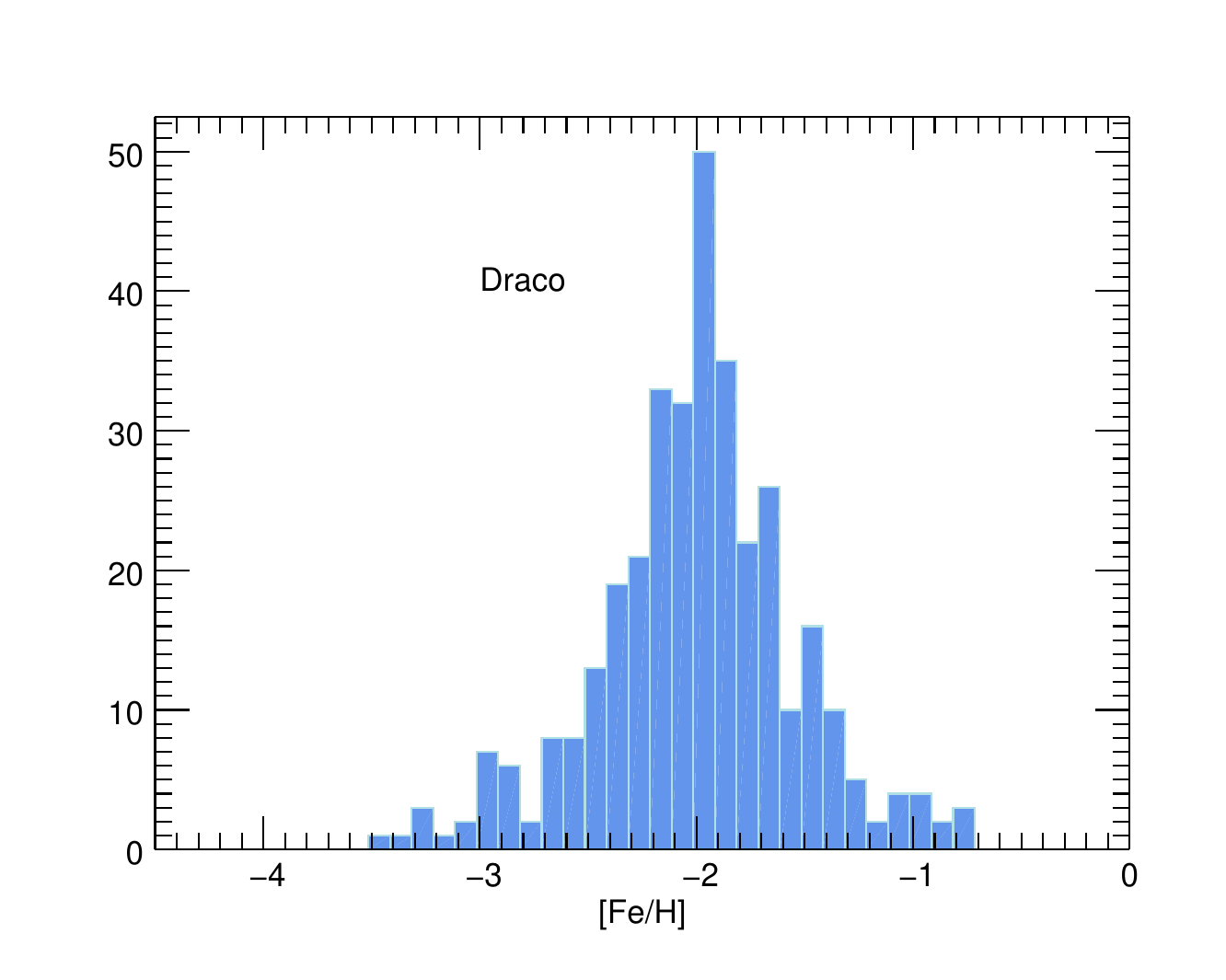}
\includegraphics[width=5.5cm,clip=true]{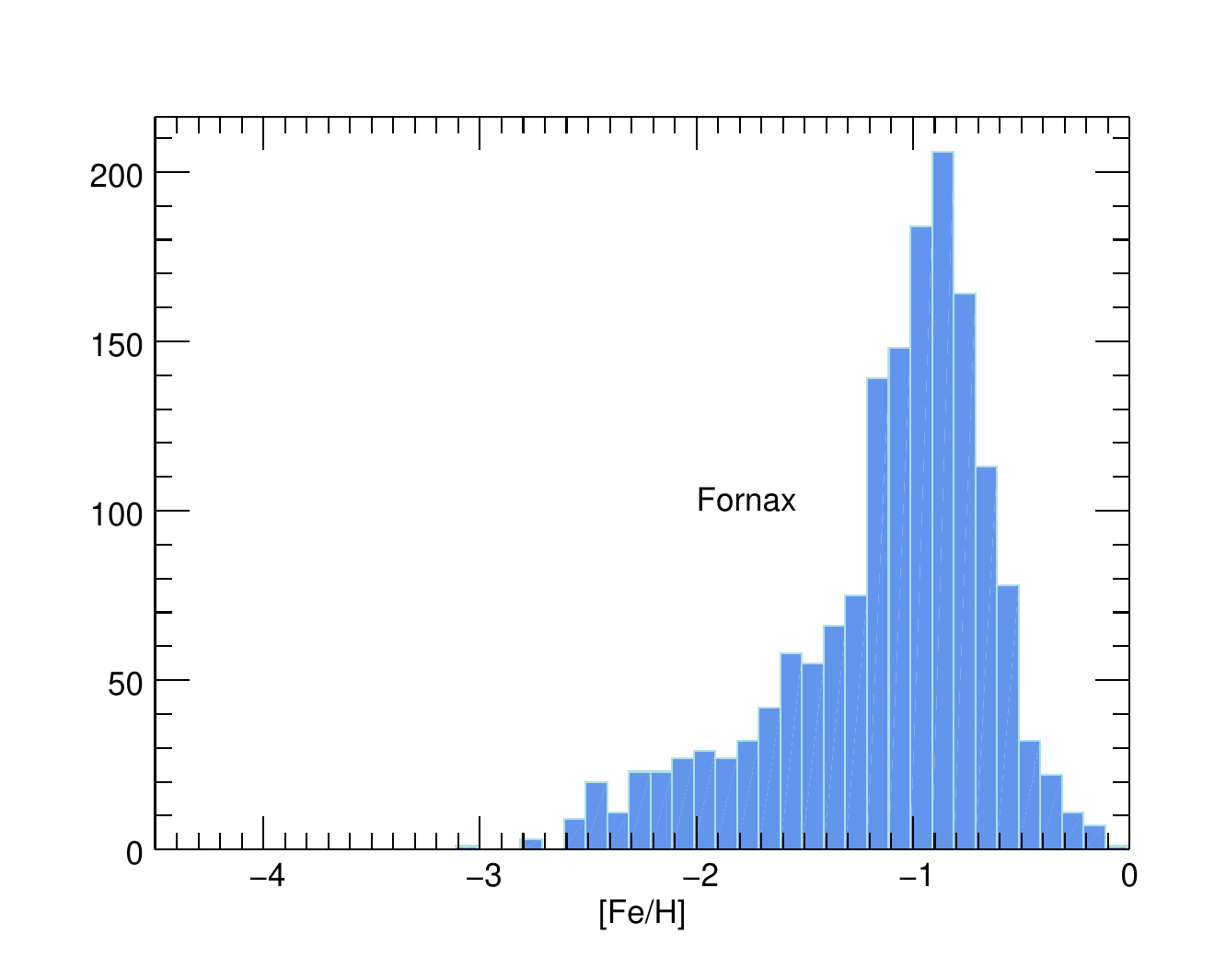}
\includegraphics[width=5.5cm,clip=true]{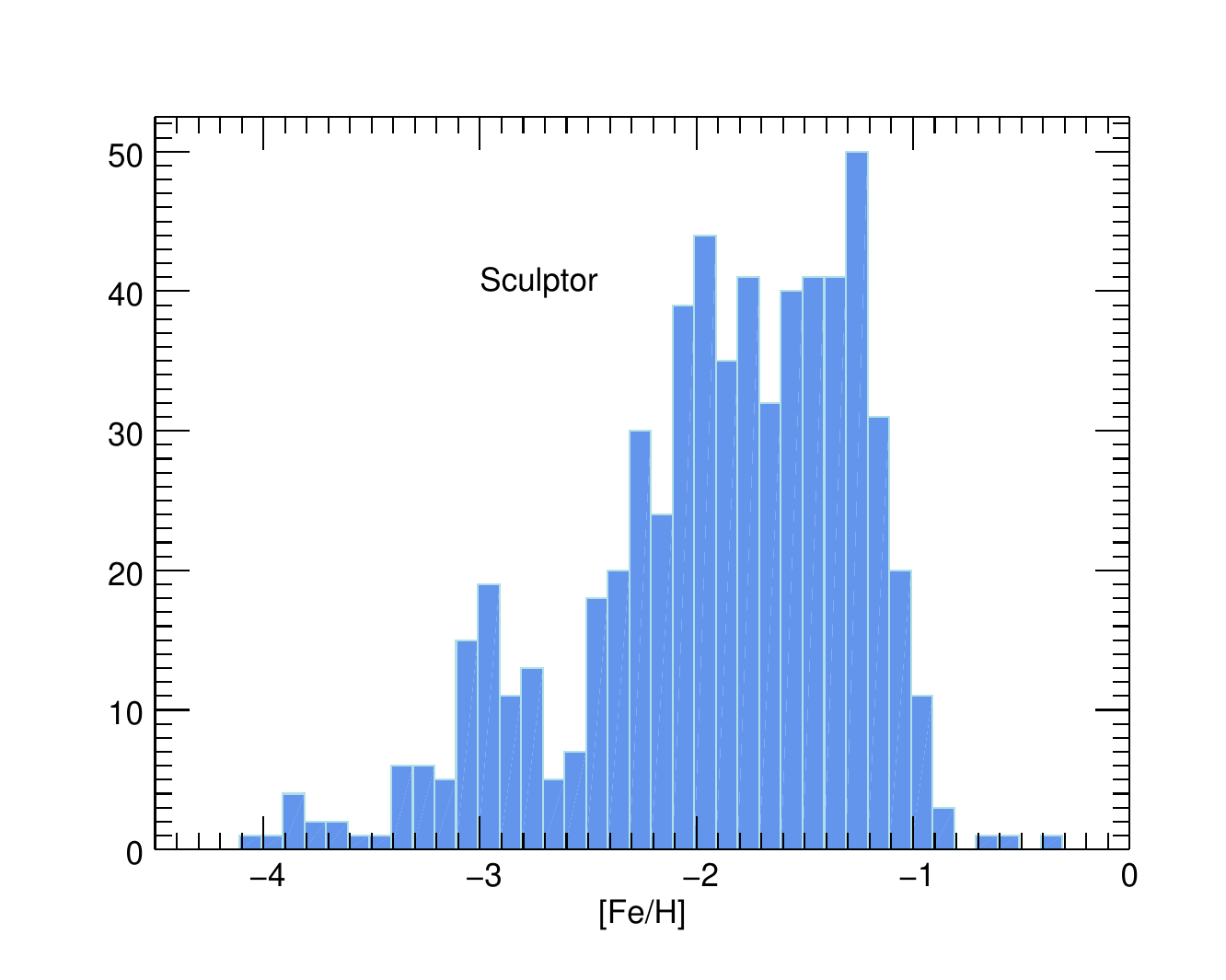}
\includegraphics[width=5.5cm,clip=true]{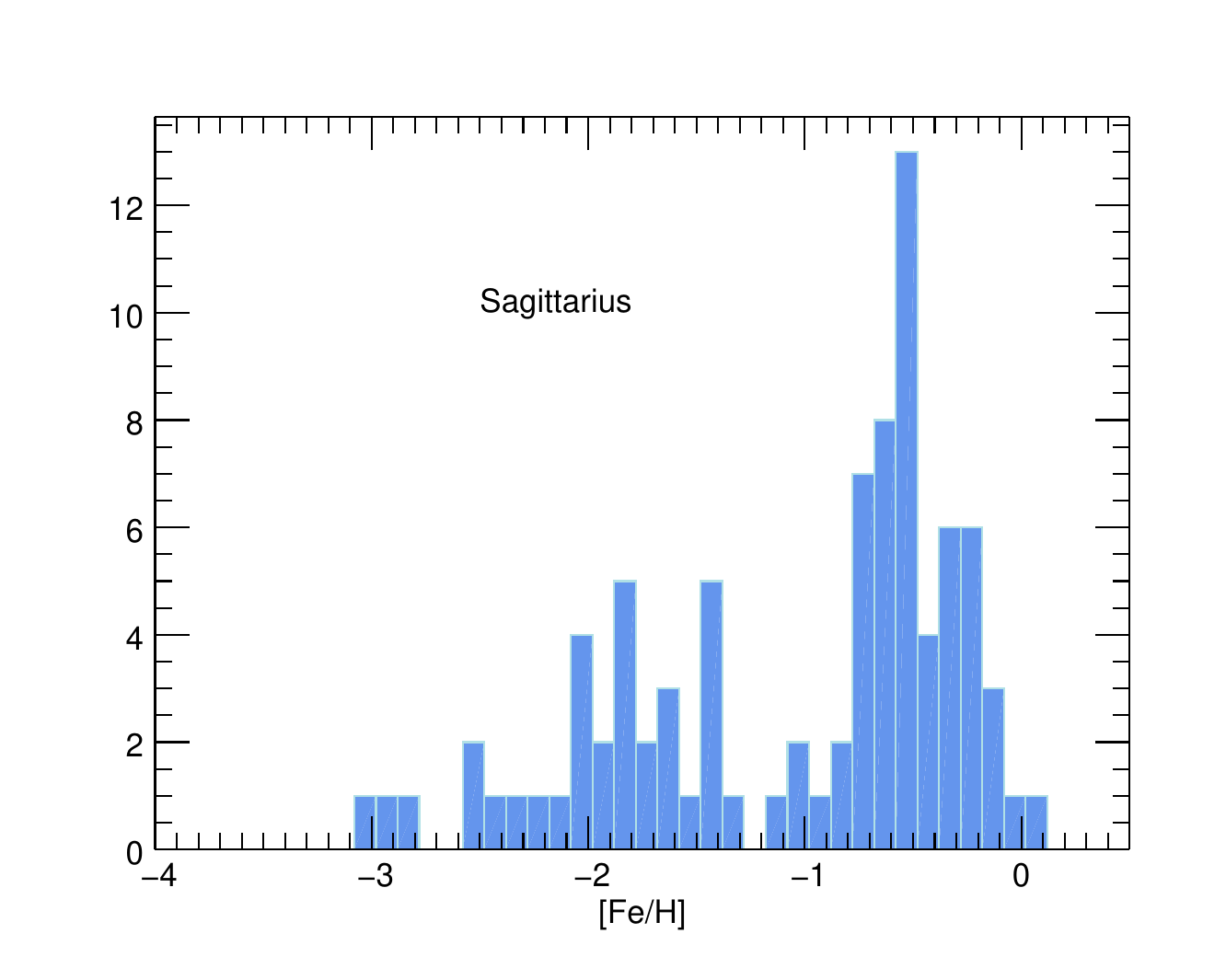}
\includegraphics[width=5.5cm,clip=true]{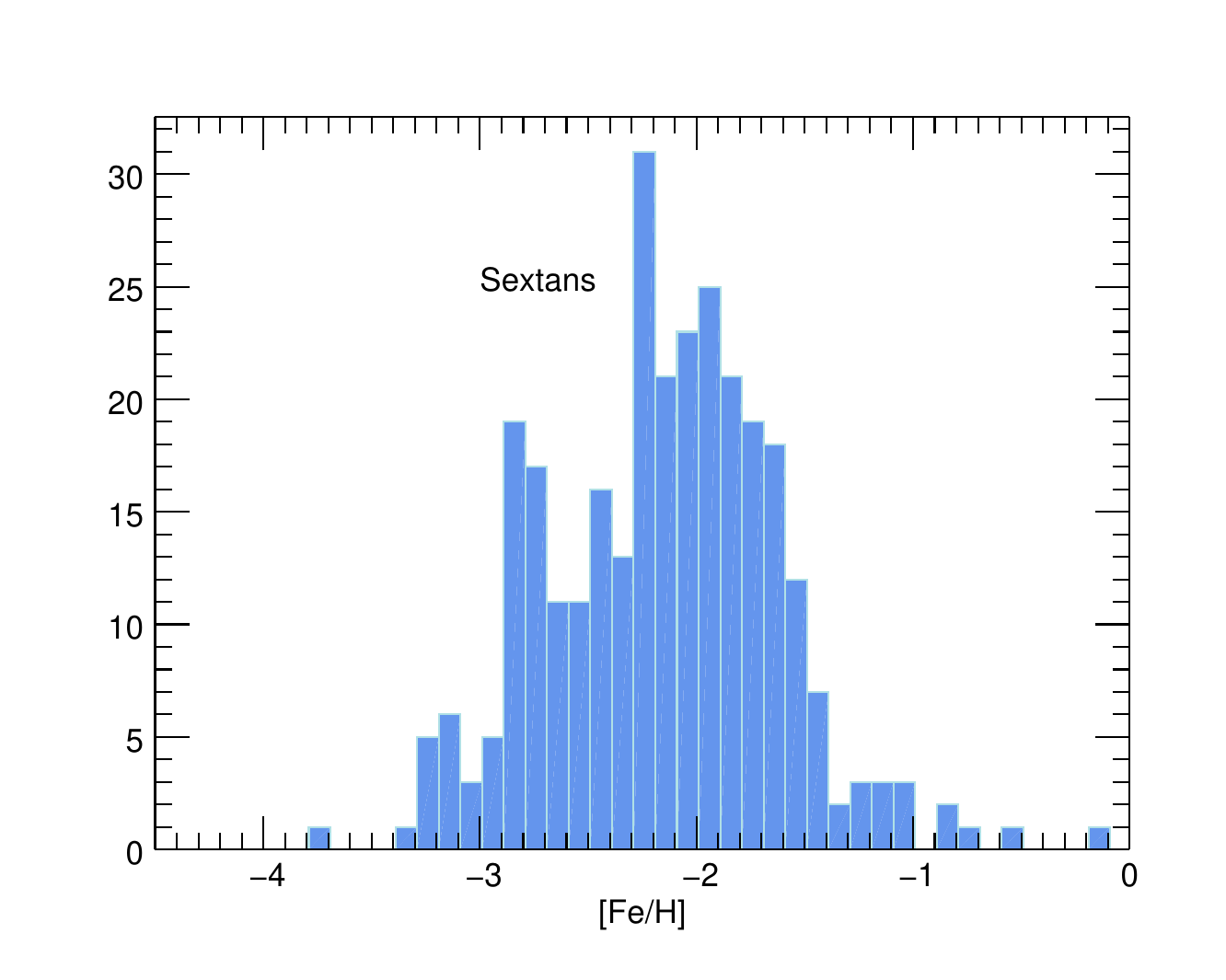}
\caption{\footnotesize Dwarf Spheroidal galaxies metallicity distribution.}
\label{Fig:dSph_mdf}
\end{figure}

   \item{}Draco \\
\citet{wilson_sculptor-type_1955} discovered the dSph galaxy Draco  by inspecting  48-inch schmidt
plates taken for the National Geographic Society-Palomar Observatory Sky Survey. 
 \citet{kirby_multi-element_2010} obtained  KECK/DEIMOS spectra of a large sample of stars in Draco. 
Based on a sample of 299 stars from Draco with metallicities ranging from $-3.6$ to $-0.83$ , they derived a mean metallicity $\rm [Fe/H] = -1.89$.  
They also measured the abundance of Mg, Ca, Si and Ti for a significant fraction of the sample. \\

 \citet{tsujimoto_enrichment_2017}  obtained high resolution spectra of 12 metal poor stars in Draco with HDS/Subaru.
 In particular, they determined the abundances of several neutron capture elements in their sample covering metallicities from $-2.5$ to $-2.0$. 
They found that these stars are separated into two groups with r-process abundances differing by one order of magnitude. \\
 
Fig.\,\ref{Fig:draco_abundance}  shows the abundance ratios of Mg, Ca and Ba as a function of [Fe/H] in the Draco dwarf galaxy. 
The grey symbols are based on  the low resolution spectroscopy made by  \citet{kirby_multi-element_2010}. The bottom plot shows the results for the [Ba/Fe] ratio as determined by  \citet{tsujimoto_enrichment_2017}. In the middle panel of the plot, we can see  that Draco exhibits  a wide range of metallicities  down to  EMP metallicities. 
The  Draco metallicity histogram based on the published data is shown in Fig. \ref{Fig:dSph_mdf}.
\\
 
 \begin{figure}
\centering
\includegraphics[width=10.0cm,clip=true]{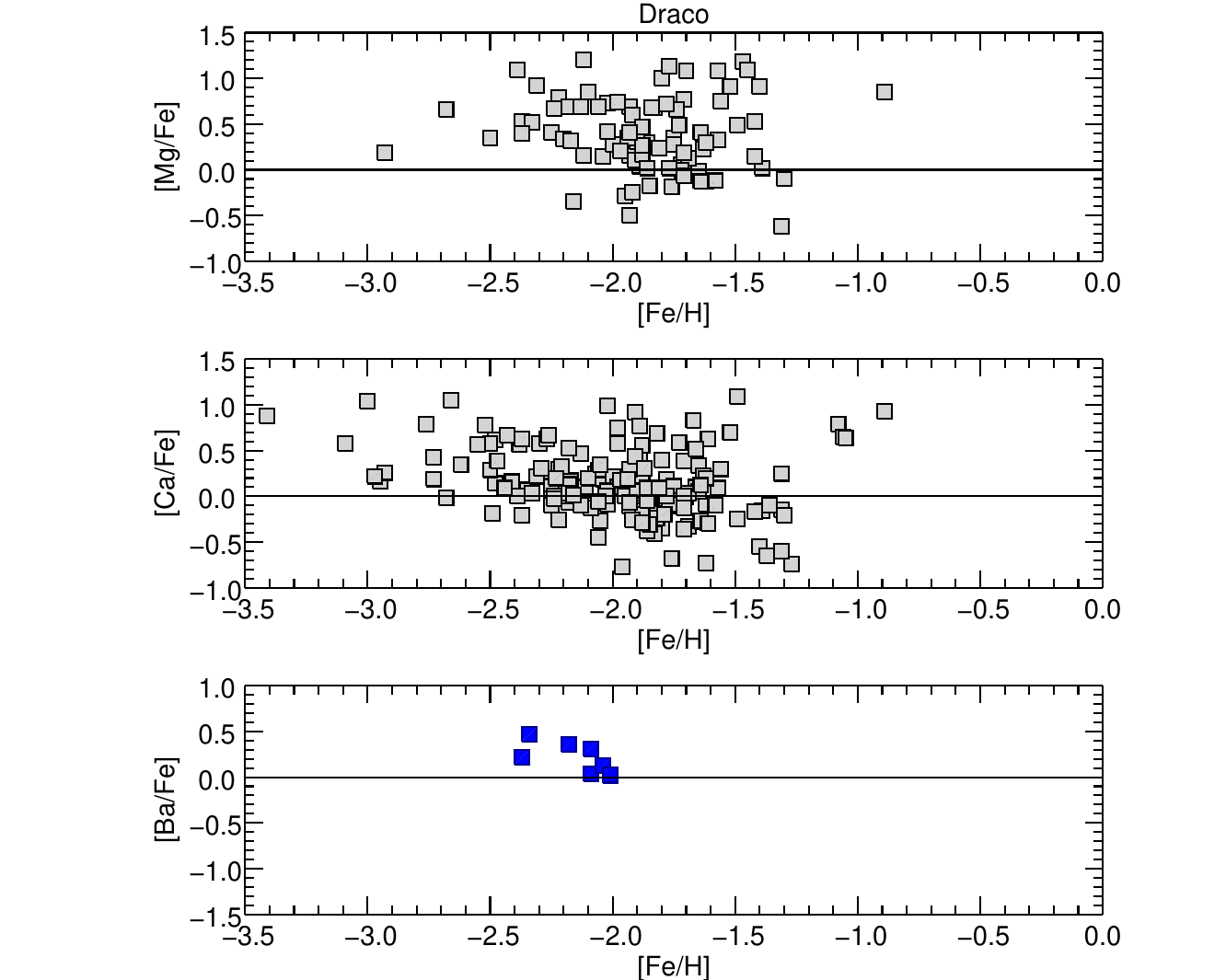}
\caption{\footnotesize Abundance ratios in Draco. The first two upper panels are the abundances of [Mg/Fe] and [Ca/Fe] as a function of [Fe/H] from low resolution spectroscopy  \citep{kirby_multi-element_2010}. The last panel presents the [Ba/Fe] vs [Fe/H] from \citet{tsujimoto_enrichment_2017} based on high resolution spectroscopy.  }
\label{Fig:draco_abundance}
\end{figure}

  \item{}Fornax \\
  Fornax has been discovered by \citet{shapley_two_1938} on photographic plates from the Harvard Boyden station in Arequipa, Peru. 
 \citet{shetrone_vltuves_2003} determined the detailed abundances of three red giants in Fornax thanks to high resolution spectra obtained with UVES/VLT. They found moderately low metallicities ranging from $-1.60$ to  $-1.21$. 
A first CaT surveys was performed by \citet{battaglia_dart_2006}  who derived metallicities
ranging for $-0.12$ to $-2.61$ in a sample of 562 stars. 

 \citet{kirby_multi-element_2010} obtained  KECK/DEIMOS spectra of a large sample of stars in Fornax. 
Based on a sample of 675 stars from Fornax with metallicities ranging from $-2.82$ to $-0.03$, they derived a mean metallicity $\rm \langle [Fe/H]\rangle = -1.05$.

In the search for EMP stars in dSph, UVES/ESO high resolution spectroscopy of one low CaT selected star in Fornax was obtained by \citet{tafelmeyer_extremely_2010}.  They found a metallicity  $\rm[Fe/H] = -3.66$  and an [$\alpha$/Fe] enhancement as found in halo field stars. 
 
\citet{lemasle_vltflames_2014}  used FLAMES/VLT  in high-resolution mode to determine the abundances of several $\alpha$, iron-peak and neutron-capture elements in a sample of 47 individual red giant branch stars in Fornax. They found metallicities ranging from --0.3  to --2.68 and [$\alpha$/Fe] following a decreasing ratio as [Fe/H] increases with a "knee" at lower [Fe/H] than for the Milky Way, a characteristic found in other dwarf  galaxies. 
 
\begin{figure}
\centering
\includegraphics[width=10.0cm,clip=true]{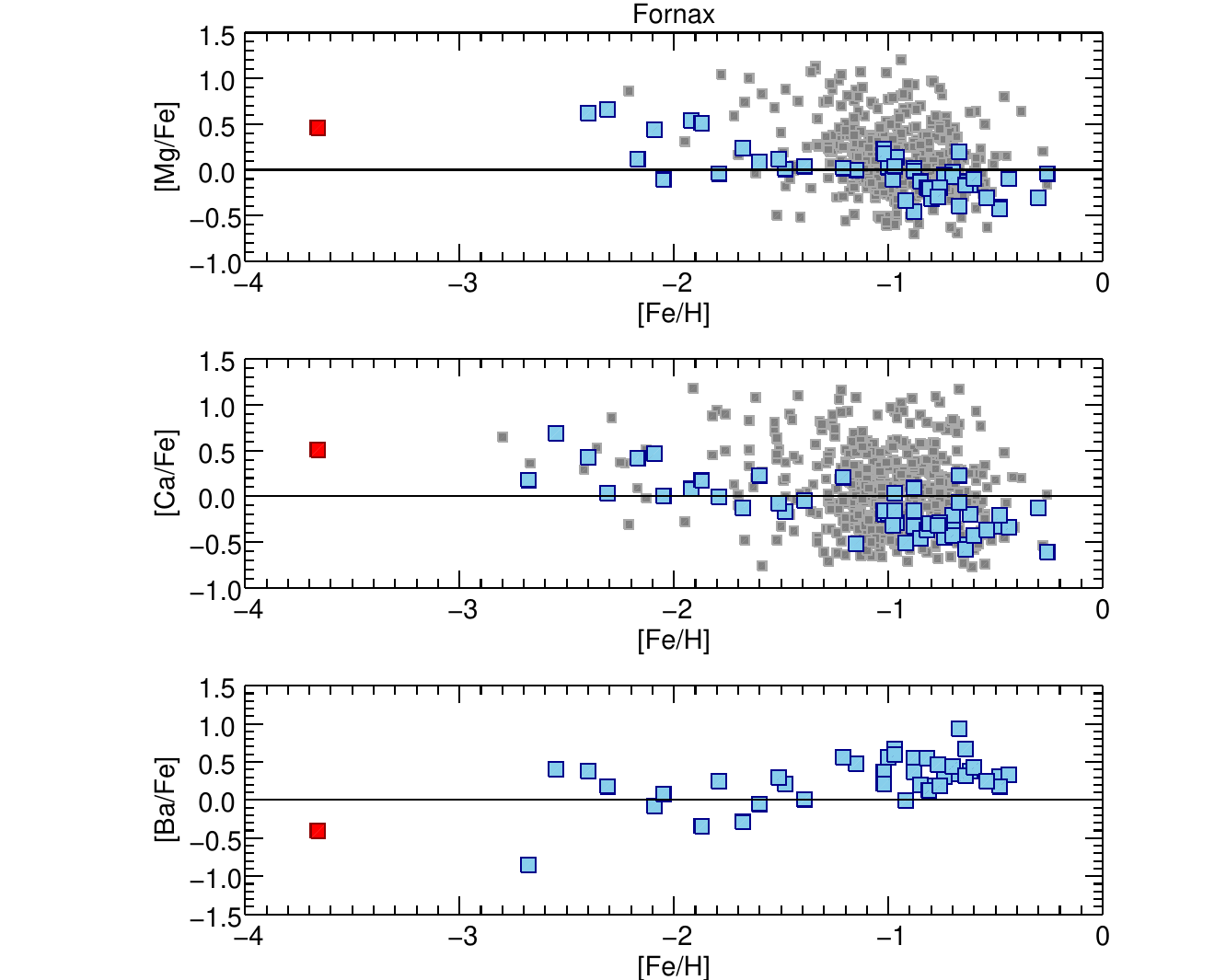}
\caption{\footnotesize Abundance ratios in Fornax. The first two upper panels are the abundances of [Mg/Fe] and [Ca/Fe] as a function of [Fe/H]. Grey symbols are results from low resolution spectroscopy  \citep{kirby_multi-element_2010} . Blue  and red symbols show results based on high resolution spectroscopy \citep{shetrone_vltuves_2003,lemasle_vltflames_2014,tafelmeyer_extremely_2010}. The last panel presents the [Ba/Fe] vs [Fe/Fe] from \citet{shetrone_vltuves_2003},  \citet[][blue symbols]{lemasle_vltflames_2014}   and \citet[][red squares]{tafelmeyer_extremely_2010}.}
\label{Fig:fornax_abundance}
\end{figure}
 Fig. \ref{Fig:fornax_abundance} show the abundance ratios [Mg/Fe], [Ca/Fe] and [Ba/Fe] ratios versus [Fe/H] in a sample of stars belonging to Fornax.  The [Mg/Fe], [Ca/Fe] vs [Fe/H] plots  show the decrease of the  [$\alpha$/Fe] as [Fe/H] increases crossing the solar 
 [$\alpha$/Fe] ratio at around [Fe/H] = --1.5. The red symbol shows the only EMP star found in Fornax discovered by \citet{tafelmeyer_extremely_2010}.
The  Fornax metallicity histogram based on the published data is shown in Fig. \ref{Fig:dSph_mdf}.\\

 \item{} dSph Galaxy Sagittarius (Sgr dSph) \\
  The Sagittarius dwarf galaxy was discovered by \citet{ibata_dwarf_1994} on automatic plate measuring (APM) scans of UK Schmidt Telescope (UKST) BJ and R plates followed by spectroscopic data  using the 3.9\,m Anglo-Australian Telescope (AAT) equipped with the multi-fibre system AUTOFIB.
Many papers  on the determination of the chemical composition of stars in Sagittarius can be found in the literature \citep[see for example][and reference therein]{mucciarelli_chemical_2017}. From these abundance analyses, the most metal poor star has a metallicity 
$\rm[Fe/H] \le -2.5$. 
\citet{bonifacio_abundances_2006} were the first to highlight the existence of EMP stars in Sgr,
\citet{hansen_ages_2018} made a high-resolution analysis  of thirteen   stars  of the main body of Sagittarius and found metallicities ranging from --1 to --3. Among them, they found an extremely  metal-poor stars with $\rm [Fe/H]\simeq -3$.  
Their abundances are similar to what is found metal-poor halo stars. Their most metal poor stars indicates a pure r-process pollution.  
\citet{sestito_pristine_2024} measured the abundances in 12 metal-poor stars in Sgr  down to a metallicity of --3.26, 
using Mike at Magellan. Their metal-poor sample
does not show any decline in [$\alpha$/Fe], implying that at these low metallicities there was no contribution from SNIa.
\citet{sbordone_wide_2020} analysed a CEMP-r/s star with [Fe/H]=--1.55 in this galaxy.
 \citet{chiti_four_2019}  used  Magellan Echellette (MagE)  to obtain low resolution spectra of four stars. They derived the  metallicity using the  Ca{\sc II} K,  Ca NIR  and MgIb strong lines.  They found moderately metal-poor metallicities between $-1.55$ and $-2.25$. 
From medium resolution spectra obtained with the MagE spectrograph on the Magellan-Baade Telescope \citet{chiti_discovery_2020} found 18 metal poor stars with metallicities ranging from $ \rm [Fe/H] = -1.47 ~to -3.06  $.
 Fig. \ref{Fig:Sagittarius_abundance} illustrates the  well known run of abundances ratios [X/Fe] as a function of [Fe/H] for three elements showing clearly the early decrease  of the [$\alpha$/Fe] ratios at much lower metallicities than what is found in the halo of the Galaxy. The variation of the [Ba/Fe] also shows a [Ba/Fe]  remaining solar down to $\rm[Fe/H] \simeq  -2.5 $, another  abundance characteristics found in dwarf galaxies. 
The  Sagittarius metallicity histogram based on the published data is shown in Fig. \ref{Fig:dSph_mdf}.
 \\
 
  \begin{figure}
\centering
\includegraphics[width=12.0cm,clip=true]{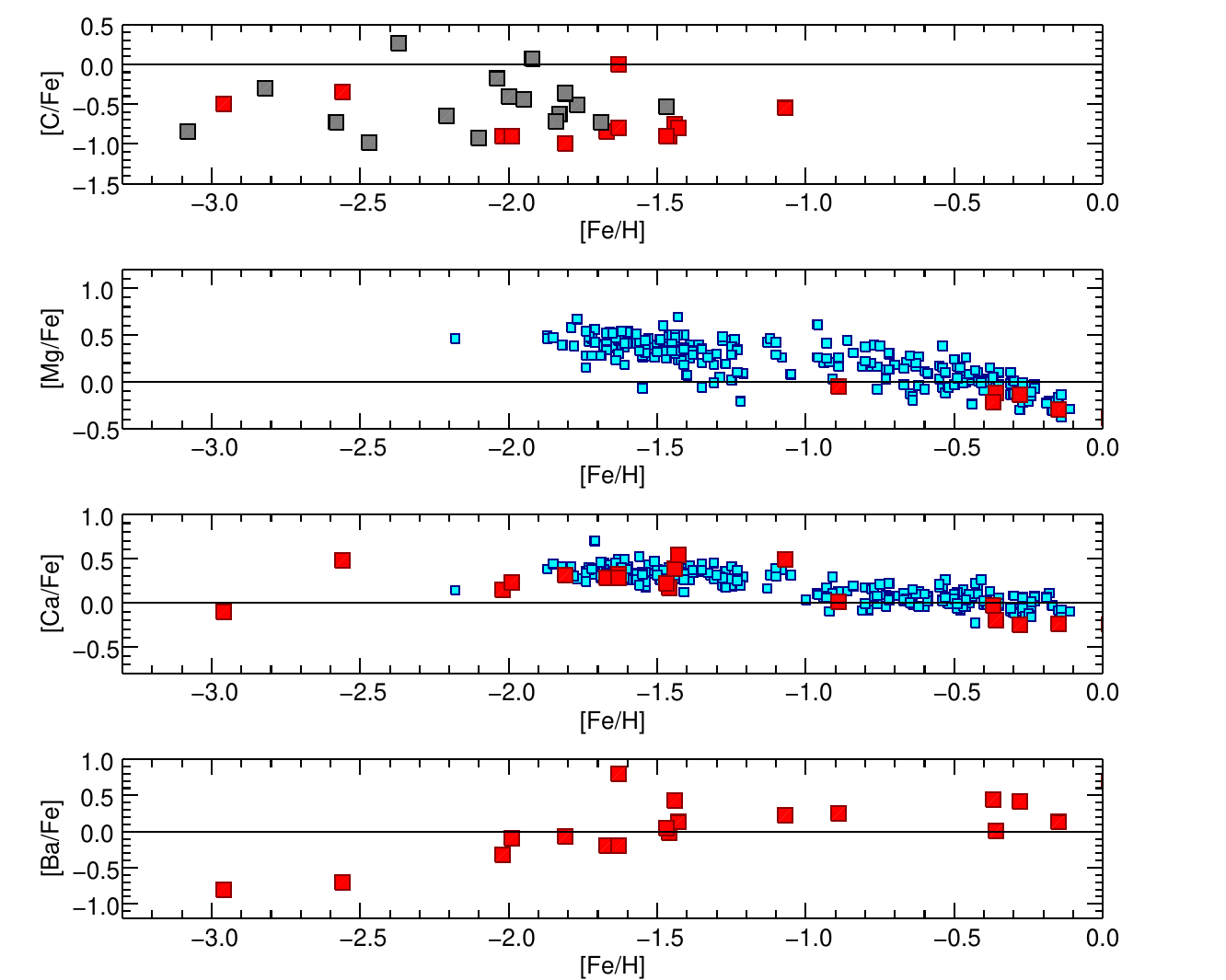}
\caption{\footnotesize Abundance ratios in Sagittarius. The  panels are the abundances of [Mg/Fe],  [Ca/Fe] and [Ba/Fe] as a function of [Fe/H]. A representative sub-sample from \citet{mucciarelli_chemical_2017} (in cyan)  of the existing literature  (medium resolution spectroscopy) covering the known metallicity range found in Sagittarius  data is shown on this plot. Red symbols show the results from \citet{hansen_ages_2018} based on high resolution spectroscopy.  Grey symbols on the upper panel are from \citet{chiti_discovery_2020}.
}
\label{Fig:Sagittarius_abundance}
\end{figure}

 \item{}  dSph Sculptor \\
This  galaxy was discovered by   \citet{shapley_stellar_1938}  on photographic plates from the Harvard Boyden station in Arequipa, Peru.
From low-resolution spectra \citet{tolstoy_using_2001} found that Sculptor's mean metallicity was $\rm[Fe/H] = -1.5$ with a 0.9\,dex metallicity spread.
 From UVES spectra of 5 stars,  \citet{shetrone_vltuves_2003}  determined  metallicities ranging from --1.1 to --1.98. 
\citet{geisler_sculptor-ing_2005}  used high-resolution UVES/VLT spectra to determine abundances of 17 elements in four red giants in Sculptor. Their  [Fe/H] values range from --2.10 to --0.97, confirming previous findings of a large metallicity spread,  and a moderate mean low metallicity.

Thanks to medium resolution Keck DEIMOS, \citet{kirby_multi-element_2009} measured the abundance of Fe, Mg, Si Ca and Ti for a large sample of 388 radial velocity member stars in Sculptor. They found a [Fe/H]  asymmetric distribution with  a mean
$\rm \langle[Fe/H]\rangle = -1.58$
and a long, metal-poor tail, indicative of a history of extended star formation.
From their analysis, they identified a significant fraction of stars with [Fe/H] $<$ --2.0.
They discovered in their sample a star with $\rm [Fe/H] = -3.80\pm 0.28$. 
From a CaT line index based selection of metal-poor stars, \citet{tafelmeyer_extremely_2010}  studied  two stars with UVES/ESO. For both stars, they  found metallicities well below $\rm[Fe/H]=  -3$. 
In particular, one giant in Sculptor at [Fe/H] = $-3.96\pm  0.06$ is  one of the most metal-poor star ever observed in an external galaxy.
\citet{frebel_linking_2010} identified another  EMP star with [Fe/H] = --3.8  thanks to MIKE spectra. 
The very low metallicity  of Sculptor was confirmed  by \citet{starkenburg_extremely_2013}
 who analysed  X-shooter spectra of seven stars finding five stars with $\rm [Fe/H]\le -3$. \\

\citet{jablonka_early_2015} observed five very metal poor candidates using high resolution UVES/ESO spectra.
Four stars appeared to be EMP stars with metallicities ranging from [Fe/H] = --3.22 to --3.88. \\
\citet{simon_chemical_2015}   analysed new and archival high resolution spectroscopy from Magellan/MIKE and VLT/UVES and determined stellar parameters and abundances in a consistent way for each star. Two of the stars in their sample, at [Fe/H] = -3.5 and [Fe/H] = -3.8, are new discoveries from their Ca K survey of Sculptor, while the other three were known in the literature. They confirm that the star Scl 07-50 is  one of the lowest metallicity star identified in an external galaxy, with a measured metallicity of  $\rm[Fe/H] = -4.1$.

\begin{figure}
\centering
\includegraphics[width=10.0cm,clip=true]{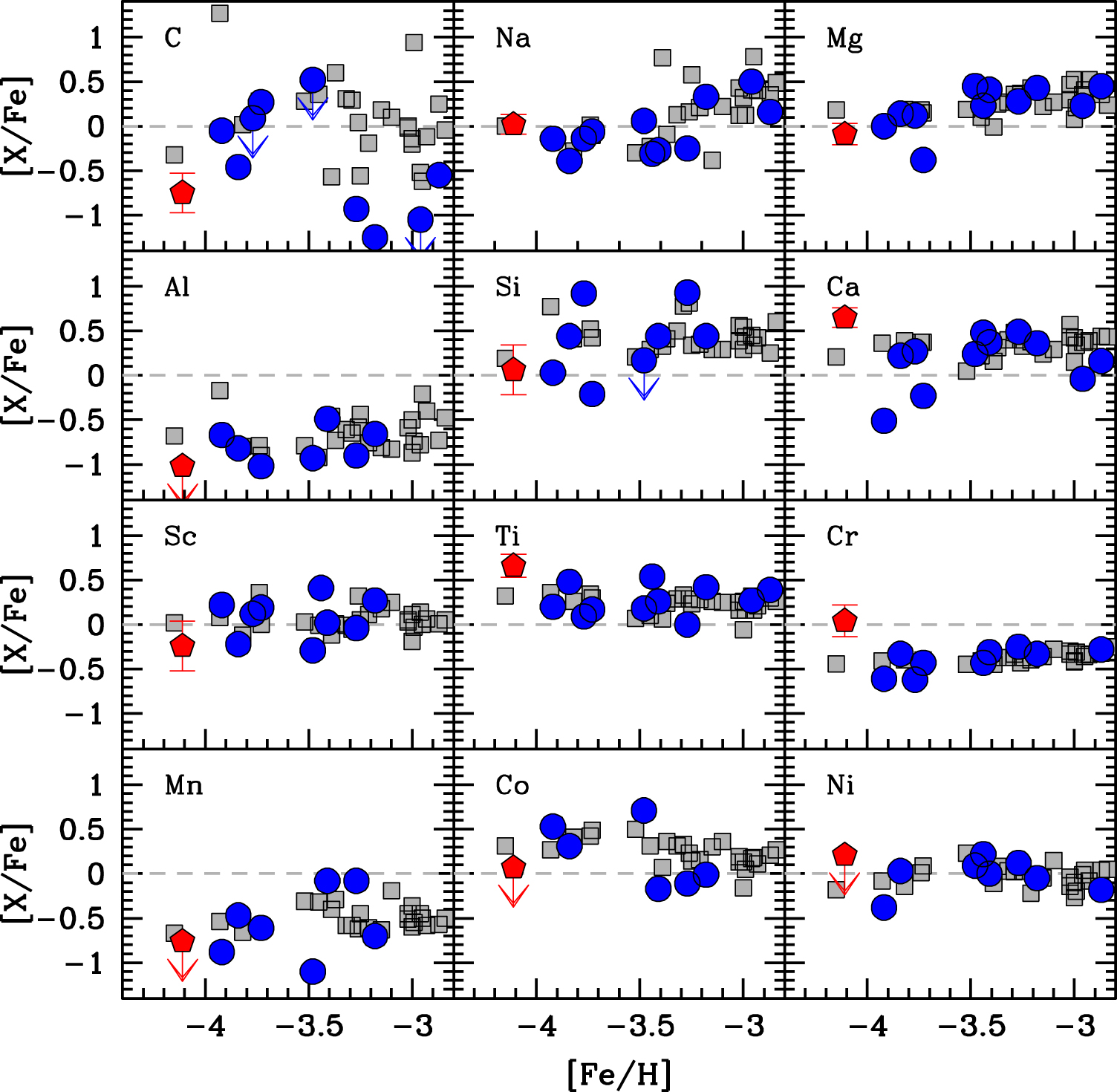}
\caption{\footnotesize  Abundance ratios in the EMP stars of 
Sculptor (from \citealt{skuladottir_zero-metallicity_2021}). }
\label{Fig:sculptor_abundances}
\end{figure}

  \citet{skuladottir_zero-metallicity_2021} reported the detection
  of an ultra-metal-poor star (AS0039) in the Sculptor dwarf spheroidal galaxy based on the analysis of X-Shooter spectra. 
With $\rm [Fe/H]_{LTE} = -4.11$, it  has been considered as   the most metal-poor star discovered in any external galaxy,  until the discovery of a star (LMC-119) with $\rm [Fe/H] =  -4.13 $ by \citet{chiti_enrichment_2024} in the LMC. Contrary to the majority of Milky Way stars at this metallicity, AS0039 is clearly not enhanced in carbon, with $\rm [C/Fe]_{LTE} = -0.75$, and A(C) = +3.60, making it the lowest detected carbon abundance in any star to date.
 Fig. \ref{Fig:sculptor_abundances} shows the abundance ratios measured in the extremely metal poor stars of Sculptor.

 The same team    \citep{skuladottir_tracing_2024} analysed a sample of new and archival data  of stars belonging to Sculptor. In particular, they made   a follow-up   on the most metal-poor star known in this (or any external) galaxy, AS0039, with high-resolution ESO VLT/UVES spectra.
 Their new analysis confirmed its low metallicity with $\rm[Fe/H]= -3.90 \pm 0.15$. They also confirmed that this star is carbon poor with A(C) = 3.6. \\

\item{} dSph Sextans \\
 The dwarf galaxy Sextans was found by \citet{irwin_new_1990} as part of a monitoring program  
 of UK Schmidt Telescope sky survey plates. 
Keck/HIRES was used by \citet{shetrone_abundance_2001} to obtain high resolution spectra of five stars. They derived metallicities ranging from [Fe/H] = --1.45 to --2.19.  They also measured a low [$\alpha$/Fe], a  characteristic already found in other dwarf galaxies.

\citet{aoki_chemical_2009} determined the chemical abundances of six extremely metal-poor ($\rm [Fe/H] < -2.5$) stars in the Sextans dwarf spheroidal galaxy thanks to  high resolution spectra obtained with HDS/Subaru. They found metallicities ranging from --2.66 to --3.10.

\citet{tafelmeyer_extremely_2010} analysed high resolution spectra of  2 stars obtained with UVES/VLT and HRS at Hobby-Eberly Telescope. They found very 
low metallicities $\rm[Fe/H]=-2.93$ and --2.94. 

From the analysis of UVES/VLT high resolution spectra, \citet{lucchesi_homogeneity_2020} identified two new extremely metal poor stars in Sextans with 
metallicities [Fe/H] = --2.95  and --3.01. They also reanalysed the stars studied by \citet{aoki_chemical_2009} and confirmed their low metallicity.  \citet{theler_chemical_2020} made a study of a sample of 81 stars using FLAMES/VLT spectra. They found a wide metallicity range from [Fe/H]= --3.2 to --1.5. From their chemical abundances derived with high accuracy on a sufficiently large number of stars, they concluded  that Sextans 
showed  a plateau in [$\alpha$/Fe] at $\simeq$ +0.4  followed by a decrease above $\rm[Fe/H] \simeq -2$.

\citet{aoki_chemical_2020} analysed three stars with Subaru HDS and two stars taken from the ESO VLT archive and determined [Fe/H]  ranging between $-2.8$ and $-3.06$. 
They found that the distribution of [$\alpha$/Fe] abundance ratios of the Sextans dwarf galaxy stars is slightly lower than the average of the values of stars in the Galactic halo.

As it can be seen in Sextans, but also in all the dwarf galaxies, stars with low [$\alpha$/Fe] have been found. These stars may be the source of low [$\alpha$/Fe] stars found 
in the Milky way halo. As suggested by \citet{bonifacio_topos_2018} the low [$\alpha$/Fe] metal-poor stars of the Galaxy  could have formed in low-mass dwarf galaxies and subsequently
been accreted to the Milky Way Halo, with a different chemical composition at a given metallicity, as has been claimed by \citet{hayes_disentangling_2018}, see also Sec.\,\ref{sec:alpha-poor-stars}.

\end{itemize}

\subsection {Ultra faint Dwarf Galaxies (UFD)}   \label{sec:UFDgal}
Ultra faint dwarf galaxies are the smallest structures known to be dominated by dark matter up to now, and they are also the oldest and most metal-poor systems \citep{simon_faintest_2019}. They play an important part in the search for the most metal poor stars.
In this section we review the current knowledge on abundance results obtained thanks to the detailed analysis of  medium and high resolution spectra, 
summarised in  Table \ref{tab:UFD_EMP}.

\begin{table}[thb]

\caption{ Main characteristics of the nearby faint dwarf galaxies. 
 $\rm \langle [Fe/H] \rangle$ is the mean metallicity, N stars is the number of stars with a metallicity estimate from spectroscopy.
  Range is, for each galaxy,  the known range in metallicity derived with spectroscopic measurements.
  The column EMP gives the number of EMP stars found in each UFD galaxy. 
 }
\centering
\begin{tabular}{llrcll}
      Name	  &   $\rm \langle [Fe/H] \rangle$	&   N     & Range     &  EMP &  Comments	     \\
\hline
 AquaII        &   -2.30	 &     9	&  -3.13 to -1.87      &  3   &              \\
 AquaIII         &   -2.61      &    10        &  -3.05 to -1.95      &  2   &              \\
 BooI           &   -2.55      &    42        &  -3.84 to -1.70      & 10   &    1 CEMP-no    \\
 BooII         &   -1.79      &    14        &  -2.93 to -1.70      &  6   &           \\
 BooIII         &   -2.10      &    10        &  -3.20 to -0.9       &  1   &  only [Fe/H]        \\
 CnVI            &         -1.98      &   184        &  -3.5 to -0.4        & 10   &    1 CEMP-no      \\
 CnVII            &        -2.21      &     8        &  -2.80 to -1.17      &  1   &  1 outlier  with high  [Sr/Fe]        \\
 CarII          &   -2.44      &     9        &  -3.53 to -2.20      &  3   &  1 outlier with low  [Sc/Fe]        \\
 CarIII          &   -1.80      &     2        &  -3.87  and  -2.27   &  1   &            \\
 CetI              &   -1.90      &    54        &  -2.81 to -0.9       &  1   &            \\
 CetII           &   2.30             &      1        &     -2.29            &  1   &   low [Sr/Fe] and [Ba/Fe]    \\
 ColI         &   -2.14      &    10        &  -3.36 to -1.65      &  1   &            \\
 Coma         &   -2.60      &    10        &  -3.38 to -2.12      &  4   &  low Sr and Ba abundances          \\
 Cra            &   -1.70      &     6        &  -2.10 to -1.59      &  0   &   GC or UFD ?        \\
 CraII            &   -1.98      &    53        &  -2.89 to -1.41      &  1   &          \\
 GrusI             &   -1.42      &     7        &  -2.55 to -0.56      &  0   &   2 stars with low Ba and Ba        \\
 GrusII            &   -2.51      &    11        &  -2.94 to -1.55      &  1   &   stars with high [Mg/Ca]         \\
 Her            &   -2.41      &    30        &  -3.17 to -1.45      &  6   &   very low Ba in several stars         \\
 HoroI         &   -2.76      &     6        &  -2.83 to -2.36      &  1   &   stars with  solar [$\alpha$/Fe]         \\
 HydI          &   -2.52      &    26        &  -3.18 to -1.40      &  9   &    1 CEMP star (+3 dex)   \\
 LeoI          &   -1.43      &   936        &  -3.13 to -0.22      &  5   &            \\
 LeoII         &   -1.62      &   307        &  -3.05 to -0.80      &  3   &            \\
 LeoIV          &   -2.54      &     5        &  -3.19 to -1.8       &  2   &            \\
 LeoVI           &   -2.20      &     9        &  -2.69 to -1.9       &  0   &            \\
 PegIV       &   -2.63      &     5        &  -3.29 to -2.00      &  3   &            \\
 PheII       &   -2.51      &     5        &  -2.89 to -2.0       &  1   &            \\
 PisII       &   -2.45      &     4        &  -2.60 to -2.10      &  0   &  1 CEMP-no with   [Ba/Fe]=-1.1    \\
 RetII         &   -2.46      &    16        &  -3.18 to -2.02      &  5   &  1 r-II star, 1 CEMP-r star      \\
 RetIII         &              &     3        &  -3.24 to -2.32      &  2   &           \\ 
 Seg 1            &   -2.72      &     8        &  -3.78 to -1.50      &  3   &  1 CEMP-s, 3 CEMP-no,    \\
 Seg 1           &              &               &                  &     &  halo like [$\alpha$/Fe]     \\
 Seg2            &   -2.22      &    10        &  -2.85 to -1.33      &  1   &           \\
 TriII            &   -2.24      &     6        &  -2.92 to -1.40      &  2   & 1 star with low Na and Ni          \\
 TucII           &   -2.23      &    11        &  -3.34 to -1.60      &  5   &     \\
 TucIII     &   -2.42      &    22        &  -2.97 to -2.15      &  2   &                     \\
 TucIV         &   -2.49      &     8        &  -3.40 to -2         &  3   &                     \\
 TucV           &   -2.17      &     4        &  -3.55 to -2.46      &  1   &  1 CEMP-no, 1 r-I         \\
 UMaI           &   -2.18      &    17        &  -2.75 to -1.14      &  0   &                       \\
 UMaII          &   -2.47      &     9        &  -3.2 to -1.04       &  2   & halo like [$\alpha$/Fe]                    \\
 UMi          &   -2.13      &   225        &  -3.64 to -0.49      & 12   &                  \\
 Wil1              &   -2.10      &     8        &  -2.10 to -0.8       &  0   &                \\

 \hline
    \end{tabular}
    \label{tab:UFD_EMP}
\end{table}

\begin{itemize}

\item{}Aquarius II   \\

Only one spectroscopic study of this galaxy can be found in the literature. 
Spectra of 12 stars  from the galaxy  Aquarius II,  a UFD galaxy recently discovered by \citet{torrealba_at_2016},  have been obtained by 
\citet{bruce_spectroscopic_2023} using IMACS at the Magellan-Baade telescope. They derived a  very low metallicity of $\rm[Fe/H] = -2.57 \pm 0.17$, with the most metal-poor of the sample at a metallicity of $\rm -3.13 \pm 0.73$. 
 The  Aquarius II metallicity histogram based on the published data is shown in Fig. \ref{Fig:Aquarius_histogram}.
\\

\begin{figure}
\centering
\includegraphics[width=5.5cm,clip=true]{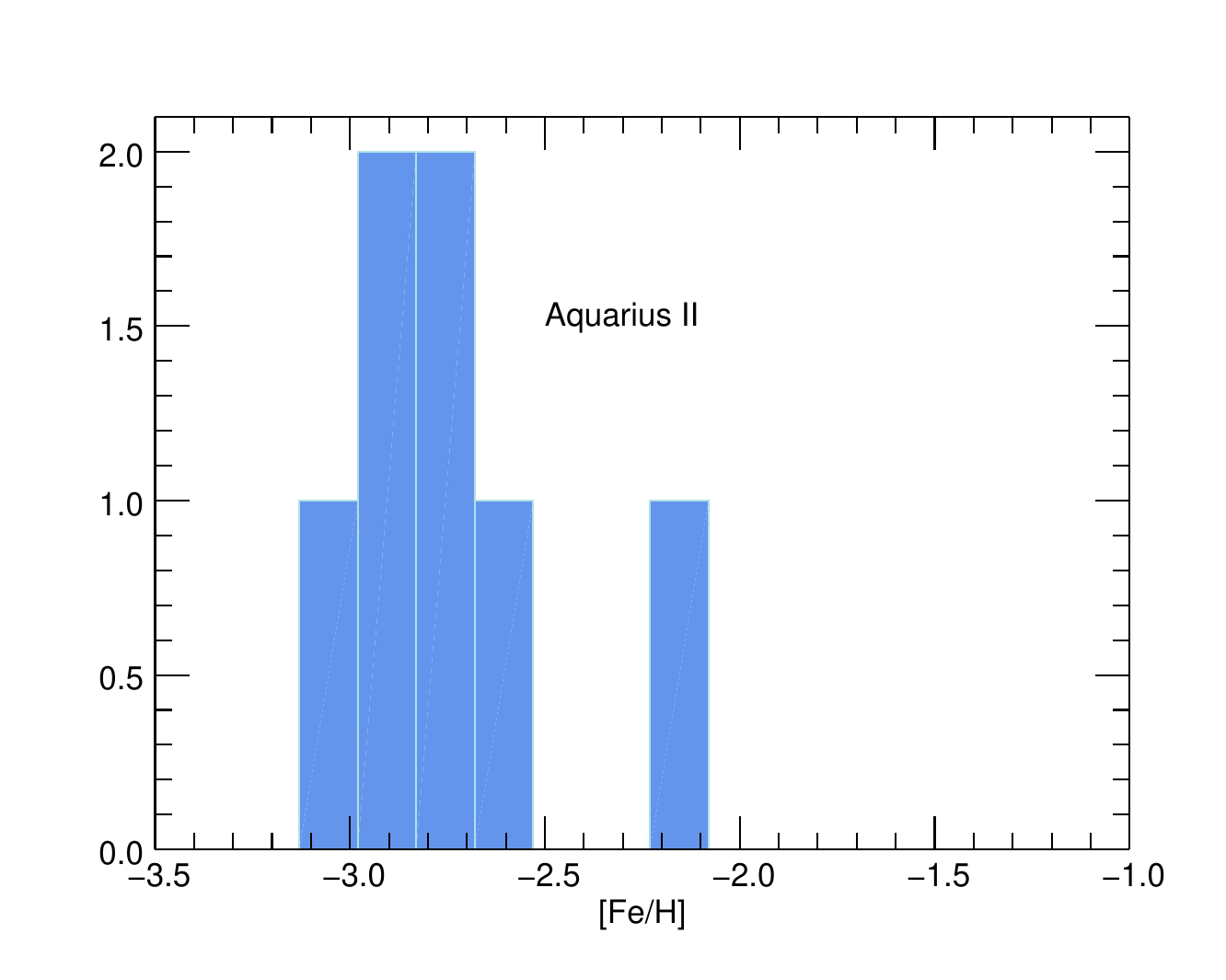}
\includegraphics[width=5.5cm,clip=true]{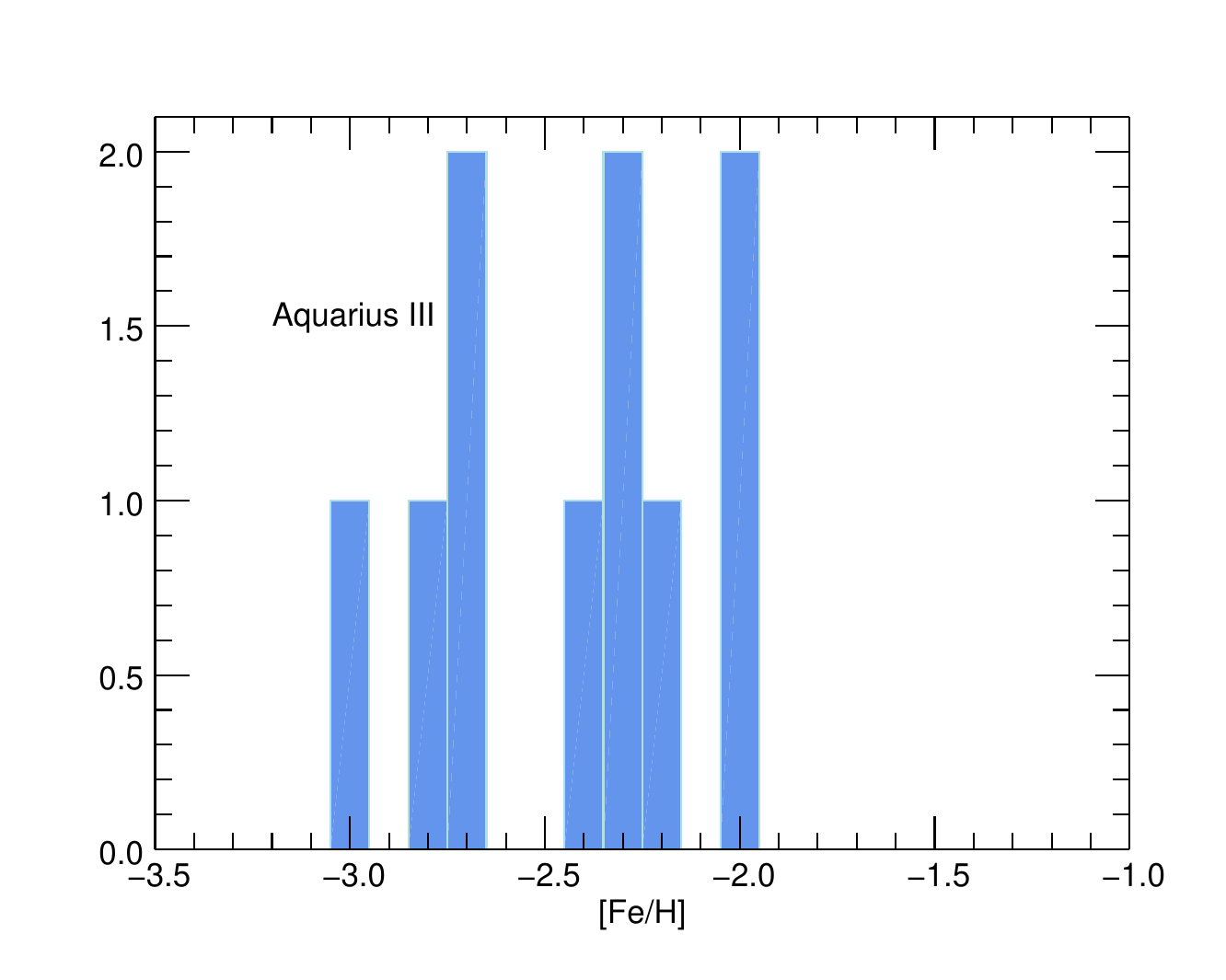}
\caption{\footnotesize Metallicity histogram of Aquarius II and Aquarius III. In these galaxies several  EMP stars have been detected.}
\label{Fig:Aquarius_histogram}
\end{figure}

\item{}Aquarius III   \\

Aquarius III has been identified as an ultra-faint Milky Way satellite galaxy in the second data release of the DECam Local Volume Exploration survey by \citet{cerny_discovery_2025}. Follow-up imaging from DECam confirmed it as an UFD galaxy. Keck DEIMOS medium resolution spectroscopy has been used to identify 11 member stars
with a low metallicity. Metallicities based on the calcium triplet strength of the six brightest members led to a low metallicity ${\rm [Fe/H] = -2.61 }$ with a metallicity spread  $\simeq 0.46$.
Higher  resolution ({$\rm R \simeq 4100$}) Magellan/MagE spectroscopy of the brightest star of the sample revealed that this star is a carbon enhanced $\alpha$-enhanced star with a metallicity  around $\rm [Fe/H] =-3.0$. 
The  Aquarius III metallicity histogram based on the published data is shown in Fig.\,\ref{Fig:Aquarius_histogram}.
\\

\item{}Bootes I \\

Bootes I has been found by \citet{belokurov_faint_2006}   in a systematic search for stellar overdensities in 
the north Galactic cap using Sloan Digital Sky Survey data. From the analysis of the  colour magnitude diagram, they concluded that the galaxy contained a single and   low metallicity stellar population of stars.  
Low resolution spectroscopy confirmed that Bootes I was very metal poor with 
metallicity of [Fe/H] $\simeq$ --2.50  \citep{martin_keckdeimos_2007}.
Medium resolution spectroscopy of 16  red giants, identified as members from radial velocity measurements have
been obtained by \citet{norris_abundance_2008} at a moderate resolution of R=5000 allowing to estimate the metallicity using the strong  \ion{Ca}{ii} K lines as a metallicity proxy. 
Assuming $\rm[Ca/Fe] \simeq +0.3$, they found a large abundance range of $\rm\Delta [Fe/H] \simeq 1.7$  in their sample,  with one star having $\rm[Fe/H] =-3.4$ .  \\
Several  high resolution spectroscopic studies have followed the medium resolution work of \citet{norris_abundance_2008} to study in detail Boo I  (\citealt{feltzing_evidence_2009, norris_chemical_2010, lai_feh_2011,gilmore_elemental_2013,ishigaki_chemical_2014, frebel_chemical_2016}). 
The first high resolution study was done by \citet{feltzing_evidence_2009} with the analysis of  7 stars of Boo I  thanks to spectra obtained 
 at the High Resolution Echelle Spectrometer (HIRES) at Keck I. 
They derived a  [Fe/H] ranging from $-1.98$ to $-2.90$.  They could also measure 
the abundances of Mg, Ca  for all their stars and Ba for 6 stars. 
 \citet{norris_chemical_2010} analysed 16 stars belonging to Bootes I and derived metallicities ranging from --1.93 to --3.66. 
\citet{lai_feh_2011},  studied 25 stars belonging to the galaxy using the low resolution imaging spectrometer (LRIS)  at Keck  Observatory. 
The spectra had a resolution of R $\simeq$ 1800. From their study, they determined a low average metallicity of $\rm \langle[Fe/H]\rangle = -2.59$.
They used  the n-SSPP  SEGUE pipeline to derive  stellar parameters together with [Fe/H],   
[$\alpha$/Fe] 
and [C/Fe] abundances.
Their sample includes   3 extremely metal-poor stars, among them  a  star  with [Fe/H]= --3.8.

 This star was also  identified as EMP by the high resolution spectroscopic study by \citet{norris_boo-1137extremely_2010} of the star Boo-1137, who found a metallicity of $\rm [Fe/H] = -3.7$ in good agreement with \citet{lai_feh_2011}.
\citet{gilmore_elemental_2013} obtained  Flames/UVES spectra of 7 giant stars with a resolution of R=47000. From their analysis,  they found that  the stars  cover [Fe/H] from $-3.7$ to $-1.9$ and include a CEMP-no star with $\rm [Fe/H] = -3.33$. 
\citet{ishigaki_chemical_2014} analysed HDS-Subaru spectra  of 6 stars  
and derived the abundances of several elements. They confirmed that the star \# Boo-094  already 
 observed by  \citet{norris_chemical_2010} had a very low metallicity with $\rm [Fe/H]= -3.18$. 

 The metallicity histogram of Boo I (Fig.\,\ref{Fig:Boo_histogram})  built from the  spectroscopy  literature data shows a peak at  a metallicity bin of $\rm [Fe/H]  = -2.5$ and  a low metallicity tail down to $\simeq -3.8$. 

\begin{figure}
\centering
\includegraphics[width=5.5cm,clip=true]{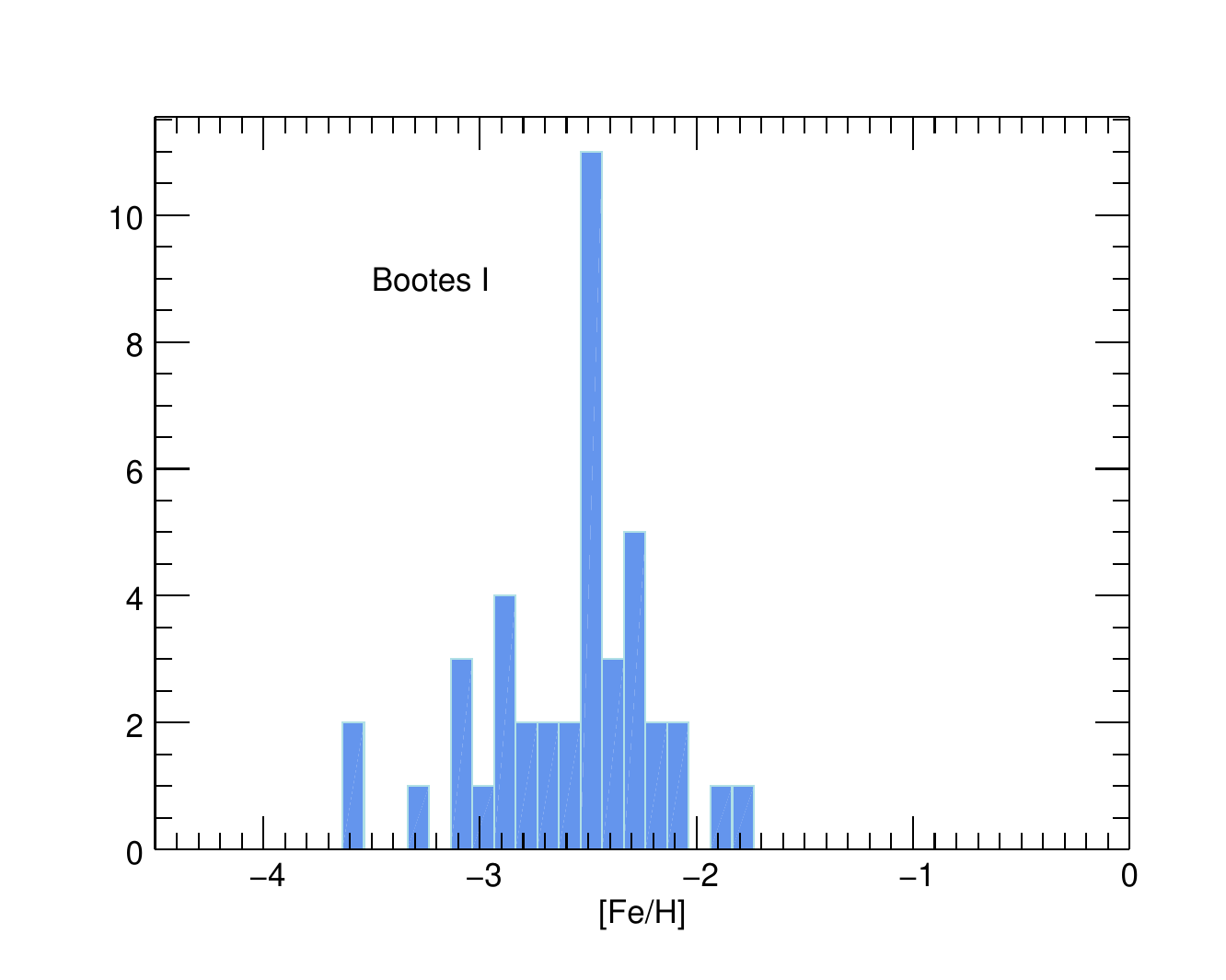}
\includegraphics[width=5.5cm,clip=true]{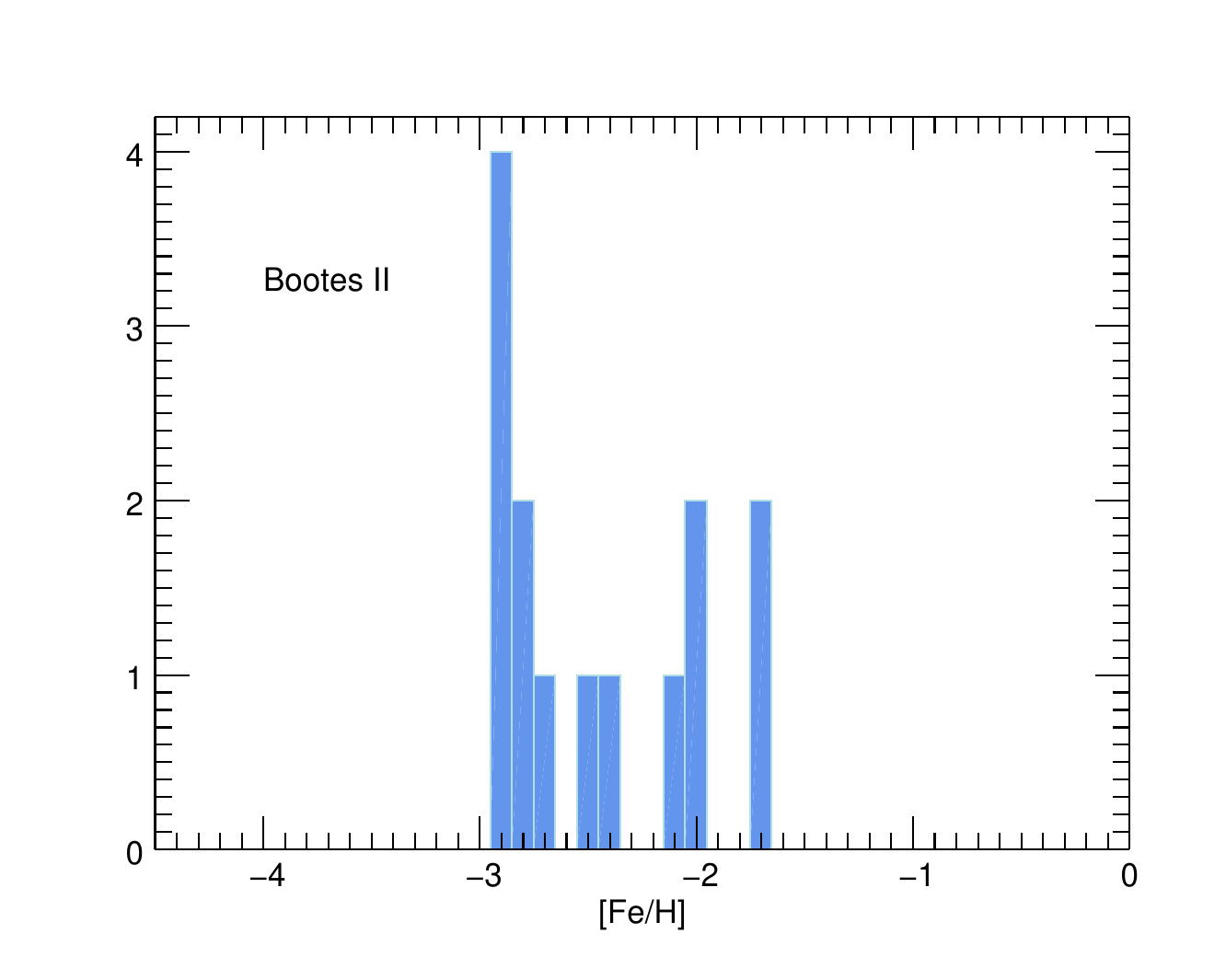}
\includegraphics[width=5.5cm,clip=true]{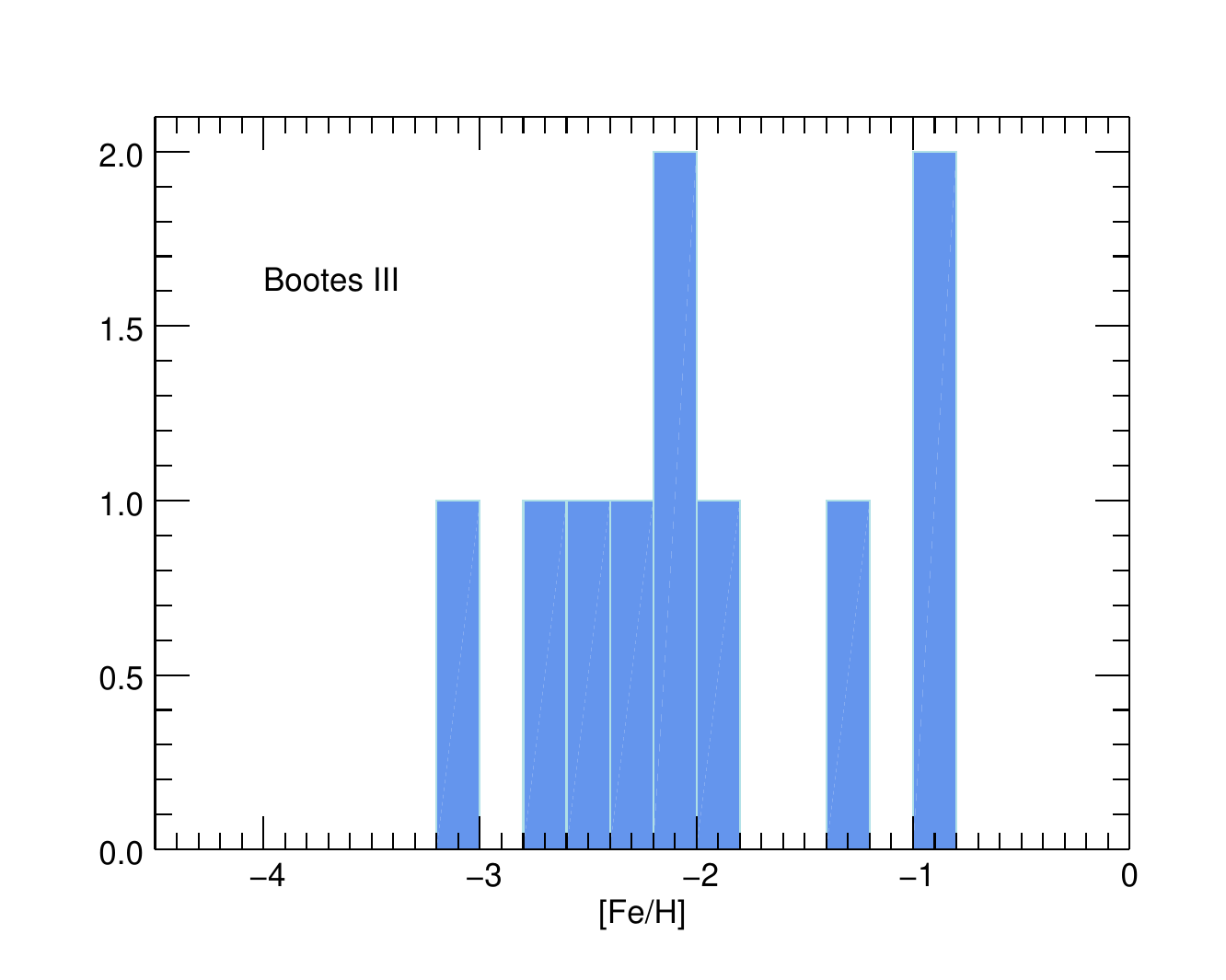}

\caption{\footnotesize Metallicity histogram of Bootes I Bootes II and Bootes III.  The Bootes I  histogram  is showing an  extended low metallicity end.}
\label{Fig:Boo_histogram}
\end{figure}

\citet{frebel_chemical_2016} analysed  2 stars from Bootes I  from high S/N high resolution  MIKE/ Magellan spectra and computed the  detailed abundance ratios of 2 stars.  
In particular, they could derive the abundance Na, Mg , Al Si,  Ca Sc, Ti, Ce, Mn, CO, Ni , Zn , Sr Ba  for the EMP star Boo-980. 

\citet{mashonkina_formation_2017}  reanalysed high resolution spectra from archive data  and computed non-local thermodynamic equilibrium (NLTE) abundances of up to 10 chemical species of 8 stars in Boo I. They confirmed the large range of [Fe/H] found in Boo I.  Their results show  also evidence for a decline in $\alpha$/Fe with increasing metallicity in Boo I  that is most probably due to the ejecta of type Ia supernovae.\\

\item{} Bootes II \\

\citet{walsh_pair_2007}  reported the discovery of a stellar overdensity in the Sloan Digital Sky Survey Data Release 5, lying at an angular distance of only 1.5 degree from  the UFD galaxy  Boo I. From the isochrone fitting of the color-magnitude diagram of the overdensity region, they conclude   that this region had the signature of an old (12 Gyr) metal-poor ($\rm[Fe/H] \simeq -2.0$) population. 
Boo II was confirmed as dwarf galaxy by \citet{koch_spectroscopic_2009} using deep photometry from INT.
They also obtained medium resolution spectra (R=3600)  for 17 stars in the field of Boo II using GMOS-N spectrograph, among them  5 stars identified as Vr members.
They interpreted these results as a spectroscopic detection of the Bootes II system. 
From the spectra centred on the near-infrared Ca{\sc II}  triplet (CaT),  they estimated  a mean stellar metallicity $\rm[Fe/H] = -1.79 $ with a dispersion of 0.14\,dex. 
This determination relied on an old calibration of the Ca {\sc II} triplet, which was later revised.
 \citet{koch_chemical_2014} made a detailed chemical analysis of the brightest confirmed member star in Boo II using Keck/HIRES and derived a
very low metallicity of $\rm [Fe/H] = -2.93$ using an updated Ca triplet calibration. 
They also found a high [$\alpha$/Fe] ratio that is compatible with the $\alpha$-enhanced plateau value of the  Galactic halo.
\citet{ji_high-resolution_2016}  obtained MIKE high resolution spectra (resolving power ranging from 22,000 to 38,000) of the 4 brightest confirmed  stars in Boo II. Due to the faintness of the targets ($\rm g \simeq 19.5$), the S/N ratio of the combined spectra at 6000\,\AA\ did not exceed 30/pixel.  They found  metallicities [Fe/H]  between --2.63 and --2.92. 

X-Shooter/VLT spectra of two members of 
Boo II were analysed by \citet{francois_abundance_2016} who confirmed its low metallicity  ([Fe/H]= --2.98 and --3.08) for both stars.
They also found an $\rm [\alpha/Fe]$ overabundance and a low [Ba/Fe] characteristic of the  Galactic halo population. [Fe/H] was determined using the measurement of the equivalent widths of several Fe I lines found in the spectra, thanks to the higher resolution of X-Shooter compared to DEIMOS and the relatively high S/N ratios of the spectra.   

  Recently, \citet{bruce_spectroscopic_2023} presented a chemical and kinematic analysis of the largest sample of member stars for Boo II. Using IMACS/Magellan R=11,000 spectra, they determined the metallicities of the stars using the equivalent widths of the CaT absorption lines and converting them into [Fe/H] following the conversion relation of \citet{carrera_near-infrared_2013}.  They obtained metallicities ranging from $\rm[Fe/H]=-2.00 \pm 0.74$ to --2.93$\pm 0.13$. Using the MCM sampler {\tt emcee} \citep{foreman-mackey_emcee_2013} they concluded that their results were compatible with a mean metallicity of  --2.71 with a low dispersion of 0.1\,dex.
    The  Bootes II metallicity histogram based on the published data is shown in Fig. \ref{Fig:Boo_histogram}.
 \\ 
  
\item{} Bootes\,III \\

Boo\,III was discovered as a stellar
overdensity, spanning $\sim 1^\circ$  on the sky, nearly coincident with the Styx stellar stream \citep{grillmair_four_2009}. 
From a low resolution spectroscopic follow-up with Hectospec at the MMT,  \citet{carlin_kinematics_2009}  found metallicities ranging from  $\rm [Fe/H]=-0.9$ to $-3.3$.   The  Bootes III metallicity histogram based on the published data is shown in Fig. \ref{Fig:Boo_histogram}.
\\ 

\item{} Canes Venatici I \\ 

Canes Venatici I was discovered in 2006 by \citet{zucker_new_2006} as a stellar overdensity in the north Galactic cap using the SDSS DR 5.
The first deep color-magnitude diagrams of the Canes Venatici I (CVn I) dwarf galaxy were provided by \citet{martin_deep_2008} from observations with the wide-field Large Binocular Camera on the Large Binocular Telescope. Interestingly, their analysis revealed a dichotomy in the stellar populations of CVn I which harbours an old ($\le$ 10 Gyr), metal-poor ([Fe/H] $\simeq -2.0$), and spatially extended population along with a much younger, more metal rich, and spatially more concentrated population.
\citet{kirby_multi-element_2010} determine the abundances of Fe (using the CaT/[Fe/H] triplet calibration) and several $\alpha$ elements in a sample of 171 stars using medium-resolution spectra (R $\simeq$ 7000) that was obtained with Keck/DEIMOS and found metallicities ranging from --1.0 to --3.3. 

Using the same Keck/DEIMOS medium-resolution spectra that were obtained by \citet{kirby_multi-element_2010}, \citet{vargas_distribution_2013} could  determine the abundance of Fe, Mg, Ca in several stars of this galaxy.  

\citet{francois_abundance_2016}  analysed X-Shooter/VLT spectra of two stars in CVn I  and derived the metallicity by a direct measurement of the equivalent widths of several Fe I lines.  
They found metallicities [Fe/H] = --2.18 and --2.52. They  could also determine the abundances of several other elements (Mg, Ca, Sr and Ba). 

\citet{yoon_identification_2020}  found a group of CEMP-no stars in CVn I. They also  identified the star SDSS 1327+3335 as the first likely  CEMP-no  with a  metallicity of 
$\rm [Fe/H] =-3.38$ in the CVn I dwarf satellite galaxy, using a classification parameter determination methodology based on maximum likelihood spectral matching on Multi-object Double Spectrographs (MODS/LBT) spectra (R=1700 and S/N of the order of 20). 
The  CVn I  metallicity histogram based on the published data is shown in Fig. \ref{Fig:CVn_histogram} revealing the existence of several EMP stars. \\

\begin{figure}
\centering
\includegraphics[width=5.5cm,clip=true]{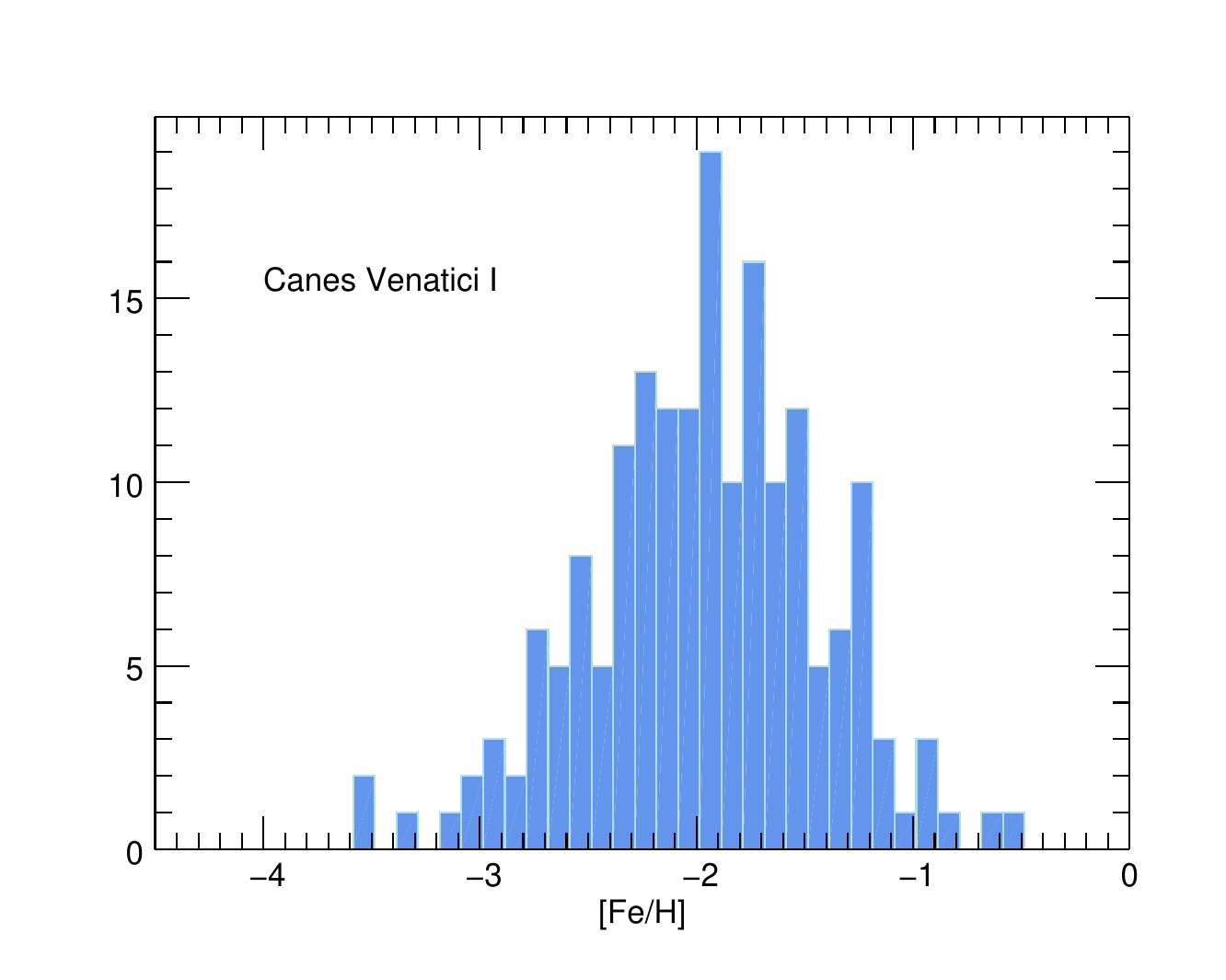}
\includegraphics[width=5.5cm,clip=true]{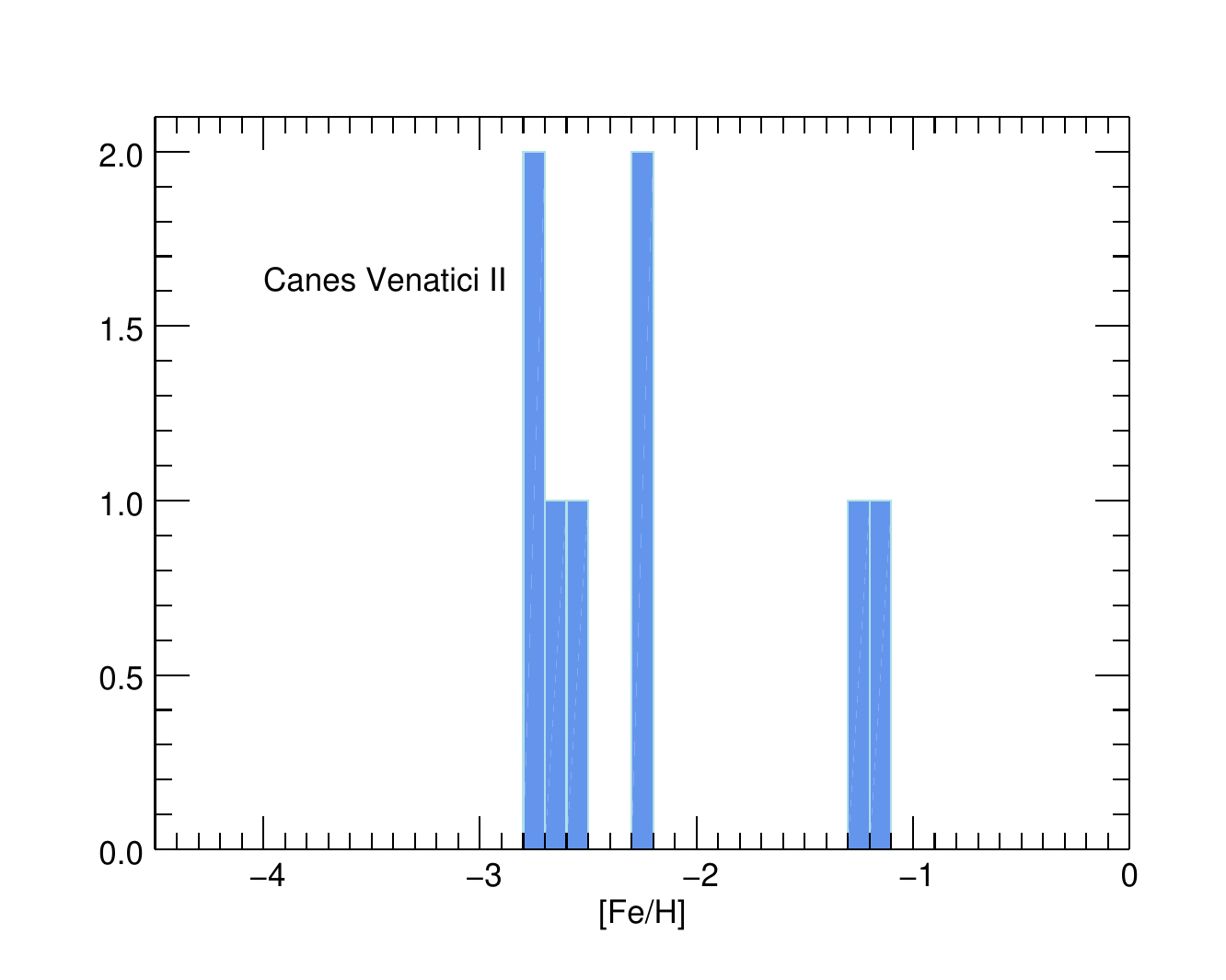}

\caption{\footnotesize Metallicity histogram of Canes Venatici Ultra faint dwarf galaxies.}
\label{Fig:CVn_histogram}
\end{figure}

\item{} Canes Venatici II \\
The galaxy Canes Venatici II (CVn II) is one of the four UFD galaxies discovered by \citet{belokurov_cats_2007} in the Sloan Digital Sky Survey. 
It has also been discovered by \citet{sakamoto_discovery_2006} in the Sloan Digital Sky Survey and named as
SDSS J1257-3419. 
Follow-up spectroscopic observations were performed in 2008 by \citet{kirby_uncovering_2008} who analysed 16 stars. They used DEIMOS on the Keck II telescope to obtain spectra at R $\simeq$ 6000 over a spectra range of roughly 6500-9000 \AA. They derived a mean metallicity of [Fe/H] = -2.19 $\pm$ 0.05  with a dispersion of 0.58\,dex. \citet{vargas_distribution_2013} computed the [$\alpha$/Fe] ratios in eight stars of this galaxy and found an increase of the [$\alpha$/Fe] as metallicity decreases with a solar ratio at [Fe/H] $\simeq -1.30$ to reach on average an [$\alpha$/Fe] overabundance of $\simeq$ 0.5\,dex  at [Fe/H] $\simeq -2.50$. The distribution of [Ca/Fe] and [Ti/Fe] abundance ratios tends to point towards the presence of a significant scatter at low [Fe/H]. The metallicity was later revised by \citet{kirby_universal_2013} who found [Fe/H] = -2.12 with a dispersion of $\simeq$ 0.59\,dex.

\citet{francois_abundance_2016} obtained X-Shooter/VLT spectra of one star in CVn II.  Using the equivalent widths of several Fe I lines,  
they could determine a metallicity of $\rm [Fe/H] = -2.58$.
They also found  in this star a very high strontium abundance of [Sr/Fe] = +1.32   and a low  upper limit $\rm [Ba/Fe] \le -1.28$.
The  CVn II  metallicity histogram based on the published data is shown in Fig. \ref{Fig:CVn_histogram}. Data are needed to better define the metallicity distribution.
\\

\item{}   Carina II  and Carina III \\

\begin{figure}
\centering
\includegraphics[width=5.5cm,clip=true]{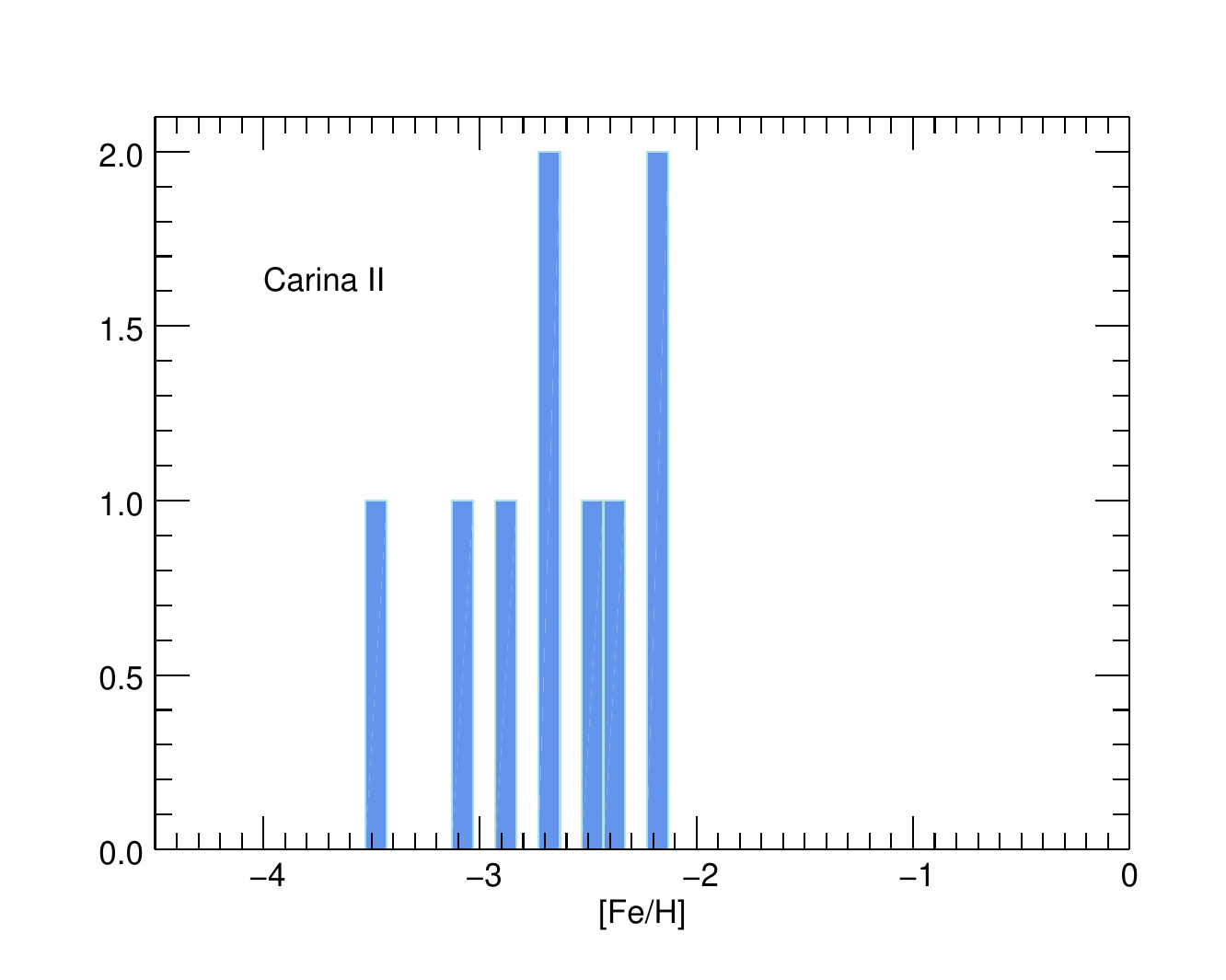}
\includegraphics[width=5.5cm,clip=true]{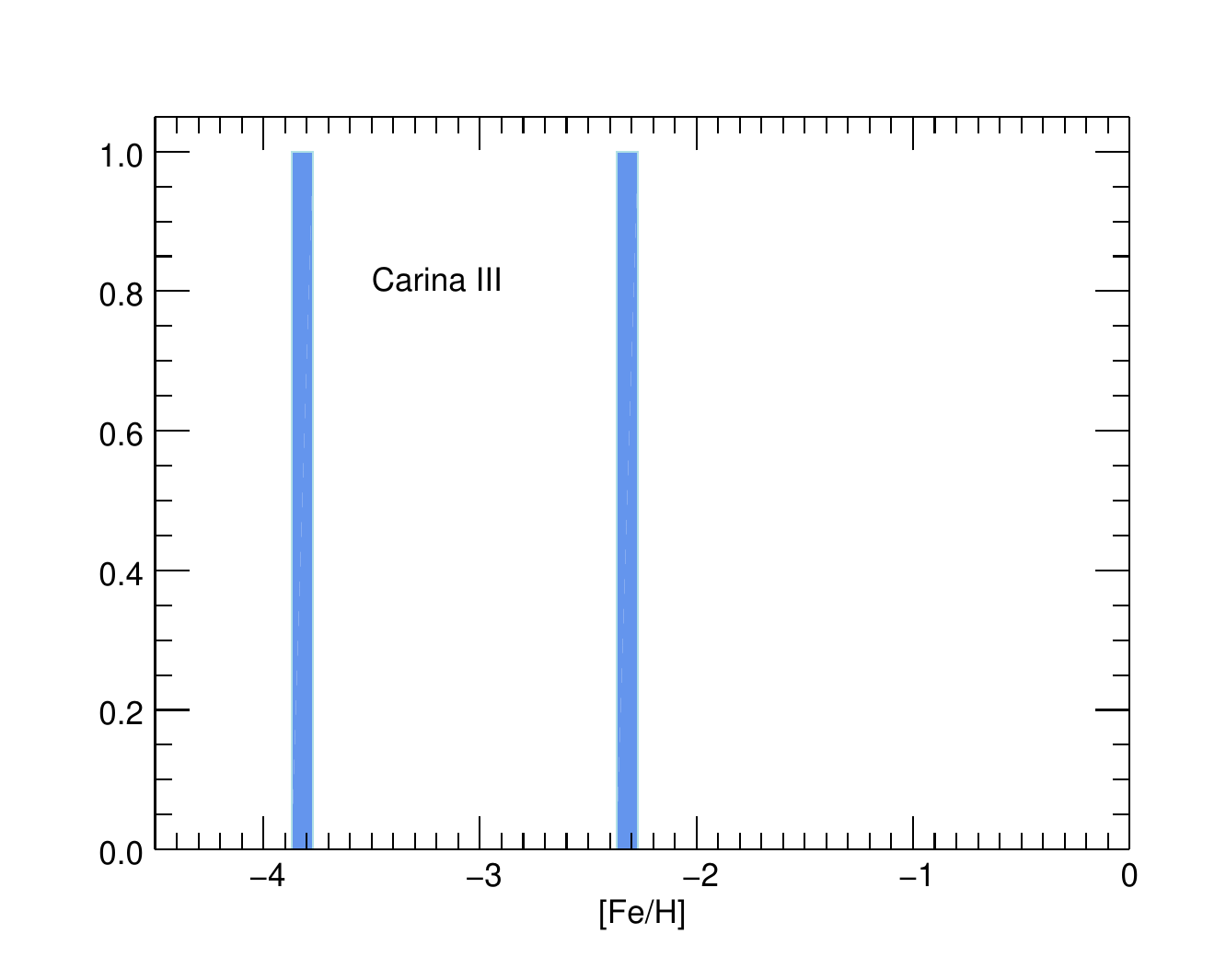}

\caption{\footnotesize Metallicity histogram of Carina II and III Ultra faint dwarf galaxies.}
\label{Fig:CarII_III_histogram}
\end{figure}

Carina II and Carina III  were discovered by \citet{torrealba_discovery_2018} in the vicinity of the Large Magellanic Cloud  in data from the Magellanic Satellite Survey (MagLiteS). MagLiteS covers a previously unexplored region of the Magellanic neighbourhood outside of the  Dark Energy Survey (DES)  footprint. The survey uses the Dark Energy Camera (DECam; \citealt{flaugher_dark_2015}) on the Blanco 4m telescope at Cerro Tololo Inter-American Observatory. 
\citet{ji_detailed_2020}  made the first detailed abundance analysis of nine stars in Car II and two stars in Car III using the high resolution spectrograph MIKE installed on the Magellan telescope. Using a combination of equivalent widths and spectral synthesis to measure the abundances of individual lines, they found [Fe/H] ranging from  --2.21  to --3.53 for the Car II sample and  [Fe/H] = --2.27  and --3.87  for the two Car III stars. 
The stars in Car II and Car III  mostly display abundance trends matching those of other similarly faint dwarf galaxies: enhanced  [$\alpha$/Fe] ratios declining at higher metallicity, iron-peak elements matching the stellar  Galactic halo, and unusually low neutron-capture element abundances.
 The  Car II and Car III  metallicity histograms based on the results  of \citet{ji_detailed_2020}  are shown in Fig. \ref{Fig:CarII_III_histogram}.
\\

\item{}   Cetus I \\
\begin{figure}
\centering
\includegraphics[width=5.5cm,clip=true]{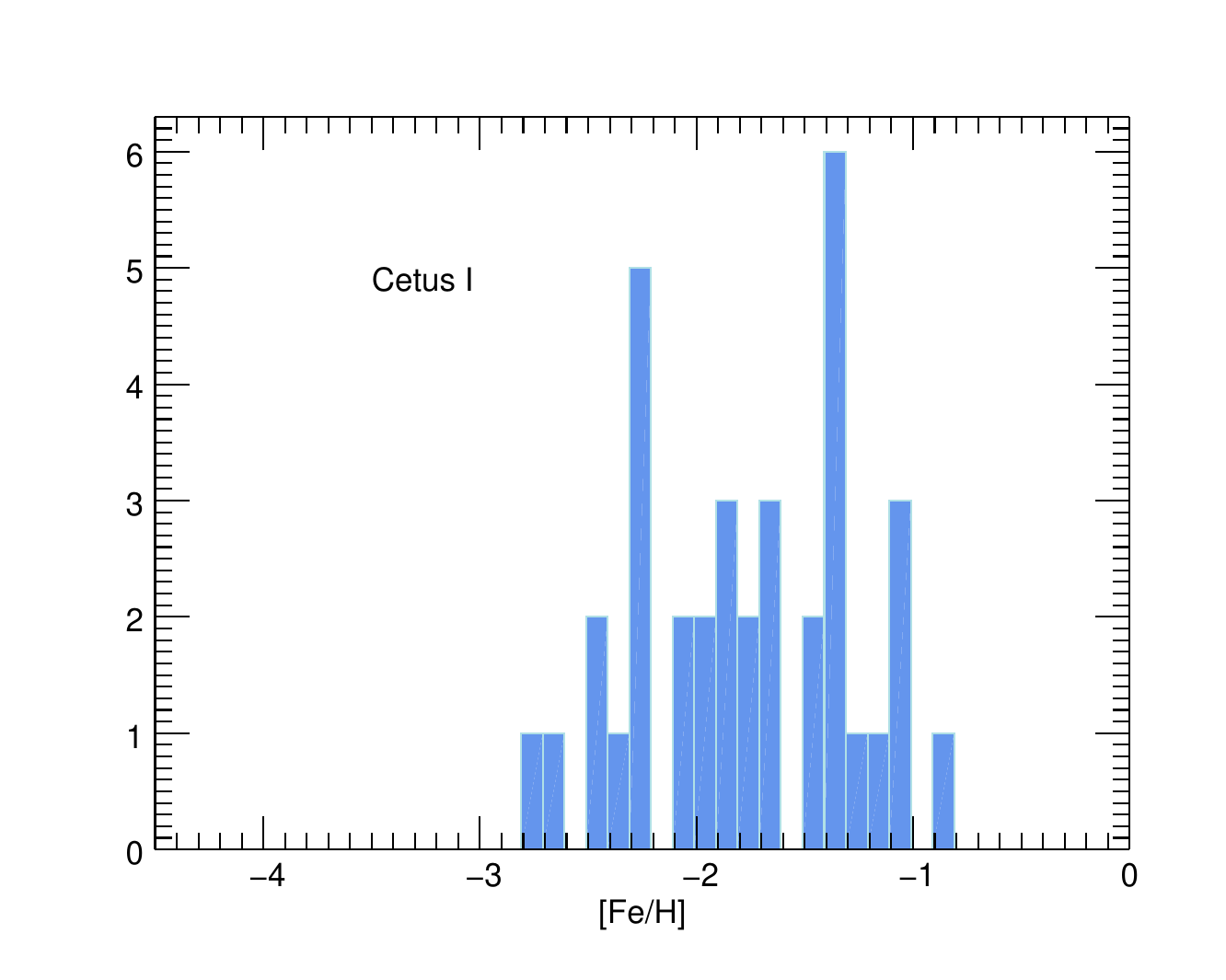}
\includegraphics[width=5.5cm,clip=true]{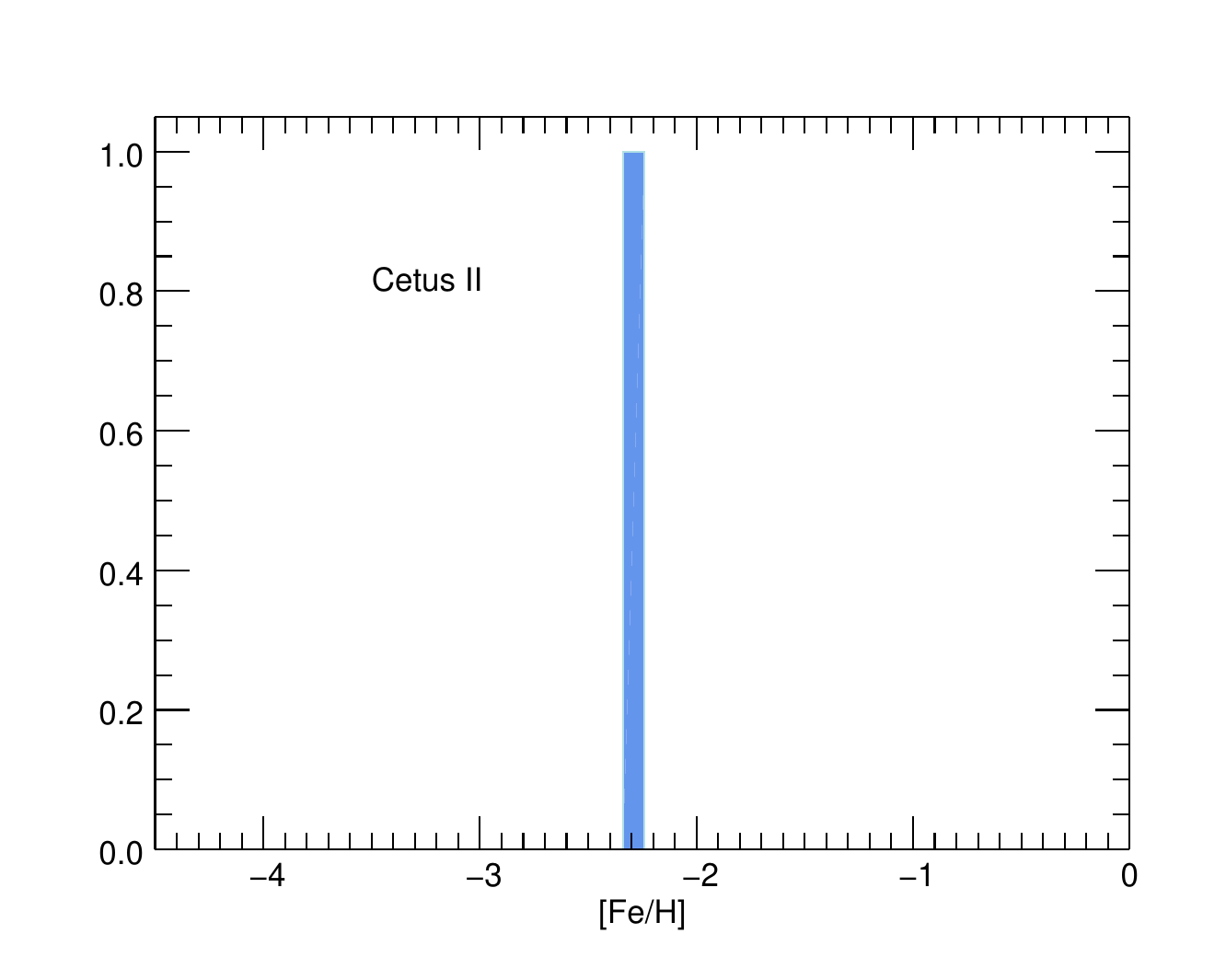}

\caption{\footnotesize Metallicity histograms of  Cetus I and Cetus II.  }
\label{Fig:Cet_histogram}
\end{figure}

Cetus I was discovered  by \citet{whiting_new_1999} by  a visual examination of the fields  covered by the ESO-SRC and SERC Equatorial surveys of the southern sky. From the analysis of the locus of the giant branch, they derived a metallicity of $\rm[Fe/H] = -1.9 \pm 0.2$. 
 The first spectroscopic study of individual stars in Cetus was
conducted by \citet{lewis_inside_2007}, using Keck/DEIMOS data of
$\simeq$70 stars selected from the RGB of the galaxy. They confirmed a population consistent with a metallicity  $\rm [Fe/H] \simeq -1.9$.

\citet{taibi_stellar_2018}  made VLT/FORS2 MXU spectroscopic observations in the region of the near infrared  Calcium triplet for a sample of 80  RGB stars. From the computed [Fe/H] values, they found   that Cetus  I had a significant metallicity spread  with a median value   $\rm [Fe/H] = -1.71$ and a standard deviation $= 0.45$\,dex.   The Cet I  metallicity histogram based on the results  of \citet{taibi_stellar_2018}  is shown in Fig. \ref{Fig:Cet_histogram}.
\\

  \item{}   Cetus II \\
  
Cetus II was discovered by \citet{drlica-wagner_eight_2015} in the analysis of the combined data set from the first two years of  the Dark Energy Survey (DES) covering 
$\rm\simeq 5000 deg^2$ of the south Galactic cap.

From deep Gemini GMOS-S g, r photometry  data of the ultra-faint dwarf galaxy candidate Cetus II,  \citet{conn_nature_2018}  concluded that their results strongly support the picture that Cetus II is not an ultra-faint stellar system in the Milky Way halo, but made up of stars from the Sagittarius tidal stream. \\
Based on medium-resolution spectroscopy of the Cet II field obtained with the Magellan/IMACS spectrograph  , \citet{webber_chemical_2023} identified a set of likely Cet II member stars centred at a velocity of $\rm V_{helio}=-82$\,km/s.
From the analysis of high-resolution spectral data  obtained for J0117, a bright giant  with the MIKE echelle spectrograph,
they  show that this star is a metal-poor $\rm [Fe/H] = -2.29$, $\alpha$-enhanced ($\simeq + 0.4$) star with low abundances of the neutron capture elements ($\rm [Sr/Fe] = -2.10$ and $\rm [Ba/Fe] = -2.23$), following the trends seen for the chemical analysis of other UFD galaxy stars. Another peculiar feature of this star is a  high [K/Fe] abundance of +0.81.  The Cet II  metallicity histogram  shown in Fig. \ref{Fig:Cet_histogram} contains the single metallicity measurement from \citet{webber_chemical_2023}.
\\

\item{}   Columba I \\

\begin{figure}
\centering
\includegraphics[width=5.5cm,clip=true]{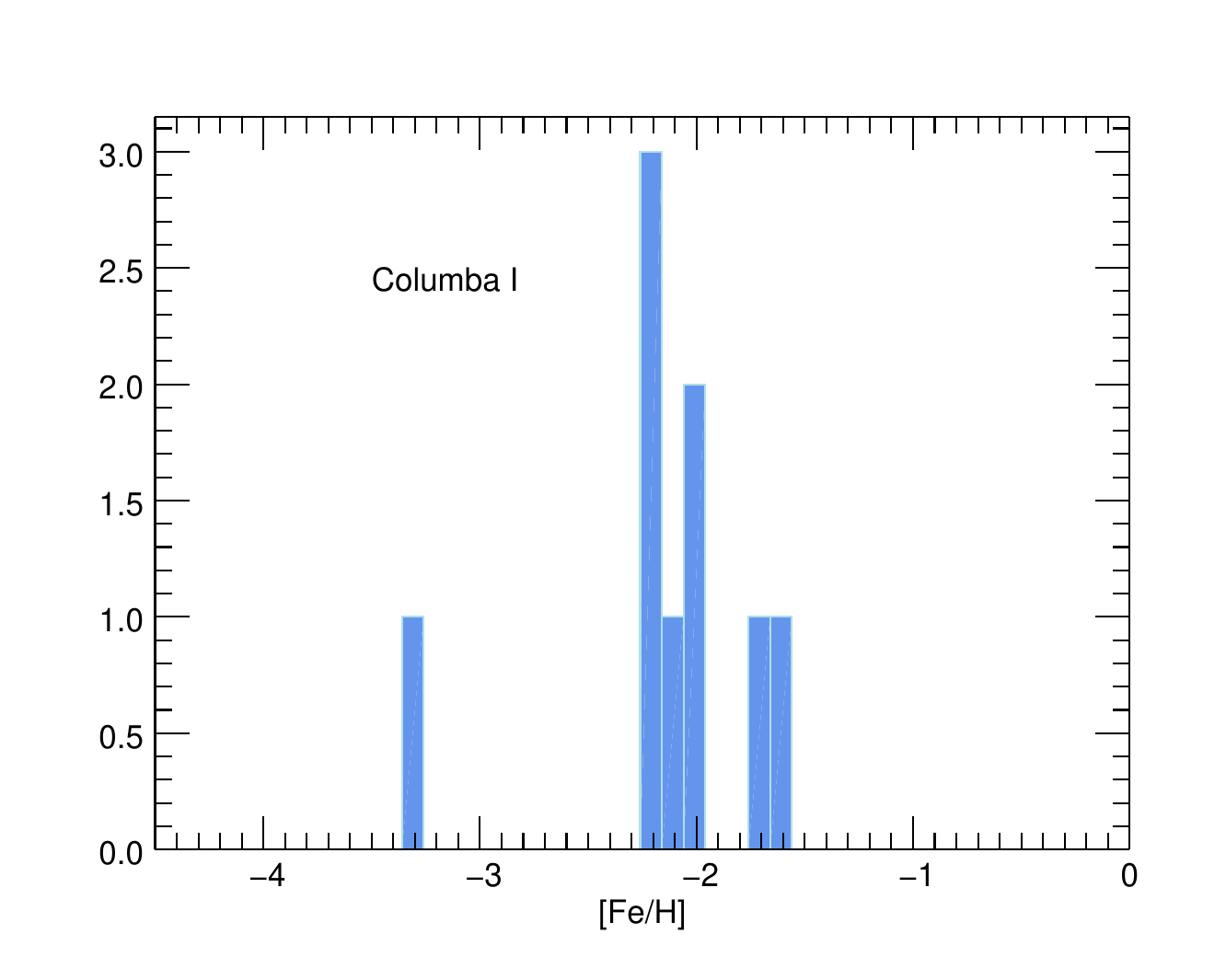}
\includegraphics[width=5.5cm,clip=true]{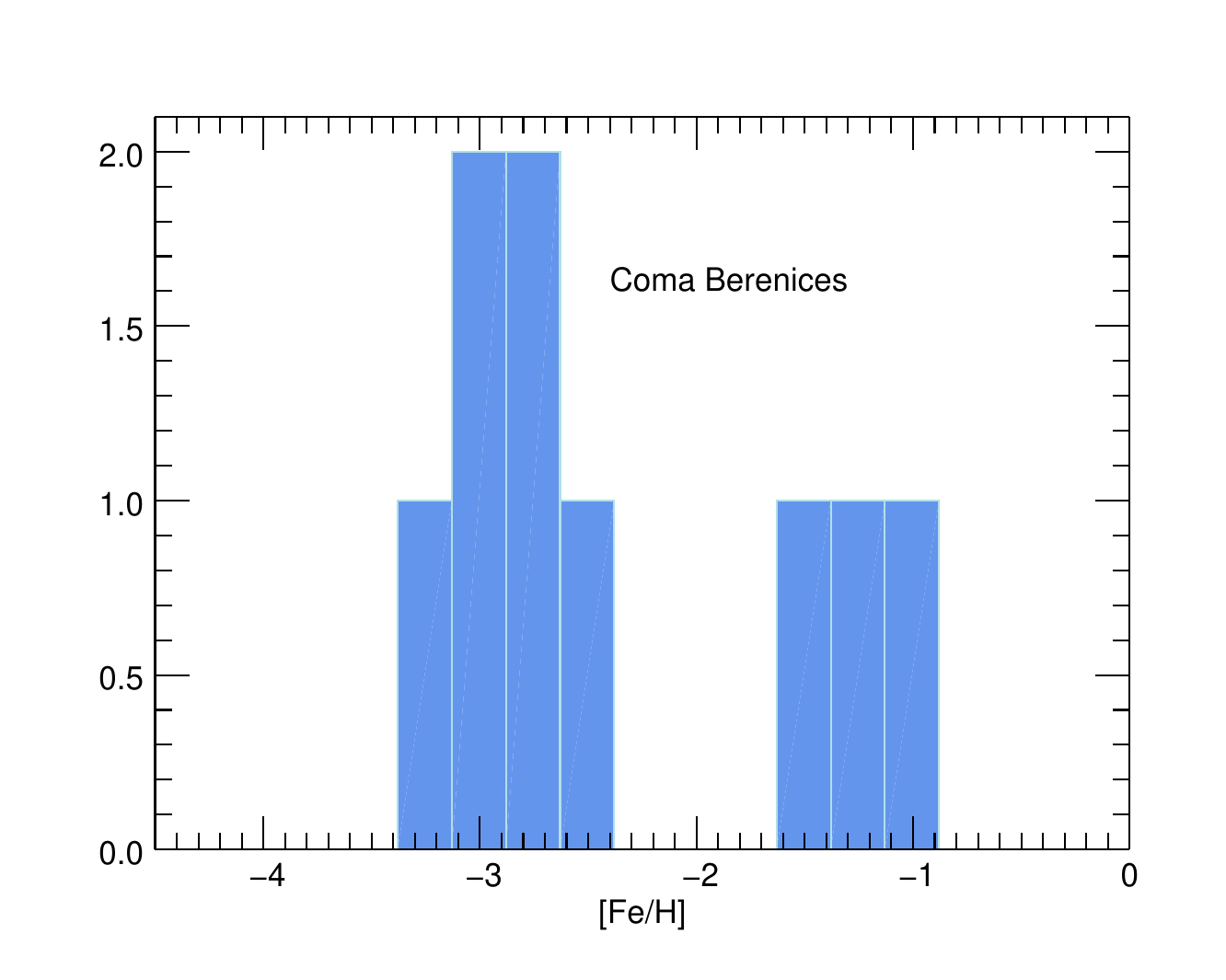}
\caption{\footnotesize Metallicity histogram of  Columba I and Coma Berenices  }
\label{Fig:ColI_histogram}
\end{figure}

Columba I has been discovered by \citet{drlica-wagner_eight_2015}  in the second year of optical imaging data from the Dark Energy Survey (DES).
The authors found that Columba I  is isolated from the other new DES systems and is likely not associated with the Magellanic system.
Using a combination of Gaia DR2 astrometric measurements, photometry, and new FLAMES/GIRAFFE intermediate-resolution spectroscopic data in the region of the near-IR Ca{\sc II} triplet lines,  \citet{fritz_gaia_2019} determined the metallicity of six likely  Col I members and found [Fe/H] ranging from --1.9 to --3.3. 
Note that the brightest target of their sample has a $g$ magnitude of 19.77, a very faint object for high resolution spectroscopy. 
The Col I  metallicity histogram   is shown in Fig. \ref{Fig:ColI_histogram}.
\\
  
 \item{}  Coma Berenices \\
 
 The galaxy Coma Berenices (Coma Ber) is one of the five  UFD galaxies discovered by \citet{belokurov_cats_2007} in the Sloan Digital Sky Survey along with follow-up
 deep  photometry made at the Subaru telescope with the Suprime-Cam camera. From the CMD, they conclude that this galaxy was metal poor compatible with a single 
 population with $\rm [Fe/H] \simeq -2$. 
Keck DEIMOS spectroscopy was done by  \citet{simon_kinematics_2007} to measure the velocity dispersion of Com Ber. They also measured the metallicity thanks to the measure of the CaT triplet absorption lines.  From the analysis of 59 member stars, they confirmed the low metallicity of the Galaxy with $\rm [Fe/H] =-2.00 \pm 0.07$.

  \citet{frebel_high-resolution_2010} obtained Keck/HIRES spectra for three stars in Coma Ber. Thanks to  a  spectral  resolving power of R = 37,000 a wavelength range from 4100 to 7200\, \AA , a S/N ratio of the order of 25-30 at 5000\,\AA, they could make a detailed abundance analysis based of many elements. They found a metallicity [Fe/H] ranging 
 from --2.31 to --2.88. From a  comparison with Milky Way  halo stars of similar metallicities, their results revealed a substantial agreement between the abundance patterns of the ultra-faint dwarf galaxies and the MW halo for the light, $\alpha$, and iron-peak elements (C to Zn).
 They found  extremely low  abundances of neutron-capture elements (Sr to Eu)  compared to the abundances found in the  Milky Way stars at the same metallicity.\\
From the analysis of  Keck/DEIMOS spectra,  \citet{vargas_distribution_2013} computed the [$\alpha$/Fe] ratios in nine stars of this galaxy and found a decreasing ratio as the metallicity of the star increases similar to what they found for CVnI.

 The Coma  metallicity histogram   is shown in Fig. \ref{Fig:ColI_histogram}.
\\

\item{Crater} \\

Crater has been  discovered by \citet{belokurov_atlas_2014} from the ATLAS ESO VST survey  \citep{shanks_vst_2013} associated to  deep imaging with the WHT 4m telescope.
It has also been discovered the same year by \citet{laevens_new_2014} from the Pan-STARRS1 Survey 2  supplemented by deep photometry with ESO/MPG 2.2 m telescope. Their estimate of the metallicity  is $\rm [Fe/H] < -1.8$  according to  \citet{belokurov_atlas_2014} and $\rm [Fe/H] =  -1.9$
according to \citet{laevens_new_2014}. Their conclusion diverged on the nature of this compact object: \citet{belokurov_atlas_2014} concluding it is an UFD  while 
\citet{laevens_new_2014} favouring the hypothesis of Crater being a globular cluster. 
\citet{bonifacio_chemical_2015} obtained X-Shooter spectra of two giant stars belonging to Crater and derived metallicities $\rm [Fe/H] = -1.73$ and $\rm [Fe/H]=-1.67$. 
They also measured the abundances of several other elements. 
From the presence of a well developed blue plume and sub-giant branch consistent
with a population of 2.2 Gyr, while the bulk of the population implies an age of 7 Gyr,
they  favoured the hypothesis of Crater being a dwarf galaxy.
This was also supported by the rather large difference in radial velocity  found for their two stars.

\begin{figure}
\centering
\includegraphics[width=5.5cm,clip=true]{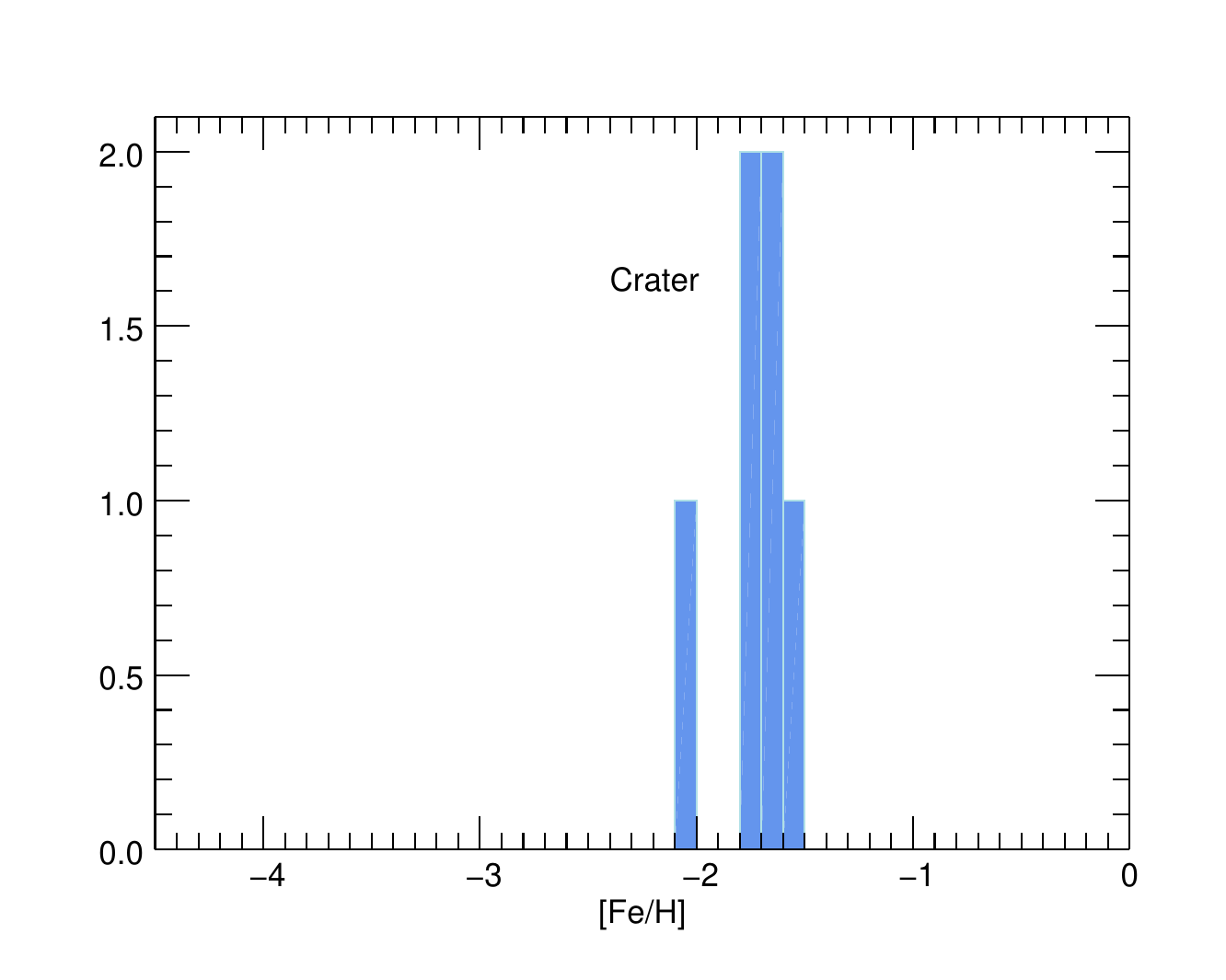}
\includegraphics[width=5.5cm,clip=true]{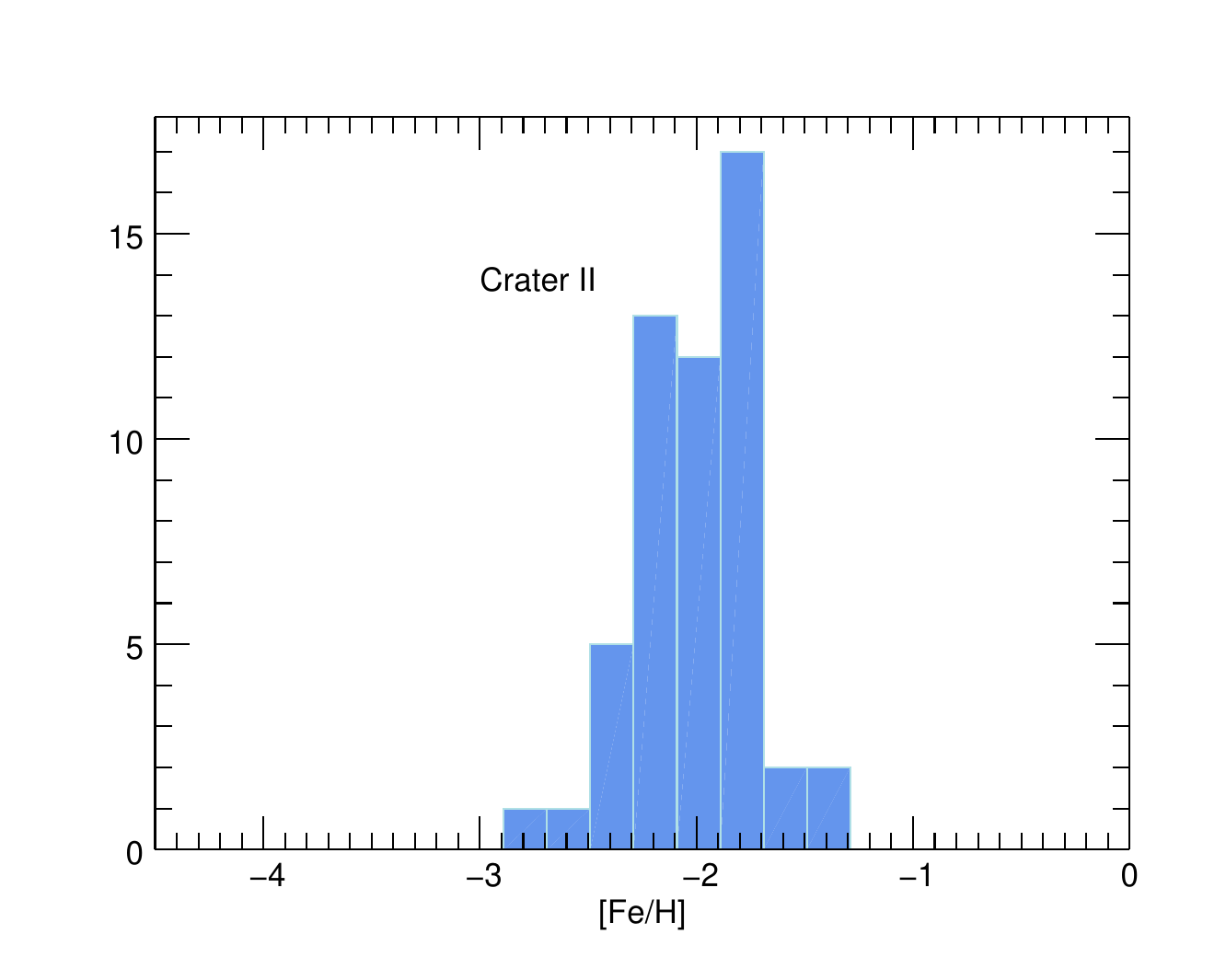}
\caption{\footnotesize Metallicity histogram of  Crater  and Crater II.   }
\label{Fig:Cra_histogram}
\end{figure}

\citet{kirby_spectroscopic_2015} measured 15 stars in Crater, of which 10 turned out to be radial velocity members,  
at medium resolution using DEIMOS at Keck,
including the two stars measured by 
\citet{bonifacio_chemical_2015}. For these two stars they found consistent metallicities but they found a radial velocity difference
between the two stars consistent with zero. They argue in favour of Crater being a globular cluster since it appears
too metal-rich for its luminosity compared to the luminosity-metallicity relation defined by other galaxies.
This is certainly true, however, a similar situation is true also for Sgr \citep[see Fig. 12 of ][]{mcconnachie_observed_2012}
and could arise if an object was more massive (and therefore luminous) in the past and has subsequently lost mass.

From the the determination of the velocity dispersion of a sample  of 26 spectroscopically confirmed member stars of Crater and a new determination of a mass/light ratio  consistent with a pure bayronic matter, \citet{voggel_probing_2016}  suggested that the presence of dark matter was not required. They  concluded that their study strongly supports Crater being  a faint intermediate-age outer halo globular cluster and not a dwarf galaxy. 

\citet{weisz_hubble_2016}, using HST photometry argued that the apparently young blue plume and sub giant branch identified by \citet{bonifacio_chemical_2015}
can be very well explained by blue stragglers and evolved blue stragglers. 
This is certainly true, one should also note, however, that while many globular clusters show a clearly defined
population of blue stragglers a sub giant branch is generally not seen. If it is a cluster, Crater is certainly exceptional, also considering
its young age.

Considering the controversial nature of Crater we did   not  include it in the list of galaxies presented in  Fig. \ref{Fig:Galaxies_South_overview}.
The Crater  metallicity histogram   is shown in Fig. \ref{Fig:Cra_histogram}.

\item{Crater II} \\

Crater II  was discovered by \citet{torrealba_feeble_2016} by a systematic search  of overdensities in  the  VST/Paranal  ATLAS ESO  public survey  \citep{shanks_vlt_2015}.
Due to its low luminosity ($M_{V} \simeq -8$), it has been classified as  UFD Galaxy. Using MMT/Hectochelle spectra of  stars in the field of view of Crater II,
\citet{caldwell_crater_2017} identified 62 members deriving a mean radial velocity of 87.5 km/s. They also measured a mean metallicity $\rm [Fe/H] = -1.98$ with a dispersion of $\simeq 0.22$\,dex.

Time series observations of 130 variables stars in Crater II were obtained by \citet{vivas_decam_2020} using DECam at CTIO. From the   the analysis  of the 98 RR Lyrae stars of the sample, they inferred a  small metallicity dispersion ($\simeq 0.17 dex)$ for the old metal poor population of Crater II.  
Combining  Southern Stellar Stream Spectroscopic Survey ($S{^5}$) medium resolution spectroscopy data and Gaia eDR3 data,  \citet{ji_kinematics_2021}  performed a  kinematical study of Crater II members. Using the  CaT near-infrared calcium triplet, they  derived a slightly  lower metallicity $\rm [Fe/H]= -2.16$  with a dispersion of $\simeq 0.24$\,dex.  No high resolution study has been made  yet on Crater II.
The Crater II metallicity histogram   is shown in Fig. \ref{Fig:Cra_histogram}.
\\

 \item{}  Grus I \\
 
 Using the publicly released Dark Energy Survey (DES) data, \citet{koposov_beasts_2015} discovered  the UFD galaxy Grus I. 
 Thanks to spectroscopic observations (R$\simeq$ 18,000, S/N ranging from 2.7 to 13)  with  the Michigan/Magellan Fiber System (M2FS) of a sample of stars in the line of sight of Grus I, \citet{walker_magellanm2fs_2016} identified 7 probable members estimating  metallicities ranging from --0.56 to --2.37.
 Note that \citet{walker_magellanm2fs_2016} increased all their [Fe/H] measurements by 0.32\,dex, which is the offset they obtained 
 from fitting twilight spectra of the Sun.  Two stars were observed at high resolution by   \citet{ji_chemical_2019} using the Magellan/MIKE spectrograph. 
 Thanks to high  resolving power (R from 22,000 to 28,000) and a better S/N ratio (S/N $\simeq$ 20), they could determine the abundance using the measure of the 
 equivalent widths of many lines.  They found metallicities $\rm [Fe/H] =-2.55$ and $-2.49$ for their stars. High $\rm [\alpha/Fe]$ ratios were found  as in halo stars. 
 They also found low [Sr/Ba] in these 2 stars, a strong evidence that Grus I is a Galaxy. 
 The Grus I  metallicity histogram  is shown in Fig. \ref{Fig:Gru_histogram}.
 \\
 
 \begin{figure}
\centering
\includegraphics[width=5.5cm,clip=true]{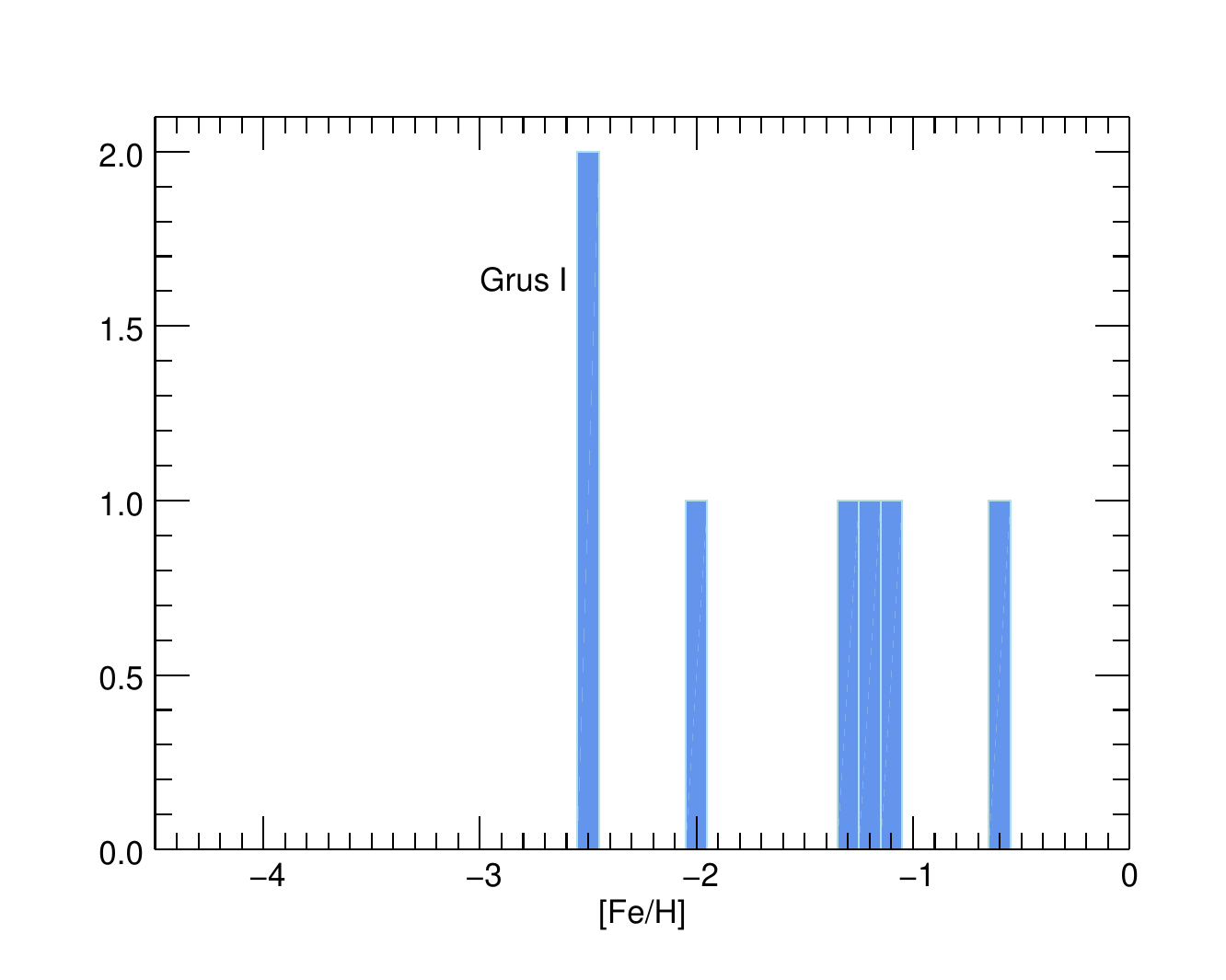}
\includegraphics[width=5.5cm,clip=true]{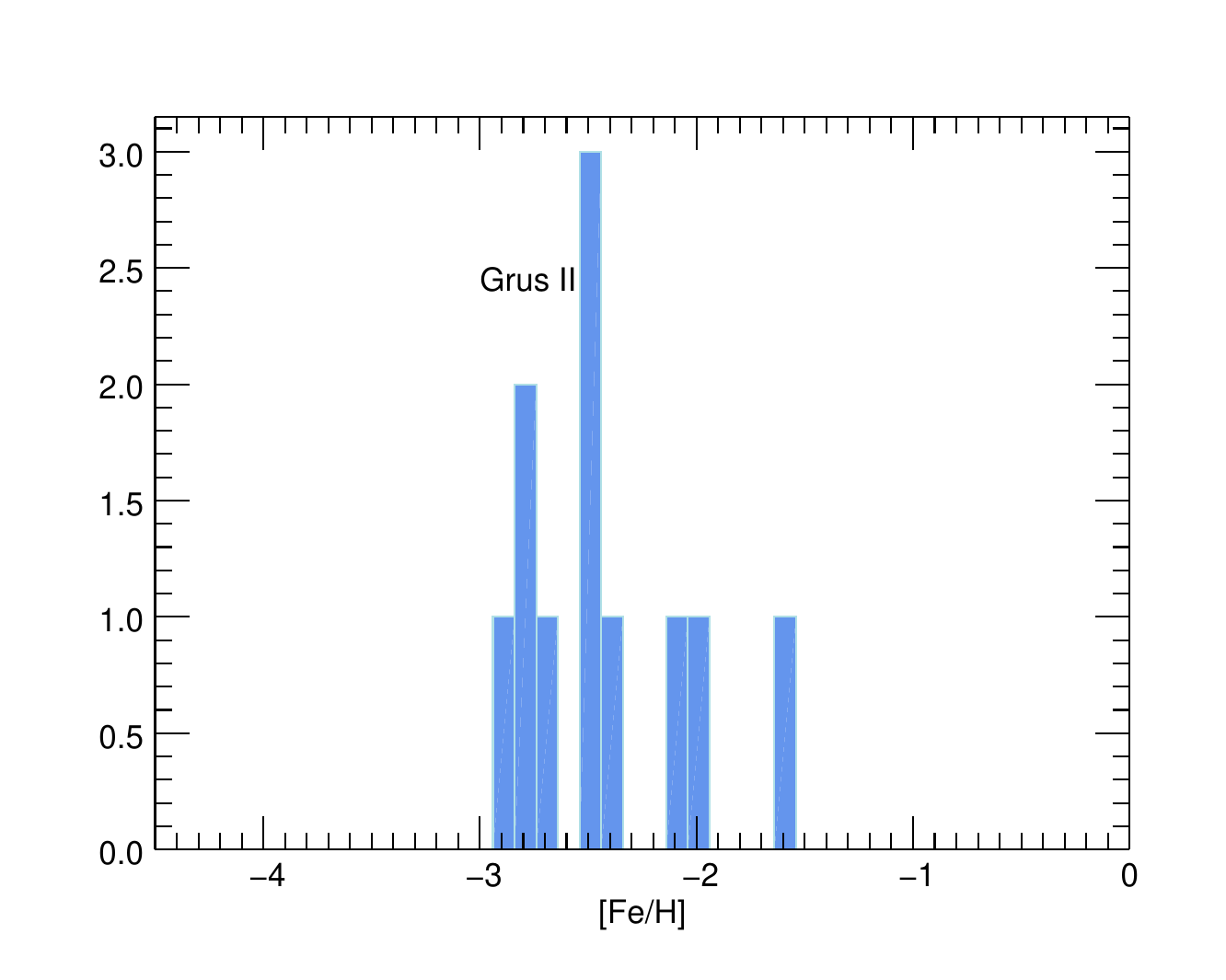}
\caption{\footnotesize Metallicity histogram of  Grus I and Grus II .   }
\label{Fig:Gru_histogram}
\end{figure}

  \item{} Grus II \\
  
Grus II was discovered by \citet{drlica-wagner_eight_2015} in the analysis of the combined data set from the first two years of  the Dark Energy Survey (DES) covering 
$\rm \simeq 5000\,deg^2$ of the south Galactic cap.

\citet{simon_birds_2020} analysed medium resolution spectra of seven stars taken with Magellan/IMACS (R= 11,000). Metallicity was measured from the equivalent width of the CaT triplet. They found  metallicities ranging from [Fe/H]=--1.55  to --3.09  with  estimated errors ranging from $\pm 0.15$ to $\pm 0.5$. 
Using the maximum likelihood method to determine the metallicity distribution of the Galaxy, they conclude that  the RGB stars in Grus II have a mean metallicity $\rm [Fe/H] = -2.51 \pm 0.21$.

In 2020,  \cite{hansen_chemical_2020} determined the detailed abundances of the three brightest members  at the top of the giant
branch of the ultra faint dwarf galaxy Grus II thanks to a series of spectra obtained with the Magellan/MIKE spectrograph. They found metallicities ranging from --2.49 to --2.94  with an error of the order of $\pm 0.3$\,dex. 
They also found that all stars exhibited a higher than expected [Mg/Ca] ratio compared to metal-poor stars in other UFD galaxies and in the Milky Way (MW) halo. Low Sr abundances have also been found in these stars.
The abundances of Grus II also revealed an enhancement in r-process elements in the most metal-rich of the three stars analysed.
The Grus II  metallicity histogram  is shown in Fig. \ref{Fig:Gru_histogram}.
\\

 \item{}  Hercules \\
 
 Hercules is a dwarf galaxy satellite of the Milky Way, found at a distance of 138 kpc. This UFD galaxy has been discovered  by \citet{belokurov_cats_2007}. 
 \cite{simon_kinematics_2007} obtained a first estimate of the metallicity using Keck/DEIMOS spectroscopy of 30 stars and find $\rm [Fe/H] \simeq -2.27$ with a dispersion of 0.31\,dex. \citet{koch_highly_2008} analysed two red giants and derived a metallicity of $\rm [Fe/H] \simeq -2.00$ with strong enhancements in Mg and O and a high deficiency in the neutron capture elements. Later, \citet{aden_abundance_2011} studied 11 stars in Hercules and obtained a metallicity spread ranging from $\rm [Fe/H] = -2.03$ to $-3.17$.
\citet{koch_neutron-capture_2013} analysed a new sample of four red giants and confirmed the high level of depletion of the neutron capture elements suggesting that the chemical evolution of Her was dominated by very massive stars.
\citet{francois_abundance_2016} analysed VLT/X-Shooter spectra of four member stars of Hercules. Their sample had metallicities that range from $-2.28$ to $-2.83$. Their results clearly showed an increase of the [$\alpha$/Fe] ratios as the metallicity decreases, as expected from classical models of chemical evolution where the impact of the contribution of type SNIa iron on the abundance ratios [$\alpha$/Fe] vs. metallicity relation is shown as a decrease of this ratio as the metallicity increases.
The Hercules  metallicity histogram  is shown in Fig. \ref{Fig:Her_histogram}.
\\
 
 \begin{figure}
\centering
\includegraphics[width=5.5cm,clip=true]{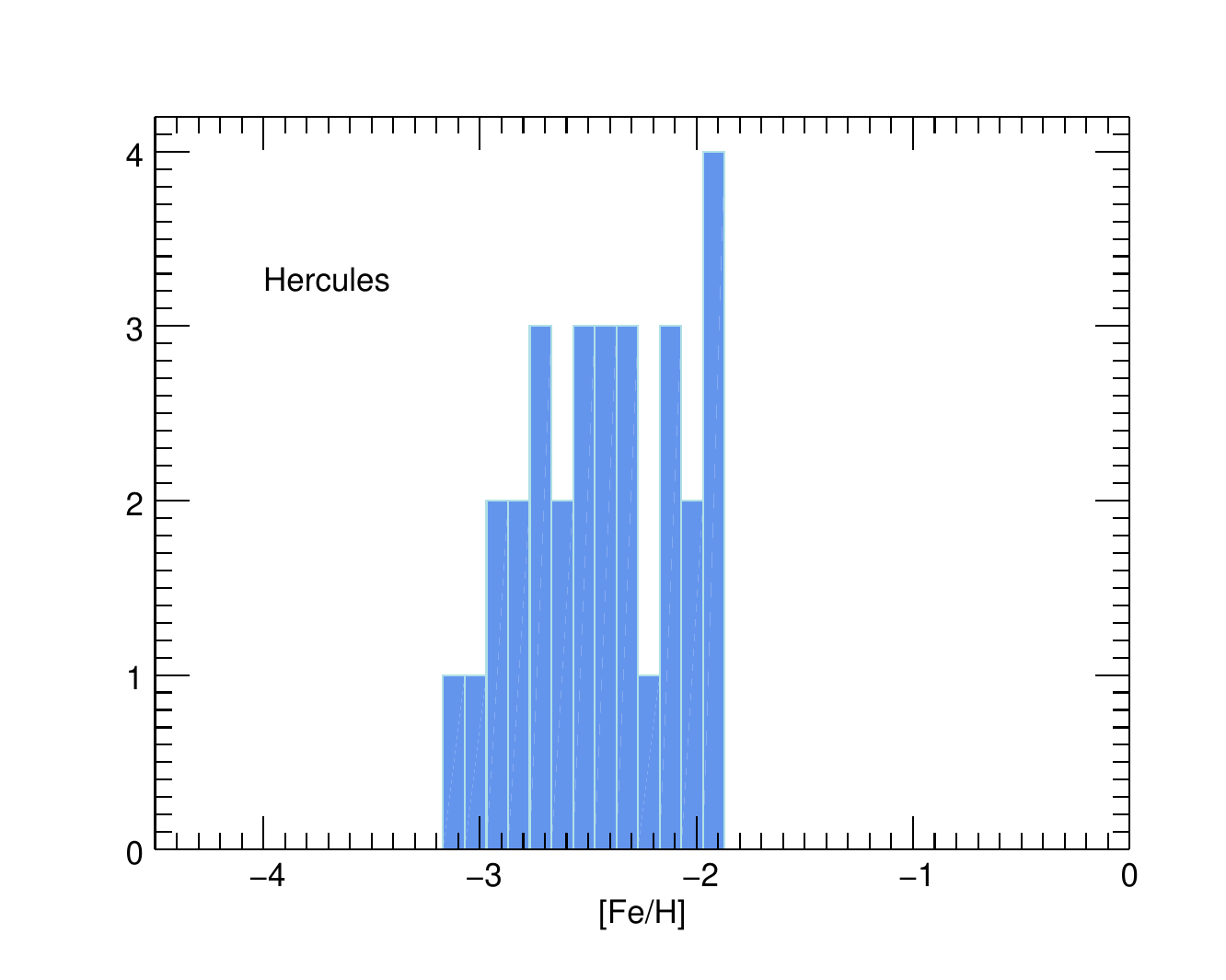}
\includegraphics[width=5.5cm,clip=true]{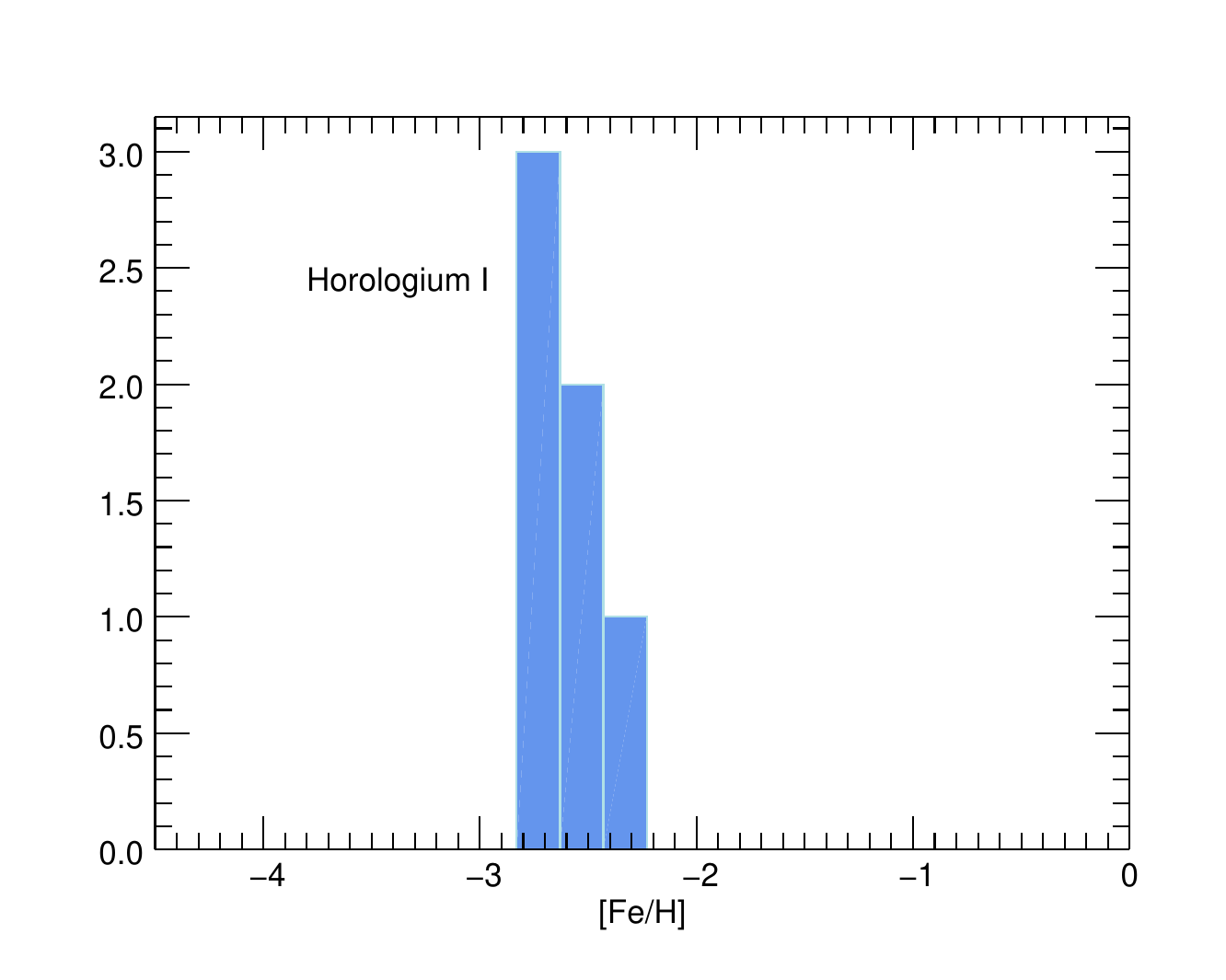}
\caption{\footnotesize Metallicity histogram of  Her and HorI .   }
\label{Fig:Her_histogram}
\end{figure}

 \item{}Horologium I \\ 
 
After the discovery  of the UFD galaxy Horologium I by \citet{koposov_beasts_2015} from  photometry and astrometry data of the Dark Energy Survey, 
follow up spectroscopic analysis  were performed by \citet{koposov_kinematics_2015} using VLT/Giraffe spectra obtained as part of the {\it Gaia }-ESO survey. From the five member stars  of Hor I identified in the Survey, they derived a mean metallicity $\rm [Fe/H]= -2.76$ with a sample ranging from --2.3 to --3.01. They also find Horologium I to have [$\alpha$/Fe] $\simeq +0.3$, consistent with the dwarf galaxy population of the Milky Way.

 \citet{nagasawa_chemical_2018}  analysed the spectra of three stars members of Horologium  I  using the UVES/VLT  and the Magellan/MIKE  
 high resolution spectrographs. They measured metallicities  [Fe/H] between --2.43 and --2.83. 
 They found the [$\alpha$/Fe] abundances to be much lower than expected when compared to other metal-deficient stars.
The Horologium I  metallicity histogram  is shown in Fig. \ref{Fig:Her_histogram}.

\item{} Hydrus I   \\  

\citet{koposov_snake_2018}   reported the discovery of a nearby dwarf galaxy in the constellation of Hydrus, between the Large (LMC) and the Small Magellanic Clouds (SMC). 
From $R\sim 18000$ spectra covering a small spectra range (513--519\,nm) obtained with M2FS on the Magellan 6.5\,m telescope 
they derived metallicities for 30 radial velocity members and found a mean metallicity of --2.5 and a scatter of about 0.4\,dex in the abundances.
One star was found extremely C enhanced, from  the Swan band they estimated [C/Fe]$\sim +3$.
Although this galaxy is rather close, no detailed abundance determination has been done so far. 
The Hydrus I  metallicity histogram  is shown in Fig. \ref{Fig:Hyd_Leo_histogram}.
\\
 
  \begin{figure}
\centering
\includegraphics[width=5.5cm,clip=true]{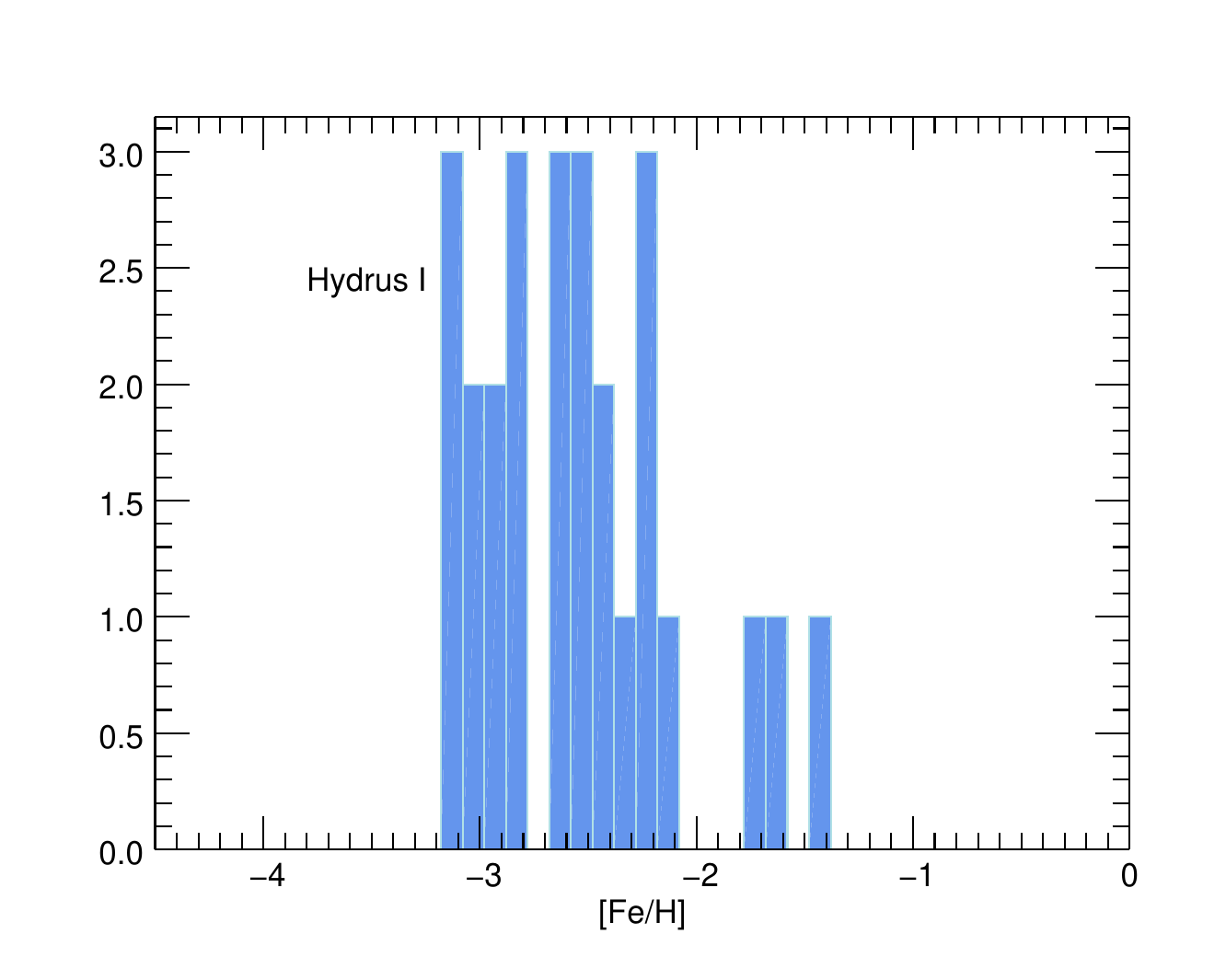}
\includegraphics[width=5.5cm,clip=true]{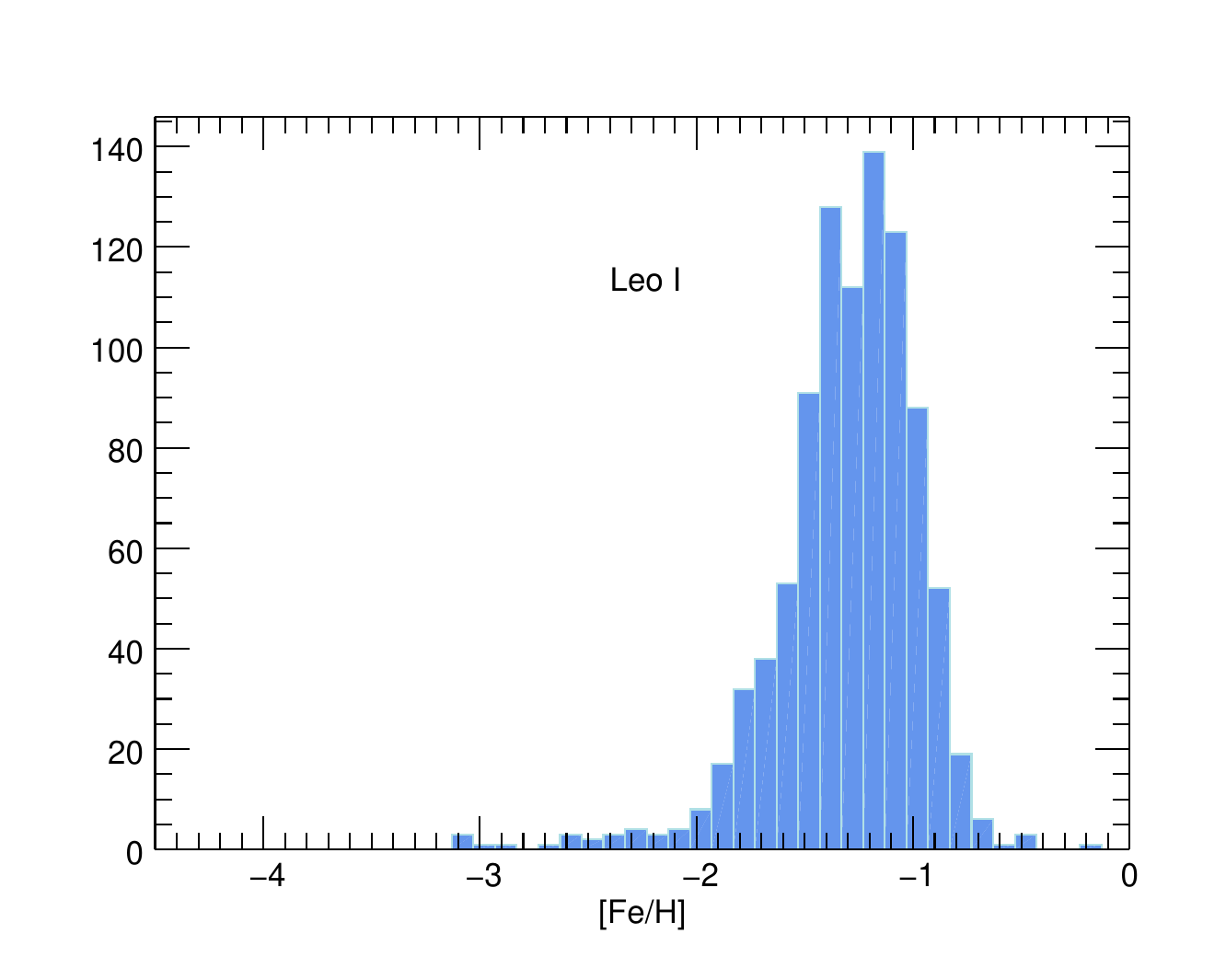}
\includegraphics[width=5.5cm,clip=true]{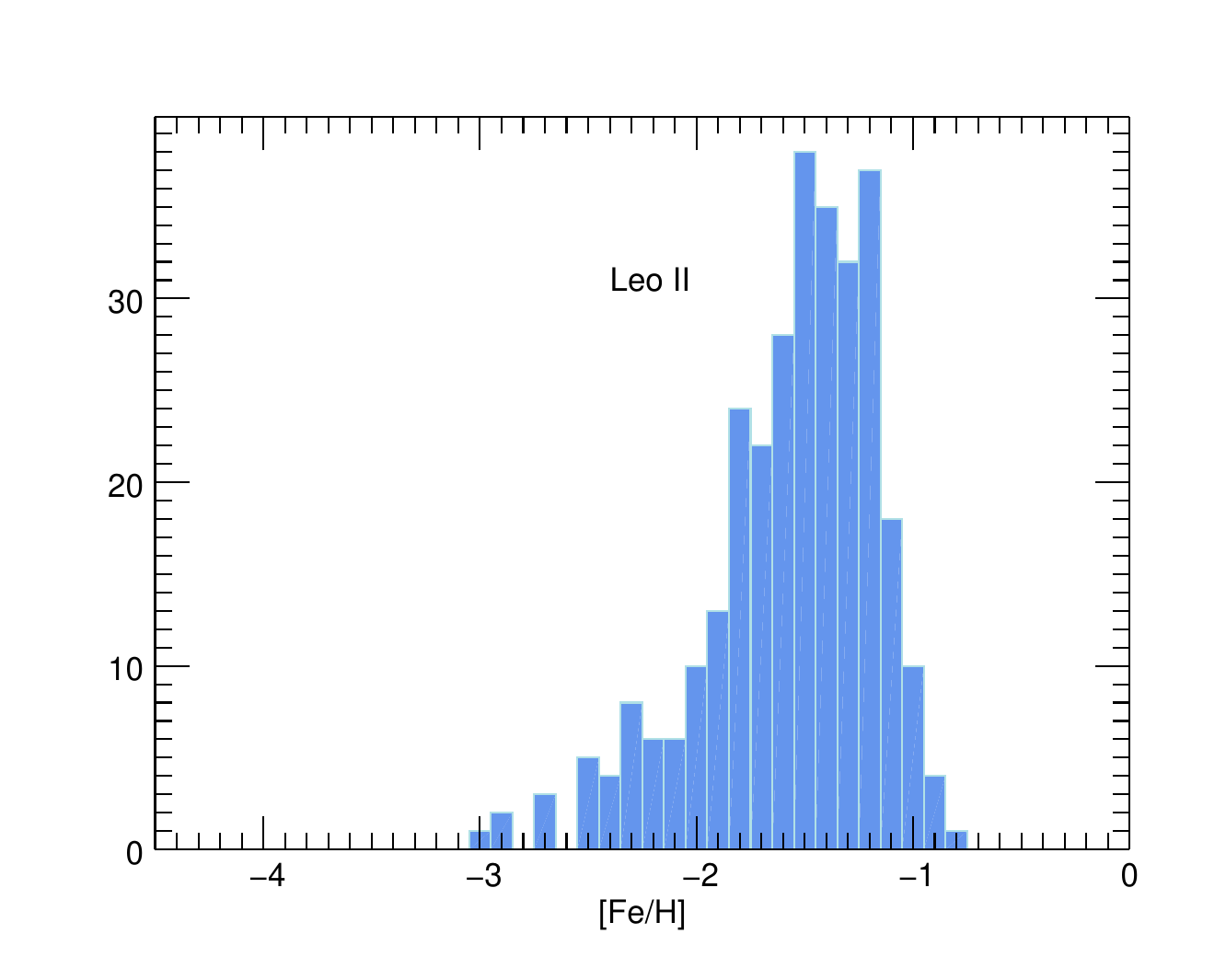}
\includegraphics[width=5.5cm,clip=true]{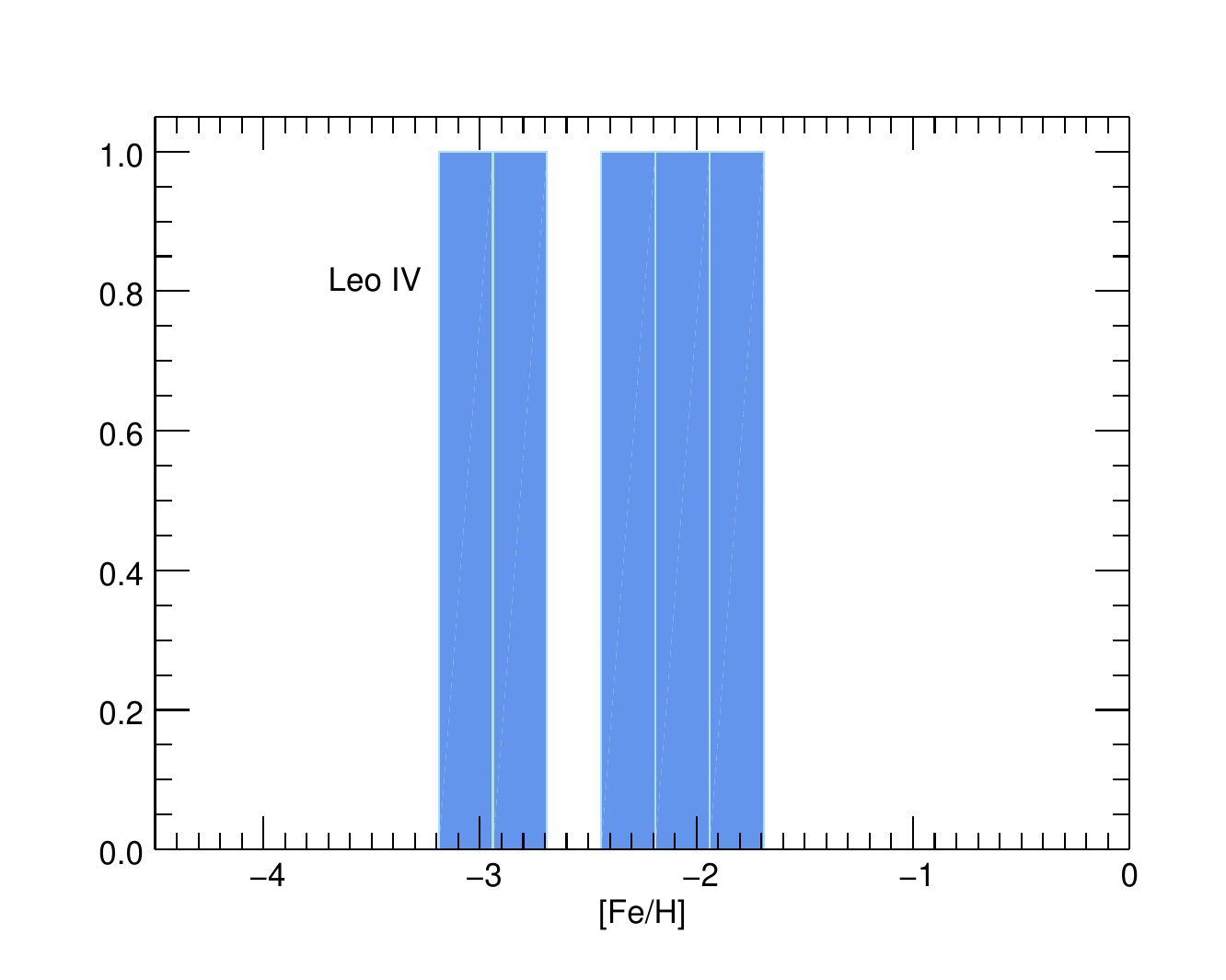}
\includegraphics[width=5.5cm,clip=true]{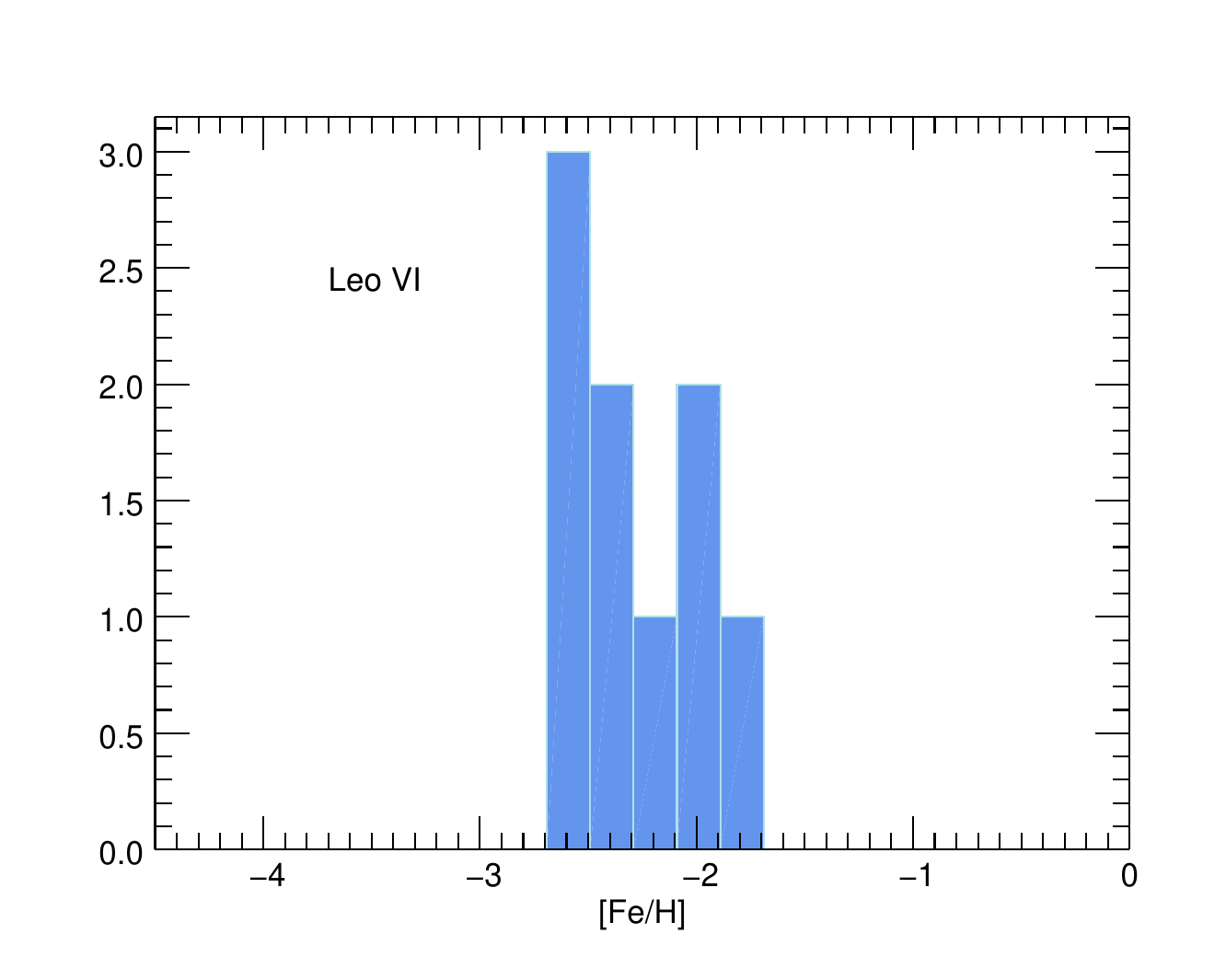}
\caption{\footnotesize Metallicity histogram of  Hydrus I, LeoI, LeoII,  LeoIV and Leo VI.  }
\label{Fig:Hyd_Leo_histogram}
\end{figure}

   \item{}   Indus I \\
   
 Indus I  is not present in the list of galaxies shown in   Fig. \ref{Fig:Galaxies_South_overview}.  It was discovered  by \citet{kim_discovery_2015}.
 With deeper observations, they  concluded that this object is likely a star cluster.\\

\item{}    Leo I and Leo II   \\

The UFD Leo I and Leo II  are among  the first UFD discovered on photographic plates taken with the 48-inch schmidt telescope for the National Geographic Society-Palomar Sky Survey \citep{harrington_two_1950}.
Almost 60 years after, the first spectroscopic analysis of Leo I  with Hectoechelle/MMT  was made by \citet{mateo_velocity_2008} to determine the  radial velocity dispersion profile  of the galaxy. 

 From high resolution  spectroscopy of two stars of Leo I,  \cite{shetrone_vltuves_2003}  measured metallicities $\rm [Fe/H] = -1.05$ and --1.56. They found that at a given metallicity the stars exhibit lower [X/Fe] abundance ratios for the $\alpha$ elements than stars in the Galactic halo.

 \citet{kirby_multi-element_2010, kirby_multi-element_2011} obtained  KECK/DEIMOS spectra of a large sample of stars in Leo I and Leo II. 
Based on a sample of 827 stars from Leo I with metallicities ranging from --3.3 to --0.39, they derived a mean metallicity $\rm \langle[Fe/H]\rangle = -1.46$. For Leo II, they obtained a mean $\rm \langle [Fe/H]\rangle = -1.7$ from a sample of 258 stars with metallicities ranging from --3.22 to --0.97.  \\

The LeoI and LeoII  metallicity histograms are shown in Fig. \ref{Fig:Hyd_Leo_histogram}.

\item{} Leo IV  \\

The galaxy Leo IV was  discovered by \citet{belokurov_cats_2007} in the Sloan Digital Sky Survey. \citet{simon_kinematics_2007} obtained low resolution Keck/DEIMOS spectra for 18 bright stars in Leo IV from which they  used  the Ca{\sc II} triplet absorption lines calibration to  derive an average metallicity $\rm\langle [Fe/H]\rangle=-2.31 \pm 0.10$ with a dispersion in [Fe/H] $\sigma = 0.15$\,dex.
Using a pixel-to-pixel matching method between observed and synthetic spectra, \citet{kirby_uncovering_2008} determined the metallicity of 12 stars in Leo IV and derived a mean $\rm [Fe/H] = -2.58\pm 0.08$ with a dispersion $\sigma = 0.75$\,dex. 
\citet{simon_high-resolution_2010}  made the detailed abundance analysis of the brightest star of Leo IV thanks to spectra obtained with  MIKE/Magellan. They measured an iron abundance of $\rm [Fe/H]=-3.2$\,dex . The star is enhanced in the $\alpha$ elements Mg, Ca, and Ti by about 0.3\,dex, very similar to the typical Milky Way halo abundance pattern.
\citet{francois_abundance_2016} reanalysed this star along with another Leo IV member star using X-Shooter/VLT spectra. 
They confirmed the low [Fe/H] abundance  found by  \citet{simon_high-resolution_2010} for  the star in common.
For the second star, they found a higher metallicity, with $\rm [Fe/H] = -2.18$, $\rm [Mg/Fe] =-0.06$ and $\rm [Ca/Fe] = -0.05$,  
in good agreement with the theoretical predictions from the galactic chemical evolution models of UFD galaxies of \citet{vincenzo_chemical_2014}.
The LeoIV  metallicity histogram is shown in Fig. \ref{Fig:Hyd_Leo_histogram}.
\\
\\

\item{} Leo VI  \\

\citet{tan_pride_2025} reported the discovery of a new ultra-faint Milky Way galaxy in the Leo  constellation identified as an overdensity in DECam data from an early version of the third data release of the DECam Local Volume Exploration (or DELVE) survey. 
Their claim is supported by  the  observations of the main characteristics of this system (Luminosity, size and distance).
Using Keck/DEIMOS low resolution spectroscopy, they identified nine 
members stars of this new UFD galaxy Leo VI  and four candidate members.  They find that the systemic spectroscopic metallicity of Leo VI is $\rm [Fe/H] = -2.39$ and a metallicity dispersion of  $\simeq \pm 0.19$\,dex. \\

\item{} Pegasus IV \\ 

\citet{cerny_pegasus_2023}  reported the discovery of Pegasus IV, an UFD galaxy found in the archival from the Dark Energy Camera processed by the DECam Local Volume Exploration Survey.  From the analysis of five  non-variable members observed with Magellan/IMACS, they measured a mean metallicity $\rm [Fe/H] = -2.63$ with stars metallicities ranging from $\rm [Fe/H] = -2.00$ to --3.29. 
The Peg IV  metallicity histogram is shown in Fig. \ref{Fig:Peg_Phe_Pisc_histogram}.\\
 
  \begin{figure}
\centering
\includegraphics[width=5.5cm,clip=true]{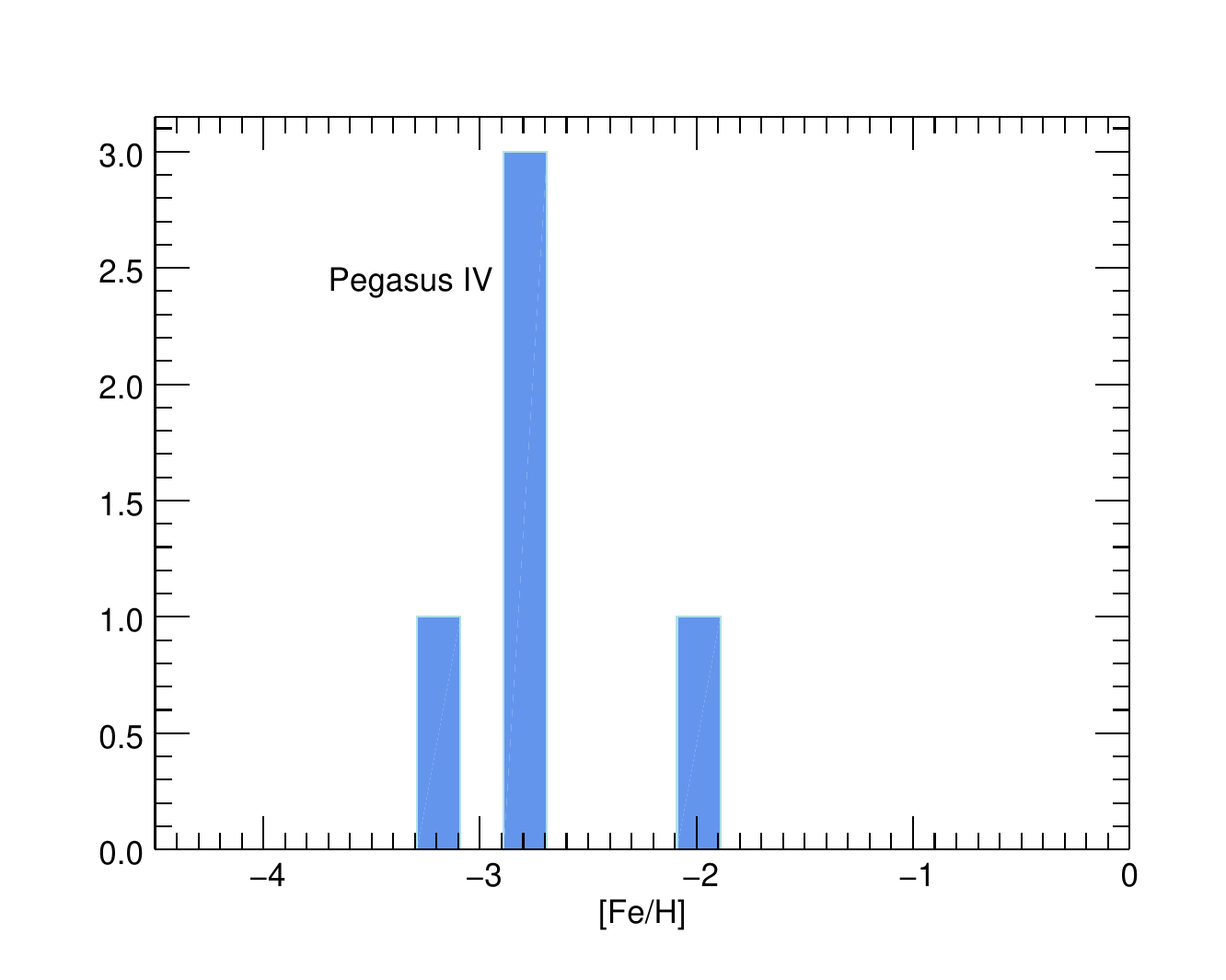}
\includegraphics[width=5.5cm,clip=true]{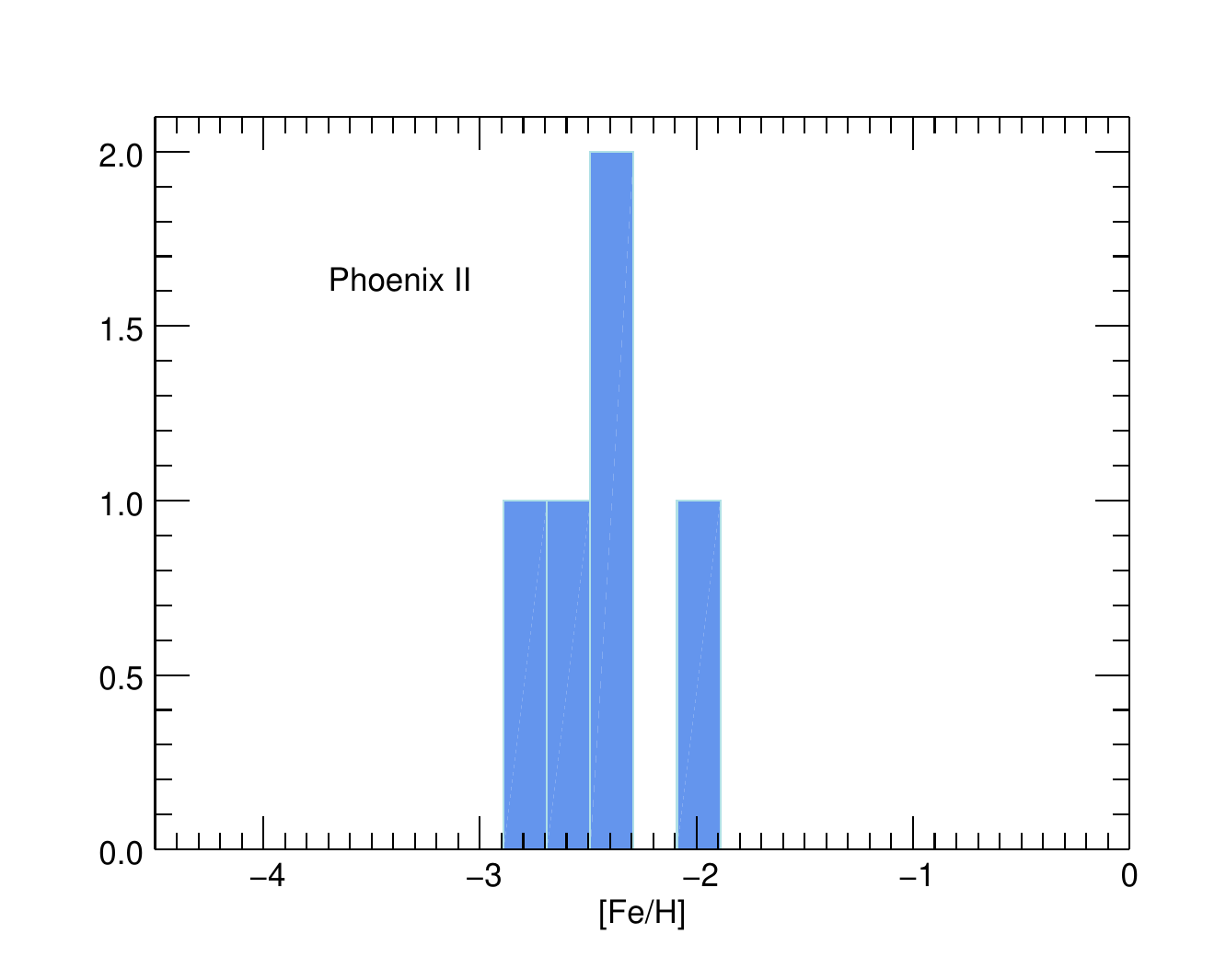}
\includegraphics[width=5.5cm,clip=true]{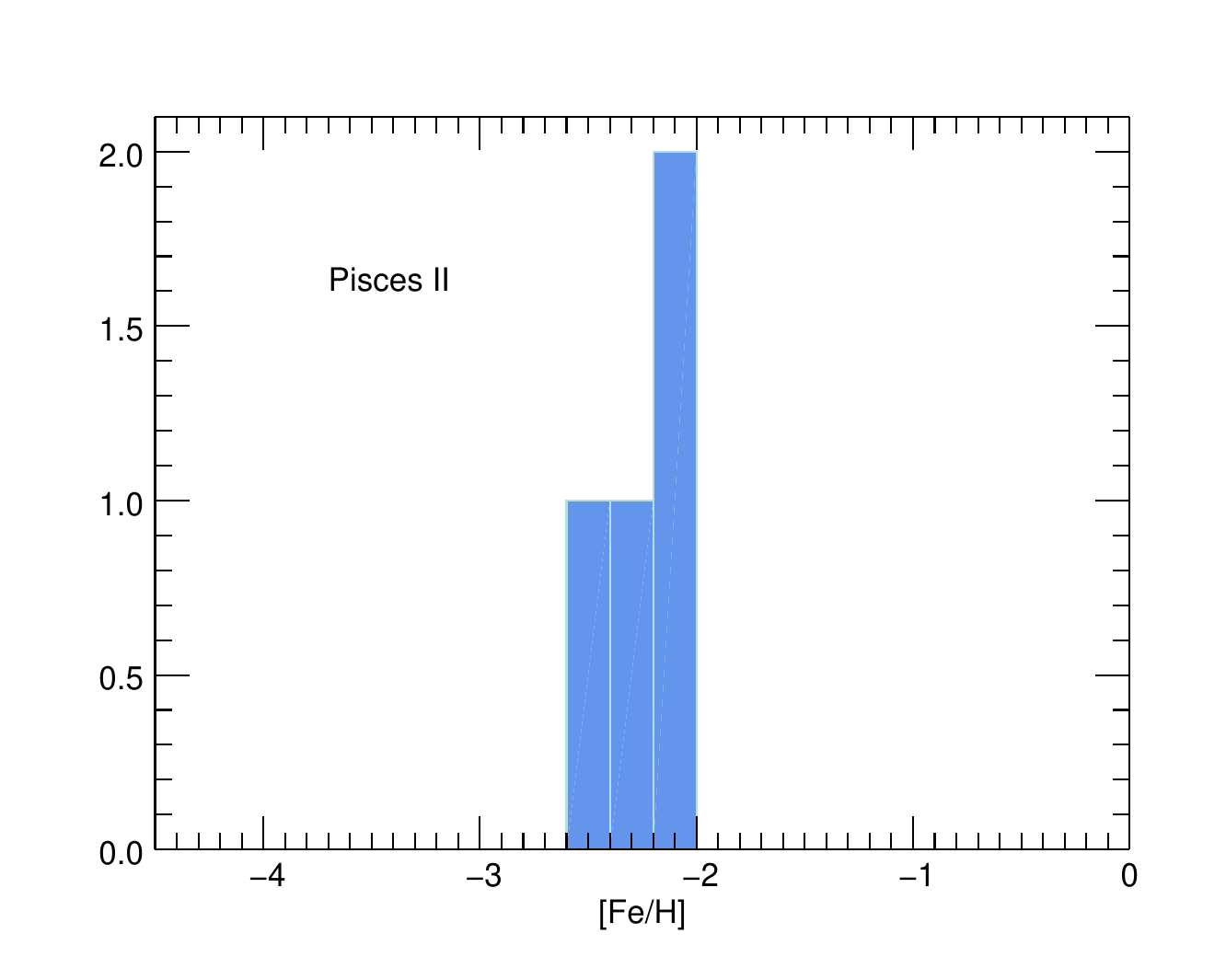}
\caption{\footnotesize Metallicity histograms of Pegasus IV, Phoenix II and Pisces II.}
\label{Fig:Peg_Phe_Pisc_histogram}
\end{figure}

\item{}  Phoenix II \\

The UFD galaxy Phoenix II (Phe II) has been discovered by \citet{koposov_beasts_2015}  from the publicly released Dark Energy Survey (DES) data.
The galaxy has also been  identified by \citet{bechtol_eight_2015} using  optical imaging data collected during the first year of the Dark Energy Survey (DES).

 \citet{fritz_gaia_2019} combined Gaia DR2 astrometric measurements, photometry, and new FLAMES/GIRAFFE
spectroscopic data centred on the Ca{\sc II} triplet.
From five confirmed members of Phoenix II,  they derived metallicities  [Fe/H] ranging from --2.00 to --2.89.
From the analysis of Magellan/Megacam photometry data \citet{mutlu-pakdil_deeper_2018} concluded  that  Phe II is more likely an UFD galaxy than a star cluster. 
 The Phoenix II metallicity histogram is shown in Fig. \ref{Fig:Peg_Phe_Pisc_histogram}.
\\

 \item{}  Pisces II   \\
 
 After its discovery as a stellar overdensity in SEGUE data by \citet{belokurov_big_2010},   
spectroscopic confirmation of the metallicity of Pisces II were performed by \citet{kirby_spectroscopic_2015} using Keck/DEIMOS spectra. 
They could determine the metallicity in four stars of their Pisc II sample and found metallicities ranging from $\rm [Fe/H] = -2.10$ to --2.70. 
One of the stars  (Pisces II 10694) of the  sample was found to be carbon rich. 
 Using X-shooter spectra (R=12,000),  \citet{spite_cemp-no_2018} determined the chemical composition of this carbon rich star.
 They  found that this star is a CEMP-no star with $\rm [Fe/H]=-2.60$, $\rm [C/Fe]=+1.23$ and a low barium abundance $\rm [Ba/Fe] = -1.10$. 
The Pisces II  metallicity histogram is shown in Fig. \ref{Fig:Peg_Phe_Pisc_histogram}.
 \\

 \item{} Reticulum II  \\
 
 As for the UFD galaxy Phoenix II, the UFD Reticulum II (Ret II) has been discovered by \citet{koposov_beasts_2015} and \citet{bechtol_eight_2015} the same year
 from the analysis of DES data. 
 Follow up spectroscopic analysis  were performed by \citet{koposov_kinematics_2015} using VLT/Giraffe spectra obtained as part of the {\it Gaia}-ESO survey. 
  They determined  the chemical composition of  16 stars of Ret II  identified in the Survey and found
 a mean iron abundance $\rm [Fe/H] = -2.46$ with a sample ranging from --1.98 to --3.19. They also found Ret II to have $\rm [\alpha/Fe] \simeq +0.4$, typical of the dwarf galaxy population of the Milky Way.
 This large range of metallicity was also found  by \citet{walker_magellan_2015} from an analysis of  Michigan/Magellan Fiber system (M2FS)  spectra (17  Ret II member stars)  obtained in high resolution mode (R$\simeq$ 18,000) and a SNR of the order of 5/pixel.

 \citet{ji_complete_2016} obtained Magellan/MIKE high resolution spectra of the nine brightest red giants members of Ret II. They confirmed the large range of iron abundance 
 ( $\rm -3.5 <  [Fe/H] < -2.0$). For seven stars of their sample, they found very high level of Europium with {$\rm [Eu/Fe] \simeq 1.7$}. 

\citet{ji_metal_2023} reported a high resolution  study of 32 spectroscopic members of Ret II with VLT/GIRAFFE and Magellan/M2FS. 
They found that most of the stars which metallicities range from  $\rm [Fe/H] = --2.02$ to $--3.18$   and are   are r-process enhanced. \\

 \citet{roederer_detailed_2016} observed the four brightest red giants in Ret II at high spectral resolution using the Michigan/Magellan Fiber System  confirming  the previous detection of high levels of r-process material in Ret II.

\citet{hayes_ghost_2023} analysed two stars in  Ret II using the recently commissioned Gemini High resolution Optical SpecTrograph (GHOST), one being already identified as a  Ret II member star (GDR3  0928 or 
Gaia DR3 4732600514724860928) but with no information on its detailed chemical composition. 
As for the other Ret II star, the analysis of the spectrum of GDR3 0928 gave a low metallicity {$\rm [Fe/H] \simeq -2.5$}. The authors also  found it is enriched in r-process elements like the majority of the giants in Ret II, however at a level that it can be classified as an r-II star. In addition, it is enriched in carbon, identifying it is a CEMP-r star. 
The Ret II metallicity histogram based on the  updated [Fe/H] from \citet{ji_metal_2023} and \citet{hayes_ghost_2023}   is shown in Fig. \ref{Fig:Ret_histogram}.
  \begin{figure}
\centering
\includegraphics[width=5.5cm,clip=true]{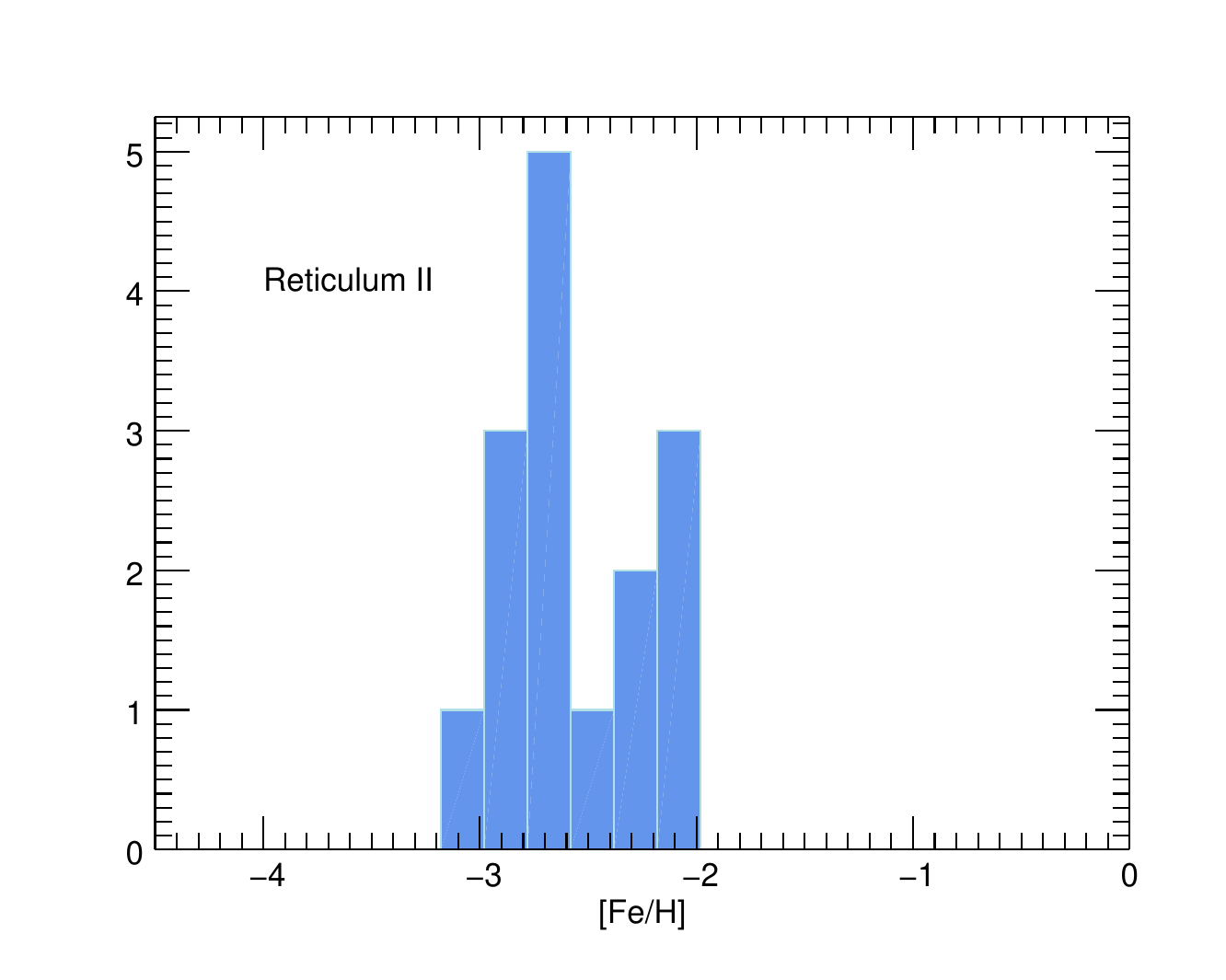}
\includegraphics[width=5.5cm,clip=true]{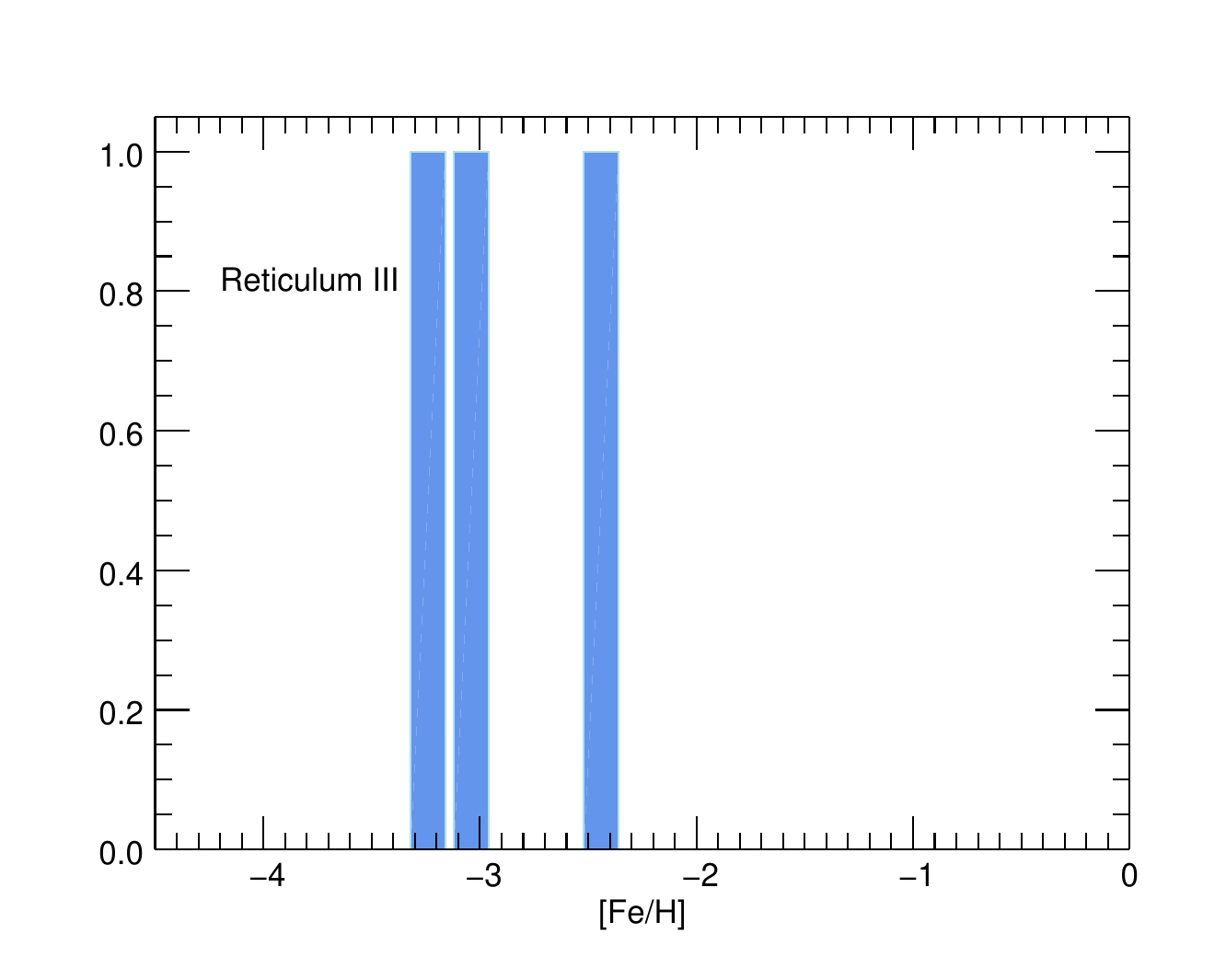}

\caption{\footnotesize Metallicity histograms of Reticulum II and Reticulum III.}
\label{Fig:Ret_histogram}
\end{figure}
\\

 \item{} Reticulum III  \\
 
 Reticulum III has been discovered by \citet{drlica-wagner_eight_2015} in DES optical imaging data. 
 From the combination of Gaia DR2 astrometric measurements, photometry, and  FLAMES/GIRAFFE intermediate-resolution spectroscopy of three stars 
 in the region of the near-IR Ca{\sc II} triplet lines, \citet{fritz_gaia_2019} determine metallicities ranging from $\rm [Fe/H] = -2.32\pm  0.15$ 
 to  $-3.24 \pm 0.15$.
 They derived a metallicity dispersion of 0.35\,dex, however, due to the faintness of the stars, they could not firmly classify Ret III as a galaxy.
The Ret III metallicity histogram  is shown in Fig. \ref{Fig:Ret_histogram}.\\

\item{}  Segue1 \\

 \begin{figure}
\centering
\includegraphics[width=5.5cm,clip=true]{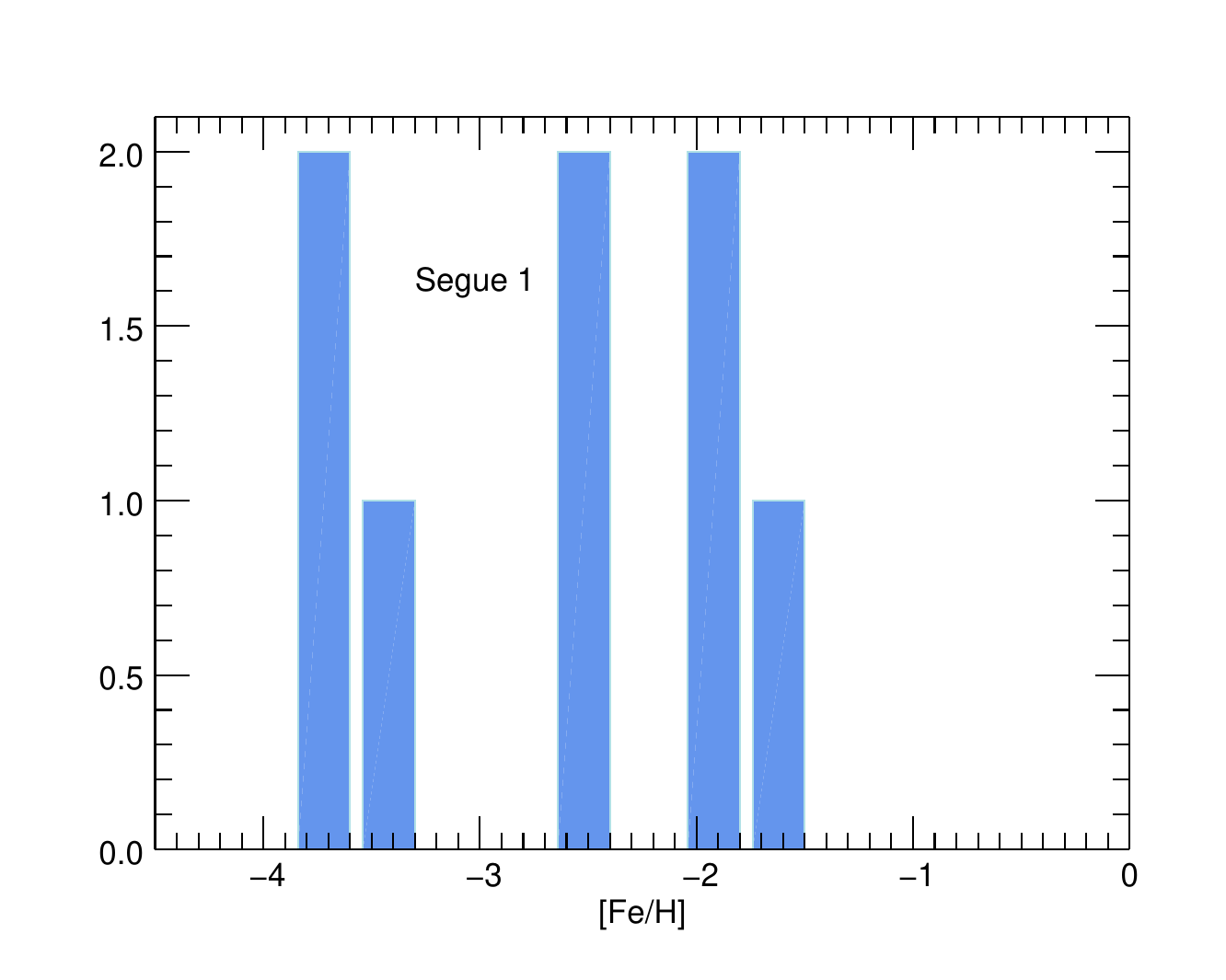}
\includegraphics[width=5.5cm,clip=true]{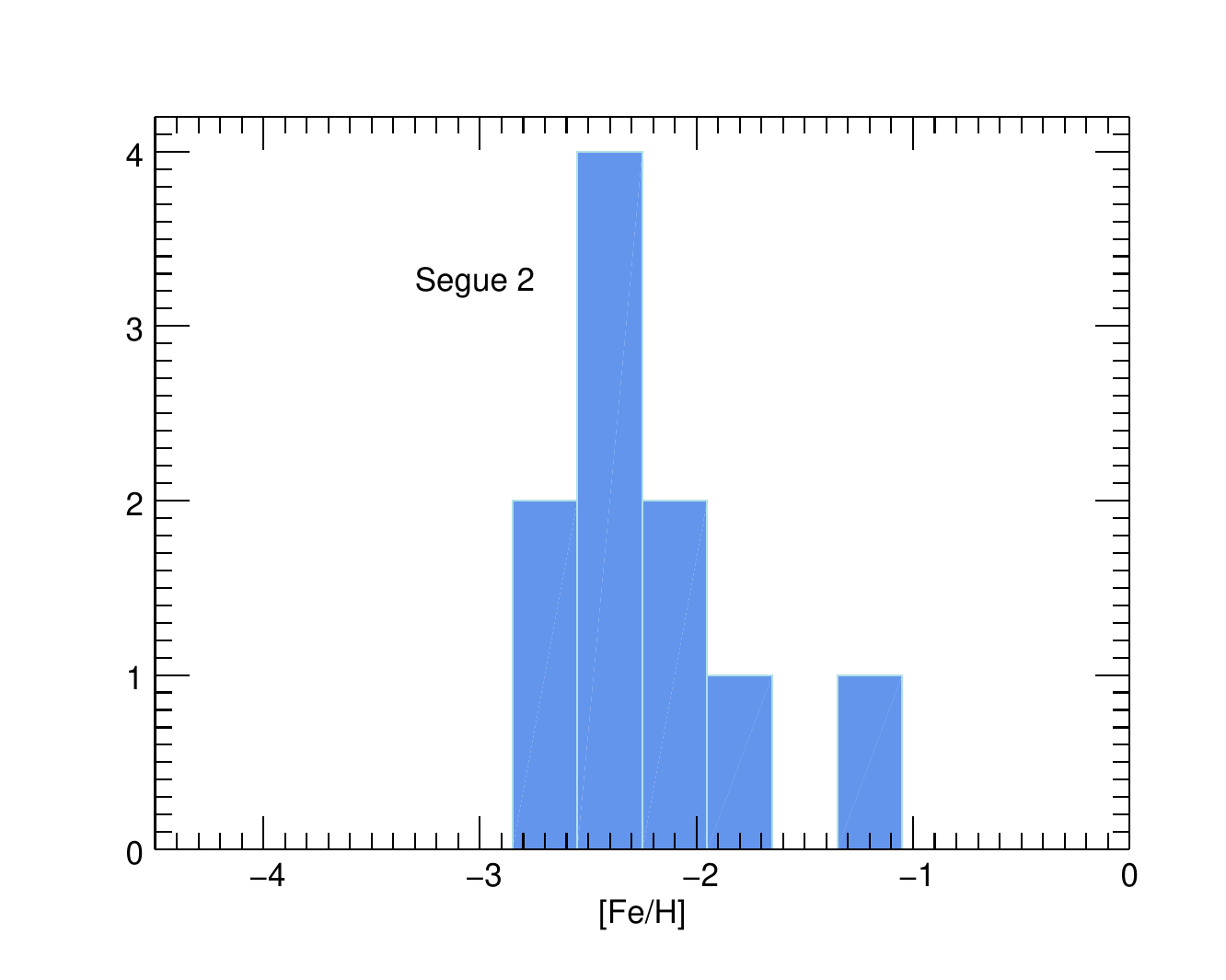}
\includegraphics[width=5.5cm,clip=true]{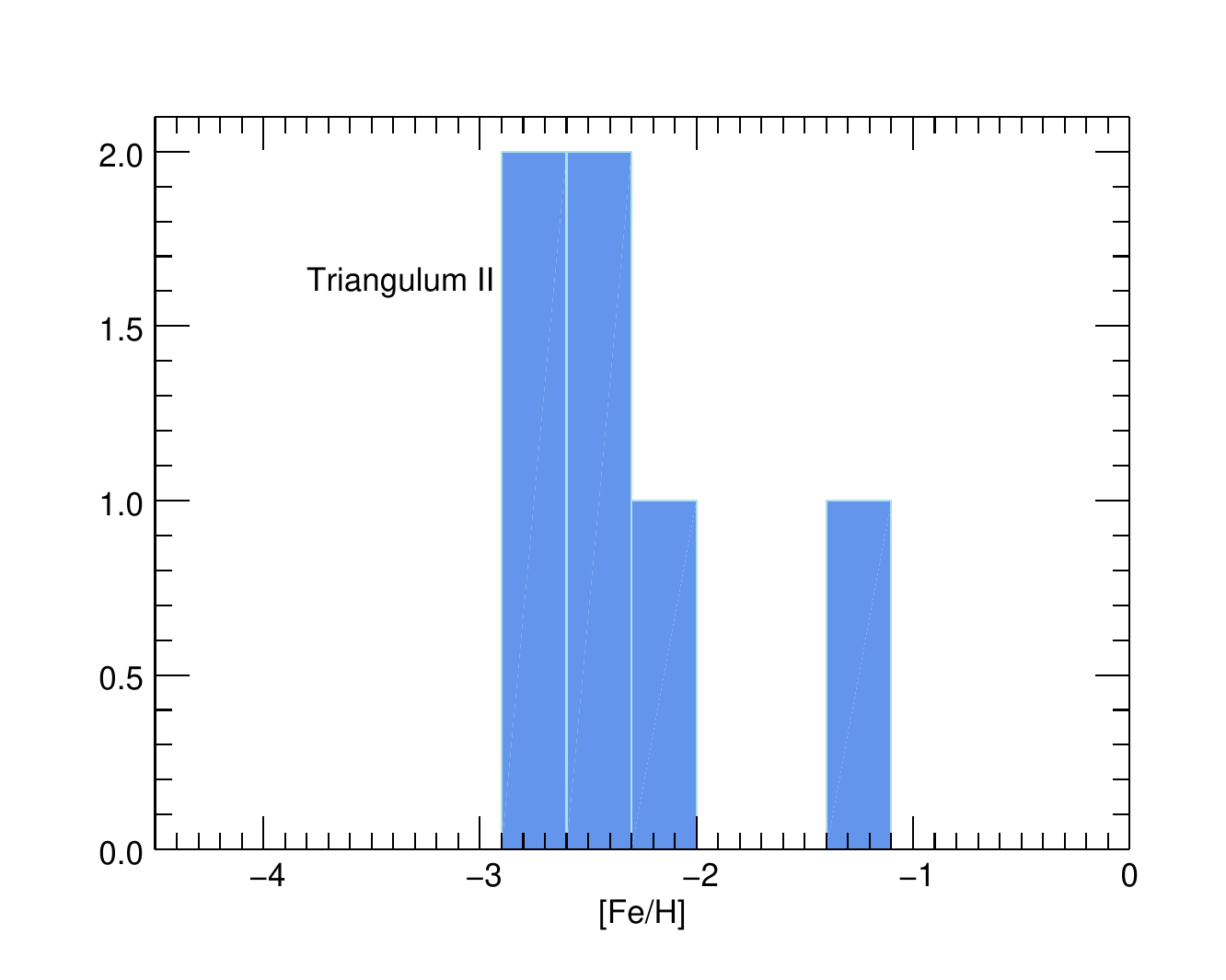}
\caption{\footnotesize Metallicity histograms of Seg1,  Seg2 and Tri II.}
\label{Fig:SegTri_histogram}
\end{figure}

Segue 1 has been discovered  by \citet{belokurov_cats_2007} from SDSS imaging data together with Subaru/Suprime-Cam deeper follow-up observations. 
They classified Segue 1 (Seg 1) as an extended globular cluster. A Keck/DEIMOS spectroscopic  survey of the Segue 1 was performed by \citet{simon_complete_2011}   who identified 71 stars as probable Seg 1 members.  They found  two extremely low metallicity stars $\rm [Fe/H] < -3$ and a large  metallicity spread leading them to the conclusion that Segue 1  is indeed a dwarf galaxy. 

Red giant members of Segue 1 were identified thanks to Anglo-Australian Telescope/AAOmega spectroscopy by \citet{norris_extremely_2010}. 
They found a significant dispersion both in iron and carbon. They measured mean $\rm\langle [Fe/H]\rangle = -2.7 \pm 0.4$ with  a [Fe/H] dispersion of $\sigma = 0.7$, and abundances spreads of
$\rm \Delta [Fe/H] = 1.6$ and $\rm\Delta [C/H] = 1.2$. They also found an  extremely metal poor carbon rich star with $\rm [Fe/H] =-3.5$  and $\rm [C/Fe] = +2.3$.

\citet{frebel_segue_2014}  analysed Magellan/MIKE and Keck/HIRES high resolution spectra of six red giant stars in the dwarf galaxy Segue 1. They found that three stars had metallicities below  $\rm [Fe/H]  = -3.5$. They confirmed  that Segue 1 stars spanned a metallicity range of more than 2 dex, from $\rm [Fe/H] = -1.4$ to $\rm [Fe/H] = -3.8$. 
  They  found a high [$\rm \alpha /Fe] \simeq +0.5$  ratio  and low  neutron-capture element abundances in all the stars. Interestingly, they did not find a 
  decline of the [$\rm \alpha/Fe]$ ratio as the metallicity of the star increases. They concluded that Segue 1 may be a surviving first galaxy that experienced a single burst of star formation. \\
\citet{sitnova_formation_2021} revisited chemical abundances in seven stars in Segue 1 using the high resolution spectra from the original paper of \citet{norris_extremely_2010} and \citet{frebel_segue_2014}. The originality of their work is that they used NLTE calculations to 
compute abundances of up to 14 elements. Details can be found in their article. 
The Seg1  metallicity histogram  is shown in Fig. \ref{Fig:SegTri_histogram}.\\
\\

\item{} Segue 2 \\

The UFD Segue 2 was discovered by \citet{belokurov_discovery_2009}  in the data of the Sloan Extension for Galactic Understanding and Exploration (SEGUE). 
Thanks  to additional deeper imaging and Hectochelle spectroscopy on the Multiple Mirror Telescope (MMT), they could measure the strength of the magnesium triplet and estimated a metallicity $\rm [Fe/H] \simeq  -2.00$. Segue 2 was later  observed by \citet{kirby_segue_2013} using Keck/DEIMOS. From the   ${\chi}^2$ comparison of observed spectra to a grid of synthetic  spectra made  for 25 Segue 2 members, they derived a mean metallicity $\rm\langle [Fe/H]\rangle = -2.22 \pm 0.13$ with a wide dispersion of the metallicity ranging from  --2.85 to --1.33. 
The Seg1  metallicity histogram  is shown in Fig. \ref{Fig:SegTri_histogram}.\\
\\

\item{}  Triangulum II \\

Triangulum II  system was found by \citet{laevens_new_2015} in  mining the Panoramic Survey Telescope And Rapid Response System (PS1) 3$\pi$ survey for localised stellar overdensities confirmed  by  additional deep imaging with the Large Binocular Cameras on the Large Binocular Telescope (LBC/LBT). \citet{laevens_new_2015} found that the color-magnitude diagram was best represented by a metal-poor ($\rm [Fe/H] = -2.19$)  old-age isochrone (13 Gyrs). 

From Keck/DEIMOS spectroscopy of Tri II member stars  \citet{martin_triangulum_2016} measured an average metallicity $\rm\langle [Fe/H]\rangle = -2.6 \pm 0.2$ based on a Ca{\sc II} triplet metallicity calibration  of \citet{starkenburg_nir_2010}.  
\citet{venn_geminigraces_2017}  determined the chemical abundance ratios and radial velocities for two stars in  Tri II from high resolution spectra obtained with  the Gemini Remote Access to CFHT ESPaDOnS Spectrograph (GRACES). They derived metallicities $\rm [Fe/H] = -2.5$ and --2.87. The detailed chemical abundances in these two stars are similar to those of similar metallicity stars in the Galactic halo, although with some anomalies like  a very low Mg in both stars.
Keck/DEIMOS spectroscopy  by \citet{kirby_triangulum_2017}  confirmed the low metallicity of Tri II. From the spectra of 6 stars, they obtain metallicities
ranging from $\rm [Fe/H] = -1.40$ to --2.86. They also found a decrease of the [$\alpha$/Fe] ratio as [Fe/H] increases,  consistent with  the   Galactic chemical evolution models where Type Ia supernovae (SNe Ia) enrichment favours the iron enrichment over the $\alpha$ elements enrichment as the galaxy evolves.
They obtained Keck/HIRES spectra  of one of the stars analysed by \citet{venn_geminigraces_2017} and  measured detailed abundances using standard high-resolution abundance analysis techniques.
They found an iron abundance $\rm [Fe/H]= -2.92$ slightly lower than the value found from KECK/DEIMOS spectra. They also found very low abundances of barium ($\rm [Ba/Fe]= -2.4$) and strontium ($\rm [Sr/Fe]= -1.5$, an abundance feature found in UFD stars. 

The two brightest members of Tri II were analysed by \citet{ji_chemical_2019} thanks to high resolution spectra acquired with GRACES. They derived metallicities $\rm [Fe/H] = -1.96$ and --2.95. In one of the two stars,  observed by \citet{venn_geminigraces_2017} and \citet{kirby_triangulum_2017}, they confirmed   a low [$\alpha$/Fe] abundance ratio. 

Although the velocity and metallicity dispersions of Tri II have not been decisive about whether it is an UFD or  a globular cluster , the authors  concluded that Tri II  is likely  an UFDs  because of the  extremely low neutron-capture element abundances found in the two stars. \\

\citet{sitnova_formation_2021} reanalysed the star S40 in Tri II  adopting observations taken by \citet{kirby_triangulum_2017} 
since they managed to measure the largest number of chemical species compared to the  studies of \citet{venn_geminigraces_2017} and 
 \citet{ji_chemical_2019}. The details of their NLTE calculations can be found in their paper \citep{sitnova_formation_2021}. 
The Triangulum II  metallicity histogram  is shown in Fig. \ref{Fig:SegTri_histogram}.\\

 \item{} Tucana II \\
 
 Tucana II  was found in the analysis of the publicly released Dark Energy Survey (DES) data by \citet{koposov_beasts_2015}.
 The galaxy was also identified by \citet{bechtol_eight_2015} using  optical imaging data collected during the first year of the Dark Energy Survey (DES).
 
From Michigan/Magellan Fiber System (M2FS) spectroscopy, \citet{walker_magellanm2fs_2016} found 9 probable members estimating  metallicities ranging from $-1.60$ to $-2.1$.
 Note that \citet{walker_magellanm2fs_2016} increased all their [Fe/H] measurements by 0.32\,dex, which is the offset they obtained 
 from fitting twilight spectra of the Sun.
\citet{ji_chemical_2016} analysed Magellan/MIKE of 4 Tuc II red giant  stars and determined  their detailed chemical abundances. 
They derived metallicities ranging for $\rm [Fe/H] = -3.2$ to $-2.6$ and low neutron-capture abundances. The stars of their sample show 
a diversity of chemical signatures  very different to what can be found for a simple 'one-shot' first galaxy as Segue 1.

Thanks to  a selection  from SkyMapper narrow-band photometry, 
 \citet{chiti_chemical_2018} determined the chemical composition of seven stars belonging to Tuc II (three of them with no previous detailed
abundanee analysis) using   high resolution Magellan/MIKE spectroscopy. They found chemical abundances that are characteristic of the UFD stellar population. Excluding one star with discrepant abundances that could be a foreground  halo star with the same systemic 
velocity as Tuc II, this galaxy could be considered as a surviving first galaxy.
\citet{chiti_detailed_2023} determined the abundance in five stars between 0.3 and 1.1 kpc from the centre 
of Tuc II   from high-resolution Magellan/MIKE spectroscopy. 
They found metallicities ranging from  $-3.6$ to $-1.9$  and  
deficiency in neutron-capture elements as is characteristic of UFD stars,  confirming their association with Tucana II.
The Tucana II  metallicity histogram  is shown in Fig. \ref{Fig:Tuc_histogram}.\\

 \begin{figure}
\centering
\includegraphics[width=5.5cm,clip=true]{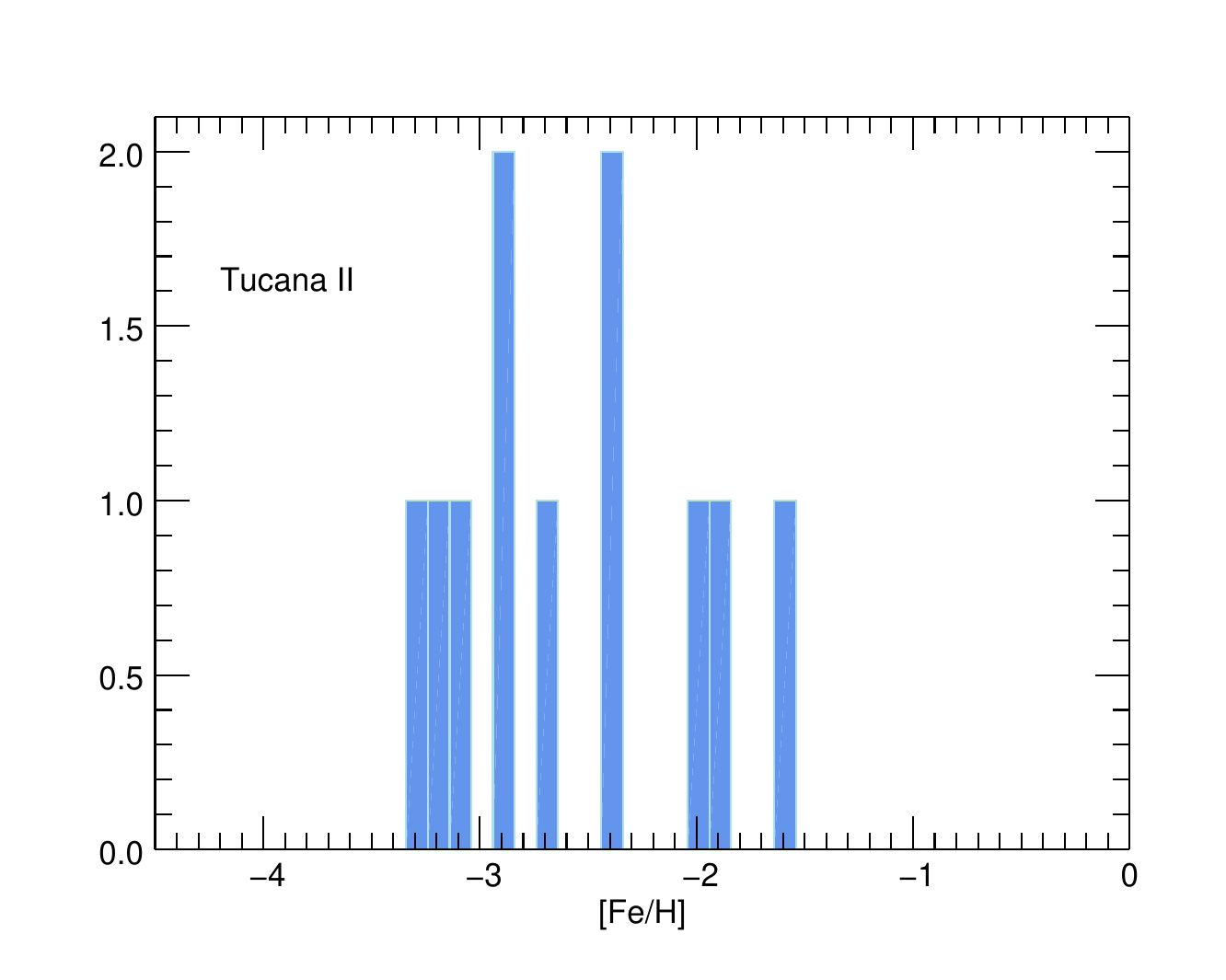}
\includegraphics[width=5.5cm,clip=true]{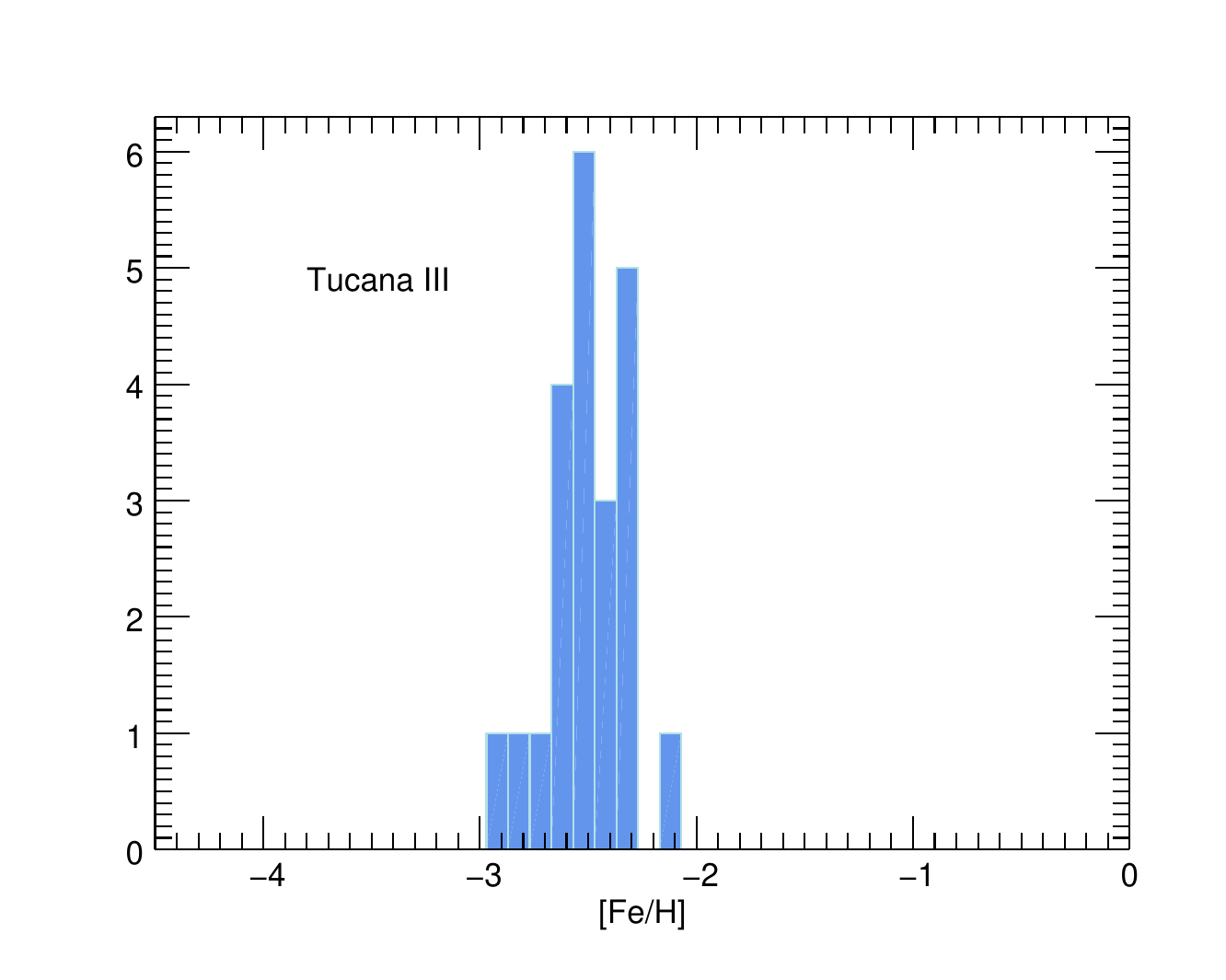}
\includegraphics[width=5.5cm,clip=true]{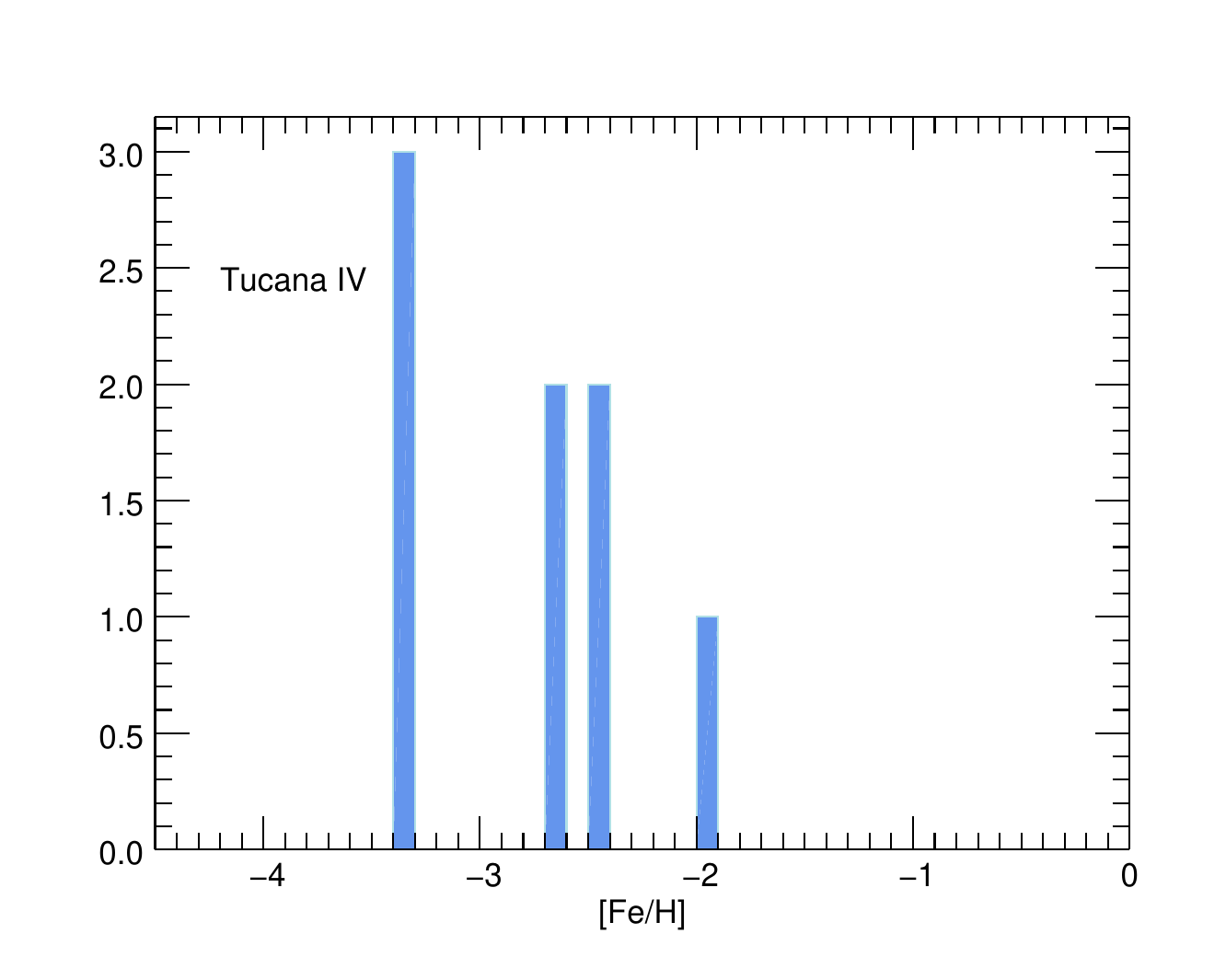}
\includegraphics[width=5.5cm,clip=true]{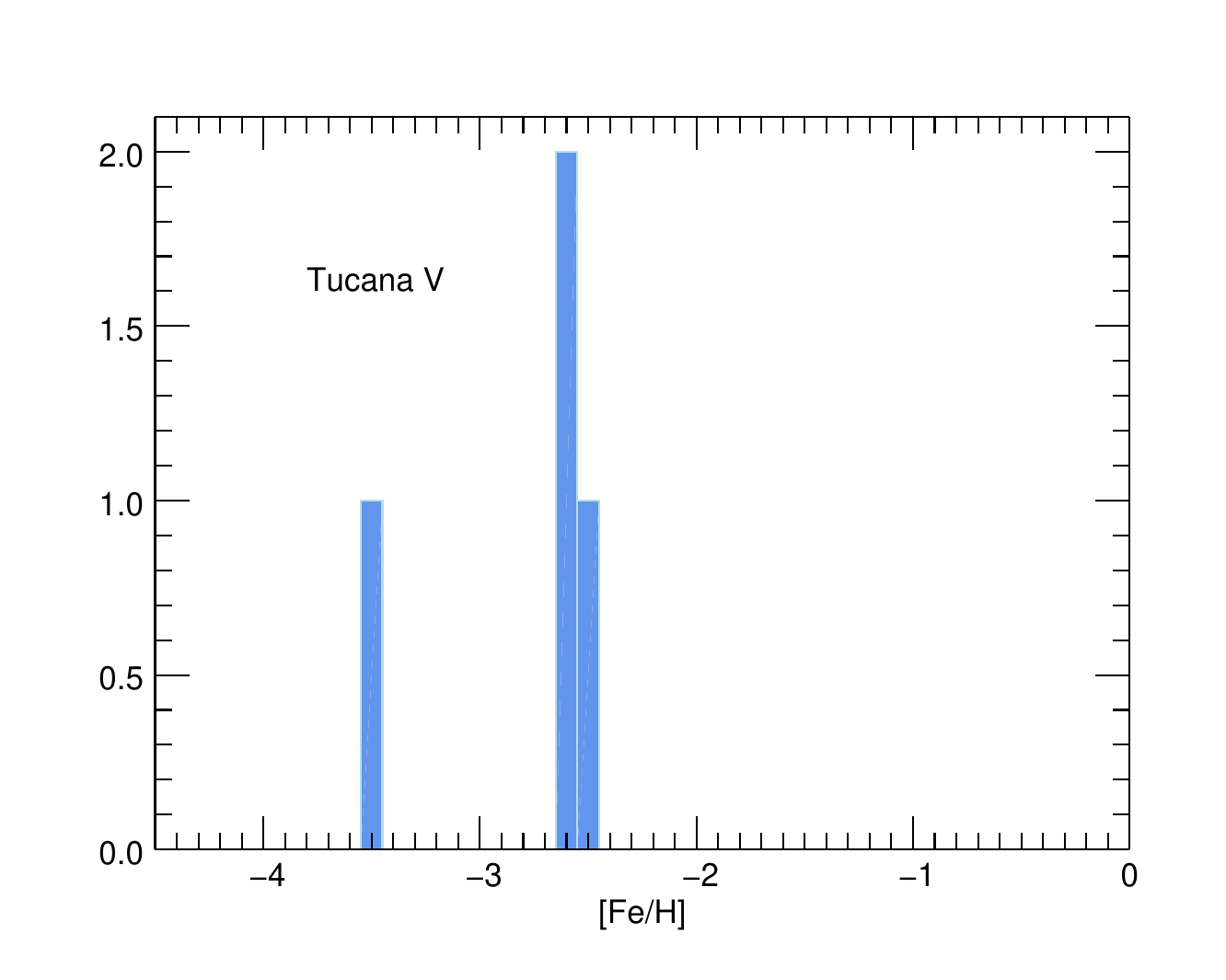}
\caption{\footnotesize Metallicity histograms of Tucana II,  Tucana III, Tucana IV and Tucana V.}
\label{Fig:Tuc_histogram}
\end{figure}

 \item{} Tucana III \\
 
 Tucana III was discovered by \citet{drlica-wagner_eight_2015} in the analysis of the combined data set from the first two years of  the Dark Energy Survey (DES) covering  $\rm\simeq 5000 deg^2$ of the south Galactic cap.
Thanks to  low resolution Magellan/IMACS spectroscopy, \citet{simon_nearest_2017} measured CaT metallicities for the RGB members of Tuc III, which range from $\rm [Fe/H] = -2.16$ to $\rm [Fe/H] = -2.58$.  Using a MCMC method \citep{foreman-mackey_emcee_2013},  they concluded that Tuc III has a mean metallicity $\rm [Fe/H]= -2.42$ with a unresolved spread $\sigma < 0.19$\,dex.  Adding that Tuc III has a low velocity dispersion, they suggested that could be the tidally stripped remnant of a dark matter-dominated dwarf galaxy. 

  As Tuc III is  located just 25 kpc away from the Sun, several bright members have been identified \citep{simon_nearest_2017}. 
  High resolution high SNR Magellan/MIKE spectra  were obtained  by \citet{hansen_r-process_2017} for a V=15.2  red giant of Tuc III. 
  From the abundance determination of 28 species (13 neutron-capture elements) they found that this star shows a mild enhancement in neutron-capture elements associated with the r-process and can be classified as an r-I star.

 \citet{marshall_chemical_2019} determined the detailed chemical abundance of four additional confirmed members of Tuc III. 
 The four stars have chemical abundance patterns consistent with the one previously studied star in Tucana III, 
 with a metallicity range of 0.44\,dex,  the expected trends in  $\alpha$-elements  and a moderate enhancement in r-process.
The Tucana III  metallicity histogram  is shown in Fig. \ref{Fig:Tuc_histogram}.\\

 \item{} Tucana IV  \&  Tucana V \\
 
 As for Tucana III, Tucana IV and Tuc V were discovered by \citet{drlica-wagner_eight_2015} in the analysis of the combined data set from the first two years of  the Dark Energy Survey (DES) covering  $\rm\simeq 5000 \, deg^2$ of the south Galactic cap.

From Magellan/IMACS spectra,   \citet{simon_birds_2020}
 found that Tuc IV has a CaT calibrated metallicity   $\rm [Fe/H]= -2.49$ and $\rm [Fe/H]= -2.17$ for Tuc V.
Based on their sizes, masses, and metallicities, they classified Tuc IV as likely dwarf galaxy, but the nature of Tuc V remained uncertain.
\citet{hansen_chemical_2024} observed three stars in Tuc V with Mike at Magellan, they found that the stars span more than 1\,dex in metallicity
from $-3.55$ to $-2.46$. One of the stars is mildly enhanced in r-process  elements (r-I, [Eu/Fe]=+0.36) and another one is
a CEMP-no star. In view of the metallicity spread and chemical diversity \citet{hansen_chemical_2024} concluded that Tuc V is
a galaxy but likely  not associated to the SMC as had been claimed by \citet{conn_nature_2018}, who preferred
``either a chance grouping of stars related to the SMC halo or a star cluster in an advanced stage of dissolution''.
The Tucana IV and Tucana V  metallicity histograms  are shown in Fig. \ref{Fig:Tuc_histogram}.\\

 \item{}   Ursa Major I \\
  
 \citet{willman_new_2005}  reported the discovery of the UFD UMa I  detected as an overdensity of red, resolved stars in Sloan Digital Sky Survey data.
  \citet{simon_kinematics_2007} obtained  low resolution Keck/DEIMOS spectra for 39 bright stars in UMa I  from which they  used the Ca{\sc II} triplet absorption lines calibration to  derive an average metallicity $\rm \langle [Fe/H]\rangle =-2.06 \pm 0.10$ with a dispersion $\rm\sigma [Fe/H] = 0.46$\,dex.
  Using a pixel-to-pixel matching method between observed and synthetic spectra on the sample of spectra obtained by \citet{simon_kinematics_2007}, \citet{kirby_uncovering_2008} 
  found an iron abundance $\rm [Fe/H] = -2.29$ with a dispersion $\sigma = 0.54$\,dex. 
  
  Thanks to   Subaru/Suprime-Cam observations  \citet{okamoto_suprime-cam_2008} derived a photometric metallicity $\rm [Fe/H] \simeq -2.00$ consistent with previous photometric and spectroscopic studies \citep{willman_new_2005,martin_keckdeimos_2007}. 
  
From the analysis of  Keck/DEIMOS spectra,  \citet{vargas_distribution_2013} computed the [$\alpha$/Fe] ratios in eleven stars of UMaI and found a decreasing [$\alpha$/Fe] ratio as the metallicity of the star increases.
The Ursa Major I  metallicity histogram is presented in Fig. \ref{Fig:Ursa_Will_histogram}. \\

 \begin{figure}
\centering
\includegraphics[width=5.5cm,clip=true]{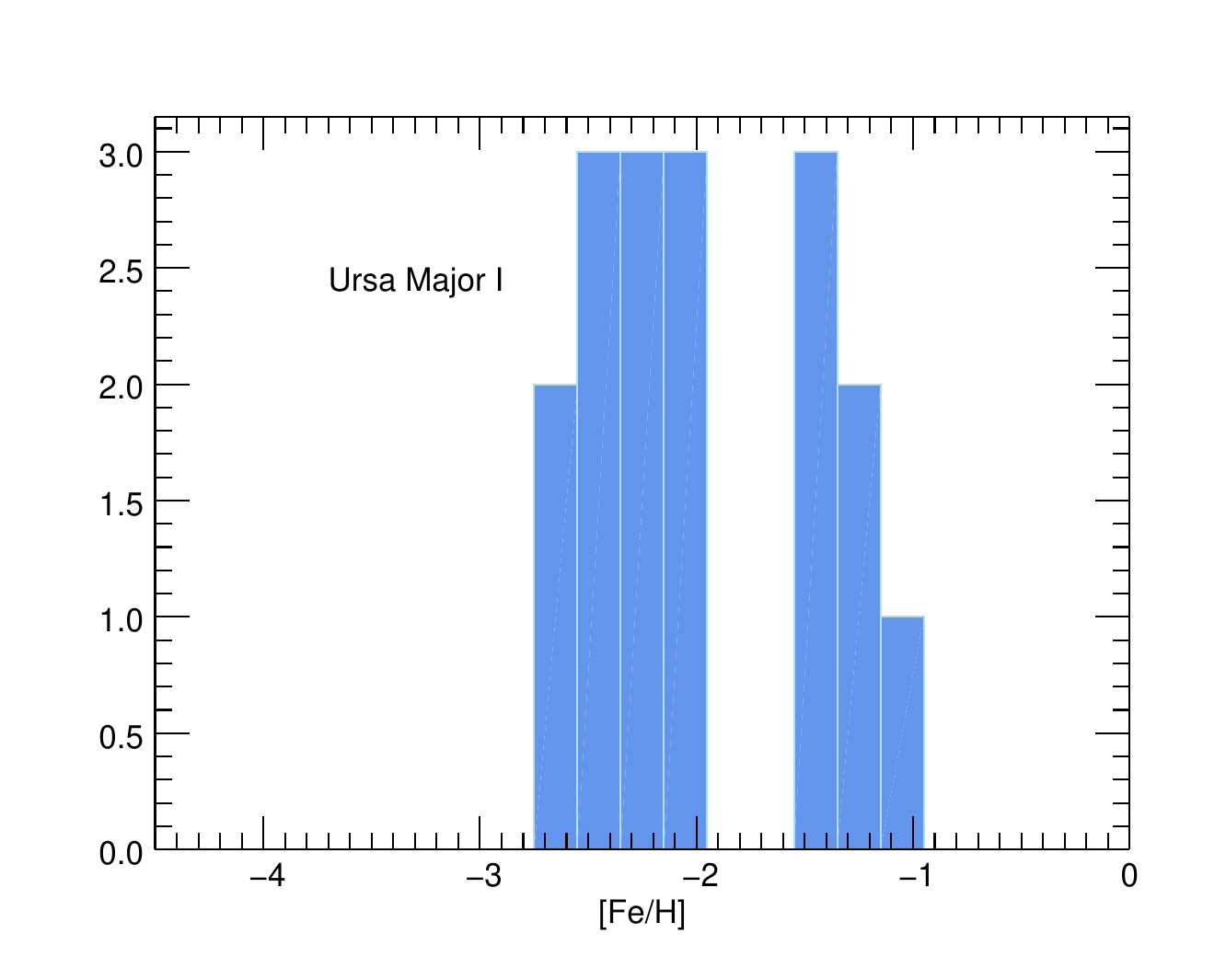}
\includegraphics[width=5.5cm,clip=true]{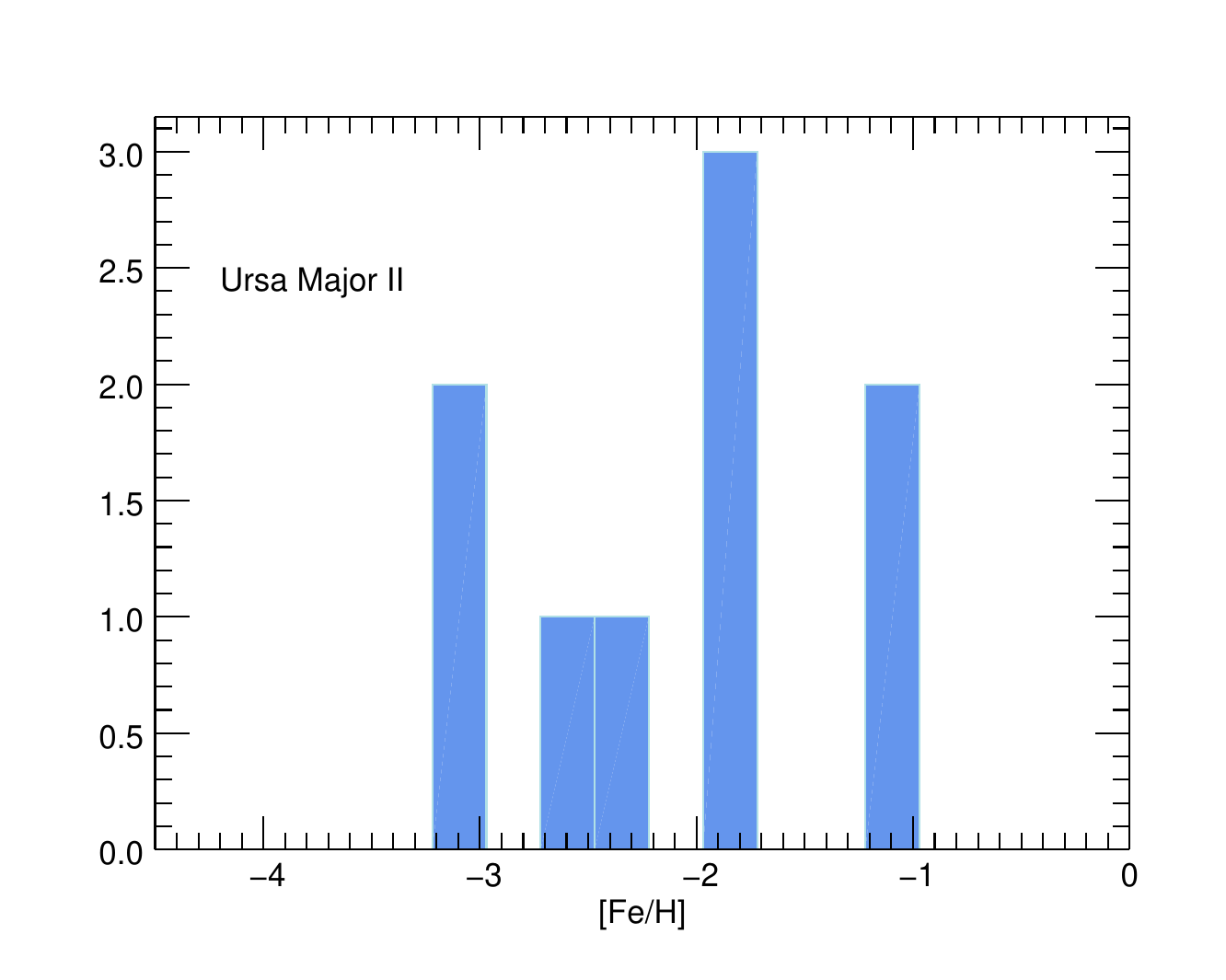}
\includegraphics[width=5.5cm,clip=true]{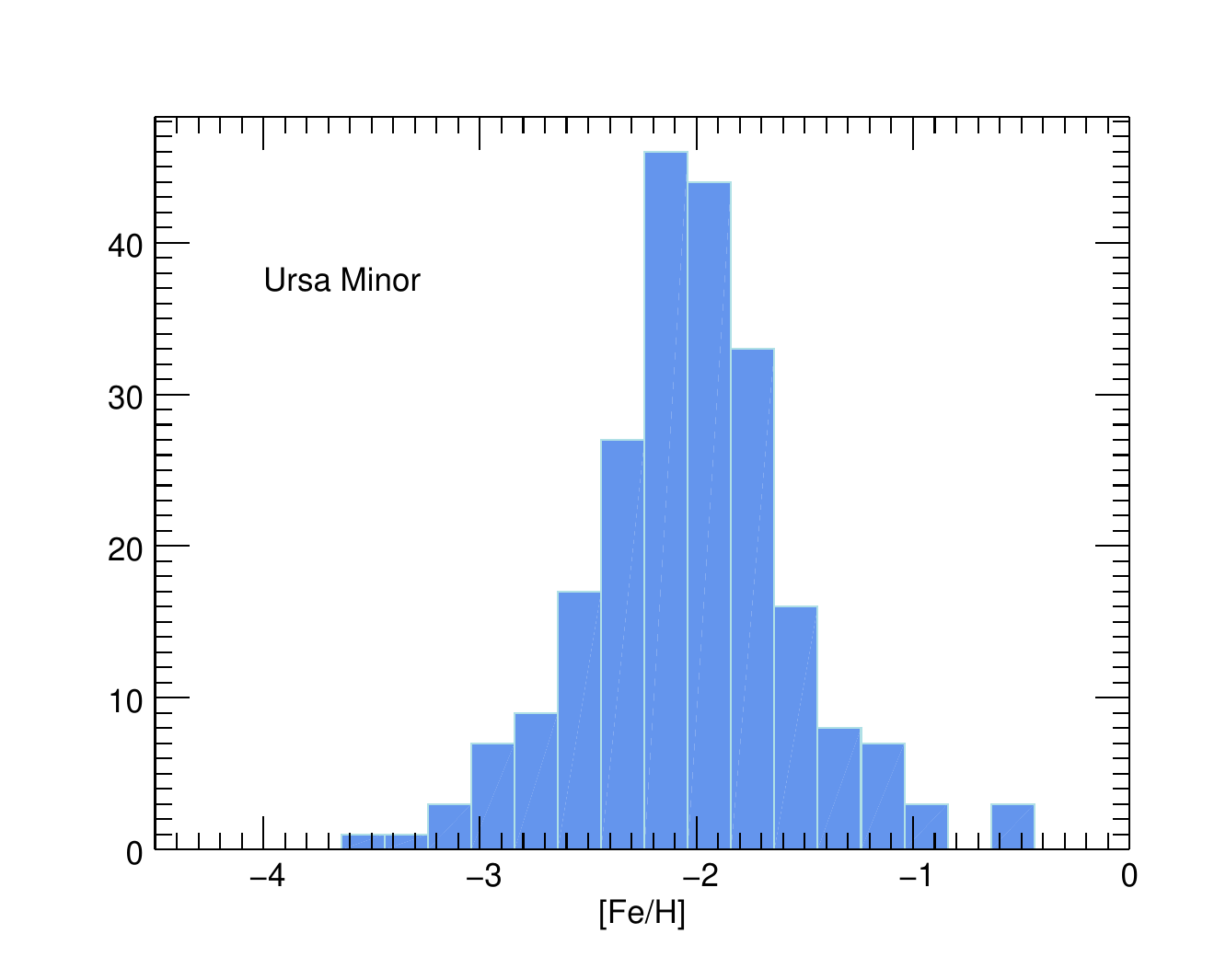}
\includegraphics[width=5.5cm,clip=true]{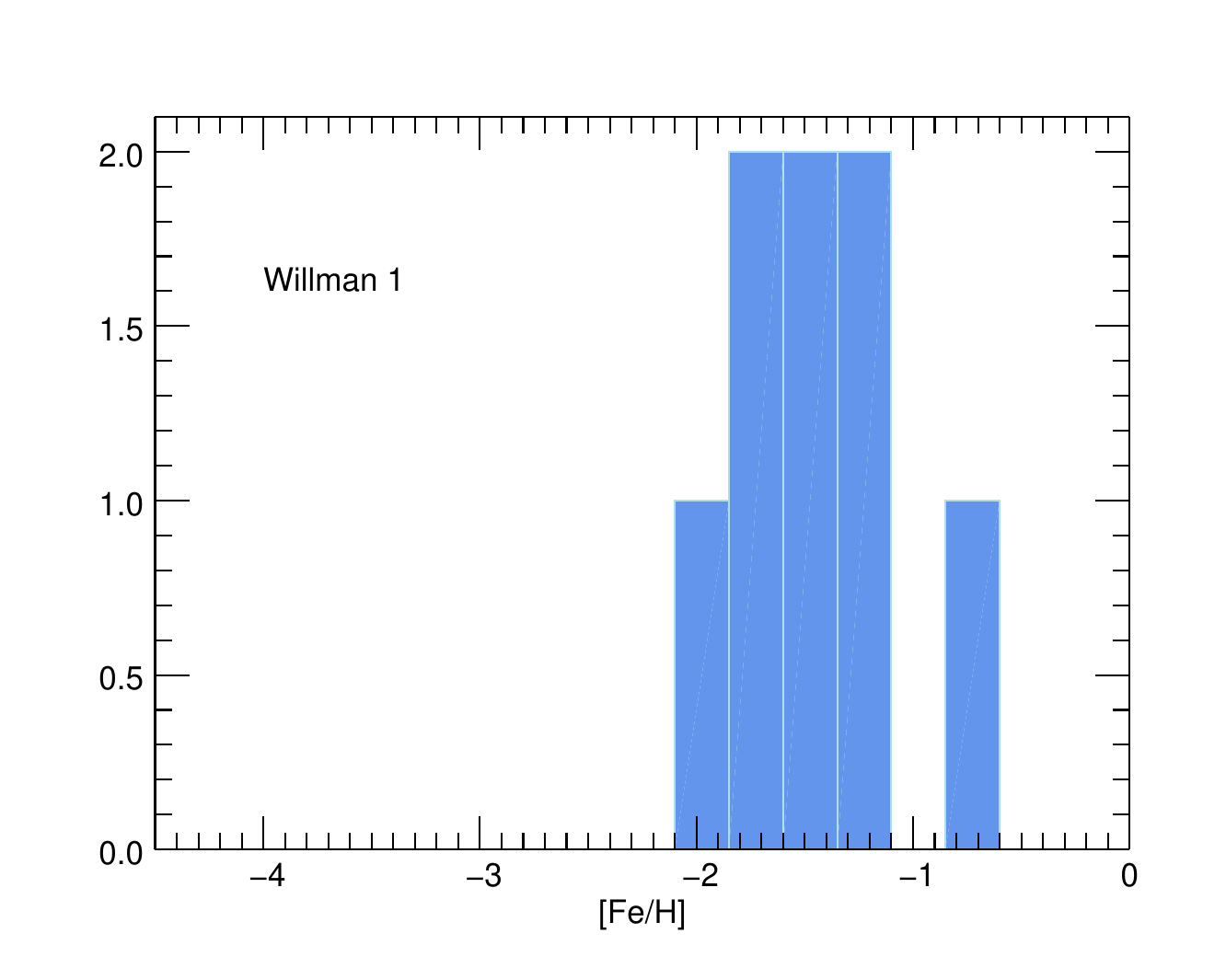}
\caption{\footnotesize Metallicity histograms of Ursa Major I, Ursa Major II, Ursa Minor and Willman 1.}
\label{Fig:Ursa_Will_histogram}
\end{figure}

\item{} Ursa Major II

\citet{zucker_curious_2006} identified UMa II as a local stellar overdensity in SDSS data  confirmed with 
 Subaru/Suprime-Cam imaging. They suggested that UMa II is a faint dwarf spheroidal galaxy on the basis of its size, structure and
 stellar population.
  \citet{simon_kinematics_2007} obtained low resolution Keck/DEIMOS spectra for 29 bright stars in UMa II.
They used the Ca{\sc II} triplet absorption lines calibration to  derive an average metallicity $\rm \langle [Fe/H]\rangle=-1.97 \pm 0.15$ with a dispersion $\rm\sigma [Fe/H] = 0.28$\,dex.
A reanalysis of their data set  by \citet{kirby_uncovering_2008} led to an iron abundance $\rm [Fe/H] = -2.44$ with a dispersion $\rm\sigma [Fe/H] = 0.57$\,dex. 
\citet{frebel_high-resolution_2010} determined the detailed chemical composition of three stars in UMa II thanks to Keck/HIRES high resolution spectra and good S/N ratio ($\simeq$ 20 to $\simeq$ 40). They found metallicities ranging from $\rm [Fe/H]= -2.34$ to $-3.23$.
They found a substantial agreement between the abundance patterns of the ultra-faint dwarf galaxies and the Milky Way halo for the light, $\alpha$, and iron-peak elements (C to Zn).
\citet{vargas_distribution_2013} analysed  Keck/DEIMOS spectra of seven stars in UMaII . All  of their [$\alpha$/Fe] average abundances measurements 
have a  rather constant value of  +0.4, spanning a large range of metallicities from [Fe/H] = -1.04  to --3.04. 
The Ursa Major II  metallicity histogram is shown in Fig. \ref{Fig:Ursa_Will_histogram}.
\\

\item{} Ursa Minor \\ 

Ursa  Minor was found by Wilson \citep{wilson_sculptor-type_1955} on 48-inch schmidt plates taken for the National Geographic Society-Palomar Observatory Sky Survey.
Using  the Keck/HIRES high resolution spectrograph, \citet{shetrone_abundance_2001} analysed six giant stars members of UMi  and derived moderately low metallicities ranging from $\rm [Fe/H]= -1.45$ to $-2.17$.  They also found a decreasing [$\alpha$/Fe] ratio as [Fe/H] increases as found in other dwarf galaxies. For example the ratio [Mg/Fe] ratio reaches a
solar value at $\rm [Fe/H]= -1.5$. 
\citet{kirby_multi-element_2010}    obtained KECK/DEIMOS spectra of a large sample of stars in UMi. Based on a sample of 255 stars  with metallicities ranging from $-3.88$ to $-0.66$, they derived a mean metallicity $\rm [Fe/H]= -2.19$. 

\citet{cohen_chemical_2010} made  an abundance analysis based on high-resolution spectra of 10 stars selected to span the full range in metallicity in UMi. They confirmed the large range of metallicities found by  \citet{kirby_multi-element_2010}. 
Among the stars , they found two EMP stars with metallicities $\rm [Fe/H]=-3.08$ and $-3.10$. 
They  found that for the UMi sample [Mg/Fe] is constant to within the uncertainties with a value $\simeq  +0.35$ for all the stars,  a high value found also in  Galactic halo stars. 
The Ursa Minor  metallicity histogram is presented in Fig. \ref{Fig:Ursa_Will_histogram}.
\\

\item{} Willman 1 \\

Willman 1 (Wil I) was identified  as an overdensity of resolved blue stars in SDSS data by \citet{willman_new_2005-1}.  
From their analysis, they could not conclude if it was a globular cluster or a  Milky Way dwarf spheroidal galaxies.
From Keck/DEIMOS spectroscopy follow-up observations,  \citet{martin_keckdeimos_2007} found metallicities (derived from the CaT triplet) ranging from $\rm [Fe/H]= -0.8$ to $-2.1$ They concluded that even though Wil 1 is very faint and small, it is probably a dwarf galaxy .

\citet{willman_willman_2011} confirmed the  spread in metallicity  supporting the scenario that Wil 1 is an ultra-low luminosity dwarf galaxy, or the remnants thereof, rather than a star cluster.
The Willman I  metallicity histogram is presented in Fig. \ref{Fig:Ursa_Will_histogram}. \\

\item{Abundance trends in UFD metal-poor stars.} \\

In Fig. \ref{Fig:UFD_MgFe}, \ref{Fig:UFD_CaFe}, \ref{Fig:UFD_SrFe} and \ref{Fig:UFD_BaFe} are collected the abundance ratios 
[Mg/Fe], [Ca/Fe], [Sr/Fe] and [Ba/Fe] determined in stars belonging to UFD galaxies.  The plus signs represent the results based on medium resolution  spectroscopy spectra while the closed squares are all results from high resolution spectroscopy.  It is important to note that 
most of the results are based on spectra with a rather low signal-to-noise ratio. Moreover, the stars are generally the brightest stars of the galaxy 
 hence with low gravity and rather low surface temperature making their analysis difficult. 
 Putting the results of all the UFD in the same diagram may be misleading as each galaxy has its own chemical history. We can however extract some general information.   Looking at the ratio of [Mg/Fe]  and [Ca/Fe] ratios as a function of metallicity, we confirm that these ratios
 are over solar for the most metal poor star sample  of each galaxy. We also confirm the decrease of these ratios as the metallicity of the star increases, 
 each galaxy following its own decreasing path. We also confirm that the [Mg/Fe]  and [Ca/Fe] ratio become sub solar at a much lower metallicity than found for stars in our galaxy.  
 The interpretation of the results for Sr and Ba is more complex as these ratios reveal the abundances of stars that have been polluted by 
 few supernovae events, thus showing the relative contamination of these events to the different UFD.  The details can be found in each section above. However, it is interesting to note the large spread in the abundance ratio  of [Sr/Fe] (4\,dex)  and [Ba/Fe] with a variation by a factor larger than 100000  at a given metallicity, a spread also found in the metal poor end of the  Galactic stars.

  \begin{figure}
\centering
\includegraphics[width=10.0cm,clip=true]{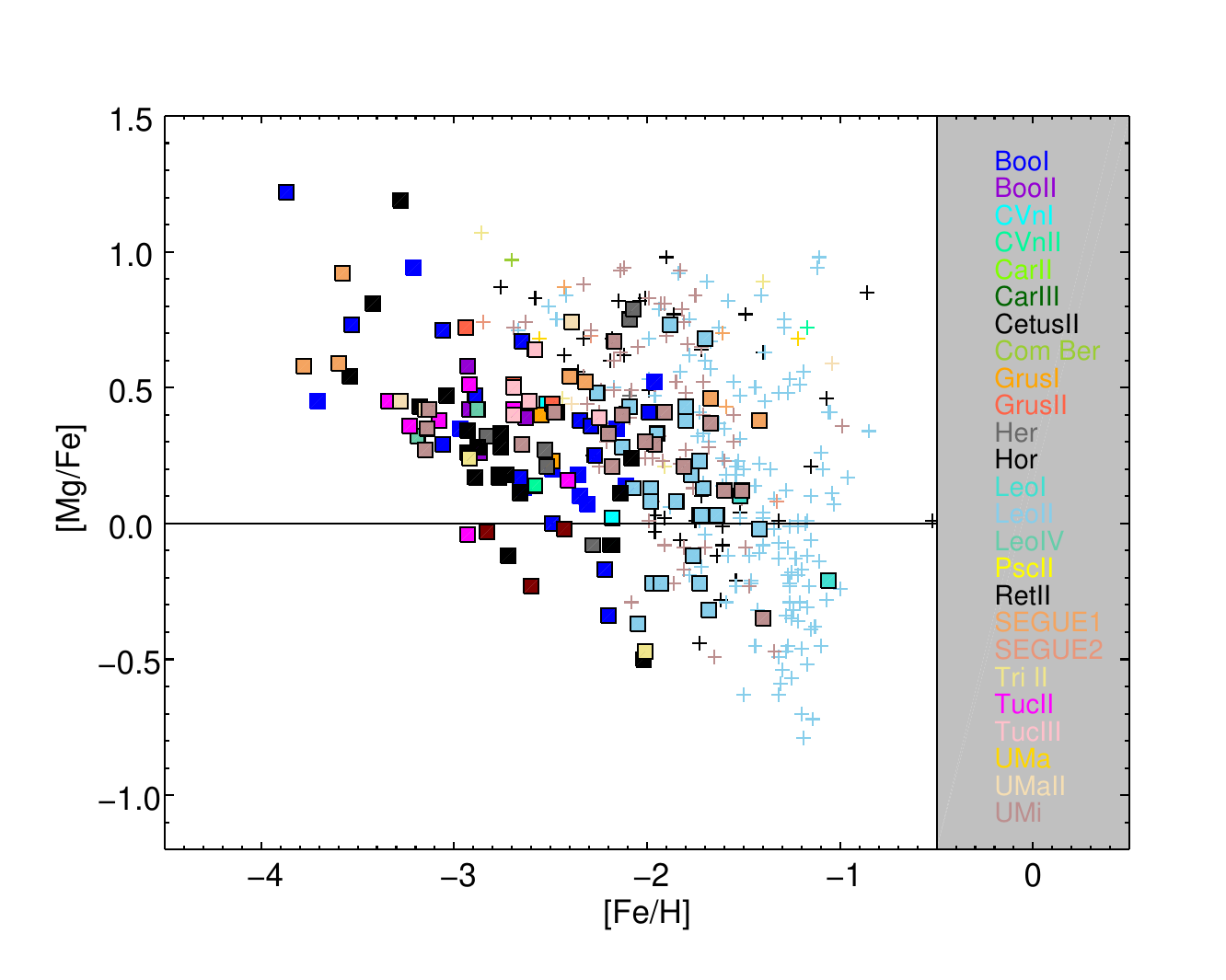}
\caption{\footnotesize  [Mg/Fe] abundances ratios versus [Fe/H] in the ultra faint dwarf spheroidal galaxies. Open symbols represent results determined using low resolution  (resolving power R$<10\,000$). Filled symbols show the results derived from medium to high resolution spectroscopy (R$>10\,000$). }
\label{Fig:UFD_MgFe}
\end{figure}

 \begin{figure}
\centering
\includegraphics[width=10.0cm,clip=true]{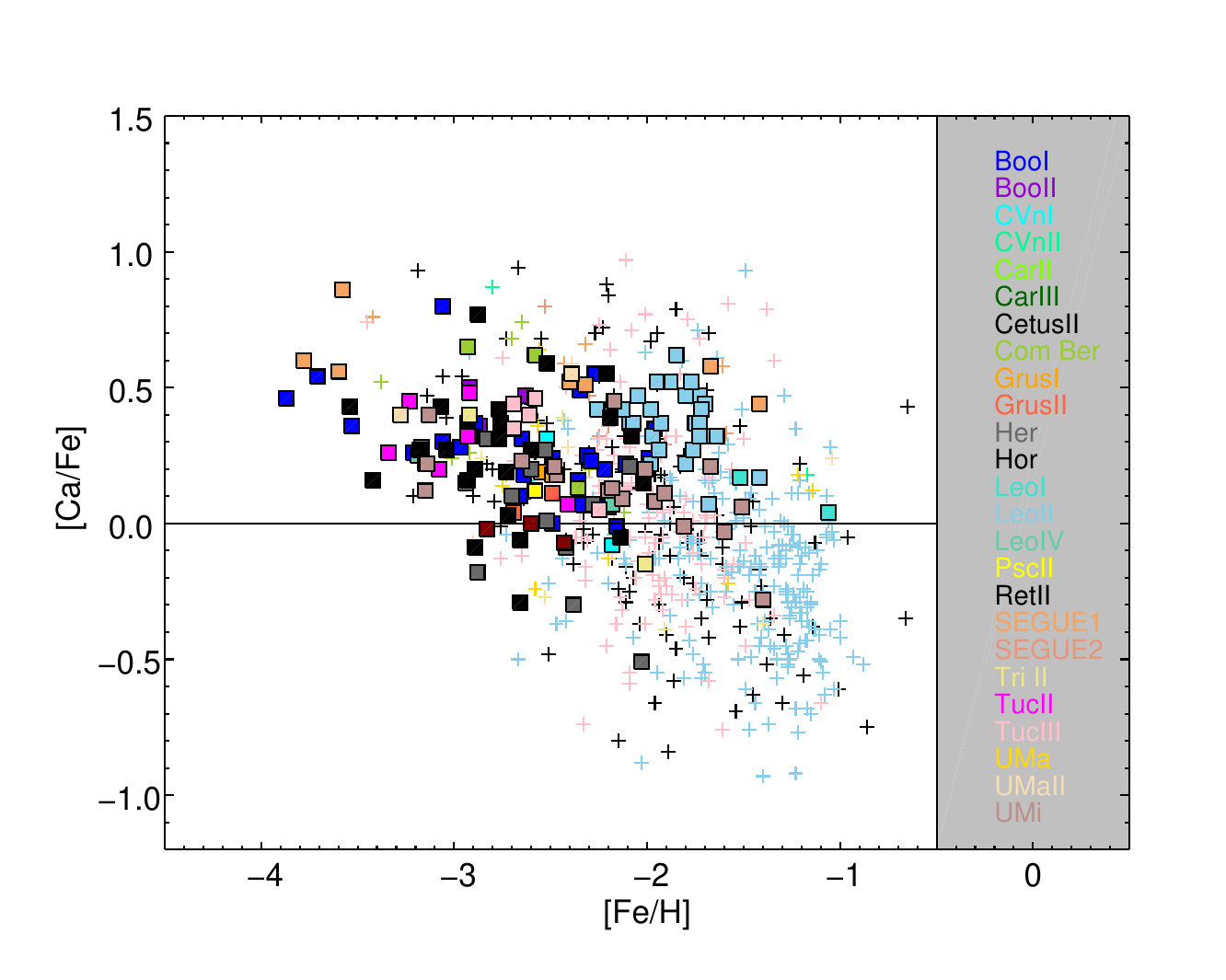}
\caption{\footnotesize[Ca/Fe]abundances ratios versus [Fe/H] in the ultra faint dwarf spheroidal galaxies. Open symbols represent results determined using low resolution (R$<10\,000$). Filled symbols show the results derived from medium to high resolution spectroscopy (R$>10\,000$).}
\label{Fig:UFD_CaFe}
\end{figure}

 \begin{figure}
\centering
\includegraphics[width=10.0cm,clip=true]{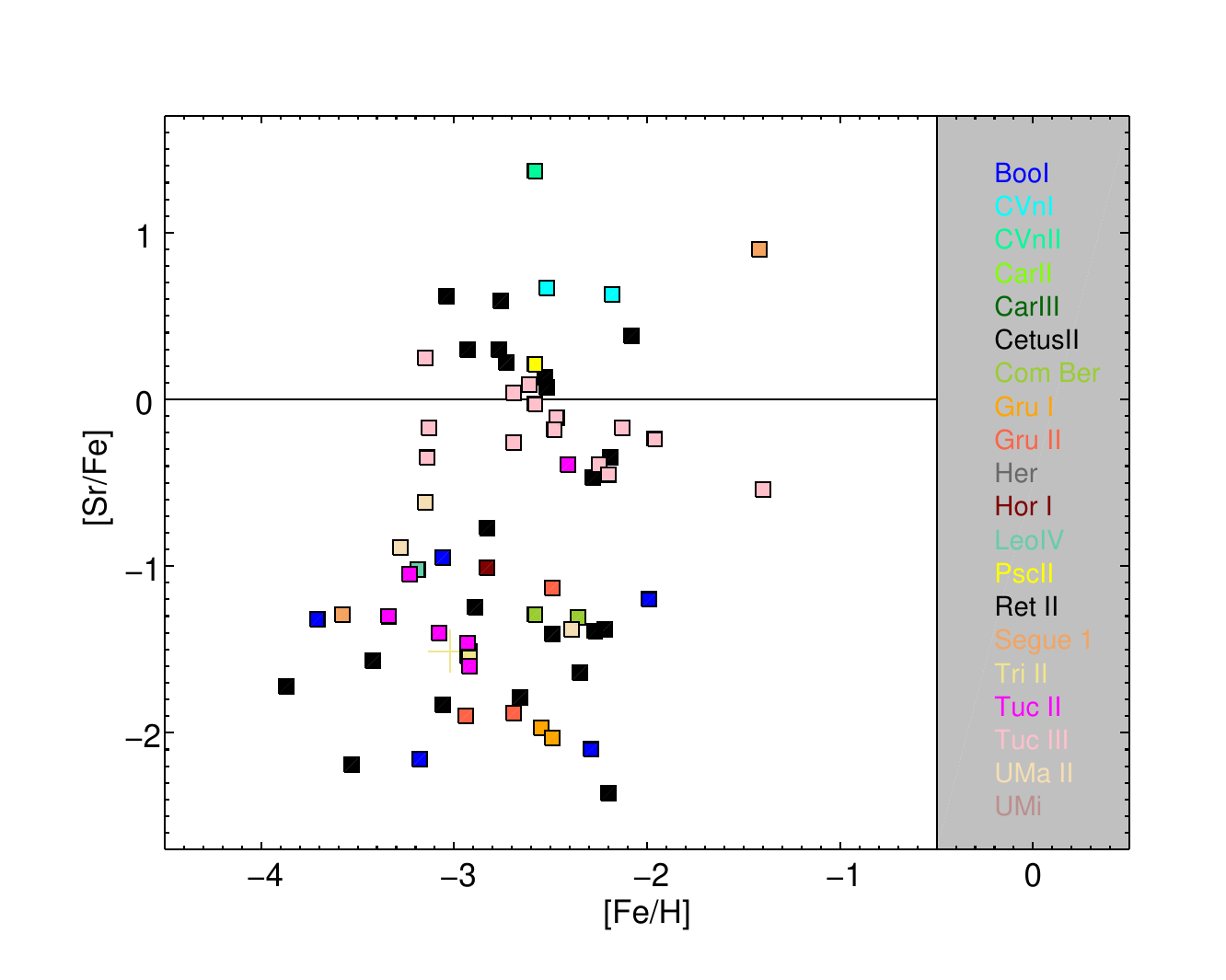}
\caption{\footnotesize [Sr/Fe] abundances ratios versus [Fe/H] in the ultra faint dwarf spheroidal galaxies. Open symbols represent results determined using low resolution (R$<10\,000$). Filled symbols show the results derived from medium to high resolution spectroscopy (R$>10\,000$). }
\label{Fig:UFD_SrFe}
\end{figure}

 \begin{figure}
\centering
\includegraphics[width=10.0cm,clip=true]{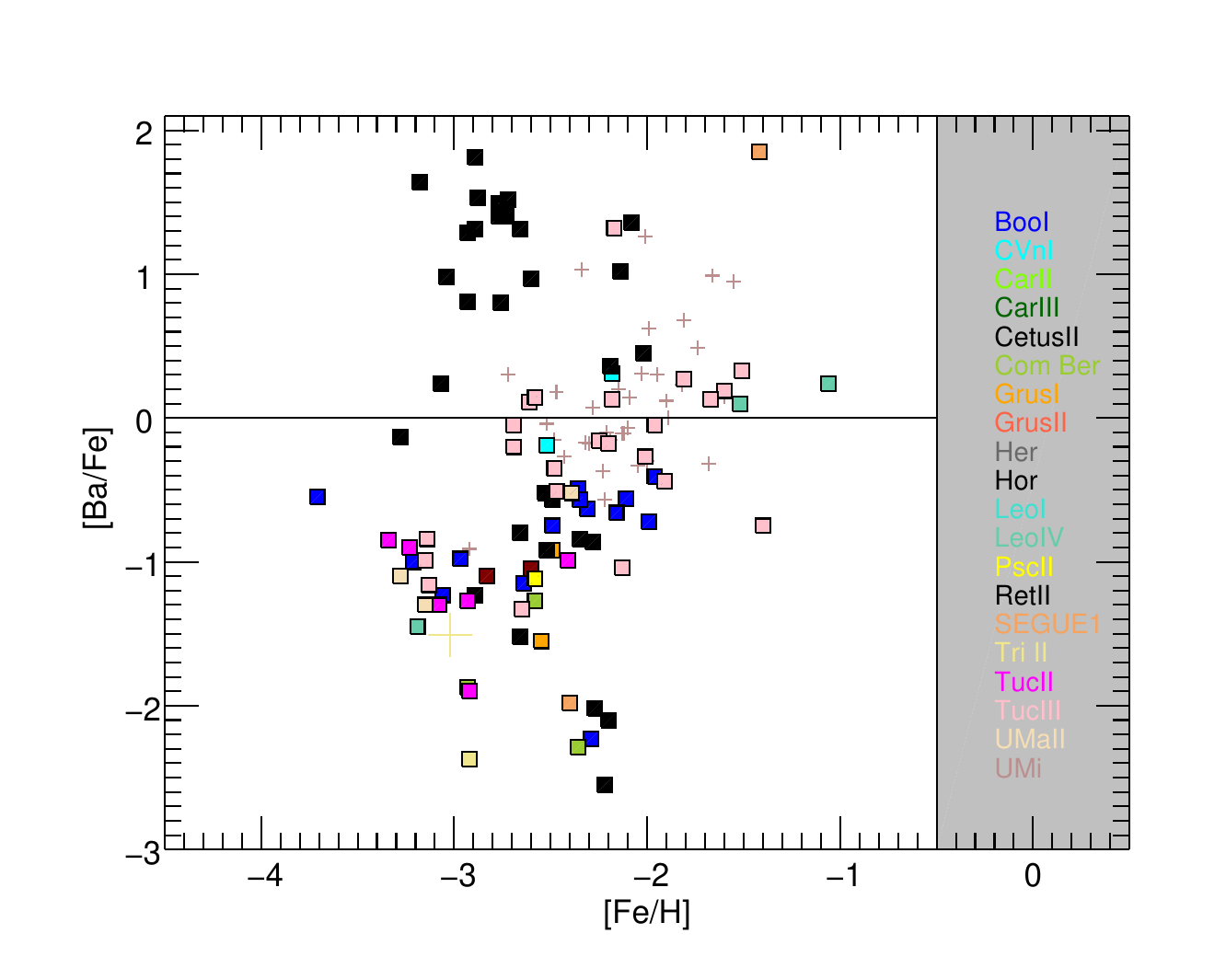}
\caption{\footnotesize [Ba/Fe] abundances ratios versus [Fe/H] in the ultra faint dwarf spheroidal galaxies. Open symbols represent results determined using low resolution (R$<10\,000$). Filled symbols show the results derived from medium to high resolution spectroscopy (R$>10\,000$). }
\label{Fig:UFD_BaFe}
\end{figure}

\end{itemize}

\section{Conclusions} 

In this article, we reported on the metallicity distribution, on the  chemical peculiarities of the EMP stars and on  the existence of these stars not only in the halo, as previously thought, but also in the bulge, in the galactic disk with disk-like orbits  and  in the external galaxies of the local group, in particular the Ultra faint dwarf spheroidal galaxies.  
Recent studies have revealed the existence of young metal-poor stars. 
In this article, we did not discuss the study of the chemical composition  of the recently discovered stellar streams, as  their vast majority do not contain EMP stars and high resolution spectroscopic studies  of the metal-poor  streams are limited at the time of writing of this article. Details can be found in \citet{martin_pristine_2022}.
Thanks to past and ongoing spectroscopic and multi-band photometric surveys (HK, HES, SDSS, DES, LAMOST, Pristine, SMSS among others), a large number of EMP stars candidates 
have been detected. Unfortunately, the time consuming  high resolution spectroscopic follow-up observations are mandatory to make a  detailed chemical analysis of these potential EMP stars, restricting the number of EMP stars with a complete chemical and kinematical analysis, leaving us with a limited picture of the early chemical epochs. 
It is therefore very important to have a high success rate in their detection, each of the surveys having its own merits  (see sec. \ref{sec:surveys}). 
In the incoming spectroscopic surveys (WEAVE, MOONS, 4MOST ) some configuration setups (wavelength range, resolution)  have been tailored to make an efficient follow-up of EMP candidates. 
The increasing number of EMP stars studied in details will help in better understanding the early chemical evolution of our Galaxy and local group galaxies.

\begin{acknowledgements}

We are grateful to M. Ishigaki for providing her data and models.
We are also grateful to R. Ezzedine, S. Lucatello, W.S. Oh, {\'A}. Sk{\'u}lad{\'o}ttir and C. Sneden
for granting us permission to reprint their figures. We thank the editor in chief of A\&A,
T. Forveille for granting us permission to reproduce our figure.
We are grateful to T. Suda and the developers of the SAGA database, that was very useful for us.
We are grateful to the two referees who provided very careful and constructive reports that helped
us to improve our paper.
PB acknowledges support   from the ERC advanced grant N. 835087 -- SPIAKID. 

\end{acknowledgements}

\addcontentsline{toc}{section}{References}
\bibliographystyle{aa}

\bibpunct{(}{)}{;}{a}{}{,} 


\end{document}